\documentclass{vldb}
\usepackage{graphics}
\usepackage{graphicx}
\usepackage{subfigure}
\usepackage{xspace}
\usepackage{amsmath}
\usepackage{amssymb}
\usepackage{mathrsfs}
\usepackage{array}
\usepackage{textcomp}
\usepackage{weiwAlgorithm}
\usepackage{url}
\usepackage[]{hyperref}
\usepackage{xcolor}
\usepackage{comment}
\usepackage{scrextend}
\usepackage{hyperref}
\usepackage{paralist}
\usepackage{multirow}

\usepackage{tkz-kiviat}
\usetikzlibrary{arrows}

\newcommand{\commentone}[1]{}  

\DeclareMathOperator*{\argmin}{arg\,min}

\newcommand{\fnurl}[1]{\footnote{\url{#1}}}

\newcommand{\topk}{top-$k$}

\newcommand{\intersect}{\cap}

\newcommand{\myparagraph}[1]{\vspace{1mm}\noindent\textbf{#1}\hspace{.5em}}

\newcommand{\Alghnsw}{\textsf{HNSW}\xspace}
\newcommand{\Algkgraph}{\textsf{KGraph}\xspace}
\newcommand{\Algflann}{\textsf{FLANN}\xspace}
\newcommand{\Algannoy}{\textsf{Annoy}\xspace}
\newcommand{\Algsw}{\textsf{SW}\xspace}

\newcommand{\Algdpg}{\textsf{DPG}\xspace}
\newcommand{\AlgOPQ}{\textsf{OPQ}\xspace}

\newcommand{\Algflannhkm}{\textsf{FLANN-HKM}\xspace}
\newcommand{\Algflannkd}{\textsf{FLANN-KD}\xspace}

\newcommand{\Algkmeans}{\textsf{k-means}\xspace}

\newcommand{\AlgNAPP}{\textsf{NAPP}\xspace}
\newcommand{\AlgRCT}{\textsf{RCT}\xspace}

\newcommand{\Algvptree}{\textsf{VP-tree}\xspace}

\newcommand{\Algsrs}{\textsf{SRS}\xspace}
\newcommand{\Algqalsh}{\textsf{QALSH}\xspace}
\newcommand{\Algfalconn}{\textsf{FALCONN}\xspace}

\newcommand{\Datanusw}{\textsf{Nusw}\xspace}
\newcommand{\Datagist}{\textsf{Gist}\xspace}
\newcommand{\Dataglov}{\textsf{Glove}\xspace}
\newcommand{\Datarand}{\textsf{Rand}\xspace}

\newcommand{\Dataaudio}{\textsf{Audio}\xspace}
\newcommand{\Databann}{\textsf{BANN}\xspace}

\newcommand{\Dataenron}{\textsf{Enron}\xspace}

\newcommand{\Datasift}{\textsf{Sift}\xspace}

\newcommand{\Datatrevi}{\textsf{Trevi}\xspace}
\newcommand{\Datauqv}{\textsf{UQ-V}\xspace}

\newcommand{\Datayout}{\textsf{Yout}\xspace}
\newcommand{\Datagauss}{\textsf{Gauss}\xspace}

\newcommand{\annbenchmark}{\textsf{ann-benchmark}}

\newcommand{\kNN}{$k$NN\xspace}
\newcommand{\KNN}{K-NN\xspace} 
\newcommand{\ANN}{ANN\xspace}

\newcommand{\dist}[2]{\parallel #1, #2 \parallel_2}
\newcommand{\set}[1]{\{\, #1 \,\}}

\let\angle\measuredangle

\pgfdeclarelayer{background}
\pgfdeclarelayer{foreground}
\pgfsetlayers{background,main,foreground}

\usetikzlibrary{snakes}

\begin{document}
\title{Approximate Nearest Neighbor Search on High Dimensional Data --- Experiments,
  Analyses, and Improvement (v1.0)}


\author{
{$^{1,3}$Wen Li, $^{1}$Ying Zhang, $^{2}$Yifang Sun, $^{2}$Wei Wang, $^{2}$Wenjie Zhang, $^{2}$Xuemin Lin} %
\vspace{1.6mm}\\
\fontsize{10}{10}
\selectfont\itshape
$^{1}$The University of Technology Sydney~~~~ $^{2}$The University of New South Wales\\
\fontsize{10}{10}
\selectfont\itshape
$^{3}$China University of Mining and Technology   \\
\fontsize{9}{9} \selectfont\ttfamily\upshape
$~$ \{Li.Wen, Ying.Zhang\}@uts.edu.au ~~ \{yifangs, weiw, zhangw, lxue\}@cse.unsw.edu.au \\}

\maketitle

\begin{abstract}
  Approximate Nearest neighbor search (ANNS) is fundamental and essential operation in
  applications from many domains, such as databases, machine learning,
  multimedia, and computer vision. Although many algorithms have been
  continuously proposed in the literature in the above domains each year, there
  is no comprehensive evaluation and analysis of their performances.

  In this paper, we conduct a comprehensive experimental evaluation of many
  state-of-the-art methods for approximate nearest neighbor search.
  Our study (1) is cross-disciplinary (i.e., including 16 algorithms in different domains, and
  from practitioners) and (2) has evaluated a diverse range of settings,
  including 20 datasets, several evaluation metrics, and different query
  workloads. %
  The experimental results are carefully reported and analyzed to understand the
  performance results. Furthermore, we propose a new method that achieves both
  high query efficiency and high recall empirically on majority of the datasets
  under a wide range of settings.
\end{abstract}

\section{INTRODUCTION}
\label{sec:intro}

Nearest neighbor search finds an object in a reference database which has the
smallest distance to a query object. It is a fundamental and essential operation
in applications from many domains, including databases, computer vision,
multimedia, machine learning and recommendation systems. 

Despite much research on this problem, it is commonly believed that it is very
costly to find the \emph{exact} nearest neighbor in high dimensional Euclidean
space, due to the \emph{curse of dimensionality}~\cite{DBLP:conf/stoc/IndykM98}.
Experiments showed that exact methods can rarely outperform the brute-force
linear scan method when dimensionality is high~\cite{DBLP:conf/vldb/WeberSB98}
(e.g., more than 20). Nevertheless, returning sufficiently nearby objects,
referred to as \emph{approximate nearest neighbor search} (ANNS), can be performed
efficiently and are sufficiently useful for many practical problems, thus
attracting an enormous number of research efforts. Both exact and approximate
NNS problems can also be extended to their \topk{} versions.

\subsection{Motivation}
\label{subsec:motivation}

There are hundreds of papers published on algorithms for (approximate) nearest
neighbor search, but there has been few systematic and comprehensive
comparisons among these algorithms. In this paper, we conduct a comprehensive
experimental evaluation on the state-of-the-art approximate nearest neighbor
search algorithms in the literature, due to the following needs:

\myparagraph{1. Coverage of Competitor Algorithms and Datasets from Different
  Areas.}%
As the need for performing ANN search arises naturally in so many diverse domains,
researchers have come up with many methods while unaware of alternative methods
proposed in another area. In addition, there are practical methods proposed by
practitioners and deployed in large-scale projects such as the music
recommendation system at \texttt{spotify.com}~\cite{misc:url/annoy}. As a
result, it is not uncommon that important algorithms from different areas are
overlooked and not compared with. For example, there is no evaluation among %
Rank Cover
Tree~\cite{DBLP:journals/pami/HouleN15} (from
Machine Learning), %
Product
Quantization~\cite{DBLP:journals/pami/JegouDS11,DBLP:journals/pami/GeHK014}
(from Multimedia), %
SRS~\cite{DBLP:journals/pvldb/SunWQZL14} (from Databases), and %
KGraph~\cite{misc:url/kgraph} (from practitioners).
Moreover, each domain typically has a \emph{small} set of commonly used datasets
to evaluate ANNS algorithms; there are very few datasets used by all these
domains.
In contrast, we conduct comprehensive experiments using carefully selected
representative or latest algorithms from different domains, and test \emph{all
  of} them on 20 datasets including those frequently used in prior studies in
different domains. %
Our study confirms that there are substantial variability of the performances of
all the algorithms across these datasets.

\myparagraph{2. Overlooked Evaluation Measures/Settings.}
An NNS algorithm can be measured from various aspects, including
\begin{inparaenum}[(i)]
\item search time complexity,
\item search quality,
\item index size,
\item scalability with respect to the number of objects and the number of
  dimensions,
\item robustness against datasets, query workloads, and parameter settings,
\item updatability, and
\item efforts required in tuning its parameters.
\end{inparaenum}
Unfortunately, none of the prior studies evaluates these measures completely and
thoroughly.

For example, most existing studies use a query workload that is essentially the
same as the distribution of the data. Measuring algorithms under different query
workloads is an important issue, but little result is known.
In this paper, we evaluate the performances of the algorithms under a wide
variety of settings and measures, to gain a complete understanding of each
algorithm (c.f., Table~\ref{tab:summary}).

\myparagraph{3. Discrepancies in Existing Results.}
There are discrepancies in the experimental results reported in some of the
notable papers on this topic.
For example, in the \annbenchmark{}~\cite{misc:url/ann-benchmark} by practitioners, \Algflann{} was shown to perform better
than \Algkgraph{},
while the study in~\cite{misc:url/kgraph} indicates otherwise.
While much of the discrepancies can be explained by the different settings,
datasets and tuning methods used, as well as implementation differences, it is
always desirable to have a maximally consistent result to reach an up-to-date
rule-of-the-thumb recommendation in different scenarios for researchers and
practitioners.

In this paper, we try our best to make a fair comparison among all methods
evaluated, and test them on all 20 datasets. For example, all search programs are
implemented in C++, and all hardware-specific optimizations are disabled
(e.g., SIMD-based distance computation). Finally, we will also publish the
source codes, datasets, and other documents so that the results can be easily
reproduced.

We classify popular \kNN{} algorithms into three categories:
\emph{LSH-based}, \emph{Space partitioning-based} and \emph{Neighborhood-based}.
The key idea of each category of the method will be introduced in Section 3-6.

\subsection{Contributions}
\label{subsec:contribution}
Our principle contributions are summarized as follows.
\begin{itemize}
\item Comprehensive experimental study of state-of-the-art ANNS methods across
  several different research areas. Our comprehensive experimental study extends
  beyond past studies by:
  \begin{inparaenum}[(i)]
  \item comparing all the methods without adding any implementation tricks,
    which makes the comparison more fair;
  \item evaluating all the methods using multiple measures; and

  \item we provide rule-of-the-thumb recommendations about how to select the
    method under different settings.
  \end{inparaenum}
  We believe such a comprehensive experimental evaluation will be beneficial to
  both the scientific community and practitioners, and similar studies have been
  performed in other areas (e.g., classification
  algorithms~\cite{DBLP:conf/icml/CaruanaKY08}).

\item We group algorithms into several categories (Section~\ref{sec:algs_LSH}, \ref{sec:algs_encoding}, \ref{sec:algs_tree} and \ref{sec:algs_proximity}), and
  then perform detailed analysis on both intra- and inter-category evaluations
  (Sections~\ref{sec:exp}). Our \emph{data-based}
  analyses provide confirmation of useful principles to solve the problem, the
  strength and weakness of some of the best methods, and some initial
  explanation and understanding of why some datasets are harder than others.
  The experience and insights we gained throughout the study enable us to engineer a new empirical algorithm, \Algdpg{} (Section~\ref{sec:dpg}), 
  that achieves both high query efficiency and high recall empirically on majority of the datasets
  under a wide range of settings.

\end{itemize}


Our paper is organised as follows. Section \ref{sec:bg} introduces the problem definition as well as some constraints in this paper. Section \ref{sec:algs_LSH}, \ref{sec:algs_encoding}, \ref{sec:algs_tree} and \ref{sec:algs_proximity} give the descriptions about some state-of-the-art ANNS algorithms that we evaluated. Section \ref{sec:dpg} presents our improved ANNS approach. The comprehensive experiments and the analyses are reports in Section \ref{sec:exp}.
Section \ref{sec:con} concludes out paper with future work. An evaluation of the parameters of the tested algorithms is reported in Appendix \ref{app:parameter}. Appendix \ref{app:final} gives some supplement results of the second round.


\section{BACKGROUND}
\label{sec:bg}

\subsection{Problem Definition}
\label{subsec:prob}

In this paper, we focus on the case where data points are $d$-dimensional
vectors in $\mathbb{R}^{d}$ and the distance metric is the Euclidean distance.
Henceforth, we will use \emph{points} and \emph{vectors} interchangeably. %
Let $\dist{p}{q}$ be the Euclidean distance between two data points $p$ and $q$
and $\mathcal{X} = \set{ x_{1}, x_{2},\ldots,x_{n}}$ be a set of reference data
points. A \kNN{} search for a given query point $q$ is defined as returning $k$
nearest neighbors $kNN(q) \in \mathcal{X}$ with respect to $q$ such that
$\left|kNN(q) \right|=k$ and $\forall x \in kNN(q)$,
$\forall x' \in \mathcal{X} \setminus kNN(q)$, $\dist{x}{q} \leq \dist{x'}{q}$.

Due to the \emph{curse of dimensionality}~\cite{DBLP:conf/stoc/IndykM98}, much
research efforts focus on the \textit{approximate} solution for the problem of
$k$ nearest neighbor search on high dimensional data. %
Let the results returned by an algorithm be
$X = \set{x_i \mid 1 \leq i \leq k}$. A common way to measure the quality of $X$
is its \emph{recall}, defined as $\frac{\left| X \intersect kNN(q) \right|}{k}$, which is
also called \emph{precision} in some papers.

\subsection{Scope}
\label{subsec:related-work}

The problem of ANNS on high dimensional data has been extensively studied in
various literatures such as databases, theory, computer vision, and machine
learning. Over hundreds of algorithms have been proposed to solve the problem
from different perspectives, and this line of research remains very active in
the above domains due to its importance and the huge challenges. To make a
comprehensive yet focused comparison of ANNS algorithms, in this paper, we
restrict the scope of the study by imposing the following constraints.

\myparagraph{Representative and competitive ANNS algorithms.} %
We consider the state-of-the-art algorithms in several domains, and omit other
algorithms that have been dominated by them unless there are strong evidences
against the previous findings.

\myparagraph{No hardware specific optimizations.} %
Not all the implementations we obtained or implemented have the same level of
sophistication in utilizing the hardware specific features to speed up the query
processing. Therefore, we modified several implementations so that no algorithm
uses multiple threads, multiple CPUs, SIMD instructions, hardware pre-fetching,
or GPUs.

\myparagraph{Dense vectors.} %
We mainly focus on the case where the input data vectors are dense, i.e.,
non-zero in most of the dimensions.

\myparagraph{Support the Euclidian distance.} %
The Euclidean distance is one of the most widely used measures on
high-dimensional datasets. It is also supported by most of the ANNS algorithms.

\myparagraph{Exact \kNN{} as the ground truth.} %
In several existing works, each data point has a label (typically in
classification or clustering applications) and the labels are regarded as the
ground truth when evaluating the recall of approximate NN algorithms. In this
paper, we use the exact \kNN{} points as the ground truth as this works for all
datasets and majority of the applications.

\paragraph*{Prior Benchmark Studies}
\label{sec:prior-benchm-stud}

There are two recent NNS benchmark studies:
~\cite{DBLP:journals/pvldb/NaidanBN15} and \annbenchmark{}~\cite{misc:url/ann-benchmark}. The former considers
a large number of other distance measures in addition to the Euclidean distance, and the latter
does not disable general implementation tricks. In both cases, their studies are
less comprehensive than ours, e.g., with respect to the number of algorithms and
datasets evaluated. More discussions on comparison of benchmarking results are
given in Section~\ref{sec:comp-with-prior}.


\section{LSH-based methods}
\label{sec:algs_LSH}

These methods are typically based on the idea of \emph{locality-sensitive
  hashing} (LSH), which maps a high-dimensional point to a low-dimensional point
via a set of appropriately chosen random projection functions. Methods in this
category enjoys the sound probabilistic theoretical guarantees on query result
quality, efficiency, and index size even in the worst case. On the flip side, it
has been observed that the data-independent methods usually are outperformed by
data-dependent methods empirically, since the latter cannot exploit the data
distribution.

Locality-sensitive hashing (LSH) is first introduced by Indyk and Motwani in~\cite{DBLP:conf/stoc/IndykM98}.
An LSH function family $\mathcal{H}$ for a distance function $f$ is defined as
$(r_1, r_2, p_1, p_2)$-sensitive iff for any two data points $x$ and $y$, there
exists two distance thresholds $r_1$ and $r_2$ and two probability thresholds
$p_1$ and $p_2$ that satisfy:
$\small
\begin{cases}
  \Pr_{h \in \mathcal{H}}[h(x) = h(y)] \geq p_1 & \text{, if } f(x, y) < r_1 \\
  \Pr_{h \in \mathcal{H}}[h(x) = h(y)] \leq p_2 & \text{, if } f(x, y) > r_2
\end{cases}\label{eq:lsh-def-1}
$.
This means, the chance of mapping two points $x$, $y$ to the same value grows as
their distance $f(x, y)$ decreases. The $2$-stable distribution, i.e., the
Gaussian/normal distribution, can be used to construct the LSH function family for
the Euclidean distance. A data point is mapped to a hash value (e.g., bucket)
based on a random projection and discritize method. A certain number of hash
functions are used together to ensure the performance guarantee, using different
schemes such as the AND-then-OR scheme~\cite{DBLP:conf/compgeom/DatarIIM04}, the
OR-then-AND scheme~\cite{DBLP:conf/sigmod/ZhaiLG11}, inverted list-based
scheme~\cite{DBLP:conf/sigmod/GanFFN12,DBLP:journals/pvldb/HuangFZFN15}, or
tree-based scheme~\cite{DBLP:journals/pvldb/SunWQZL14}.
In this category, we evaluate two most recent LSH-based methods with theoretical
guarantees: \textbf{\Algsrs}~\cite{DBLP:journals/pvldb/SunWQZL14} and
\textbf{\Algqalsh}~\cite{DBLP:journals/pvldb/HuangFZFN15}. Both of them can work with any $c>1$. Note that we exclude several recent work because either there is no known practical implementation (e.g., \cite{DBLP:conf/stoc/AndoniR15}) or they do not support the Euclidean distance (e.g., \Algfalconn{}~\cite{DBLP:journals/corr/AndoniILRS15}). There are also a few empirical LSH-based methods that loses the theoretical guarantees and some of
them will be included in other categories.

\subsection{SRS}%

SRS projects the dataset with high dimensionality into a low-dimensional space (no more than 10) to do exact $k$-NN search. We refer the distance between the query $q$ and any point $o$ in original space as $dist(o)$, and the distance in projected space as $\Delta(o)$. The key observation of \textbf{\Algsrs} is for any point $o$, $\frac{\Delta(o)^{2}}{dist(o)^{2}}$ follows the standard $\chi^{2}(m)$ distribution. Hence, The basic method solve a $c$-ANN problem in two steps: (1) obtain an ordered set of candidates by issuing a $k$-NN query with $k=T'$ on the $m$-dimensional projections of data points; (2) Examine the distance of these candidates in order and return the points with the smallest distance so far if it satisfies the early-termination test(if there exists a point that is a c-ANN point with probability at least a given threshold) or the algorithm has exhausted the $T'$ points. By setting $m = O(1)$, the algorithm guarantees that the returned point is no further away than $c$ times the nearest neighbor distance with constant probability; both the space and time complexities are linear in $n$ and independent of $d$.

\subsection{QALSH}%

Traditionally, LSH functions are constructed in a query-oblivious manner in the sense that buckets are partitioned before any query arrives. However, objects closer to a query may be partitioned into different buckets, which is undesirable. \textbf{\Algqalsh} introduces the concept of query-aware bucket partition and develop novel query-aware LSH functions accordingly. Given a pre-specified bucket width $w$, a hash function $h_{\vec{a},b}(o)=\vec{a} \cdot \vec{o}+b$ is used in previous work, while QALSH uses the function $h_{\vec{a}}(o)=\vec{a} \cdot \vec{o}$ that first projects object $o$ along the random line $\vec{a}$ as before.When a query $q$ arrives, the projection of $q$ (i.e., $h_{\vec{a}}(q)$) is computed and the query projection (or simply the query) is applied as the ``anchor'' for bucket partition.. This approach of bucket partition is said to be query-aware. In the pre-processing step, the projections of all the data objects along the random line $\vec{a}$ are calculated, and all the data projections are indexed by a $B^{+}$-tree. When a query object $q$ arrives, \textbf{\Algqalsh} computes the query projection and uses the $B^{+}$-tree to locate objects falling in the interval $[h_{\vec{a}}(q)- \frac{w}{2},h_{\vec{a}}(q) + \frac{w}{2}]$. Moverover, \textbf{\Algqalsh} can gradually locate data objects even farther away from the query, just like performing a $B^{+}$-tree range search.


\section{Encoding-based Methods}
\label{sec:algs_encoding}

A large body of works have been dedicated to learning hash functions from the
data distribution so that the nearest neighbor search result in the hash coding
space is as close to the search result in the original space as possible. Please
refer to~\cite{DBLP:journals/corr/WangSSJ14,DBLP:journals/pieee/WangLKC16} for a
comprehensive survey.

Some example methods we evaluated in this category include Neighbor Sensitive
Hashing~\cite{DBLP:journals/pvldb/ParkCM15}, Selective
Hashing~\cite{DBLP:conf/kdd/GaoJOW15}, Anchor Graph
Hashing~\cite{DBLP:conf/icml/LiuWKC11}, Scalable Graph
Hashing~\cite{DBLP:conf/ijcai/JiangL15}, Neighborhood APProximation index~\cite{DBLP:journals/pvldb/NaidanBN15}, and Optimal Product
Quantization \cite{DBLP:journals/pami/GeHK014}.

We exclude a recent work~\cite{DBLP:journals/pvldb/AndreKS15} optimizing the
performance of product quantization-based method as it is mainly based on
utilizing the hardware-specific features.

\subsection{Anchor Graph Hashing}

The most critical shortcoming of the existing unsupervised hashing methods is the need to specify a global distance measure. On the contrary, in many real-world applications data lives on a low-dimensional manifold, which should be taken into account to capture meaningful nearest neighbors. For these, one can only specify local distance measures, while the global distances are automatically determined by the underlying manifold. AGH is  a graph-based hashing method which automatically discovers the neighborhood structure inherent in the data to learn appropriate compact codes in an unsupervised manner.

The first step of the graph-hashing methods is building a neighborhood graph with all the data points. In order to compute the neighborhood graph effectively, AGH builds an approximate neighborhood graph using Anchor Graphs, in which the similarity between a pair of data points is measured with respect to a small number of anchors.The resulting graph is constructed in $O(n)$ time complexity and is sufficiently sparse with performance approaching the true kNN graph as the number of anchors increase.

AGH uses the anchor graph to approximate the neighborhood graph, and accordingly uses the graph Laplacian over the anchor graph to approximate
the graph Laplacian of the original graph. Then the eigenvectors can be fastly computed. It also uses a hierarchical hashing to address the boundary issue.

\subsection{Scalable Graph hashing}

AGH and some representative unsupervised hashing methods directly exploit the similarity (neighborhood structure) to guide the hashing learning procedure, which are considered as a class of graph  hashing. Generally, graph hashing methods are expected to achieve better performance than non-graph based hashing methods if the learning algorithms are effective enough. However, graph hashing methods need to compute the pairwise similarities between any two data points. Hence, for large-scale datasets, it is memory-consuming and time-consuming or even intractable to learn from the whole similarity graph. Existing methods have to adopt approximation or sub-sampling methods for graph hashing on large-scale datasets. However, the accuracy of approximation cannot be guaranteed.

SGH adopts a feature transformation method to effectively approximate the whole graph without explicitly computing the similarity graph matrix. Hence, the $O(n^2)$ computation cost and storage cost are avoided in SGH. The aim of SGH is approximate the similarity matrix $S$ by the learned hashing codes, which resulting in the following objective function:
\[
min_{\{b_{l}\}_{l=1}^{n}}\sum_{i,j=1}^{n}(\tilde{S_{ij}}-\frac{1}{c}b_{i}^{T}b_{j})^2
\]
where $\tilde{S_{ij}}=2S_{ij}-1$. Integrating the idea of kernelized locality-sensitive hashing, the hash function for the $k$-th bit of $b_{i}$ is defined as follow:
\[
h_{k}(x_{i})=sgn(\sum_{j=1}^{m}W_{kj}\phi(x_{i},x_{j})+bias_{k})
\]
where $W\in R^{c\times m}$ is the weight matrix, $\phi(x_{i},x_{j})$ is a kernel function.

SGH applies a feature transformation method to use all the similarities without explicitly computing $\tilde{S_{ij}}$. The discrete $sgn(.)$ function makes the problem very difficult to solve. SGH uses a sequential learning strategy in a bit-wise manner, where the residual caused by former bits can be complementarily captured in the following bits.

\subsection{Neighbor Sensitive Hashing}

For hashing-based methods, the accuracy is judged by their effectiveness in preserving the kNN relationships (among the original items) in the Hamming space. That is, the goal of the optimal hash functions is that the relative distance of the original items are ideally linear to relative Hamming distance. To preserve the original distances, existing hashing techniques tend to place the separators uniformly, while those are more effective for rNN searching.

The goal of existing methods is to assign hashcodes such that the Hamming distance between each pair of items is as close to a linear function of their original distance as possible. However, NSH changes the shape of the projection function which impose a larger slope when the original distance between a pair of items is small, and allow the Hamming distance to remain stable beyond a certain distance.

Given a reference point $p$, NSH changes the shape of the projection function which impose a larger slope when the original distance to $p$ is small, and allow the projection distance to remain stable beyond a certain distance. Hence, when using the traditional function under the projection space, it is more likely to be assigned with different hash codes for the points close to $p$. If the distances between the query $q$ and $p$ is smaller than a threshold, the above property also apply to $q$. For handling all possible queries, NSH selects multiple pivots and limits the maximal average distances between a pivot and its closest neighbor pivot to ensure that there will be at least one nearby pivot for any novel query.

\subsection{NAPP}
permutation methods are dimensionality-reduction approaches, which assess similarity of objects based on their relative distances to some selected reference points, rather than the real distance values directly. An advantage of permutation methods is that they are not relying on metric properties of the original distance and can be successfully applied to non-metric spaces.
The underpinning assumption of permutation methods is that most nearest neighbors can be found by retrieving a small fraction of data points whose pivot rankings are similar to the pivot ranking of
the query.

A basic version of permutation methods selects $m$ pivots $\Pi_{i}$ randomly from the data points. For each data point $x$, the pivots are arranged in the order of increasing distance from $x$. Such a permutation of pivots is essentially a low-dimensional integer-valued vector whose $i$-th element is the ranking order of the $i$-th pivot in the set of pivots sorted by their distances from $x$. For the pivot closest to the data point, the value of the vector element is one, while for the most distance pivot the value is $m$.

In the searching step, a certain number of candidate points are retrieved whose permutations are sufficiently close to the permutation of the query vector. Afterwards, the search method sorts these candidate data points based on the original distance function.

This basic method can be improved in several ways. For example, one can index permutations rather than searching them sequentially: It is possible to employ a permutation prefix tree, an inverted index, or an index designed for metric spaces, e.g., a VPtree.

Neighborhood APProximation index(NAPP) was implemented by Bilegsaikhan. First, the algorithm selects $m$ pivots and computes the permutations induced by the data points. For each data point, $m_{i} \le m$ most closest pivots are indexed in a inverted file. At query time, the data points that share at least $t$ nearest neighbor pivots with the query are selected. Then, these candidates points are compared directly against the query based on the real distance.

\subsection{Selective Hashing}


LSH is designed for finding points within a fixed radius of the query point, i.e., radius search. For a k-NN problem (e.g., the first page results of search engine), the corresponding radii for different query points may vary by orders of magnitude, depending on how densely the region around the query point is populated.

Recently, there has been increasing interest in designing learning-based hashing functions to alleviate the limitations of LSH. If the data distributions only have global density patterns, good choices of hashing functions may be possible. However, it is not possible to construct global hashing functions capturing diverse local patterns. Actually, k-NN distance (i.e., the desired search range) depends on the local density around the query.

Selective Hashing(SH) is especially suitable for k-NN search which works on the top of radius search algorithms such as LSH. The main innovation of SH is to create multiple LSH indices with different granularities (i.e., radii). Then, every data object is only stored in one selected index, with certain granularity that is especially effective for k-NN searches near it.

Traditional LSH-related approaches with multiple indexes include all the data points in every index, while selective hashing builds multiple indices with each object being placed in only one index. Hence, there is almost no extra storage overhead compared with one fixed radius. Data points in dense regions are stored in the index with small granularity, while data points in sparse regions are stored in the index with large granularity. In the querying phase, the algorithm will push down the query and check the cell with suitable granularity.

\subsection{Optimal Product Quantization (\AlgOPQ)}
\label{subsubsec:opq}

Vector quantization (VQ)~\cite{DBLP:journals/tit/GrayN98} is a popular and
successful method for ANN search, which could be used to build inverted index
for non-exhaustive search. A vector $x \in \mathbb{R}^d$ is mapped to a codeword
$\mathbf{c}$ in a \emph{codebook} $\mathbf{C}=\set{ \mathbf{c}_j }$ with $i$ in
a finite index set. The mapping, termed as a \emph{quantizer}, is denoted by:
$x \rightarrow \mathbf{c}(i(x))$, where the function $i(\cdot)$ and
$\mathbf{c}(\cdot)$ is called \emph{encoder} and \emph{decoder}, respectively.
The $k$-means approach has been widely used to construct the codebook
$\mathbf{C}$ where $i(x)$ is the index of the nearest \emph{mean} of $x$ and
$\mathbf{c}(i(x))$ is corresponding \emph{mean}. 
An inverted index can be subsequently built to return all data vectors that are
mapped to the same codeword.

A product quantizer~\cite{DBLP:journals/pami/JegouDS11} is a solution to VQ when
a large number of codewords are desired. The key idea is to decompose the
original vector space into the Cartesian product of $M$ lower dimensional
subspaces and quantize each subspace separately. Suppose $M=2$ and the
dimensions are evenly partitioned. Each vector $x$ can be represented as the
concatenation of $2$ sub-vectors: $[x^{(1)}, x^{(2)}]$, Then we have
$\mathbf{C} = \mathbf{C^{(1)}} \times \mathbf{C^{(2)}}$ where $\mathbf{C}^{(1)}$
(resp.\ $\mathbf{C}^{(2)}$) is constructed by applying the $k$-means algorithm
on the subspace consisting of the first (resp.\ second) half dimensions. As
such, there are $k \times k$ codewords, and $x$'s nearest codeword $\mathbf{c}$
in $\mathbf{C}$ is the concatenation of the two nearest sub-codewords $c^{(1)}$
and $c^{(2)}$ based on its sub-vectors $x^{(1)}$ and $x^{(2)}$, respectively.
Given a query vector $q = [q^{(1)}, q^{(2)}]$, the search algorithm
in~\cite{DBLP:conf/cvpr/BabenkoL12} first retrieves $L$ closest sub-codedwords
$\set{\mathbf{C^{i}_j}}_{i \in \set{1, 2}, j \in \set{1, 2, \ldots L}}$ in each
subspaces based on $q^{(1)}$ and $q^{(2)}$, respectively, and then a multi-sequence algorithm based on a priority queue is applied to traverse the set of pairs $\set{\mathbf{C^{1}_j},\mathbf{C^{2}_j}}$ in increasing distance to achieve candidate codeword set $\mathcal{C}$, finally, the distances
of points belonging to one of the code word in $\mathcal{C}$ are examined via the inverted
index.

Recently, the Optimal Product Quantization (\AlgOPQ) method is
proposed~\cite{DBLP:journals/pami/GeHK014}, which optimizes the index by
minimizing the quantization distortion with respect to the space decompositions
and the quantization codewords.

\section{Tree-based Space Partition Methods}
\label{sec:algs_tree}

Tree-based space partition has been widely used for the problem of exact and
approximate NNS. Generally, the space is partitioned in a hierarchically manner
and there are two main partitioning schemes: \textit{pivoting} and
\textit{compact} partitioning schemes.
 Pivoting methods partition the vector space relying on the distance from the data point to pivots
 while compact partitioning methods either divide the data points into clusters, approximate Voronoi partitions or random divided space.
In this subsection, we introduce two representative compact partitioning
methods, \Algannoy~\cite{misc:url/annoy} and
\Algflann~\cite{DBLP:journals/pami/MujaL14}.
as they are used in a commercial recommendation system and widely used in the
Machine Learning and Computer Vision communities. In addition, we also evaluate
a classical pivoting partitioning method, \textit{Vantage-Point}
tree~\cite{DBLP:conf/soda/Yianilos93} (VP-tree), in the experiments.

\subsection{\Algflann}
\label{subsubsec:flann}

\Algflann is an automatic nearest neighbor algorithm configuration method which
select the most suitable algorithm from \textit{randomized
  kd-tree}~\cite{DBLP:conf/cvpr/Silpa-AnanH08}, \textit{hierarchical k-means
  tree}~\cite{DBLP:journals/tc/FukunagaN75}\footnote{Also known as
  GNAT~\cite{DBLP:conf/vldb/Brin95} and vocabulary
  tree~\cite{DBLP:conf/cvpr/SchindlerBS07}.}, and \textit{linear scan} methods
for a particular data set. Below, we briefly introduce the two modified
tree-based methods and the algorithm selection criteria of \Algflann based
on~\cite{DBLP:journals/pami/MujaL14} and Version~1.8.4 source code.

\myparagraph{Randomized kd-trees}%

The main differences from \Algflann{}'s randomize kd-trees with the traditional
kd-tree are:
\begin{inparaenum}[(i)]
\item The data points are recursively split into two halves by first choosing a
  splitting dimension and then use the perpendicular hyperplane centered at the
  \emph{mean} of the dimension values of all input data points.
\item The splitting dimension is chosen at random from top-5 dimensions that
  have the largest sample variance based on the input data points.
\item Multiple randomized kd-trees are built as the index.
\end{inparaenum}

To answer a query $q$, a depth-first search prioritized by some heuristic scoring function are used to search multiple randomized kd-trees with a shared priority queue and candidate result set. The scoring function always favors the child node that is closer to the query point, and lower bounding distance is maintained across all the randomized trees.

\myparagraph{Hierarchical k-means tree}%
is constructed by partitioning the data points at each level into $K$ regions
using K-means clustering. Then the same method is applied recursively to the
data points in each region. The recursion is stopped when the number of points
in each leaf node is smaller than $K$. The tree is searched by initially
traversing the tree from the root to the closest leaf, during which the
algorithm always picks up the the inner node closest to the query point, and
adding all unexplored branches along the path to a priority queue.


\myparagraph{\Algflann}%
carefully chooses one of the three candidate algorithms by minimizing a cost
function which is a combination of the search time, index building time and
memory overhead to determine the search algorithm.
The cost function is defined as follows:
\begin{displaymath}
c(\theta)=\frac{s(\theta)+\omega_{b}b(\theta)}{min_{\theta \in \Theta}(s(\theta)+\omega_{b}b(\theta))}+\omega_{m}m(\theta)
\end{displaymath}
where $s(\theta)$, $b(\theta)$ and $m(\theta)$ represent the search time, tree build time and memory overhead for the trees constructed and queried with parameters $\theta$.

Flann use random sub-sampling cross-validation to generate the data and the query points when we run the optimization. The optimization can be run on the full data set for the most accurate results or using just a fraction of the data set to have a faster auto-tuning process.


\subsection{\Algannoy}
\label{subsubsec:annoy}

\Algannoy is an empirically engineered algorithm that has been used in the
recommendation engine in \url{spotify.com}, and has been recognized as one of
the best \ANN{} libraries.

\myparagraph{Index Construction}. %

In earlier versions, \Algannoy{} constructs multiple random projection
trees~\cite{DBLP:conf/stoc/DasguptaF08}. In the latest version (as of March
2016), it instead constructs multiple hierarchical 2-means trees. %
Each tree is independently constructed by recursively partitioning the data
points as follows. At each iteration, two centers are formed by running a
simplified clustering algorithm on a subset of samples from the input data
points. The two centers defines a partition hyperplane which has equidistant
from the centers. Then the data points are partitioned into two sub-trees by the
hyperplane, and the algorithm builds the index on each sub-trees recursively.

\myparagraph{Search}. %

The search process is carried out by traveling tree nodes of the multiple RP
trees. Initially, the roots of the RP trees are pushed into a maximum priority
queue with key values infinity. For a given tree node $n_i$ with parent node
$n_p$, if the query $q$ falls into the subtree of $n_i$, the key of $n_i$ is the
minimum of its parent node and the distance to the hyperplane; otherwise, the
key is the minimum of its parent node and the negative value of the distance to
the hyperplane. At each iteration, the node with the maximum key is chosen for
exploration.%

\subsection{VP-tree}

VP-tree is a classic space decomposition tree that recursively divides the space with respect to a randomly chosen pivot $\pi$. For each partition, a median value R of the distance from $\pi$ to every other point in the current partition was computed. The pivot-centered ball with the radius $R$ is used to partition the space: the inner points are placed into the left subtree, while the outer points are placed into the right subtree (points that are exactly at distance $R$ from $\pi$ can be placed arbitrarily). Partitioning stops when the number of points falls below the threshold $b$.

In classic VP-tree, the triangle inequality can be used to prune unpromising partitions as follows: imagine that $r$ is a radius of the query and the query point is inside the pivot-centered ball (i.e., in the left subtree). If $R - d(\pi,q) > r$, the right partition cannot have an answer and the right subtree can be safely pruned. The nearest-neighbor search is simulated as a range search with a decreasing radius: Each time we evaluate the distance between $q$ and a data point, we compare this distance with $r$. If the distance is smaller, it becomes a new value of $r$. In \cite{DBLP:journals/pvldb/NaidanBN15}, a simple polynomial pruner is employed. More specifically, the right partition can be pruned if the query is in the left partition and $(R-d(\pi,q))^{\beta}\alpha_{left} > r$. The left partition can be pruned if the query is in the right partition and $(d(\pi,q)-R)^{\beta}\alpha_{left} > r$.

\section{Neighborhood-based Methods}
\label{sec:algs_proximity}

In general, the neighborhood-based methods build the index by retaining the neighborhood information for each individual data point towards other data
points or a set of pivot points. Then various greedy heuristics are proposed to navigate the proximity graph for the given query. In this subsection, we
introduce the representative neighborhood-based methods, namely Hierarchical Navigable Small World(\Alghnsw~\cite{DBLP:journals/corr/MalkovY16}) and    \Algkgraph~\cite{DBLP:conf/www/DongCL11}. In addition, we also evaluate other two representative
methods including Small World (\Algsw~\cite{DBLP:journals/is/MalkovPLK14}), and Rank Cover Tree (\AlgRCT~\cite{DBLP:journals/pami/HouleN15}) \footnote{Though the tree structure is employed by \AlgRCT, the key idea of the \AlgRCT relies on the neighborhood information during the index construction.}.

\subsection{\Algkgraph}
\label{subsec:kgraph}

\Algkgraph~\cite{misc:url/kgraph} is the representative technique for \KNN{}
graph construction and nearest neighbor searches.

\myparagraph{Index Construction}.%
\KNN{} graph is a simple directed graph where there are $K$ out-going edges for each node, pointing to its $K$ nearest data points. Since the exact computation
of \KNN graph is costly, many heuristics have been proposed in the literature~\cite{DBLP:conf/www/DongCL11,DBLP:conf/kdd/AoyamaSSU11}. In~\cite{DBLP:conf/www/DongCL11}, the construction of \KNN graph relies on a simple principle: \emph{A neighbor's neighbor is probable also a neighbor}.
Starting from randomly picked $k$ nodes for each data point, it iteratively improves the approximation by comparing each data point against its current neighbors' neighbors, including both \KNN and reverse \KNN points, and stop when no improvement can be made. Afterwards, four optimizations are applied (local join, incremental search, sampling and early termination) to reduce the redundant computation and speed up the indexing phrase to stop at an acceptable accuracy.

\myparagraph{Search}.%
A greedy search algorithm is employed in~\cite{DBLP:conf/www/DongCL11} to discover the $k$ nearest neighbor from a query point $q$. Starting from $p$ randomly chosen nodes (i.e.,data points), the search maintains a node list to store the current best $k$ nodes sorted by their distances to the query, and recursively computes the distances from the query $q$ to each neighbor point (by following the graph edges) of first unexplored data point of the node list. The node list is updated by the neighbor points that are closer to the query. This greedy algorithm stops when each data point $x$ at the node list is closer to the query than any of its neighbor.

\subsection{Small World}
\label{subsec:SW}
A small world(SW) method is a variant of a navigable small world graph data structure. The small world graph contains an approximation of the Delaunay graph and has long-range links together with the small-world navigation property.SW uses the same searching method with K-NNG. The greatest difference between K-NNG and SW is the connection structure of nodes. K-NNG strives to obtain the local approximate optimal solution for each node, but SW explore the current optimal links for the inserted points by a incremental construction strategy. Besides, SW keeps two-way connection for each edge, but K-NNG only preserve the kNN neighbors of each node.

\textbf{Index Construction} The construction of SW is a bottom-top procedure that insert all the data points consecutively. For every new incoming point,they find the set of its closest neighbors from the graph and the undirected edges are created to connect the set and the point. As more and more elements are inserted into the graph, links that previously served as short-range links now become long-range links making a navigable small world.

\textbf{Search} The nearest neighbor search algorithm is, thus, a greedy search procedure that carries out several sub-searches. A sub-search starts at a random node and proceeds to expanding the set of traversed nodes which are not visited by following neighboring links. The sub-search stops when it cannot find points that are closer than already found $M$ nearest points ($M$ is a search parameter).

\subsection{Hierarchical Navigable Small World}
\label{subsec:HNSW}

The key idea of the Hierarchical Navigable Small World(HNSW) algorithm is to separate the links according to their length scale. In this case, the average connections per element in all of the layers can be limited.

\textbf{Index Construction} HNSW can be seen as a multi-layer and multi-resolution variant of a proximity graph. A ground (zero-level) layer includes all data points and higher layer has fewer points. Similar to SW, HNSW is constructed by incrementially inserting data points, one by one. For each data point, a maximum level $m$ is selected randomly and the new point is added to all the layers starting from layer $m$ down to the layer zero. The insertion process can be divided into two phrases. The first phrase starts from the top layer to $m+1$ by greedily traversing the graph in order to find the closest neighbor in the layer, which is used as the enter point to continue the search in the next layer.
The second phrase starts from layer $m$ down to zero. $M$ nearest neighbors are found and connected with the new point. The searching quality is controlled by the parameter $ef$, which is the number of the enter points and plays the similar role with $p$ of KGraph. In first phrase, the number of $ef$ is set as 1.

\textbf{Search} The searching algorithm is roughly equivalent to the insertion algorithm from an item with maximum level $m=0$. The closest neighbors found at the ground layer are returned as the search result.

\subsection{Rank Cover Tree}

Rank cover tree (RCT) ~\cite{DBLP:journals/pami/HouleN15} is a probabilistic data structure for similarity search, which entirely avoids the use of numerical constraints such as triangle inequality. The searching algorithm is based on the ranks of the points with respect to the query, and returns a correct result in time that depends competitively on a measure of the intrinsic dimensionality of the data set.

The structure of RCT blends some of the design features of SASH ~\cite{DBLP:conf/icde/HouleS05} and Cover Tree ~\cite{DBLP:conf/icml/BeygelzimerKL06}. It is a hierarchical tree in which each node in the bottom level (level 0) is associated with a data point, and the nodes in level $j$ are randomly selected from the set of level $j-1$ with certain probability. The index construction of RCT is performed by inserting the nodes from high levels to low levels. If one node $x$ in level $j$ appears in the RCT tree, the indexing algorithm will search its nearest neighbor $v$ in level $j+1$, and then link $x$ to $v$. Otherwise, the node links to its copy in level $j+1$.

Searching of RCT starts from the root of the tree, and on each level $j$, only a subset of nodes $V_{j}$ is kept, as the nodes in $V_{j}$ are the most similar to the query. The search algorithm iteratively perform the above procedure until reaching the bottom level.

\section{Diversified Proximity Graph}
\label{sec:dpg}

The experience and insights we gained from this study enable us to engineer a
new method, Diversified Proximity Graph (\Algdpg{}), which constructs a different
neighborhood graph to achieve better and more robust search performance.


\subsection{Motivation}
\label{sec:motivation}

\begin{figure}[htbp]
\centering
\begin{tikzpicture}[scale = 1.5]
  \begin{small}
    \draw [color=black] (4.02,2.3) circle (0pt);
    \draw [fill=black] (3.92,2.3) circle (1.5pt);
    \draw [color=black] (3.92,2.3) node (p) {};
    \draw[color=black] (4.,2.12) node {$p$};
    \draw [fill=black] (3.1,2.82) circle (1.5pt);
    \draw [fill=black] (3.1,2.82) node (a1) {};
    \draw[color=black] (3.2,3.0) node {$a_4$};
    \draw [fill=black] (2.94,2.36) circle (1.5pt);
    \draw [fill=black] (2.94,2.36) node (a2) {};
    \draw[color=black] (2.8,2.25) node {$a_3$};
    \draw [fill=black] (2.9,2.8) circle (1.5pt);
    \draw[color=black] (2.8,2.98) node {$a_1$};
    \draw [fill=black] (3.22,1.74) circle (1.5pt);
    \draw [fill=black] (2.94,2.6) circle (1.5pt);
    \draw[color=black] (2.8,2.58) node {$a_2$};
    \draw [fill=black] (5.2,2.8) circle (1.5pt);
    \draw [fill=black] (5.2,2.8) node (c) {};
    \draw[color=black] (5.3,3.0) node {$b$};

    \draw [color=black] (3.9,3.2) node (s) {};

    \draw [color=black] (5.1,2.22)-- ++(-1.5pt,-1.5pt) -- ++(2.0pt,2.0pt) ++(-2.0pt,0) -- ++(2.0pt,-2.0pt);
    \draw[color=black] (5,2.) node (q) {$q$};
    \draw [->,dash pattern=on 3pt off 3pt] (p) -- (c);
    \draw [->] (p) -- (a1);
    \draw [->,red,snake=zigzag,segment amplitude=3pt,segment length=6pt] (s) -- (p);
    \draw [->] (p) -- (a2);
    \draw [fill=black] (2.44,2.84) circle (1.5pt);
    \draw [fill=black] (1.78,2.44) circle (1.5pt);
    \draw [fill=black] (2.42,2.36) circle (1.5pt);
    \draw [fill=black] (2.2,1.92) circle (1.5pt);
    \draw [fill=black] (5.84,2.44) circle (1.5pt);
    \draw [fill=black] (5.7,3.) circle (1.5pt);
    \draw [fill=black] (6.4,2.68) circle (1.5pt);
    \draw [fill=black] (6.14,2.02) circle (1.5pt);
  \end{small}
\end{tikzpicture}
\vspace{-2mm}
\caption{Toy Example of a $2$-d dataset, $K=2$}
\vspace{0mm}
\label{fig:div_motiv}
\end{figure}
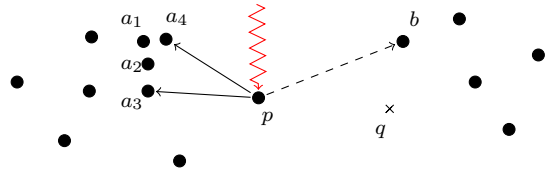

In \KNN graph construction, we only consider the distances of neighbors for each
data point. But intuitively we should also consider the \emph{coverage} of the
neighbors. As shown in Figure ~\ref{fig:div_motiv}, the two closest neighbors of
the point $p$ are $a_3$ and $a_4$, and hence in the $2$-NN graph $p$ cannot lead the
search to the NN of $q$ (i.e., the node $b$) although it is close to $b$. Since
$a_1,\ldots,a_4$ are clustered, it is not cost-effective to retain both
$a_3$ and $a_4$ in the \KNN list of $p$. This motivates us to consider the
direction diversity (i.e., angular dissimilarity) of the \KNN list of $p$ in
addition to the distance, leading to the diversified \KNN graph. Regarding the
example, including $a_3$ and $b$ is a better choice for the \KNN list of $p$.

Now assume we have replaced edge $(p, a_4)$ with the edge $(p, b)$ (i.e., the
dashed line in Figure~\ref{fig:div_motiv}), but there is still another problem.
As we can see that there is no incoming edge for $p$ because it is relatively
far from two clusters of points (i.e., $p$ is not $2$-NN of these data points).
This implies that $p$ is isolated, and two clusters are disconnected in the
example. This is not uncommon in high dimensional data due to the phenomena of
``hubness''\cite{DBLP:journals/jmlr/RadovanovicNI10} where a large portion of
data points rarely serve as K-NN of other data points, and thus have no or only
a few incoming edges in the \KNN graph. This motivates us to also use the reverse
edges in the diversified \KNN graph; that is, we keep an bidirected diversified
\KNN graph as the index, and we name it \emph{Diversified Proximity Graph}
(\Algdpg).

\subsection{Diversified Proximity Graph}
\label{sec:using-divers-kgraph}

The construction of DPG is a diversification of an existing \KNN graph, followed
by adding reverse edges.

Given a reference data point $p$, the similarity of two points $x$ and $y$ in
$p$'s \KNN{} list $\mathcal{L}$ is defined as the angle of $\angle xpy$, denoted
by $\theta(x,y)$. We aim to choose a subset of $\kappa$ data points, denoted by
$\mathcal{S}$, from $\mathcal{L}$ so that the average angle between two points
in $\mathcal{S}$ is maximized; or equivalently,
$\mathcal{S} = \argmin_{\mathcal{S} \subseteq \mathcal{N}, |\mathcal{S}|=\kappa
} \sum_{o_i, o_j \in \mathcal{S}} \theta(o_i, o_j)$.

The above problem is NP-hard~\cite{kuo1993analyzing}. Hence, we design a simple
greedy heuristic. Initially, $\mathcal{S}$ is set to the closest point of $p$ in
$\mathcal{L}$. In each of the following $\kappa-1$ iterations, a point is moved
from $\mathcal{L} \setminus \mathcal{S}$ to $\mathcal{S}$ so that the average
pairwise angular similarity of the points in $\mathcal{S}$ is minimized. Then
for each data point $u$ in $\mathcal{S}$, we include both edges $(p, u)$ and
$(u, p)$ in the diversified proximity graph. The time complexity of the
diversification process is $O(\kappa^2 K n)$ where $n$ is the number of data
points, and there are totally at most $2\kappa n$ edges in the diversified
proximity graph. %

It is critical to find a proper $K$ value for a desired $\kappa$ in the
diversified proximity graph as we need to find a good trade-off between
diversity and proximity. In our empirical study, the \Algdpg{} algorithm usually
achieves the best performance when $K=2\kappa$. Thus,
we set $K=2\kappa$ for the diversified proximity graph construction. %
%
Although the angular similarity based \Algdpg{} algorithm can achieve very good
search performance, the diversification time of $O(\kappa^2 K n)$ is still
costly. In our implementation, we use another heuristic to construct the
diversified K-NN graph as follows. We keep a counter for each data point in the
\KNN list $\mathcal{L}$ of the data point $p$. For each pair of points $u$, $v$
$\in \mathcal{L}$, we increase the counter of $v$ by one if $v$ is closer to $u$
than to $p$; that is, $\dist{v}{u}$ $ < \dist{v}{p}$. Then, we simply keep the
$\kappa$ data points with lowest count frequencies for $p$ because, intuitively,
the count of a data point is high if there are many other points with similar
direction. This leads to the time complexity of $O(K^2 n)$ for diversification.
Our empirical study shows that we
can achieve similar search performance, while significantly reduce the
diversification time. We also demonstrate that both diversification and reverse
edges contribute to the improvement of the search performance.

Note that the search process of the \Algdpg{} is the same as that of
\Algkgraph{} introduced in Section~\ref{subsec:kgraph}.



\section{EXPERIMENTS}
\label{sec:exp}

In this section, we present our experimental evaluation.

\subsection{Experimental Setting}
\label{subsec:exp_setting}

\subsubsection{Algorithms Evaluated}
\label{subsubsec:exp_algs}

We use $15$ representative existing NNS algorithms from the three categories and
our proposed diversified proximity graph (\Algdpg{}) method. All the source codes are
publicly available. Algorithms are implemented in
C++ unless otherwise specified. We carefully go through all the
implementations and make necessary modifications for fair comparisons. For
instance, we re-implement the search process of some algorithms in C++. We also
disable the multi-threads, SIMD instructions, fast-math, and hardware
pre-fetching technique. All of the modified source codes used in this paper are public available on GitHub~\cite{misc:url/nns-benchmark}.

\myparagraph{(1) LSH-based Methods.}
We evaluate Query-Aware LSH~\cite{DBLP:journals/pvldb/HuangFZFN15}
(\textbf{\Algqalsh}\footnote{\url{http://ss.sysu.edu.cn/~fjl/qalsh/qalsh_1.1.2.tar.gz}},
PVLDB'15) and SRS~\cite{DBLP:journals/pvldb/SunWQZL14}
(\textbf{\Algsrs}\footnote{\url{https://github.com/DBWangGroupUNSW/SRS}},
PVLDB'14).

\myparagraph{(2) Encoding-based Methods.}
We evaluate Scalable Graph Hashing~\cite{DBLP:conf/ijcai/JiangL15}
(\textbf{SGH}\footnote{\url{http://cs.nju.edu.cn/lwj}}, IJCAI'15), Anchor Graph
Hashing~\cite{DBLP:conf/icml/LiuWKC11}
(\textbf{AGH}\footnote{\url{http://www.ee.columbia.edu/ln/dvmm/downloads}},
ICML'11), Neighbor-Sensitive Hashing~\cite{DBLP:journals/pvldb/ParkCM15}
(\textbf{NSH}\footnote{\url{https://github.com/pyongjoo/nsh}}, PVLDB'15). We use
the hierarchical clustering trees in \Algflann{} to index the resulting binary
vectors to support more efficient search for the above algorithms. Due to increased sparsity for the Hamming space with more bits, precision within Hamming radius 2 drops significantly when longer codes are used. Therefore, we check the real distances of at most $N$ data points to achieve the tradeoff between the search quality and search speed.

We also evaluate the Selective Hashing~\cite{DBLP:conf/kdd/GaoJOW15}
(\textbf{SH}\footnote{\url{http://www.comp.nus.edu.sg/~dsh/index.html}}, KDD'15)
, Optimal Product Quantization\cite{DBLP:journals/pami/GeHK014}
(\textbf{\AlgOPQ}\footnote{\url{http://research.microsoft.com/en-us/um/people/kahe}},
TPAMI'14) and Neighborhood APProximation
index~\cite{DBLP:journals/pvldb/NaidanBN15}
(\textbf{\AlgNAPP{}}\footnote{\label{first_ref}\url{https://github.com/searchivarius/nmslib}},
PVLDB'15). Note that we use the inverted multi-indexing
technique\footnote{\url{http://arbabenko.github.io/MultiIndex/index.html}}~\cite{DBLP:conf/cvpr/BabenkoL12} to perform
non-exhaustive search for \AlgOPQ{}.

\myparagraph{(3) Tree-based Space Partition Methods.}
We evaluate
\textbf{\Algflann}\footnote{\url{http://www.cs.ubc.ca/research/flann}}
(\cite{DBLP:journals/pami/MujaL14}, TPAMI'14),
\textbf{\Algannoy}\footnote{\url{https://github.com/spotify/annoy}}, and an
advanced Vantage-Point tree~\cite{DBLP:conf/nips/BoytsovN13}
(\textbf{VP-tree}\footref{first_ref}, NIPS'13).

\myparagraph{(4) Neighborhood-based Methods.}
We evaluate Small World Graph~\cite{DBLP:journals/is/MalkovPLK14}
(\textbf{\Algsw}\footref{first_ref}, IS'14), Hierarchical Navigable Small World~\cite{DBLP:journals/corr/MalkovY16}(\textbf{HNSW}\footref{first_ref}, CoRR'16), Rank Cover
Tree~\cite{DBLP:journals/pami/HouleN15}
(\textbf{\AlgRCT{}}\footnote{\label{nnsbench}\url{https://github.com/DBWangGroupUNSW/nns_benchmark}},
TPAMI'15), \KNN graph~\cite{DBLP:conf/www/DongCL11,misc:url/kgraph}
(\textbf{\Algkgraph}\footnote{\url{https://github.com/aaalgo/kgraph}}, WWW'11),
and our diversified proximity graph (\textbf{\Algdpg{}}\footref{nnsbench}).

\myparagraph{Computing Environment.}
All C++ source codes are complied by g++ $4.7$, and MATLAB source codes (only for index construction of some algorithms)
are compiled by MATLAB $8.4$. All experiments are conducted on a Linux server
with Intel Xeon $8$ core CPU at $2.9$GHz, and $32$G memory.

\subsubsection{Datasets and Query Workload}
\label{subsubsec:exp_data}

We deploy $18$ real datasets used by existing works which cover a wide range of
applications including image, audio, video and textual data. We also use two
synthetic datasets. Table~\ref{tab:dataset} summarizes the characteristics of the
datasets including the number of data points, dimensionality, \textit{Relative
  Contrast} (RC~\cite{DBLP:conf/icml/HeKC12}), \textit{local intrinsic
  dimensionality} (LID~\cite{DBLP:conf/kdd/AmsalegCFGHKN15}), and data type
where RC and LID are used to describe the hardness of the datasets.

\textbf{Relative Contrast}  evaluates the influence of several crucial data characteristics such as dimensionality, sparsity, and database size simultaneously in arbitrary normed metric spaces.

Suppose $X=\{x_{i},i=1,...,n\}$ and a query $q$ where $x_{i}$, $q \in R^{d}$ are i.i.d  samples from an unknown distribution $p(x)$.  Let $D_{min}^{q}= \min_{i=1,...,n}{D(x_{i},q)}$ is the distance to the nearest database sample, and $D_{mean}^{q}=E_{x}[D(x,q)]$ is the expected distance of a random database sample from the query $q$. The relative contrast of the dataset X for a query $q$ is defined as $C_{r}^{q}=\frac{D_{mean}^{q}}{D_{min}^{q}}$. The relative contrast for the dataset X is given as $C_{r}=\frac{E_{q}[D_{mean}^{q}]}{E_{q}[D_{min}{q}]}=\frac{D_{mean}}{D_{min}}$.

Intuitively, $C_{r}$ captures the notion of difficulty of NN search in $X$. Smaller the $C_{r}$, more difficult the search. If $C_{r}$ is close to 1, then on average a query $q$ will have almost the same distance to its nearest neighbor as that to a random point in $X$.

We also can define Relative Contrast for $k$-nearest neighbor setting as $C_{r}^{k}=\frac{D_{mean}}{D_{knn}}$, where $D_{knn}$ is the expected distance to the $k$-th nearest neighbor. Smaller RC value implies harder datasets.

\textbf{Local Intrinsic Dimensionality} evaluates the rate at which the number of encountered objects grows as the considered range of distances expands from a reference location. We employ RVE(estimation using regularly varying funtions) to compute the value of LID. It applies an ad hoc estimator for the intrinsic dimensionality based on the characterization of distribution tails as regularly varying functions. The dataset with higher LID value implies harder than others.

We mark the first four datasets in Table~\ref{tab:dataset} with asterisks to indicate that they are ``hard'' datasets compared with others according to their RC and LID values.

Below, we describe the datasets used in the experiments.

\vspace{0.5mm}
\noindent \textbf{Nusw}\footnote{\url{http://lms.comp.nus.edu.sg/research/NUS-WIDE.htm}} includes around $2.7$ million web images,
each as a $500$-dimensional bag-of-words vector.

\vspace{0.5mm}
\noindent \textbf{Gist}\footnotemark[15] is an image dataset which contains about 1 million data points with 960 dimensions.

\vspace{0.5mm}
\noindent \textbf{Random} contains $1$M randomly chosen points in a unit
hypersphere with dimensionality $100$.

\vspace{0.5mm}
\noindent \textbf{Glove} \footnote{\url{http://nlp.stanford.edu/projects/glove/}} contains 1.2 million 100-d word feature vectors extracted from Tweets.

\vspace{0.5mm}
\noindent  \textbf{Cifar} \footnote{\url{http://www.cs.toronto.edu/~kriz/cifar.html}} is a labeled subset of \textbf{TinyImage} dataset, which consists of 60000 32 $\times$ color images in 10 classes, with each image represented by a 512-d GIST feature vector.

\vspace{0.5mm}
\noindent  \textbf{Audio} \footnote{\url{http://www.cs.princeton.edu/cass/audio.tar.gz}} has about 0.05 million 192-d audio feature vectors extracted by Marsyas library from DARPA TIMIT audio speed dataset.

\vspace{0.5mm}
\noindent \textbf{Mnist} \footnote{\url{http://yann.lecun.com/exdb/mnist/}} consists of 70k images of hand-written digits, each as a 784-d vector concatenating all pixels. 

\vspace{0.5mm}
\noindent \textbf{Sun397} \footnote{\url{http://groups.csail.mit.edu/vision/SUN/}} contains about 0.08 million 512-d GIST features of images.

\vspace{0.5mm}
\noindent \textbf{Enron} origins from a collection of emails. yifang et. al. extract bi-grams and form feature vectors of 1369 dimensions.

\vspace{0.5mm}
\noindent \textbf{Trevi}  \footnote{\url{http://phototour.cs.washington.edu/patches/default.htm}} consists of 0.4 million $\times$ 1024 bitmap(.bmp) images, each containing a 16 $\times$ 16 array of image patches. Each patch is sampled as 64 $\times$ 64 grayscale, with a canonical scale and orientation. Therefore, Trevi patch dataset consists of around 100,000 4096-d vectors.

\vspace{0.5mm}
\noindent \textbf{Notre} \footnote{\url{http://phototour.cs.washington.edu/datasets/}} contains about 0.3 million 128-d features of a set of Flickr images and a reconstruction.

\vspace{0.5mm}
\noindent \textbf{Youtube\_Faces} \footnote{\url{http://www.cs.tau.ac.il/~wolf/ytfaces/index.html}}
contains 3,425 videos of 1,595 different people. All the videos were downloaded from YouTube. 0.3 million vectors are extracted from the frames , each contains 1770 features. 

\vspace{0.5mm}
\noindent \textbf{Msong}\footnote{\url{http://www.ifs.tuwien.ac.at/mir/msd/download.html}}
is a collection of audio features and metadata for a million contemporary popular music tracks with $420$ dimensions.

\vspace{0.5mm}
\noindent \textbf{Sift}\footnote{\url{http://corpus-texmex.irisa.fr}} consists of 1 million 128-d SIFT vectors.

\vspace{0.5mm}
\noindent \textbf{Deep}\footnote{\url{https://yadi.sk/d/I_yaFVqchJmoc}} dataset contains deep neural codes of natural images obtained from the activations of a convolutional neural network, which contains about 1 million data points with 256 dimensions.

\vspace{0.5mm}
\noindent \textbf{UKbench} \footnote{\url{http://vis.uky.edu/~stewe/ukbench/}} contains about 1 million 128-d features of images.

\vspace{0.5mm}
\noindent \textbf{ImageNet} \footnote{\url{http://cloudcv.org/objdetect/}} is introduced and employed by ``The ImageNet Large Scale Visual Recognition Challenge(ILSVRC)'', which contains about 2.4 million data points with 150 dimensions dense SIFT features.

\vspace{0.5mm}
\noindent \textbf{Gauss} is generated by randomly choosing $1000$ cluster centers with in space $[0,10]^{512}$,
and each cluster follows the a Gaussian distribution with deviation $1$ on each dimension.

\vspace{0.5mm}
\noindent \textbf{UQ\_Video} is video dataset and the local features based on some keyframes are extracted which include 256 dimensions.

\vspace{0.5mm}
\noindent \textbf{Bann}\footnotemark[15] is used to evaluate the scalability of the algorithms,
where $1$M, $2$M, $4$M, $6$M, $8$M, and $10$M data points are sampled from $128$-dimensional SIFT descriptors extracted from natural images.

\begin{table}[htb]
  \small
  \centering
  \begin{tabular}{|c|r|r|r|r|r|r|r|l|}
    \hline \textbf{Name} & \textbf{$n$ $(\times 10^3)$} & \textbf{$d$} & \textbf{RC} & \textbf{LID} & \textbf{Type} \\
    \hline
    \hline Nus*   & 269                          & 500          & 1.67        & 24.5         & Image         \\
    \hline Gist*  & 983                          & 960          & 1.94        & 18.9         & Image         \\
    \hline Rand*  & 1,000                        & 100          & 3.05        & 58.7         & Synthetic     \\
    \hline Glove* & 1,192                        & 100          & 1.82        & 20.0         & Text          \\
    \hline Cifa   & 50                           & 512          & 1.97        & 9.0          & Image         \\
    \hline Audio  & 53                           & 192          & 2.97        & 5.6          & Audio         \\
    \hline Mnist  & 69                           & 784          & 2.38        & 6.5          & Image         \\
    \hline Sun    & 79                           & 512          & 1.94        & 9.9          & Image         \\
    \hline Enron  & 95                           & 1,369        & 6.39        & 11.7         & Text          \\
    \hline Trevi  & 100                          & 4,096        & 2.95        & 9.2          & Image         \\
    \hline Notre  & 333                          & 128          & 3.22        & 9.0          & Image         \\
    \hline Yout   & 346                          & 1,770        & 2.29        & 12.6         & Video         \\
    \hline Msong  & 922                          & 420          & 3.81        & 9.5          & Audio         \\
    \hline Sift   & 994                          & 128          & 3.50        & 9.3          & Image         \\
    \hline Deep   & 1,000                        & 128          & 1.96        & 12.1         & Image         \\
    \hline Ben    & 1,098                        & 128          & 1.96        & 8.3          & Image         \\
    \hline Imag   & 2,340                        & 150          & 2.54        & 11.6         & Image         \\
    \hline Gauss  & 2,000                        & 512          & 3.36        & 19.6         & Synthetic     \\
    \hline UQ-V   & 3,038                        & 256          & 8.39        & 7.2          & Video         \\
    \hline BANN   & 10,000                       & 128          & 2.60        & 10.3         & Image         \\
    \hline

  \end{tabular}                                                                                                                    \\
  \vspace{-2mm}
  \caption{\small Dataset Summary} 
  \label{tab:dataset}
\end{table}

\vspace{0.5mm}
\noindent \textbf{Query Workload}.
Following the convention, we randomly remove $200$ data points as the query points for each datasets.
The average performance of the $k$-NN searches is reported.
The $k$ value varies from $1$ to $100$ in the experiments with default value $20$.
In this paper, we use Euclidean distance for ANNS.

\subsection{Evaluation Metrics}
\label{subsubsec:metrics}

Since exact \kNN{} can be found by a brute-force linear scan algorithm (denoted
as $\mathit{BF}$), we use its query time as the baseline and define the \textbf{speedup}
of Algorithm $A$ as $\frac{t_{\mathit{BF}}}{t_{A}}$, where $t_{x}$ is the average search
time of Algorithm $x$. %

The search quality of the $k$ returned points is measured by the standard
\textbf{recall} against the \kNN{} points (See Section~\ref{subsec:prob}).

All reported measures are averaged over all queries in the query workload. We
also evaluate other aspects such as index construction time, index size, and
scalability.

Note that the same algorithm can achieve different combination of speedup and
recall (typically via using different threshold on the number of points
verified, i.e., $N$).


\subsection{Parameter Tunning}

Below are the \emph{default} settings of the key parameters of the algorithms in
the second round evaluation in Section~\ref{subsec:exp_final}.
\begin{itemize}
\item \Algsrs. The number of projections ($m$) is set to $8$.
\item \AlgOPQ. The number of subspaces is $2$, and each subspace can have $2^{10}$
  codewords (i.e., cluster centers) by default.
\item \Algannoy. The number of the \Algannoy{} trees, $m$, is set to $50$.
\item \Algflann. We let the algorithm tune its own parameters.
\item \Alghnsw. The number of the connections for each point, $M$, is set to $10$. 
\item \Algkgraph. By default, we use $K=40$ for the \KNN{} graph index. 
\item \Algdpg. We use $\kappa = \frac{K}{2} = 20$ so that the index size of
  \Algdpg{} is the same as that of \Algkgraph{} in the worst case.
\end{itemize}

More details about how to tune each algorithm and the comparisons about our counting\_based DPG and angular\_based DPG were reported in Appendix A.

\subsection{Comparison with Each Category}
\label{subsec:exp_team}
In this subsection, we evaluate the trade-offs between speedup and recall of all
the algorithms in each category. Given the large number of algorithms
in the space partition-based category, we evaluate them in the encoding-based
and tree-based subcategories separately. The goal of this round of evaluation is
to select several algorithms from each category as the representatives in the
second round evaluation (Section~\ref{subsec:exp_final}).

\subsubsection{LSH-based Methods}
\label{subsubsec:exp_team_di}

\begin{figure}[tbh]
\centering
\subfigure[\small sift ]{
\includegraphics[width=.48\linewidth]{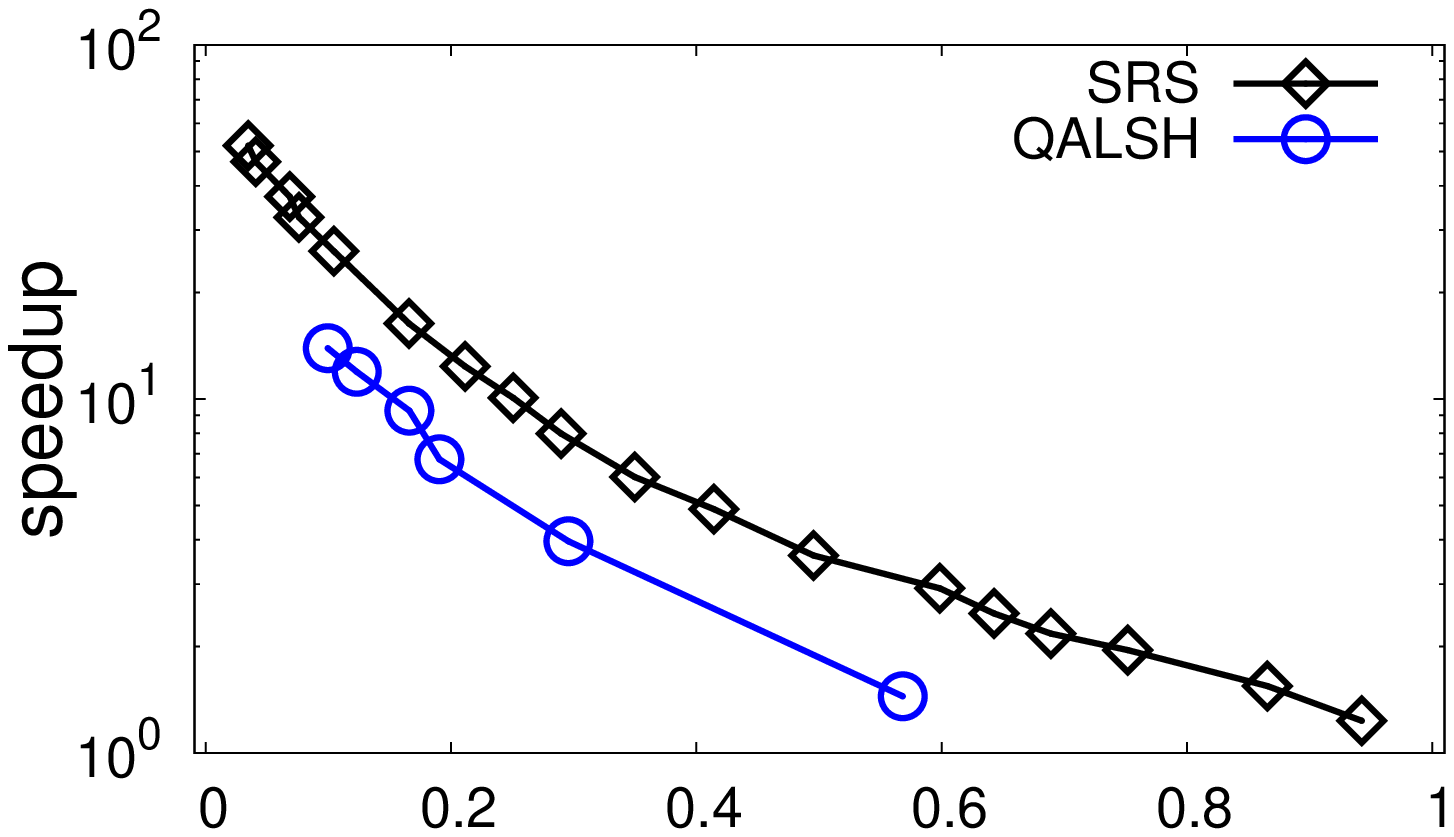}}
\vspace{-3mm}
\subfigure[\small audio ]{
\includegraphics[width=.48\linewidth]{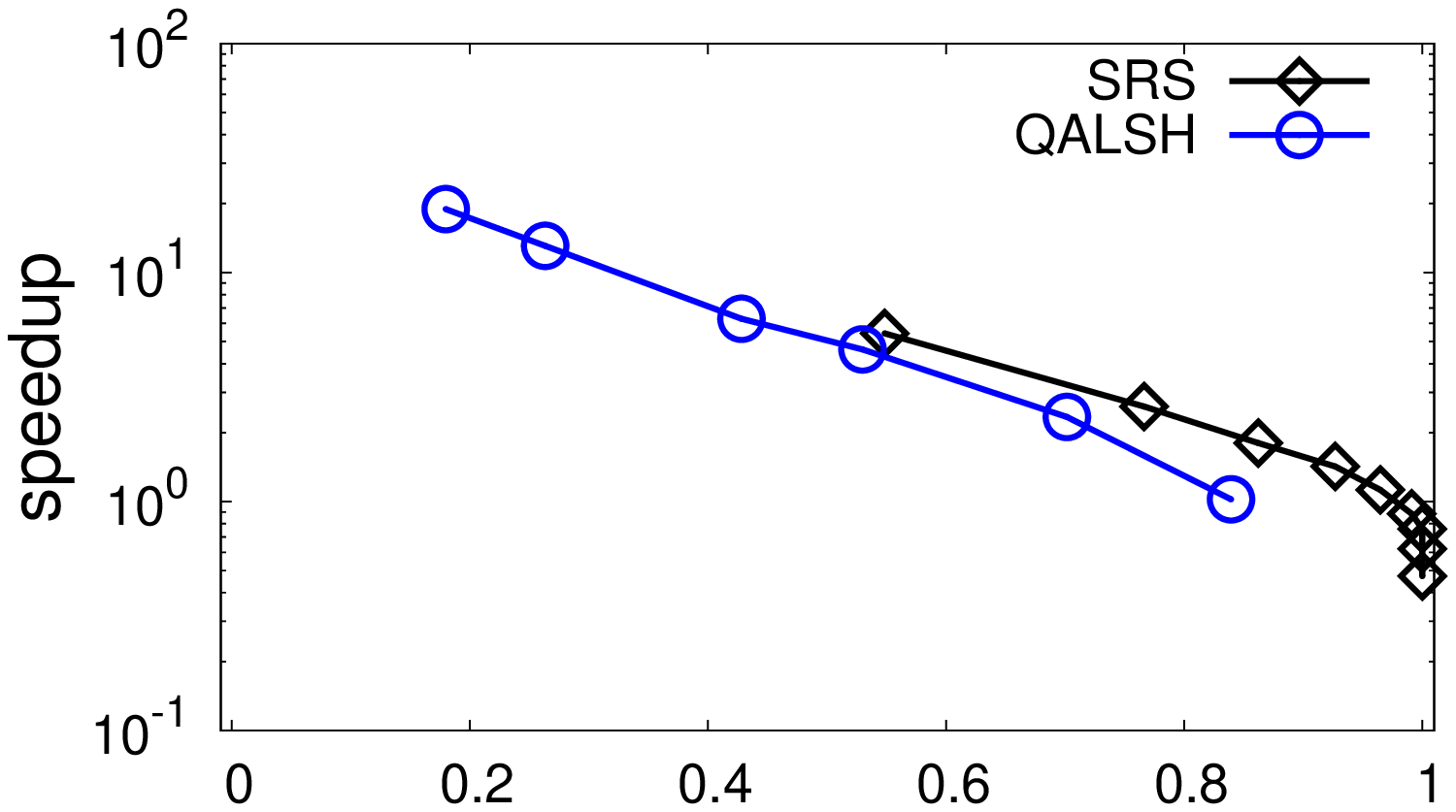}}
\vspace{-3mm}
\caption{\small Speedup vs Recall (Data-independent)}
\label{fig:exp_team_DI}
\end{figure}

Figure~\ref{fig:exp_team_DI} plots the trade-offs between the speedup and recall of two most recent data-independent algorithms
\Algsrs{} and \Algqalsh{} on \Datasift{} and \Dataaudio{}.
As both algorithms are originally external memory based approaches, we evaluate the speedup by means of
the total number of pages of the dataset divided by the number of pages accessed during the search.
It shows that \Algsrs{} consistently outperforms \Algqalsh{}. Table ~\ref{tab:exp_team_DI_IS} shows the construction time and index size of SRS and QALSH, we can see that SRS has smaller index size than QALSH (at least 5 times larger than SRS).
Thus, \Algsrs{} is chosen as the representative in the second round evaluation where a cover-tree based in-memory implementation will be used.

\begin{table}[htb]
\centering
\caption{index size and construction time (Data-independent)}
\begin{tabular}{|c|l|l|l|l|}
\hline \multirow{2}{*}{Dataset} & \multicolumn{2}{c|}{SRS} & \multicolumn{2}{c|}{QALSH} \\
\cline{2-5} & size(MB) &  time(s) &  size(MB) & time(s) \\
\hline Audio  & 2.8  & 26.5   & 14.1  & 27.3  \\
\hline Sift   & 46.6 & 980    & 318   & 277  \\
\hline Notre  & 15.5 & 253.9  & 98.1  & 95.2 \\
\hline Sun    & 4.1  & 45.1   & 21.5  & 67.2\\
\hline
\end{tabular}\\
\vspace{0mm}
\label{tab:exp_team_DI_IS}
\end{table}

\subsubsection{Encoding-based Methods}
\label{subsubsec:exp_dh}

\begin{figure*}[tbh]
\centering
\begin{minipage}[b]{0.8\linewidth}
\centering
\includegraphics[width=1.0\linewidth]{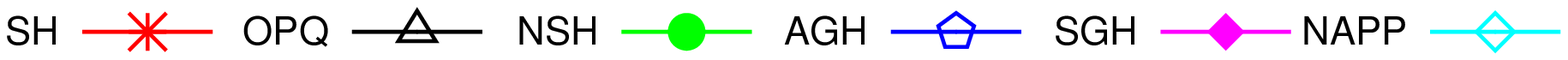}%
\vspace{-3mm}
\end{minipage}
\centering
\begin{minipage}[t]{1.0\linewidth}
\subfigure[\small Sift]{
\includegraphics[width=.234\linewidth]{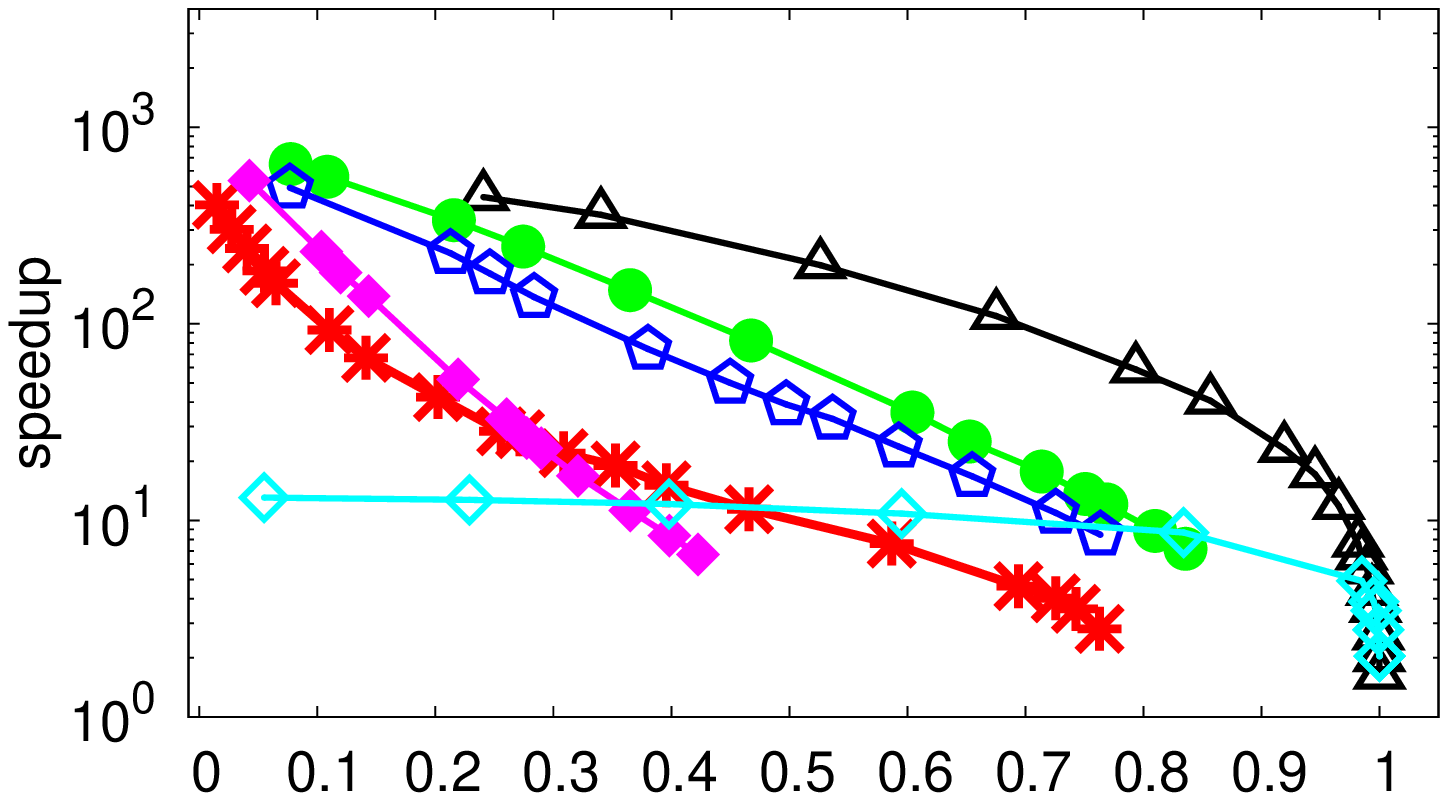}}
\centering
\subfigure[\small Trevi]{
\includegraphics[width=.234\linewidth]{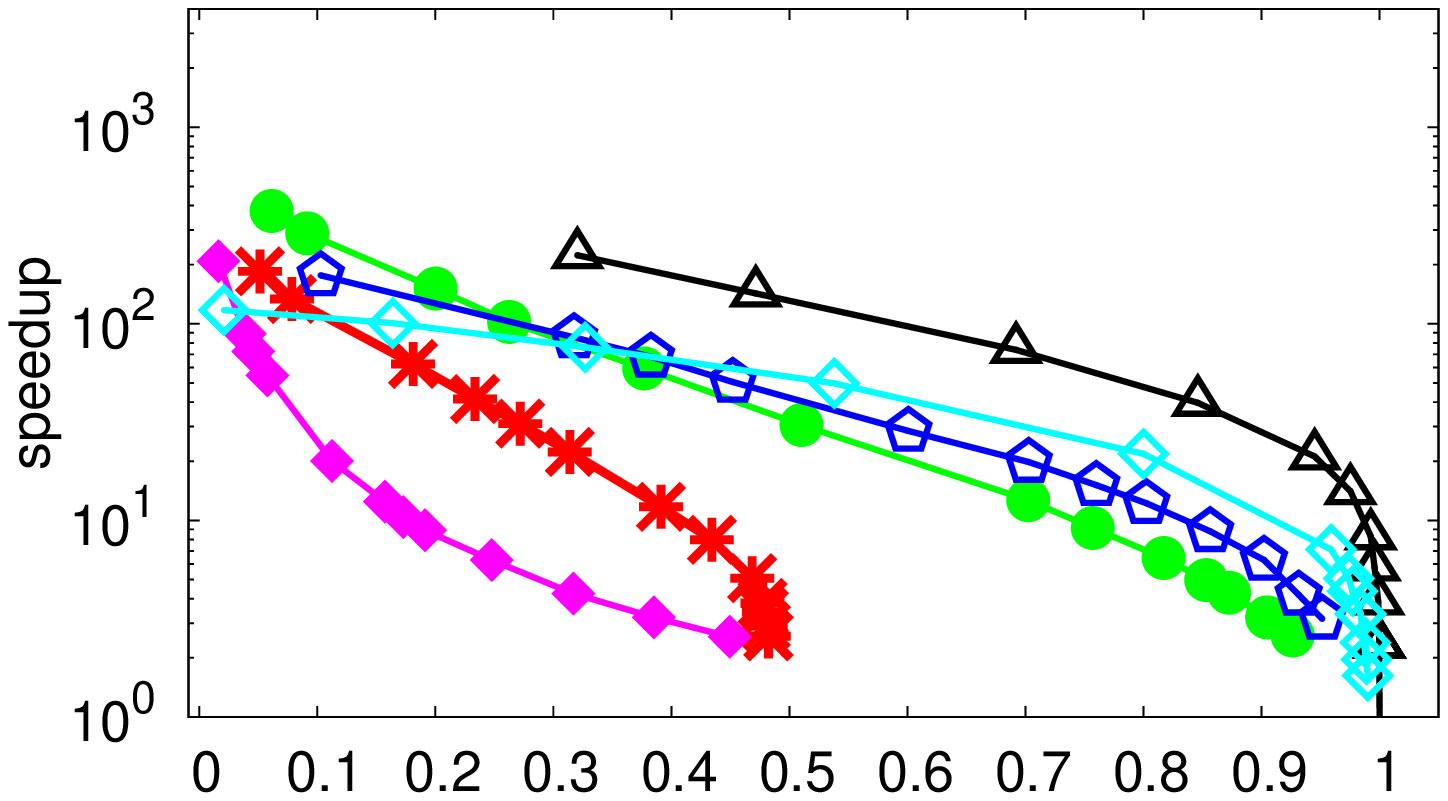}}
\centering
\subfigure[\small Mnist]{
\includegraphics[width=.234\linewidth]{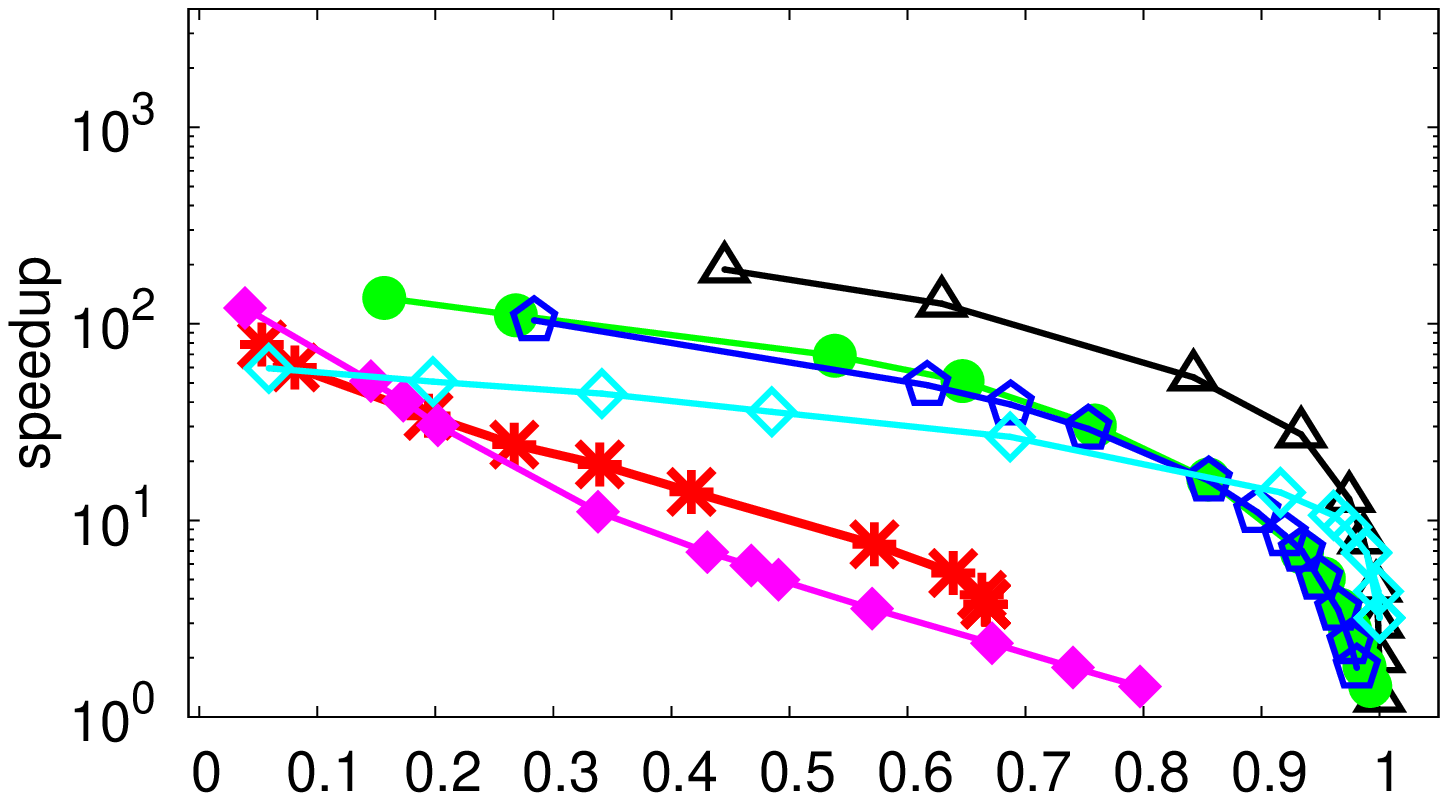}}
\centering
\subfigure[\small Glove]{
\includegraphics[width=.234\linewidth]{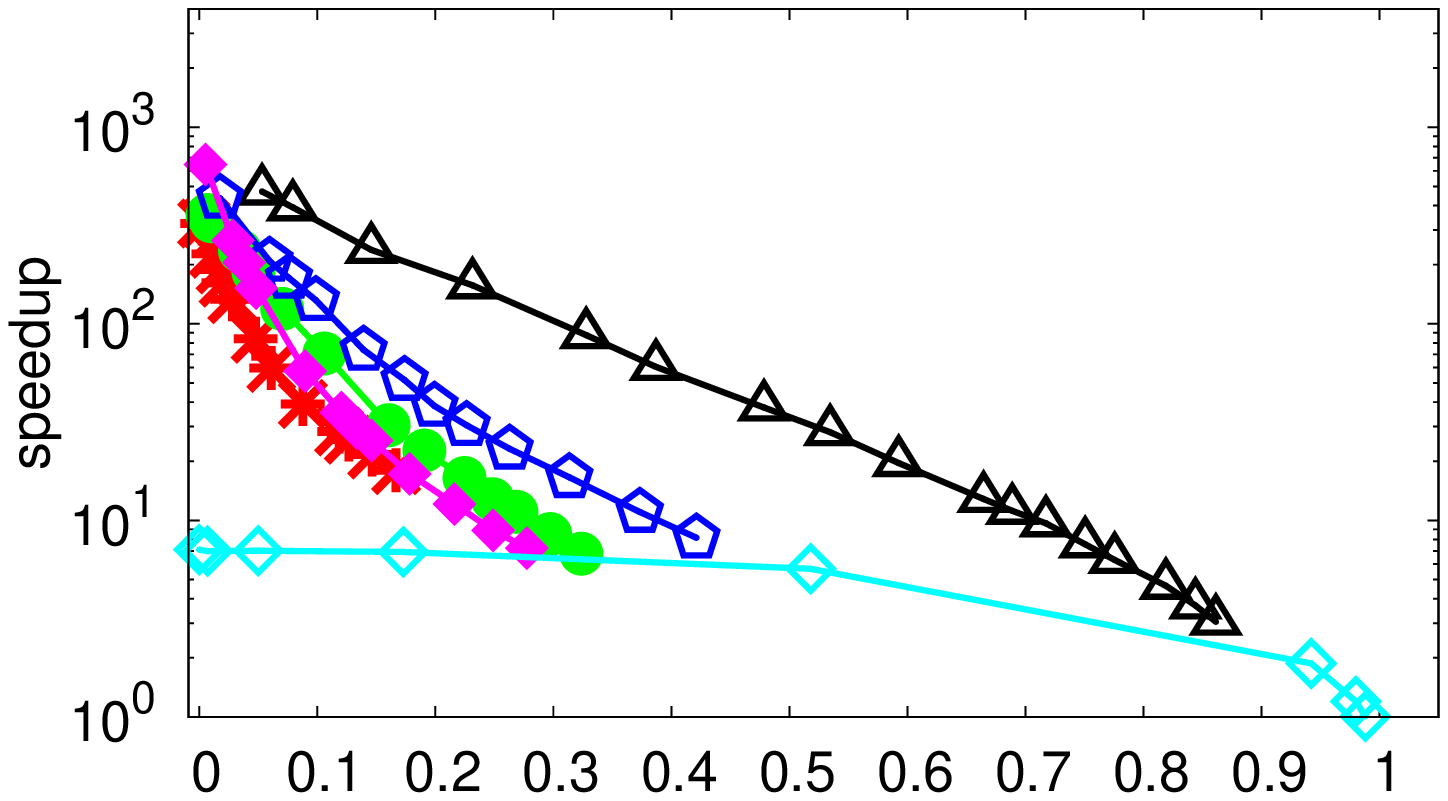}}
\end{minipage}
\vfill
\begin{minipage}[t]{1.0\linewidth}
\centering
\subfigure[\small Notre ]{
\includegraphics[width=.234\linewidth]{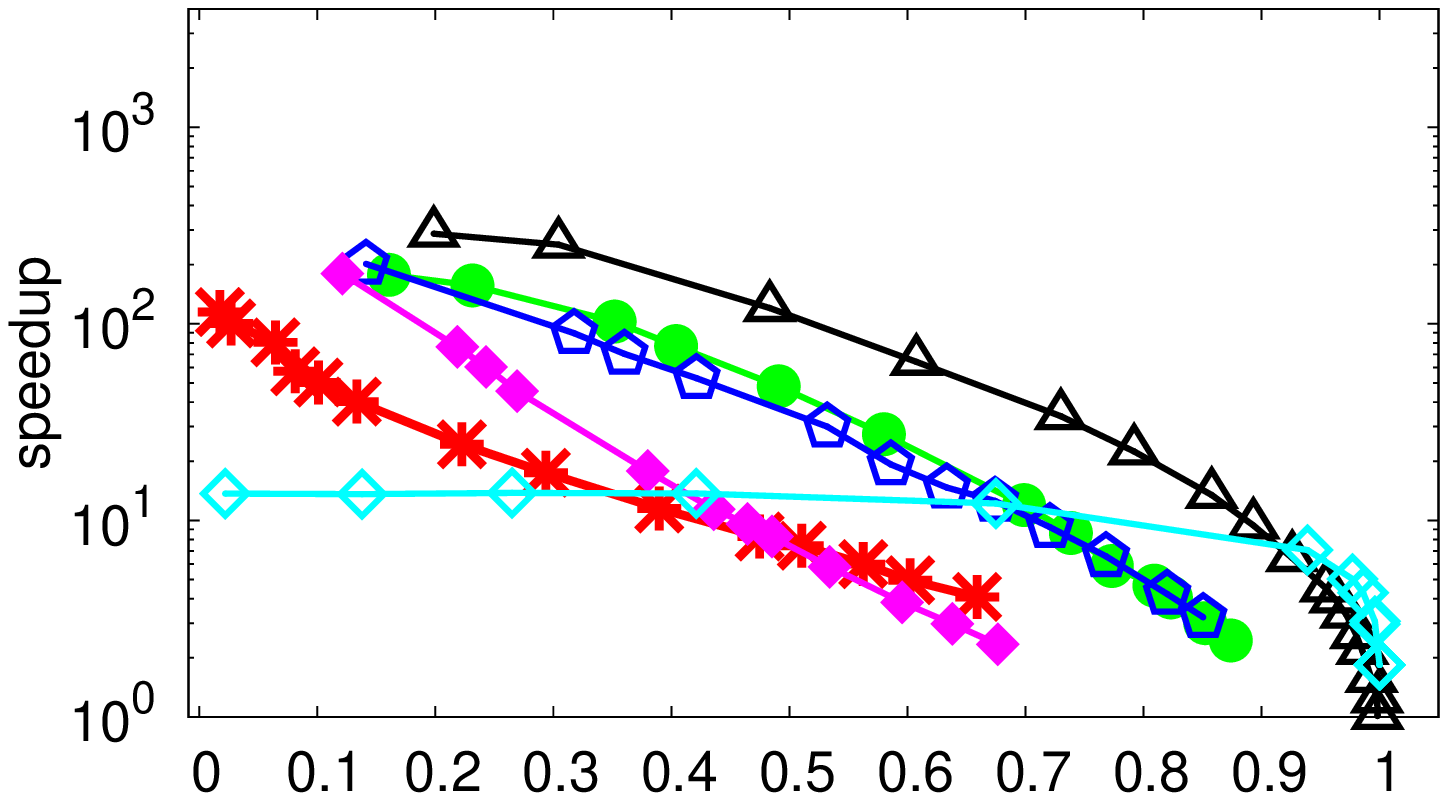}}
\centering
\subfigure[\small Ben ]{
\includegraphics[width=.234\linewidth]{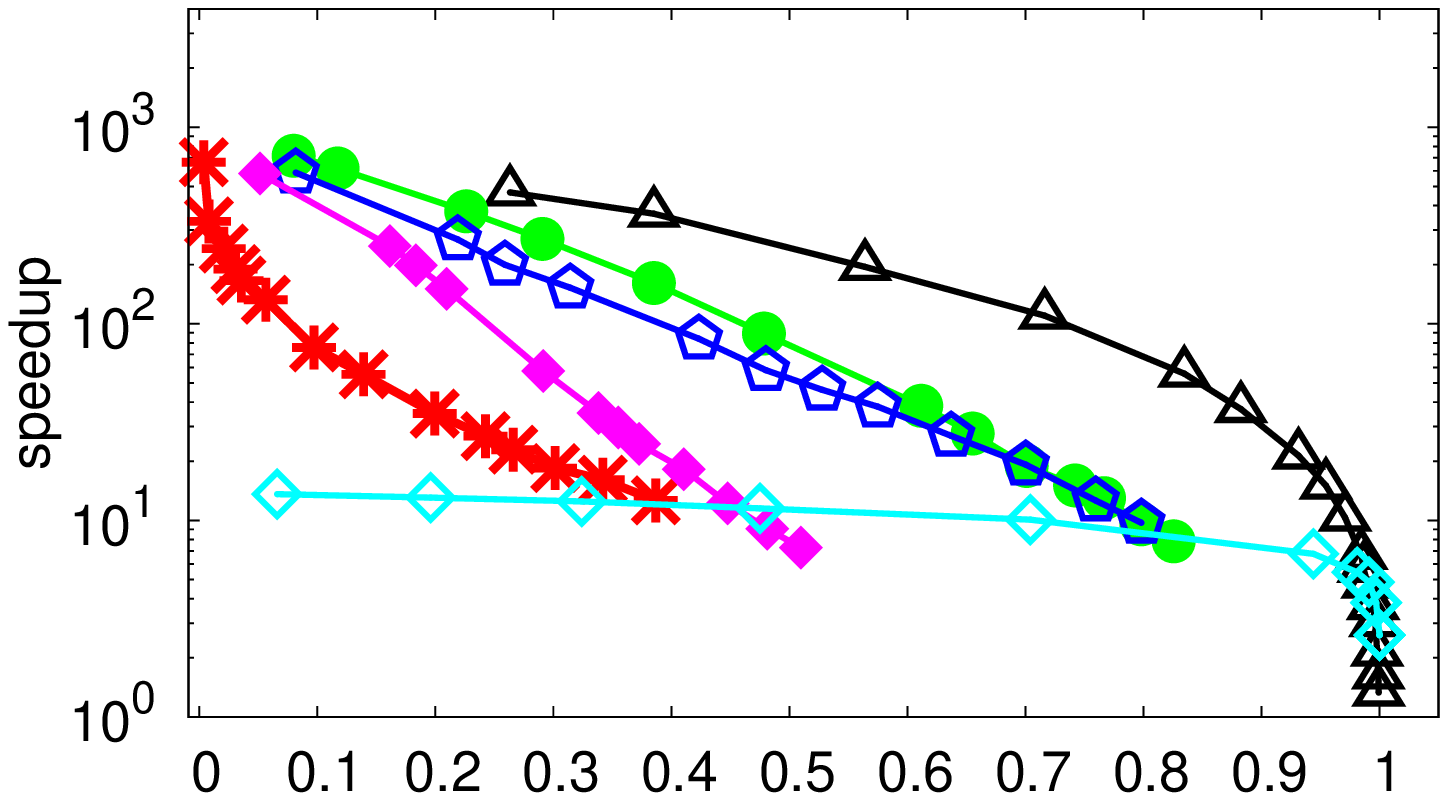}}
\centering
\subfigure[\small Audio ]{
\includegraphics[width=.234\linewidth]{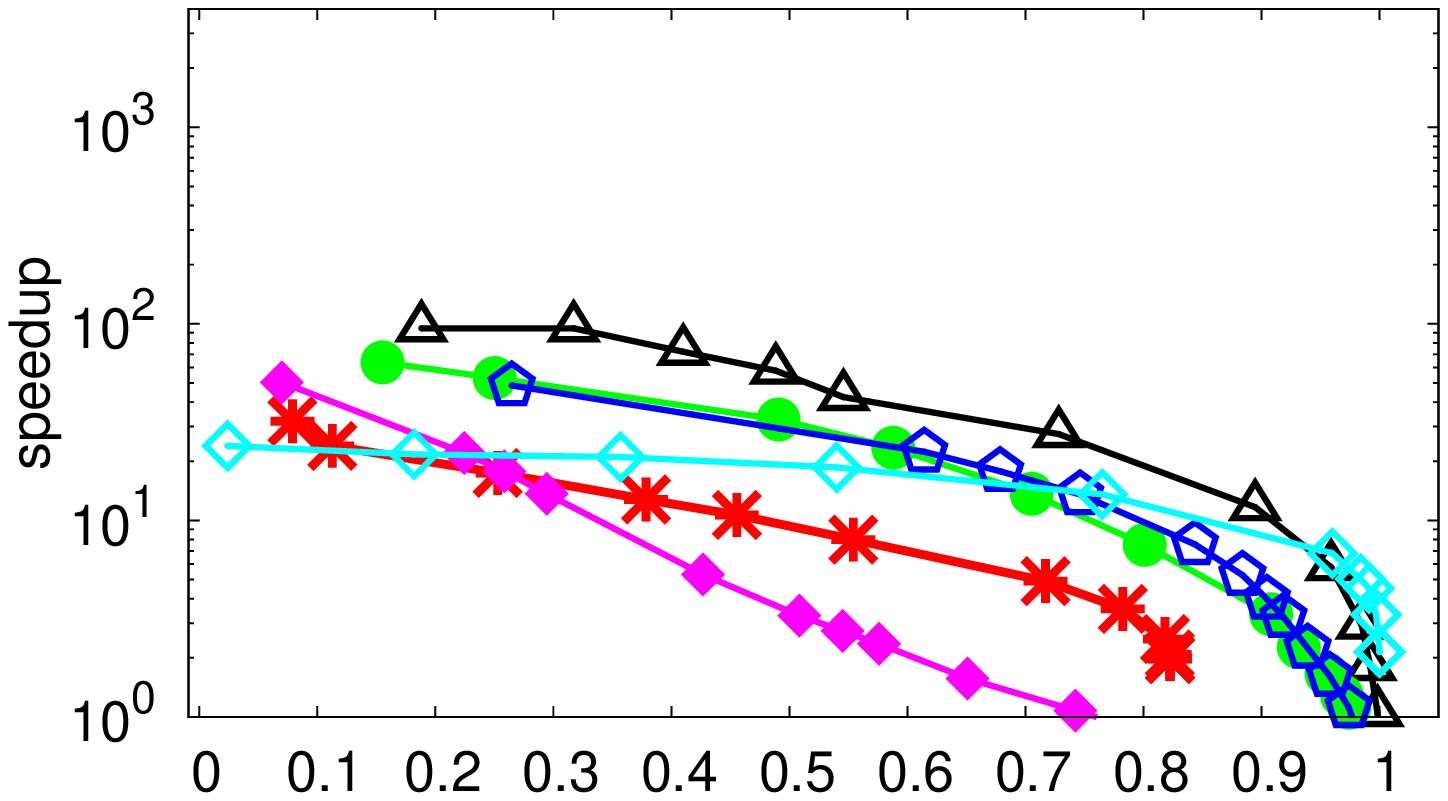}}
\centering
\subfigure[\small Cifa ]{
\includegraphics[width=.234\linewidth]{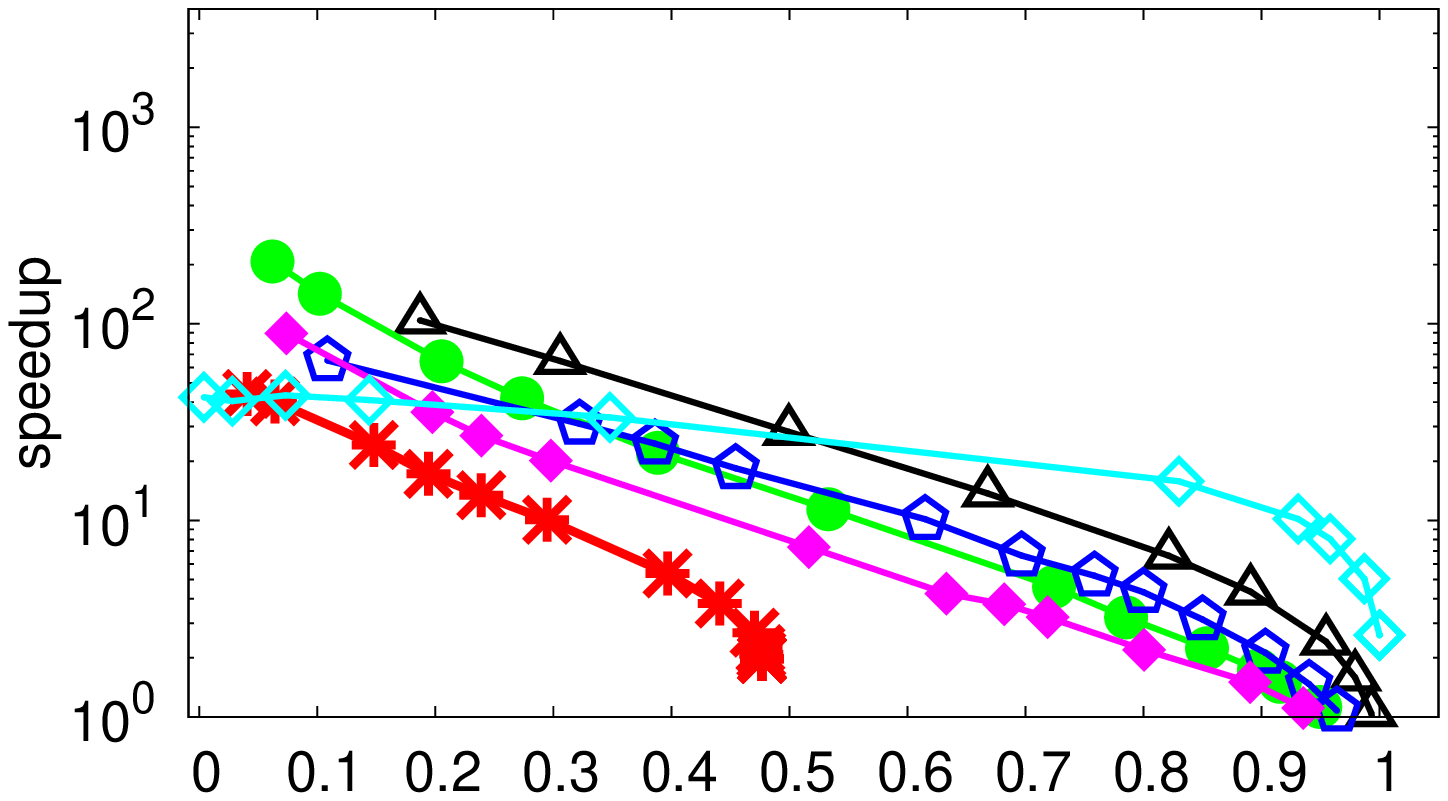}}
\end{minipage}
\vspace{-3mm}
\caption{\small Speedup vs Recall (Encoding)}
\vspace{-3mm}
\label{fig:exp_team_DSH}
\end{figure*}

We evaluate six encoding based algorithms including \AlgOPQ{}, \AlgNAPP{},
\textsf{SGH}, \textsf{AGH}, \textsf{NSH} and \textsf{SH}.
Figure~\ref{fig:exp_team_DSH} demonstrates that, of all methods, the search
performance of \AlgOPQ{} beats other algorithms by a
big margin on most of the datasets.

%

Table \ref{tab:exp_team_DSH_IS} reports the construction time (second) and index size (MB) of encoding-based methods. For most of datasets, NSH has the smallest index size, followed by AGH, SGH and OPQ. Selective Hashing has the largest index size because it requires long hash table to reduce the points number in a bucket and multiple hash tables to achieve high recall.

NSH and AGH spend relatively small time to build the index. The index time value of OPQ has a strong association with the length of the sub-codeword and dimension of the data point. Nevertheless, the index construction time of OPQ still turns out to be very competitive compared with other algorithms in the second round evaluation. Therefore, we choose \AlgOPQ as the representative of the encoding based methods.

\begin{table*}[htb]
\centering
\caption{index size and construction time (Encoding)}
\begin{tabular}{|c|l|l|l|l|l|l|l|l|l|l|l|l|}
\hline \multirow{2}{*}{Dataset} & \multicolumn{2}{c|}{SH} & \multicolumn{2}{c|}{OPQ} & \multicolumn{2}{c|}{NAPP} & \multicolumn{2}{c|}{NSH} & \multicolumn{2}{c|}{AGH} & \multicolumn{2}{c|}{SGH} \\
\cline{2-13} & size &  time &  size & time  &  size & time  &  size & time  &  size & time  &  size & time \\
\hline Sift   & 729 & 320    & 65   & 788.7    & 119   & 122     & 34.1 & 35.06     & 65   & 28     & 65.2  & 221  \\
\hline Trevi  & 185 & 132    & 158  & 10457    & 12    & 262     & 1.9  & 242.9     & 19.4 & 133    & 19.4  & 246  \\
\hline Mnist  & 165 & 34.5   & 11   & 152.4    & 8.3   & 35      & 2.4  & 96.4      & 4.6  & 26.9   & 6.9   & 358  \\
\hline Glove  & 850 & 375    & 43   & 697.7    & 143   & 119     & 77.3 & 105.9     & 78   & 20.6   & 41.3  & 89.4 \\
\hline Notre  & 325 & 107.2  & 17   & 138.4    & 40    & 39      & 16.5 & 28.2      & 11.9 & 5.3    & 22.3  & 206  \\
\hline Ben    & 792 & 352.3  & 68   & 844.6    & 131   & 134     & 37.7 & 38        & 38.2 & 16.8   & 71.9  & 447  \\
\hline Audio  & 155 & 18.1   & 9    & 45.8     & 6.4   & 8.4     & 1.8  & 24.8      & 3    & 8.1    & 4.6   & 360  \\
\hline Cifa   & 153 & 21.5   & 9    & 92.6     & 6     & 18.3    & 0.8  & 18.8      & 3.7  & 22     & 1.9   & 3    \\
\hline
\end{tabular}\\
\vspace{-2mm}
\label{tab:exp_team_DSH_IS}
\end{table*}

\subsubsection{Tree-based Space Partition Methods}
\label{subsubsec:exp_tree}

\begin{figure*}[tbh]
\centering
\begin{minipage}[b]{0.8\linewidth}
\centering
\includegraphics[width=1.0\linewidth]{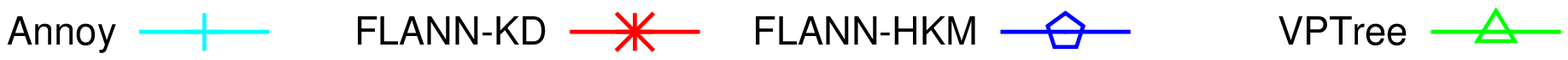}%
\vspace{-3mm}
\end{minipage}
\centering
\begin{minipage}[t]{1.0\linewidth}
\subfigure[\small sift ]{
\includegraphics[width=.234\linewidth]{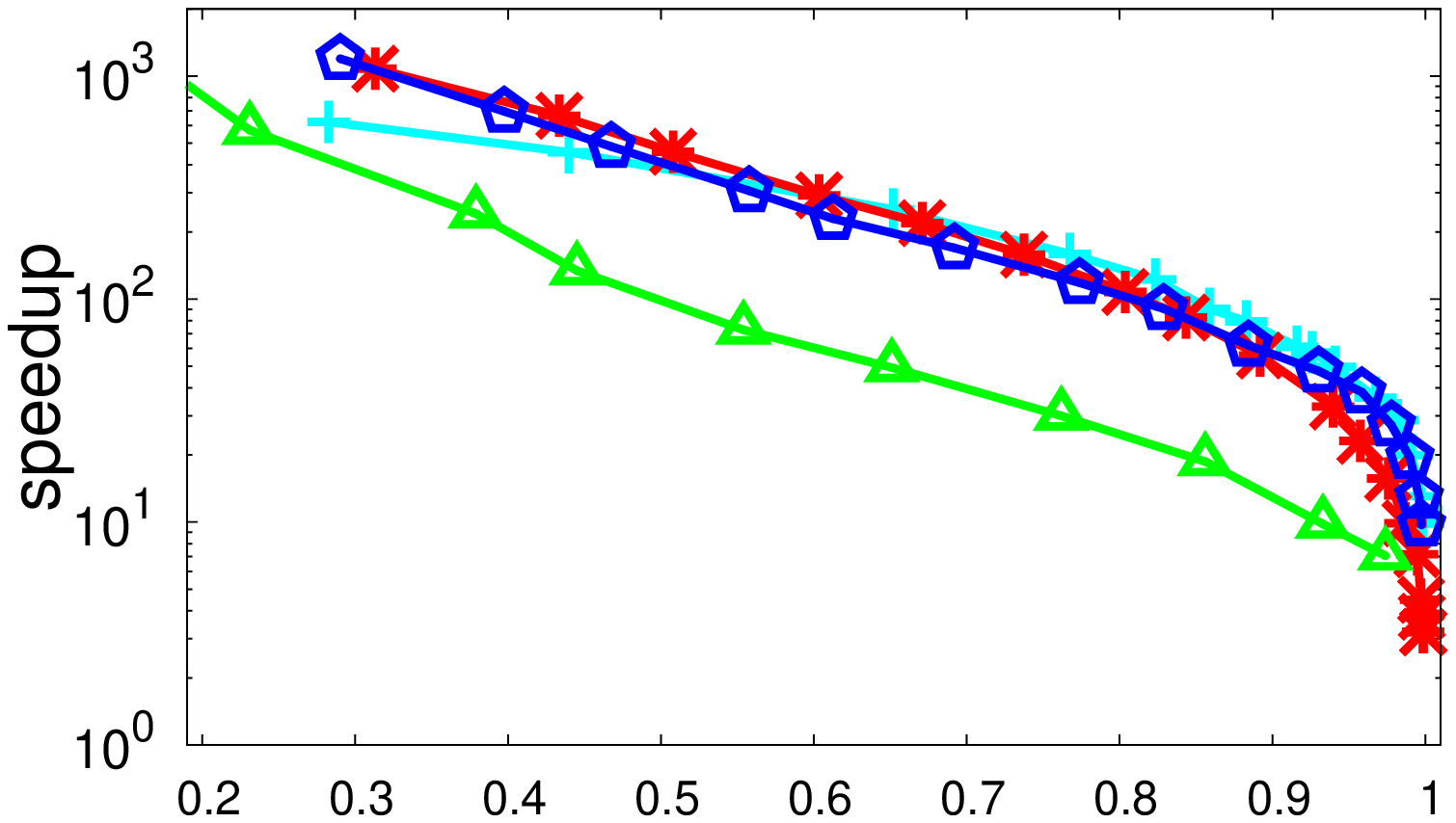}}
\centering
\subfigure[\small deep ]{
\includegraphics[width=.234\linewidth]{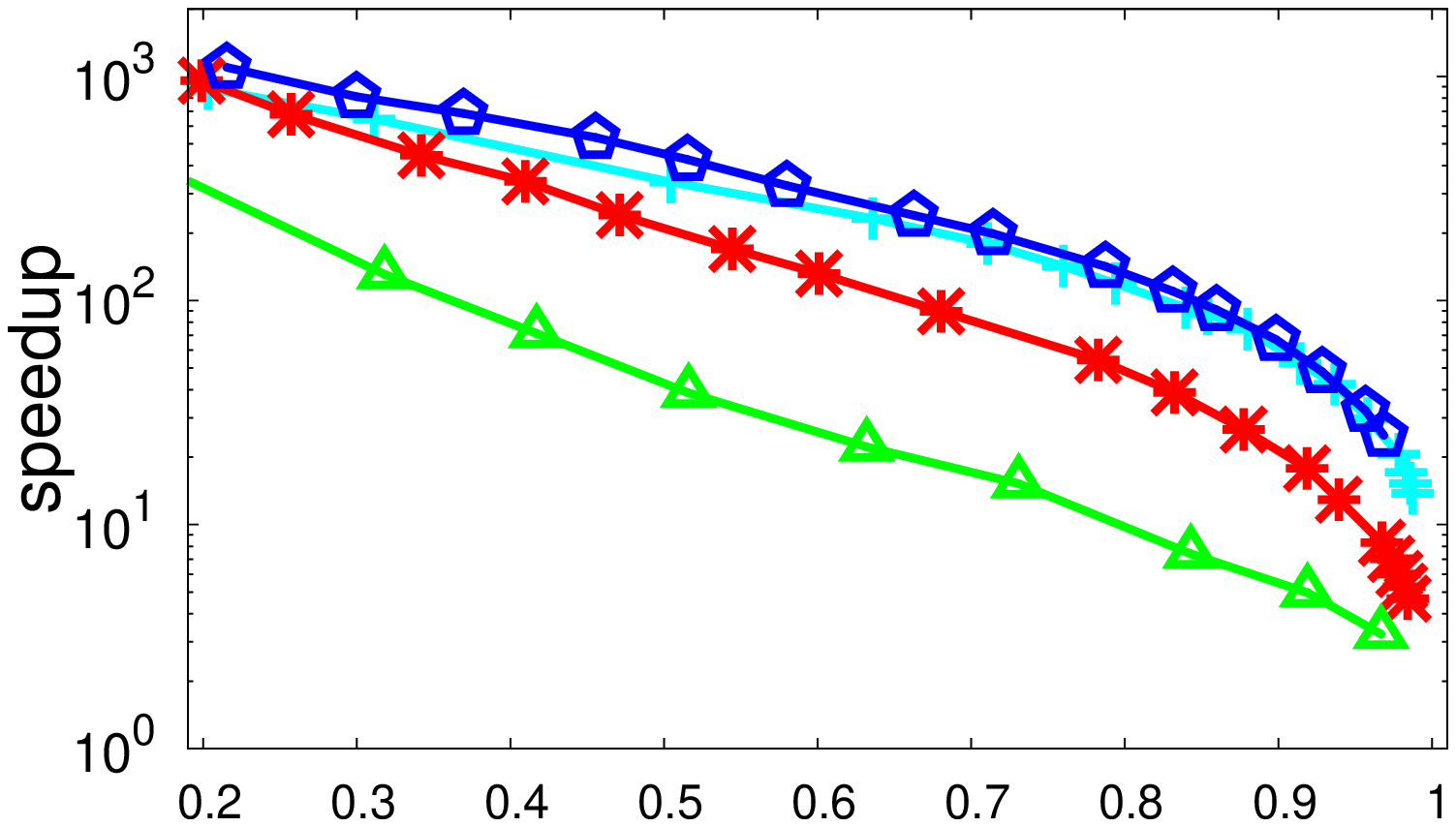}}
\centering
\subfigure[\small Enron ]{
\includegraphics[width=.234\linewidth]{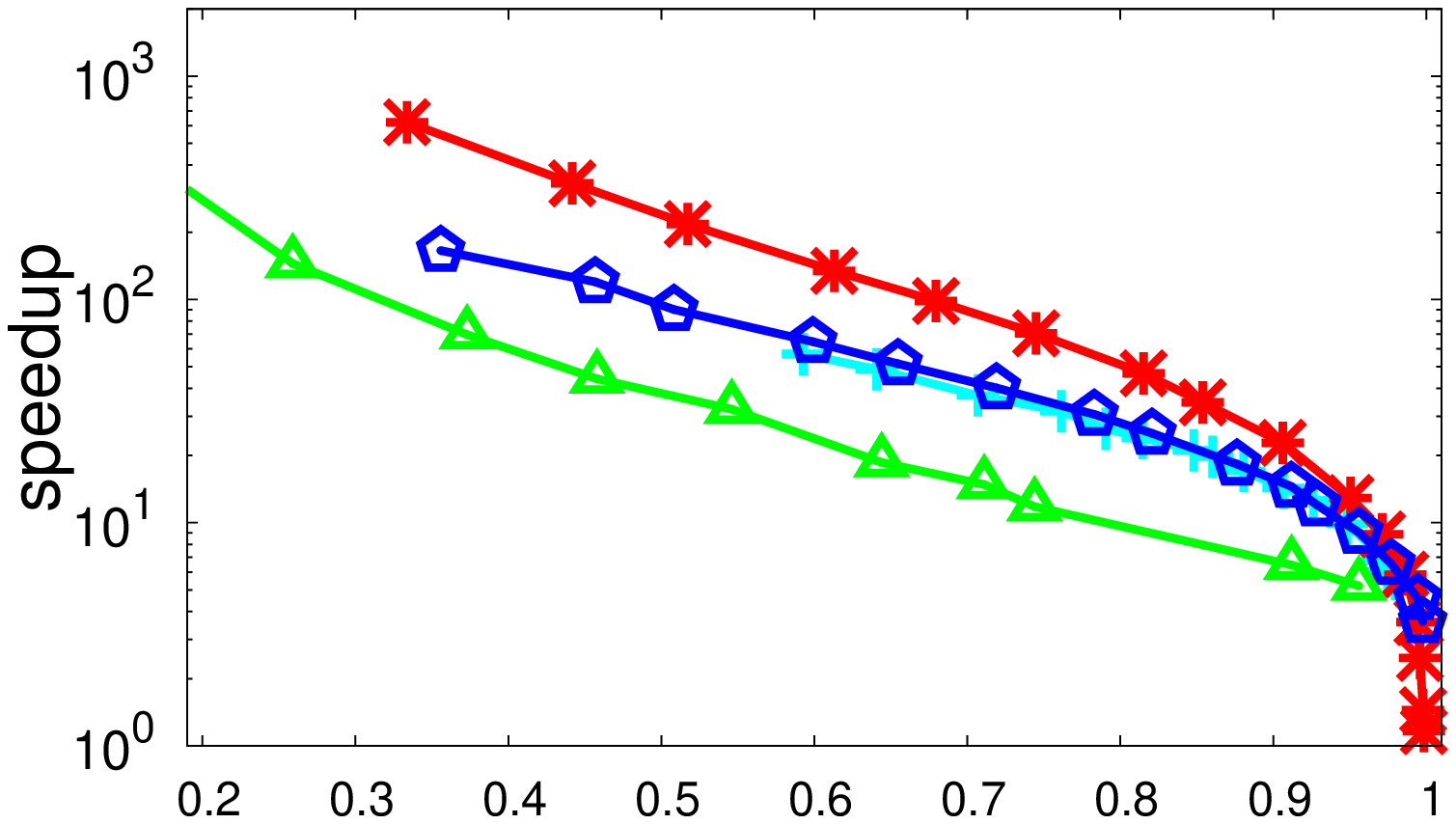}}
\centering
\subfigure[\small Gist ]{
\includegraphics[width=.234\linewidth]{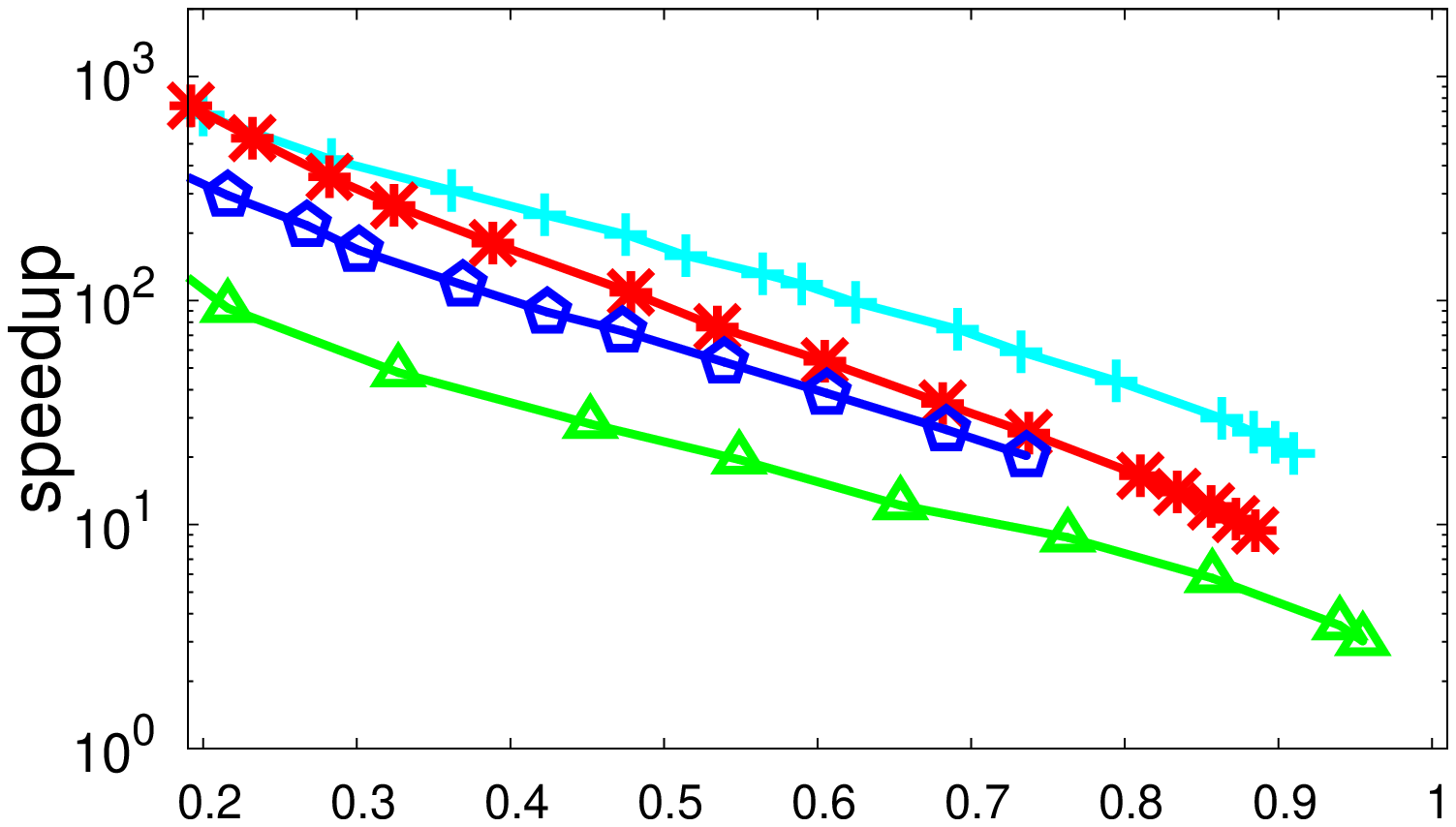}}
\end{minipage}
\vfill
\begin{minipage}[t]{1.0\linewidth}
\centering
\subfigure[\small UQ-V ]{
\includegraphics[width=.234\linewidth]{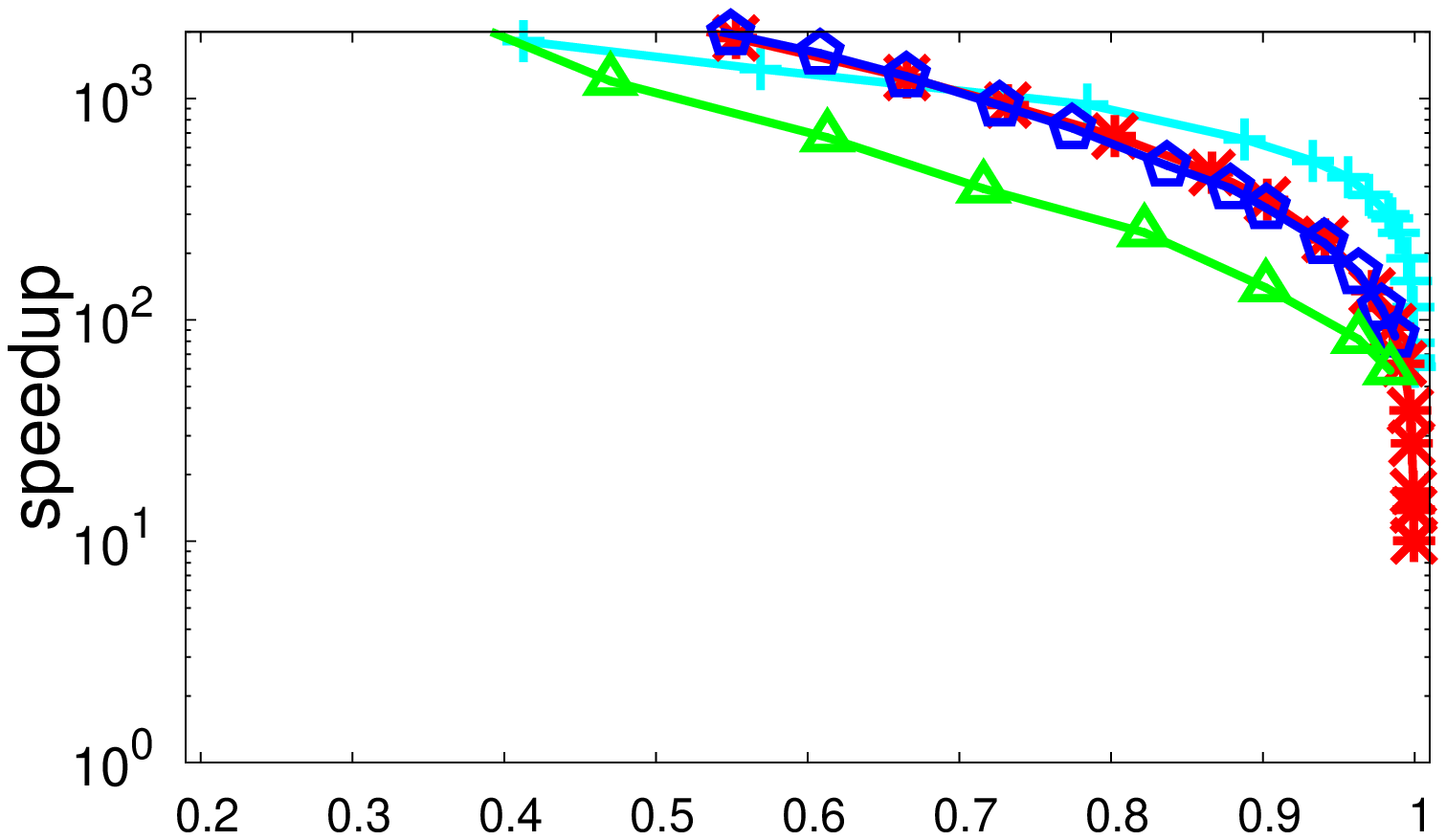}}
\centering
\subfigure[\small Sun ]{
\includegraphics[width=.234\linewidth]{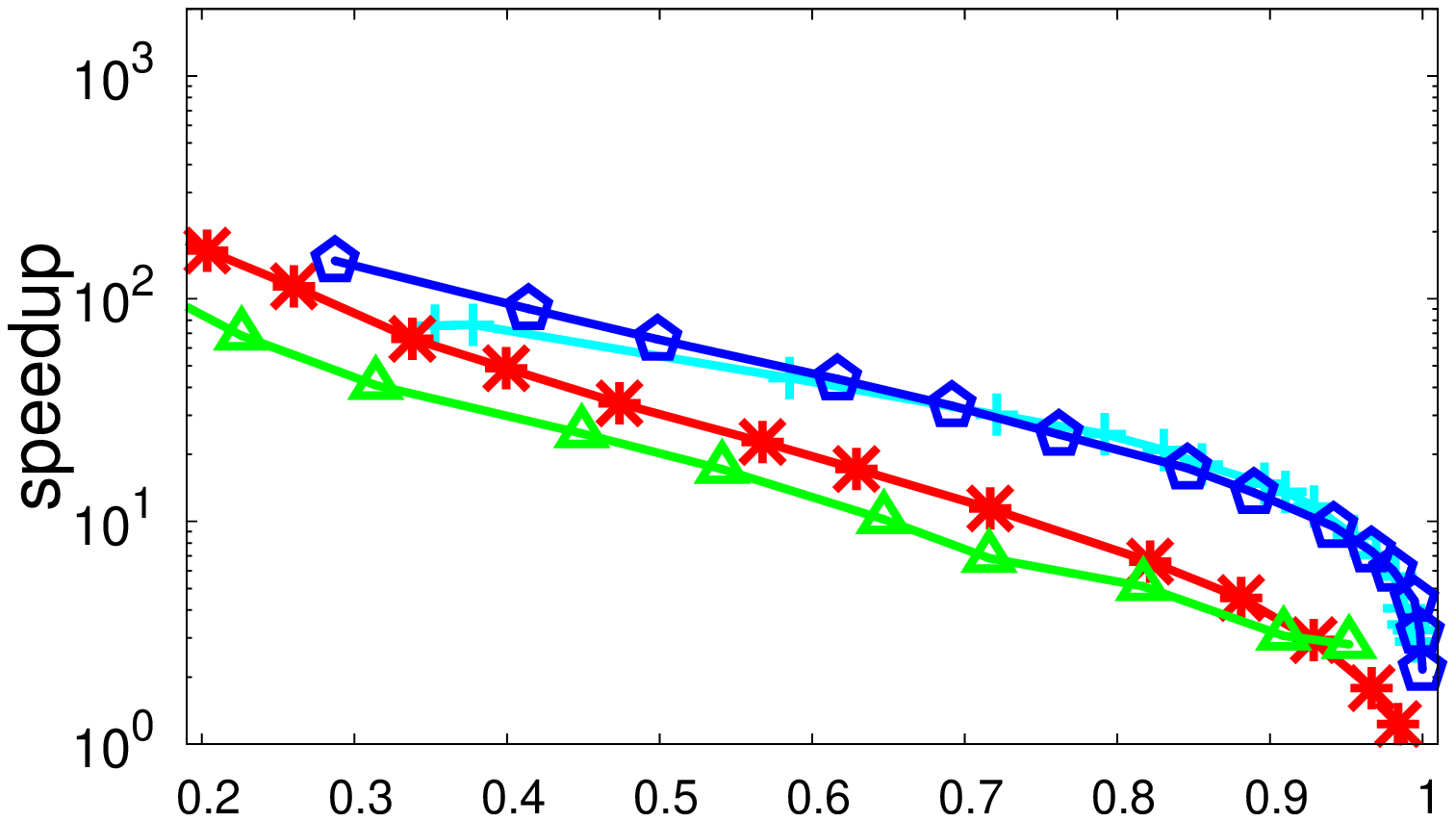}}
\centering
\subfigure[\small Audio ]{
\includegraphics[width=.234\linewidth]{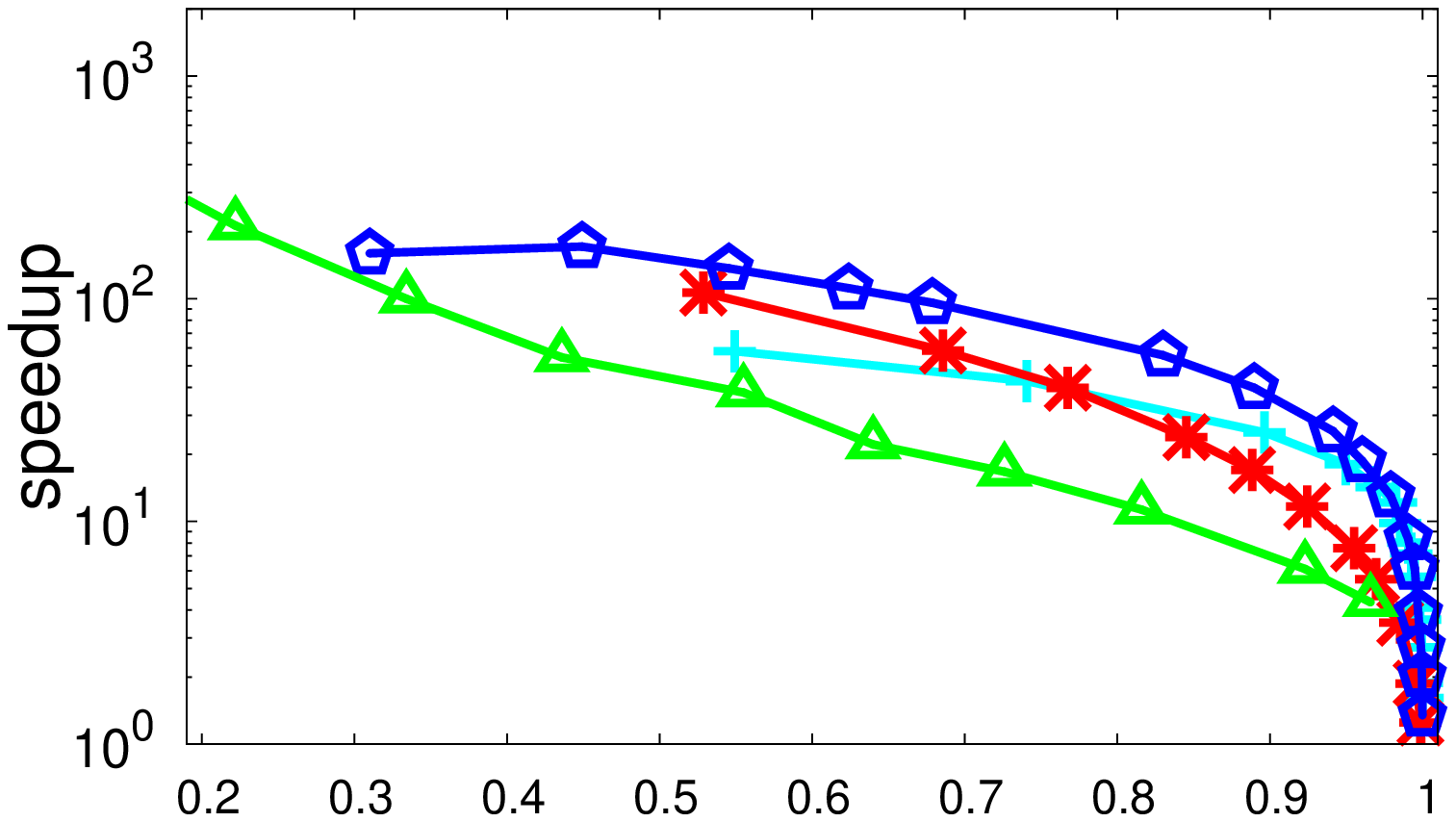}}
\centering
\subfigure[\small Cifar ]{
\includegraphics[width=.234\linewidth]{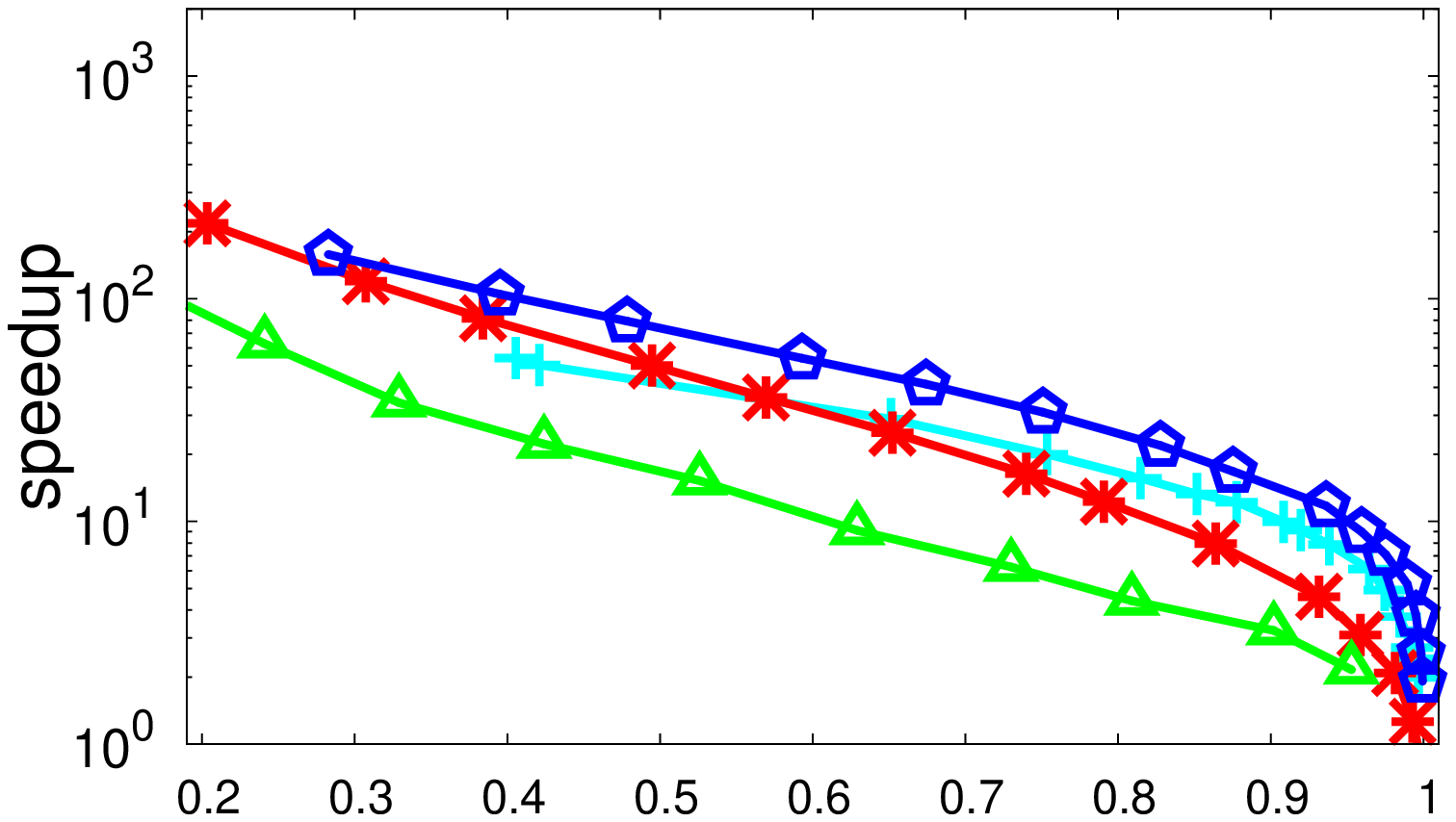}}
\end{minipage}
\caption{\small Speedup vs Recall (Tree-based)}
\vspace{-1mm}
\label{fig:exp_team_tree}
\end{figure*}

We evaluate three algorithms in this category: \Algflann{}, \Algannoy{} and \textsf{VP-tree}.
To better illustrate the performance of \Algflann{}, we report the performance of both randomized kd-trees
and hierarchical $k$-means tree, namely \Algflannkd{} and \Algflannhkm{}, respectively.
Note that among $20$ datasets deployed in the experiments, the randomized kd-trees method (\Algflannkd{})
is chosen by \Algflann{} in five datasets: \Dataenron{}, \Datatrevi{}, \Datauqv{}, \Databann{} and \Datagauss{}.
The linear scan is used in the hardest dataset: \Datarand{}, and the hierarchical $k$-means tree (\Algflannhkm) is
employed in the remaining $14$ datasets.
Figure~\ref{fig:exp_team_tree} shows that \Algannoy{}, \Algflannhkm{} and \Algflannkd{} can obtain the highest performance in different datasets.

Note that the index size of \Algvptree{} is the memory size used during the indexing. From table \ref{tab:exp_team_tree}, we can see VP-Tree almost has the largest index time, which is because \Algvptree{} spend long time to tune the parameters automatically. While the search performance of \Algvptree{} is not competitive compared to \Algflann and \Algannoy under all settings, it is excluded from the next round evaluation.

\begin{table*}[htb]
\centering
\caption{index size and construction time (Tree-based)}
\begin{tabular}{|c|l|l|l|l|l|l|l|l|}
\hline \multirow{2}{*}{Dataset} & \multicolumn{2}{c|}{Annoy} & \multicolumn{2}{c|}{Flann-KD} & \multicolumn{2}{c|}{Flann-HKM} & \multicolumn{2}{c|}{VP-tree} \\
\cline{2-9} & size(MB) &  time(s) &  size(MB) & time(s)  &  size(MB) & time(s)  &  size(MB) & time(s)  \\
\hline Sift  & 572  & 144.6     & 274.0   &45.5     &230.0  &27.4      &573.1  & 131    \\
\hline Deep  & 571  & 224       & 550.0   &148.6    &74.0   &1748.1    &1063   & 474     \\
\hline Enron & 56   & 51.1      & 27.0    &30.0     &74.0   &412.9     &500.7  & 954      \\
\hline Gist  & 563  & 544.9     & 540.0   &462.9    &1222.0 &775.5     &4177.6 & 1189     \\
\hline UQ-V  & 1753 & 801.8     & 1670.0  &462.0    &99.0   &3109.2    &3195.5 & 15.9     \\
\hline Sun   & 46   & 18.4      & 44.0    &20.6     &58.0   &43.0      &161    & 124       \\
\hline Audio & 31   & 7.08      & 15.0    &3.0      &20.0   &10.6      &45     & 151       \\
\hline Cifa  & 28   & 10.8      & 14.0    &7.0      &9.0    &23.7      &101    & 131      \\
\hline
\end{tabular}\\
\vspace{-2mm}
\label{tab:exp_team_tree}
\end{table*}

\subsubsection{Neighborhood-based Methods}
\label{subsubsec:exp_nb}

\begin{figure*}[tbh]
\centering
\begin{minipage}[b]{0.8\linewidth}
\centering
\includegraphics[width=0.8\linewidth]{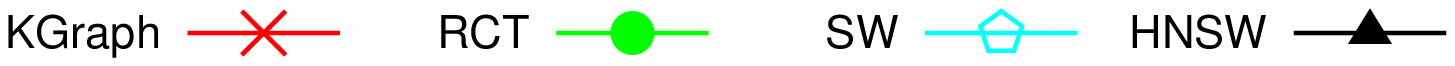}%
\vspace{-3mm}
\end{minipage}
\centering
\begin{minipage}[t]{1.0\linewidth}
\subfigure[\small Sift ]{
\includegraphics[width=.234\linewidth]{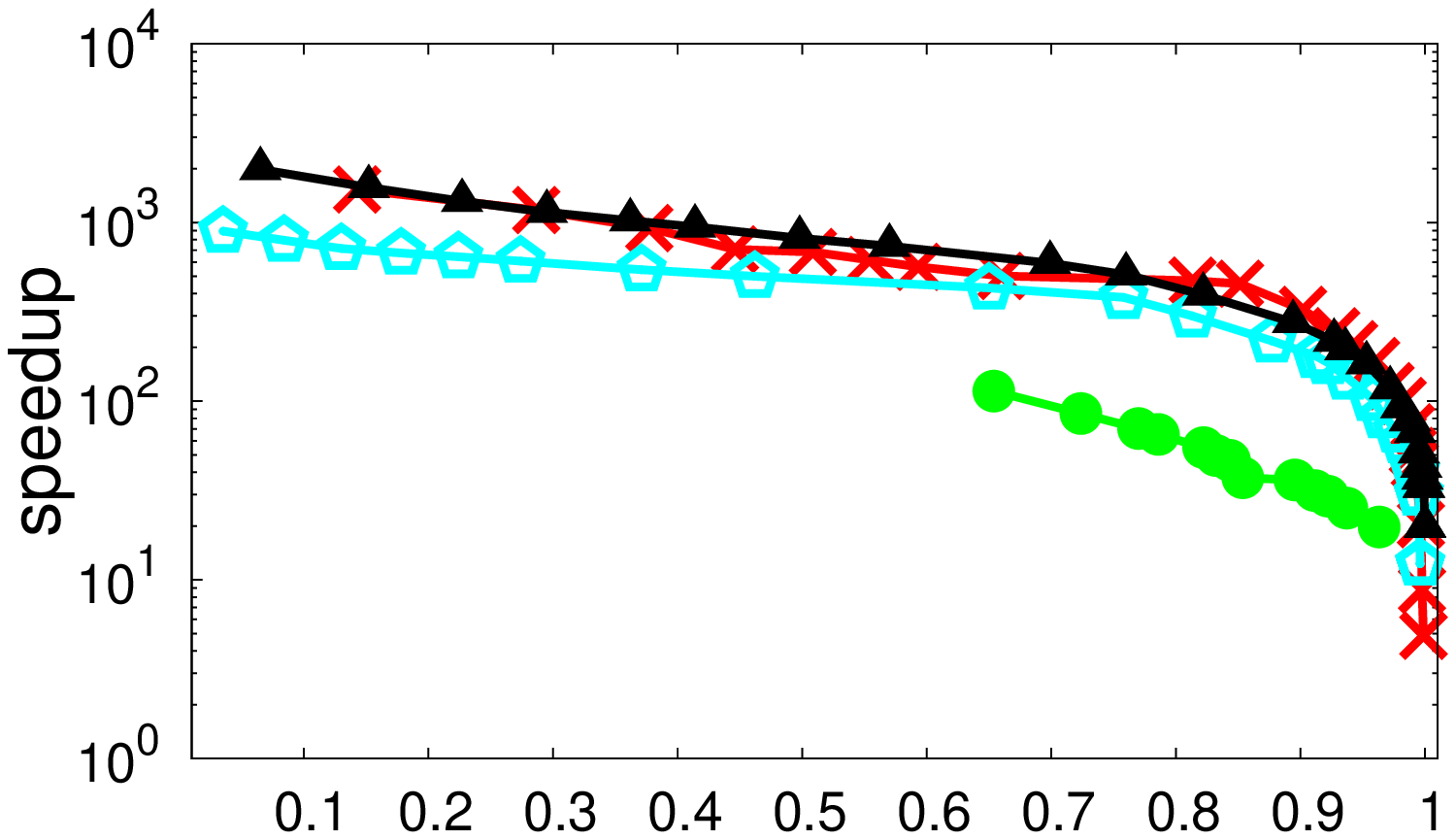}}
\centering
\subfigure[\small Nusw ]{
\includegraphics[width=.234\linewidth]{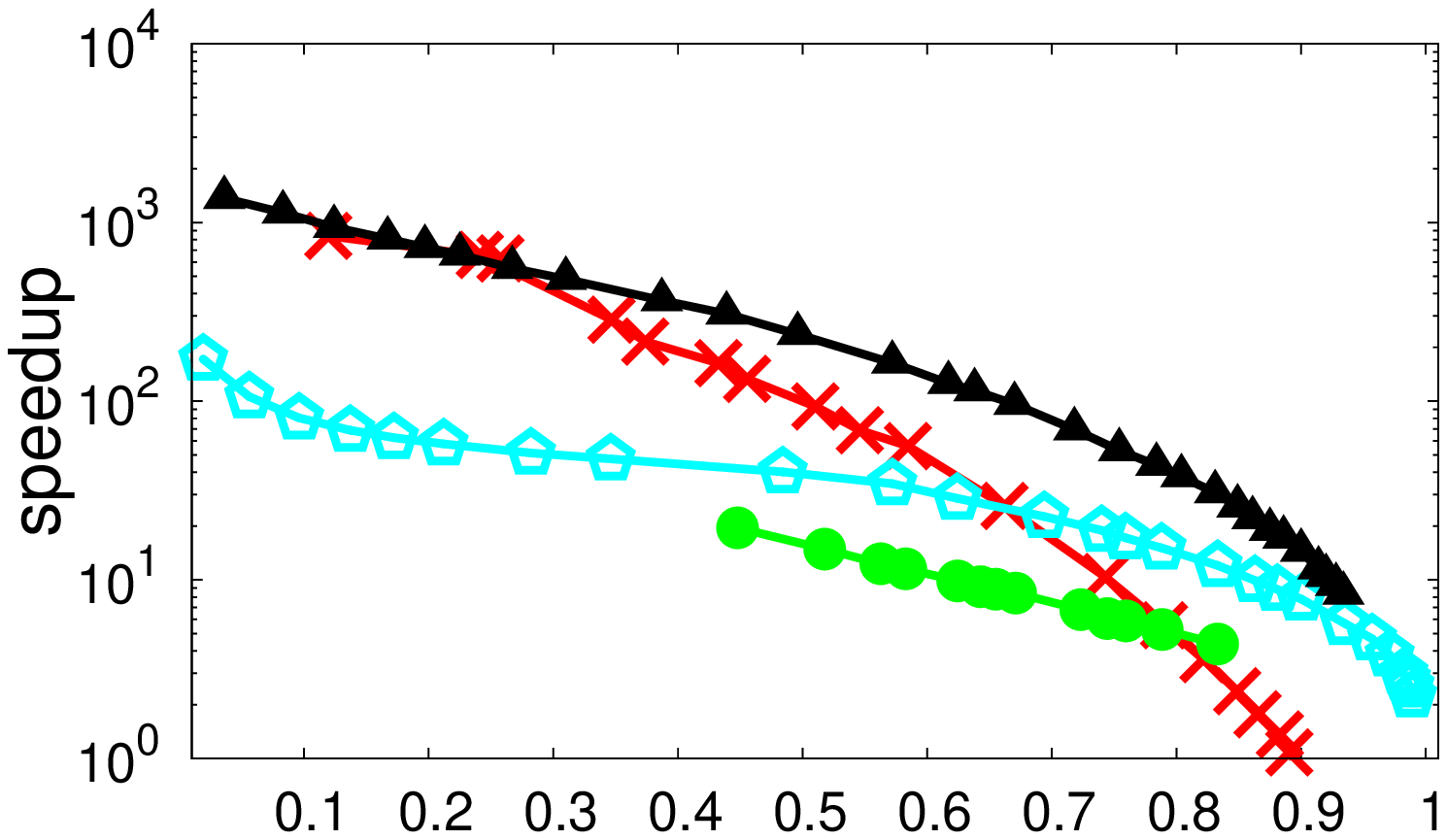}}
\centering
\subfigure[\small Audio ]{
\includegraphics[width=.234\linewidth]{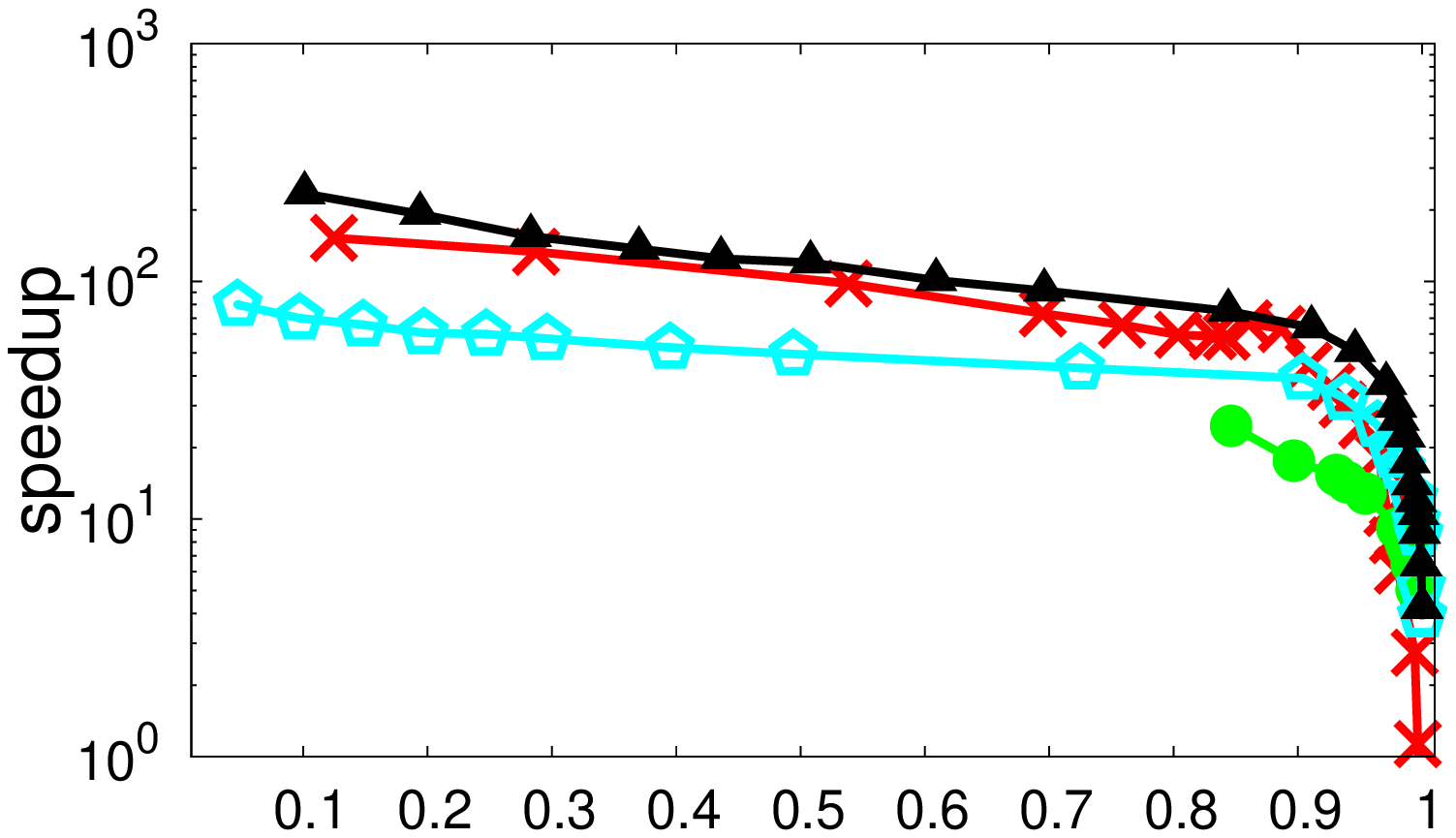}}
\centering
\subfigure[\small Gist ]{
\includegraphics[width=.234\linewidth]{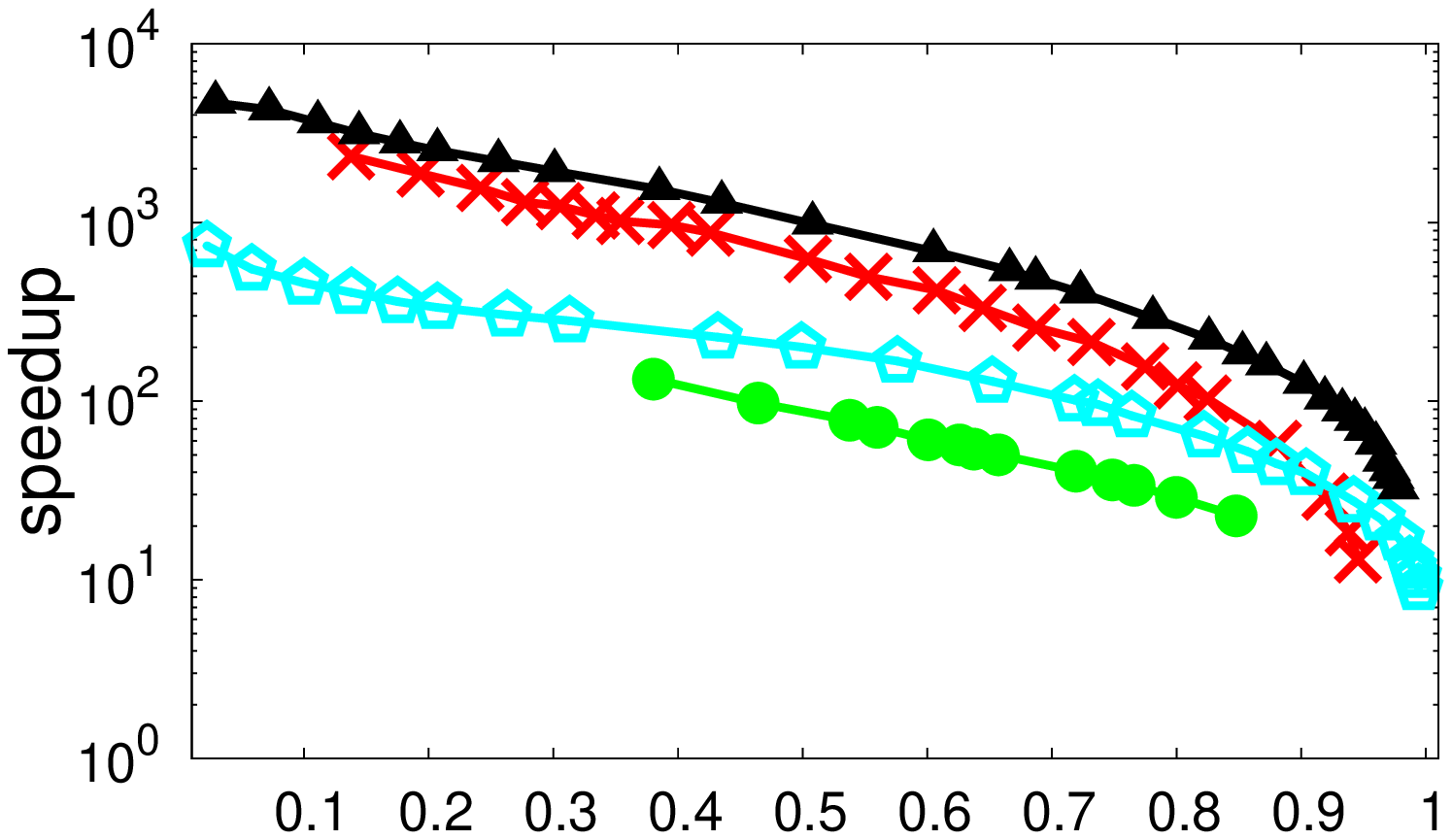}}
\end{minipage}
\vfill
\begin{minipage}[t]{1.0\linewidth}
\centering
\subfigure[\small Deep ]{
\includegraphics[width=.234\linewidth]{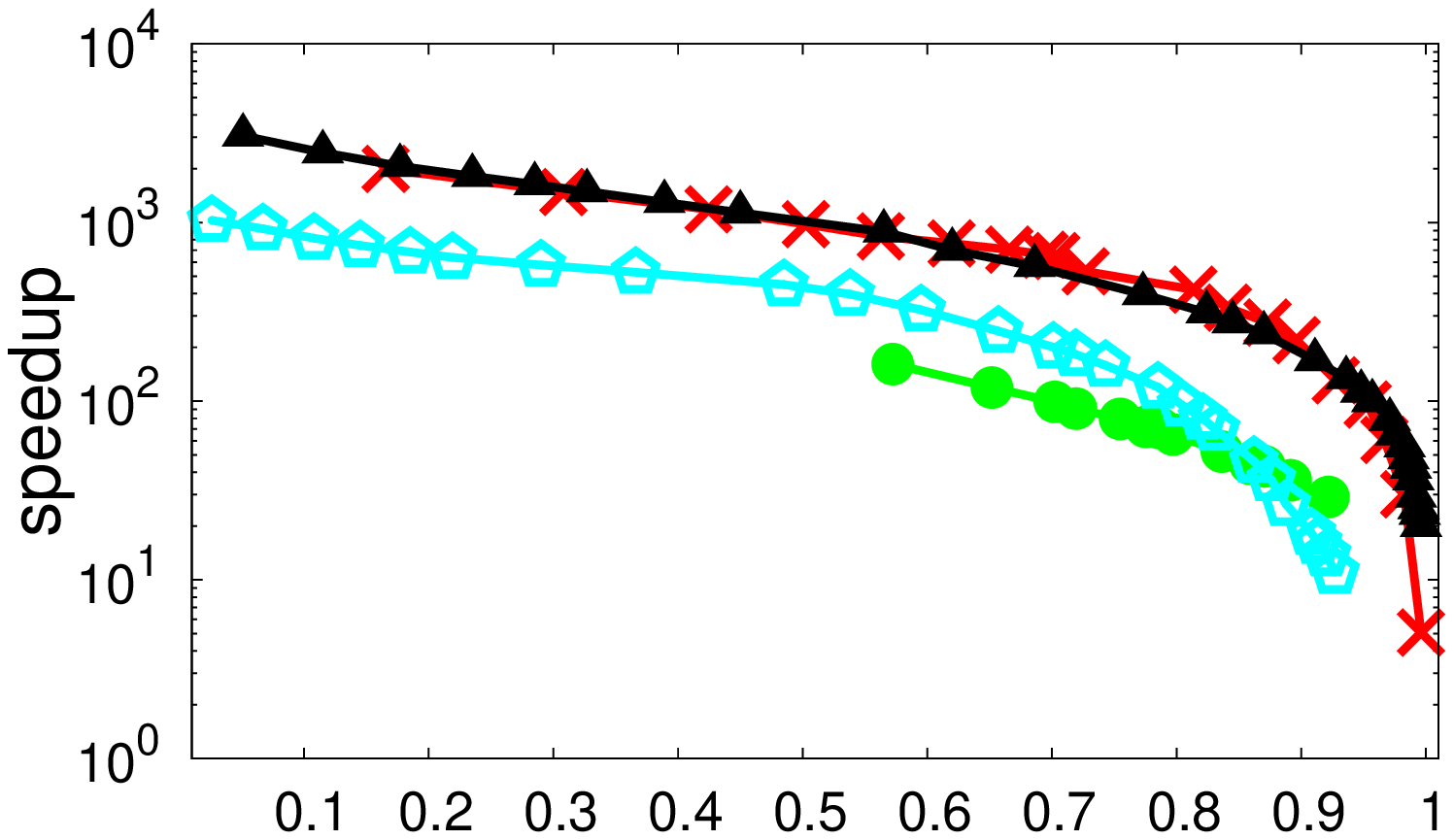}}
\centering
\subfigure[\small Cifar ]{
\includegraphics[width=.234\linewidth]{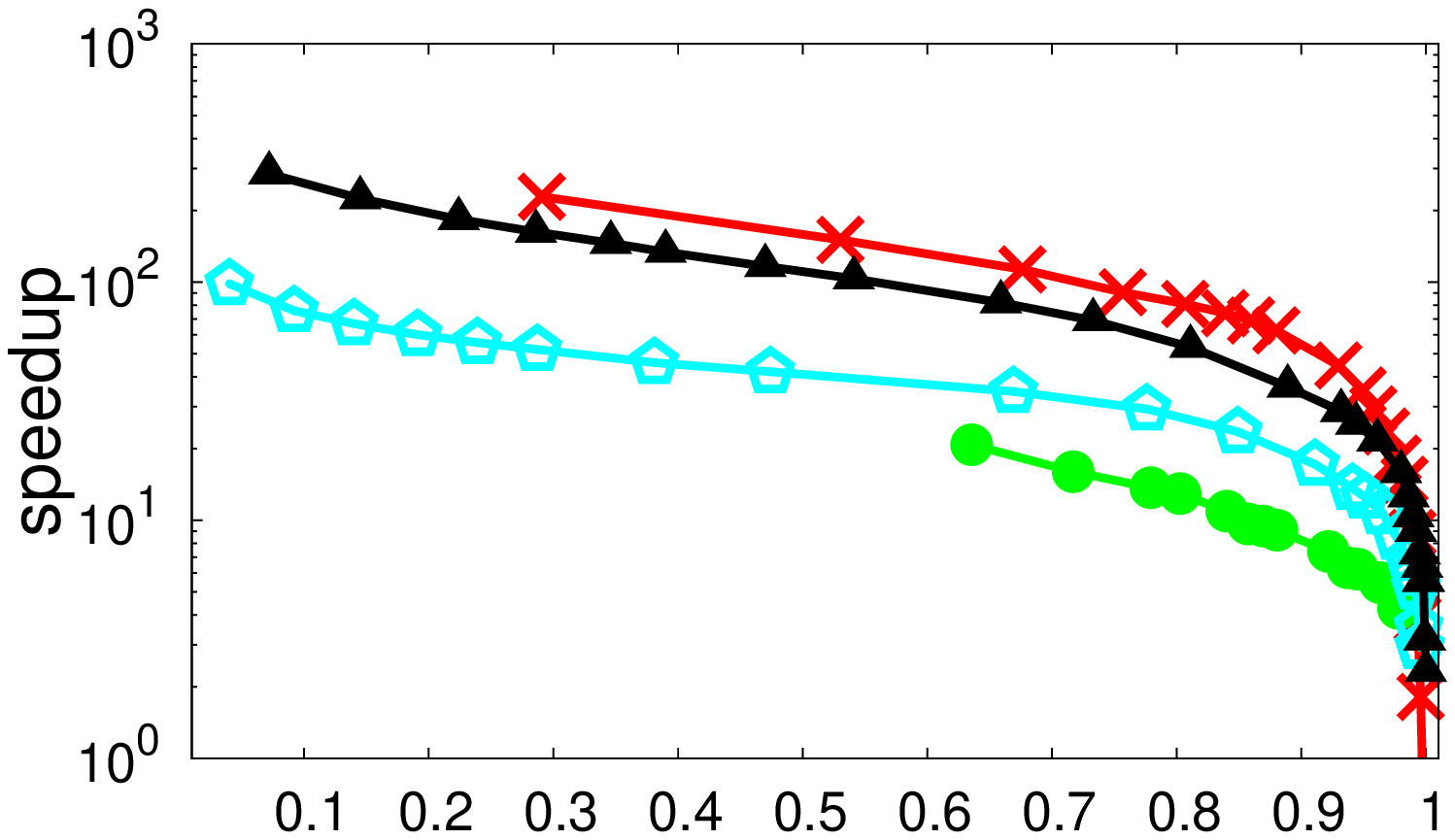}}
\centering
\subfigure[\small Trevi ]{
\includegraphics[width=.234\linewidth]{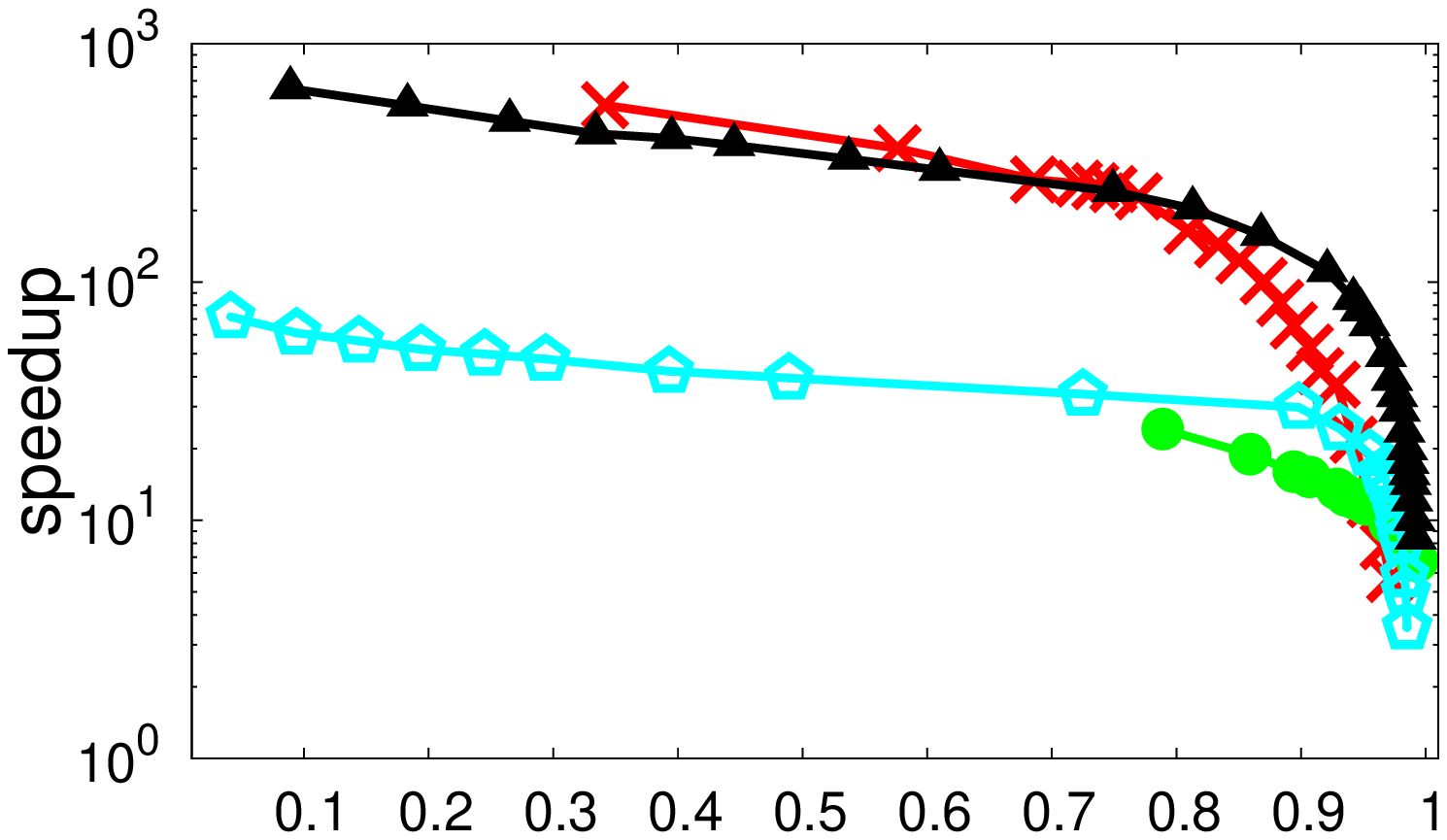}}
\centering
\subfigure[\small Ben ]{
\includegraphics[width=.234\linewidth]{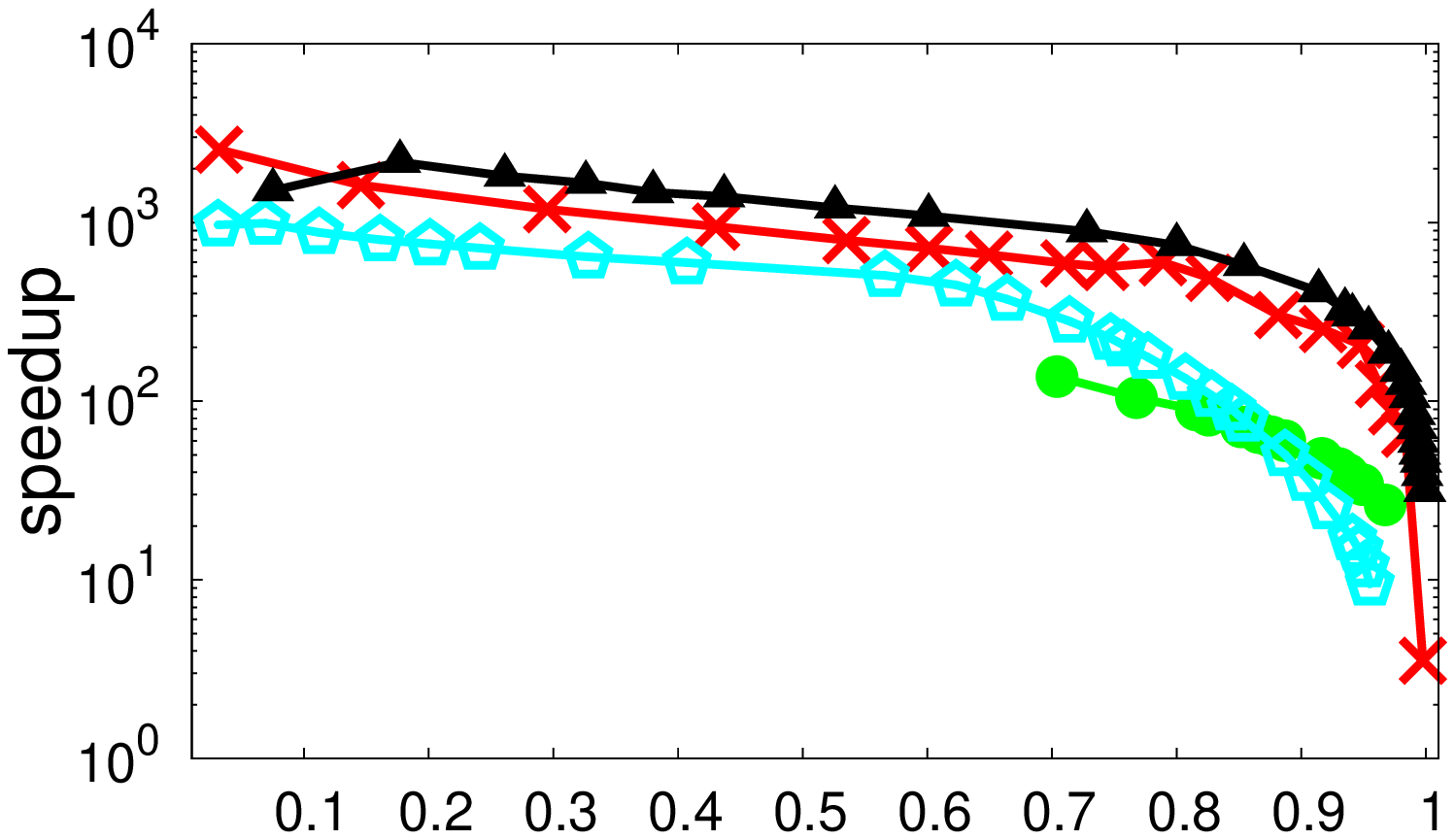}}
\end{minipage}
\vspace{-3mm}
\caption{\small Speedup vs Recall (Neighborhood-based) }
\vspace{-1mm}
\label{fig:exp_team_NB}
\end{figure*}

In the category of neighborhood-based methods, we evaluate four existing techniques: \Algkgraph{}, \Algsw{}, \Alghnsw{} and \AlgRCT{}.
Figure~\ref{fig:exp_team_NB} shows that the search performance of \Algkgraph{} and \Alghnsw{} substantially outperforms
that of other two algorithms on most of the datasets.

\AlgRCT{} has the smallest index size and the construction time of \Algkgraph{} and \Alghnsw{} are relatively large.
Due to the outstanding search performance of \Algkgraph{} and \Alghnsw{}, we choose them as the representatives of the neighborhood-based methods.
Note that we delay the comparison of \Algdpg{} to the second round evaluation.

\begin{table*}[htb]
\centering
\caption{index size and construction time (Neighborhood-based)}
\begin{tabular}{|c|l|l|l|l|l|l|l|l|}
\hline \multirow{2}{*}{Dataset} & \multicolumn{2}{c|}{HNSW} & \multicolumn{2}{c|}{KGraph} & \multicolumn{2}{c|}{SW} & \multicolumn{2}{c|}{RCT}\\
\cline{2-9}   & size(MB) &  time(s) & size(MB) &  time(s) &  size(MB) & time(s)  &  size(MB) & time(s) \\
\hline Sift   & 589   & 1290               & 160 & 815.9        & 236   & 525          & 50   & 483     \\
\hline Nusw   & 541   & 701                & 44  & 913          & 64    & 605          & 22.1 & 813.9   \\
\hline Audio  & 45    & 36.9               & 9   & 38.2         & 13    & 8.1          & 2.4  & 11.8    \\
\hline Gist   & 3701  & 5090               & 158 & 6761.3       & 233   & 2994         & 50.3 & 3997     \\
\hline Deep   & 1081  & 1800               & 161 & 1527         & 236.5 & 372.1        & 51.2 & 815     \\
\hline Cifar  & 103   & 99.8               & 8   & 97.5         & 5.9   & 25.6         & 2.3  & 28      \\
\hline Trevi  & 1572  & 1292               & 16  & 1800.8       & 12.2  & 1003         & 4.5  & 649      \\
\hline Ben    & 651   & 1062               & 176 & 863          & 260   & 225          & 57   & 482      \\
\hline
\end{tabular}\\
\vspace{-2mm}
\label{tab:exp_team_DI}
\end{table*}


\begin{figure*}[tbh]
\centering
\includegraphics[width=1.0\linewidth]{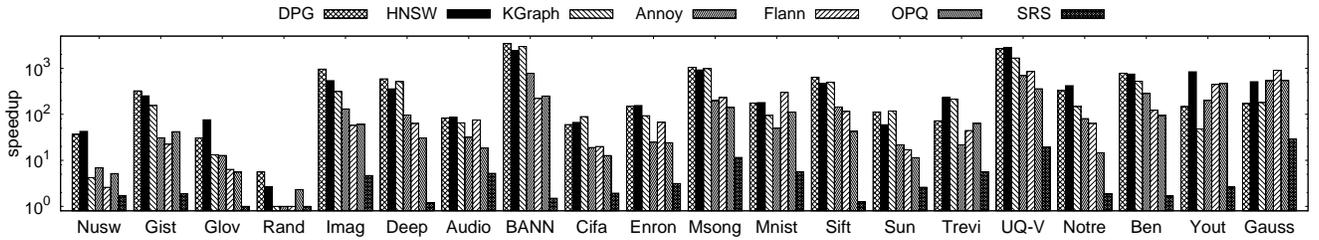}
\caption{\small Speedup with Recall of $0.8$}
\vspace{-3mm}
\label{fig:exp_bar_speedup}
\end{figure*}

\begin{figure*}[tbh]
\centering
\includegraphics[width=1.0\linewidth]{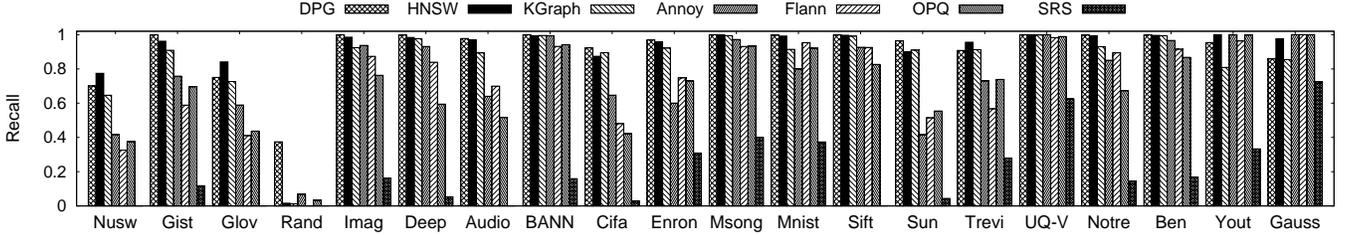}
\caption{\small Recall with Speedup of $50$}
\vspace{-3mm}
\label{fig:exp_bar_recall}
\end{figure*}
\begin{figure*}[htb]
\centering%
\begin{minipage}[b]{0.8\linewidth}
\centering
\includegraphics[width=1.0\linewidth]{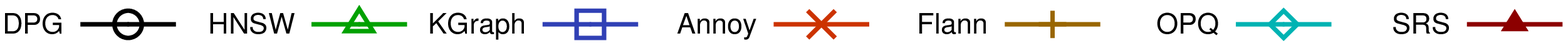}%
\vspace{-3mm}
\end{minipage}
\begin{minipage}[t]{1.0\linewidth}
\centering
      \subfigure[\small Nusw]{
      \label{fig:exp_recall_nusw} 
      \includegraphics[width=0.234\linewidth]{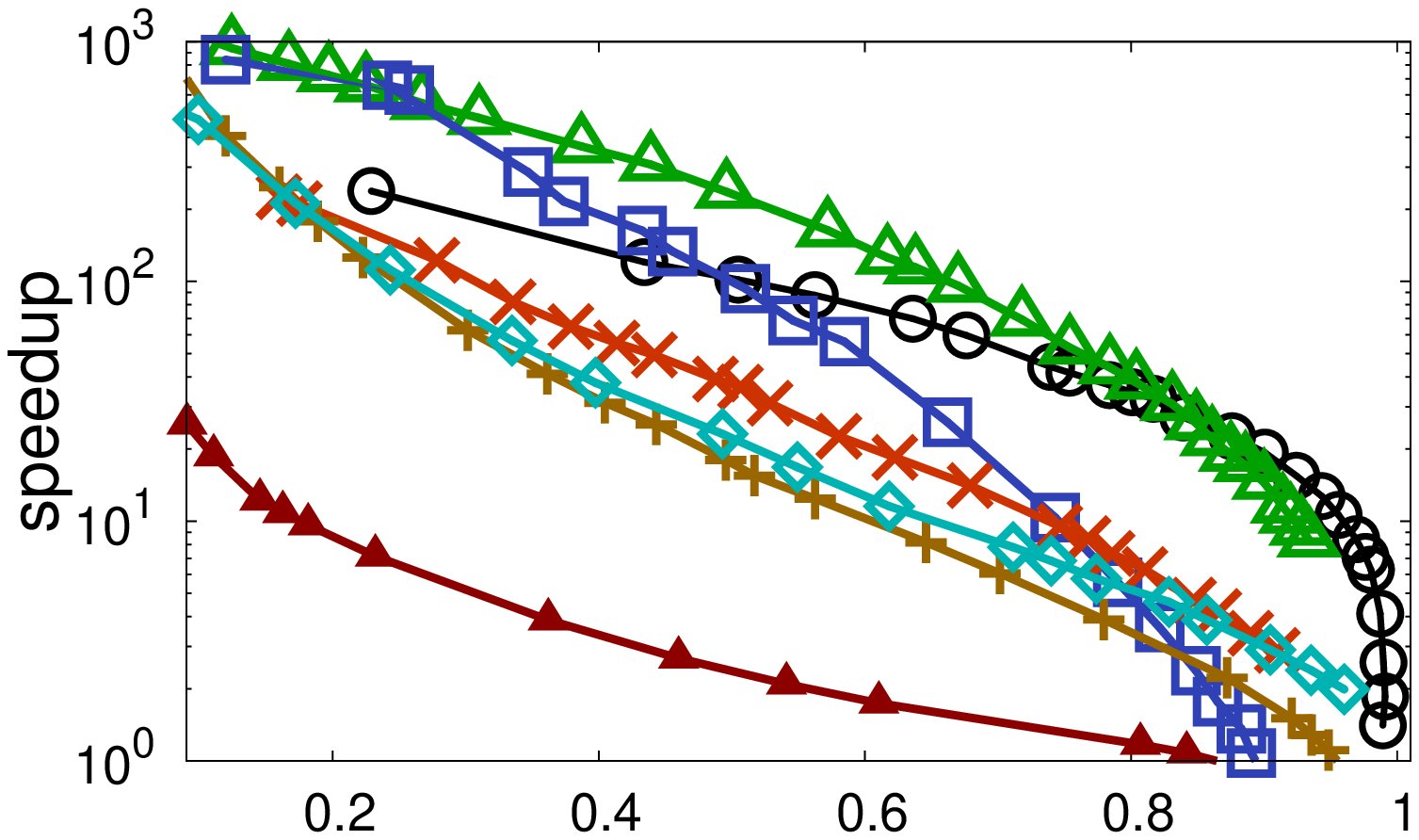}}
      \subfigure[\small Gist]{
      \label{fig:exp_recall_gist} 
      \includegraphics[width=0.234\linewidth]{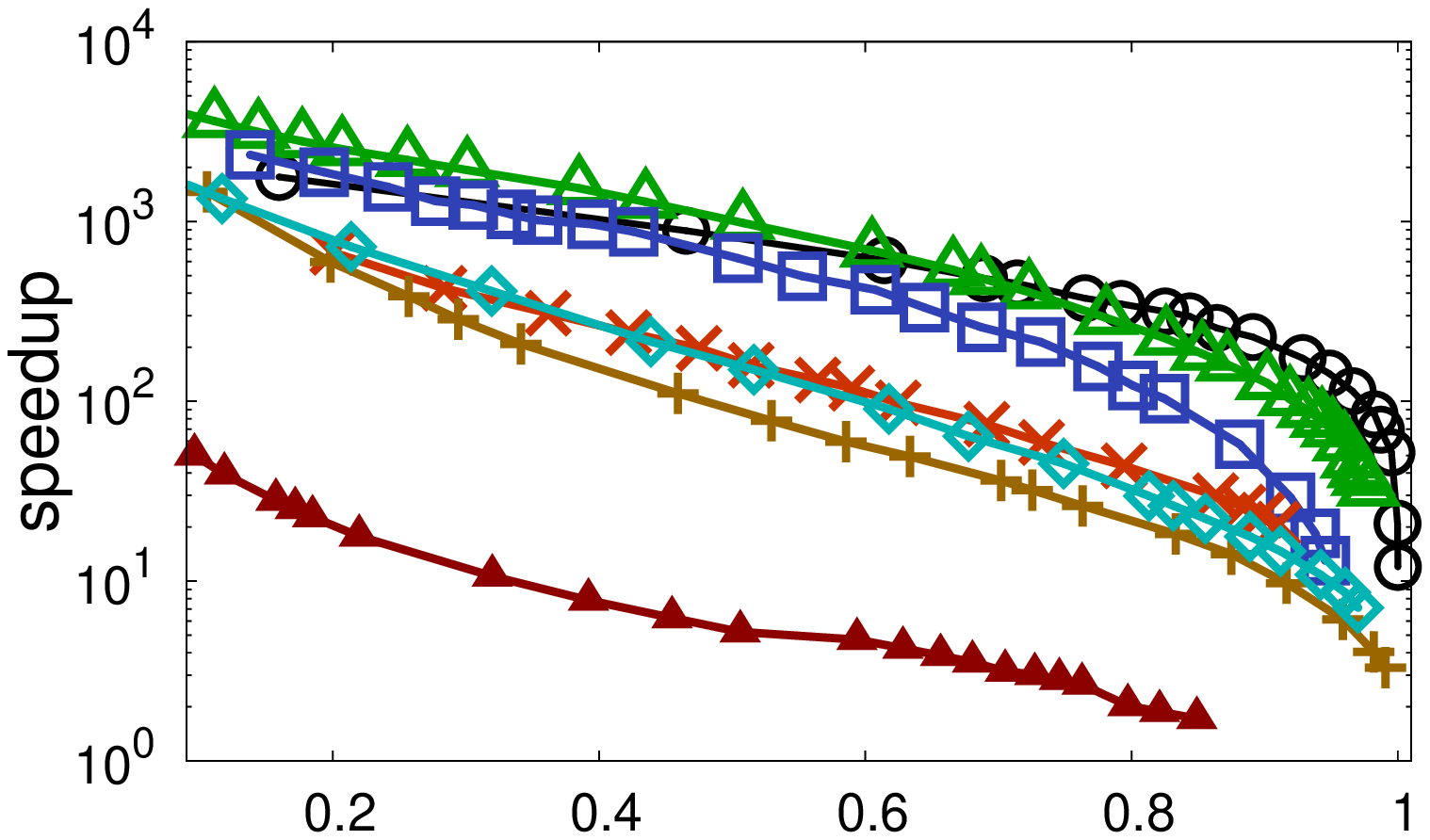}}
      \subfigure[\small Glov ]{
      \label{fig:exp_recall_glov} 
      \includegraphics[width=0.234\linewidth]{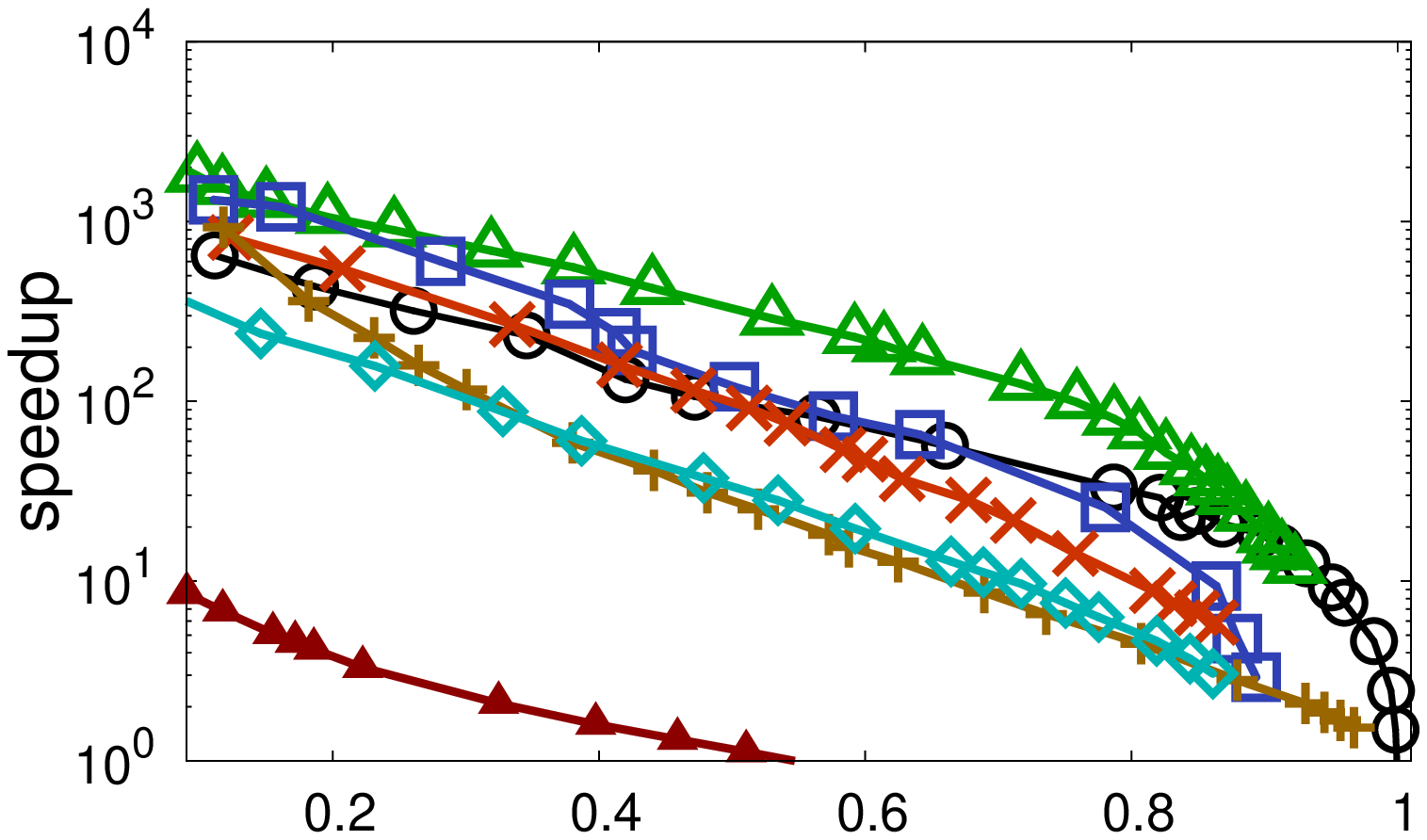}}
      \subfigure[\small Rand]{
      \label{fig:exp_recall_rand} 
      \includegraphics[width=0.234\linewidth]{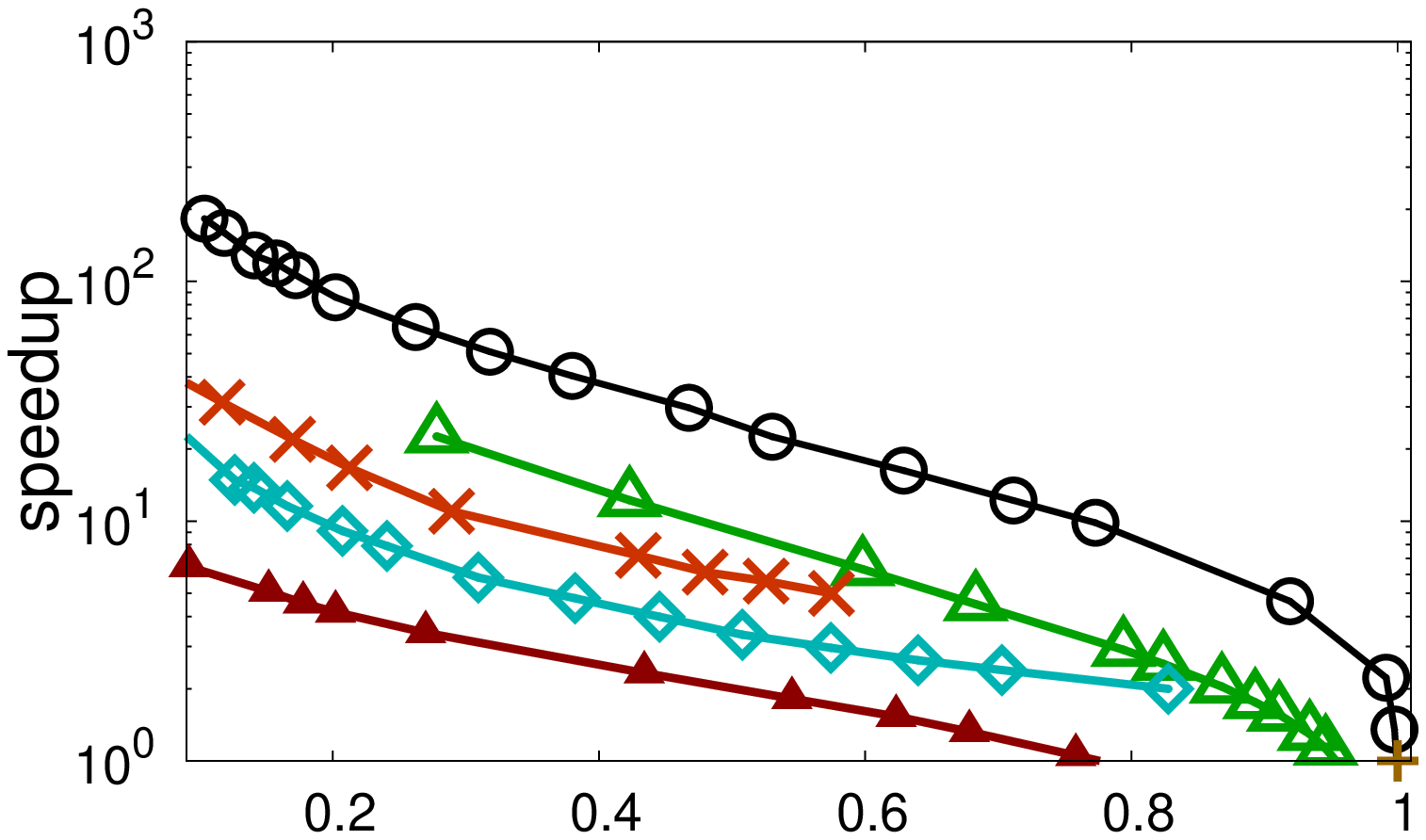}}
\end{minipage}%
\vfill
\begin{minipage}[t]{1.0\linewidth}
\centering
      \subfigure[\small Msong ]{
      \label{fig:exp_recall_msong} 
      \includegraphics[width=0.234\linewidth]{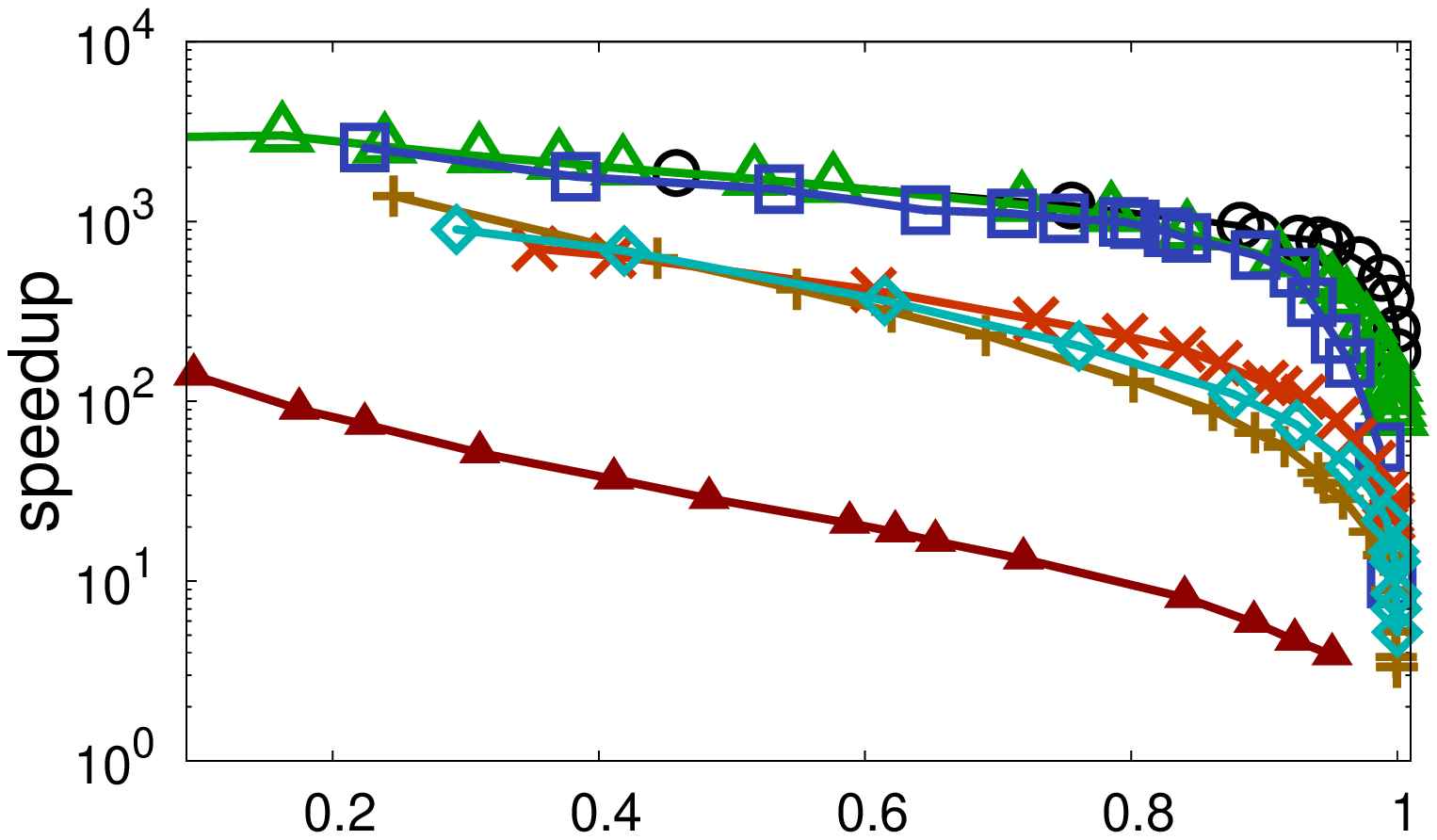}}
      \subfigure[\small Deep]{
      \label{fig:exp_recall_deep} 
      \includegraphics[width=0.234\linewidth]{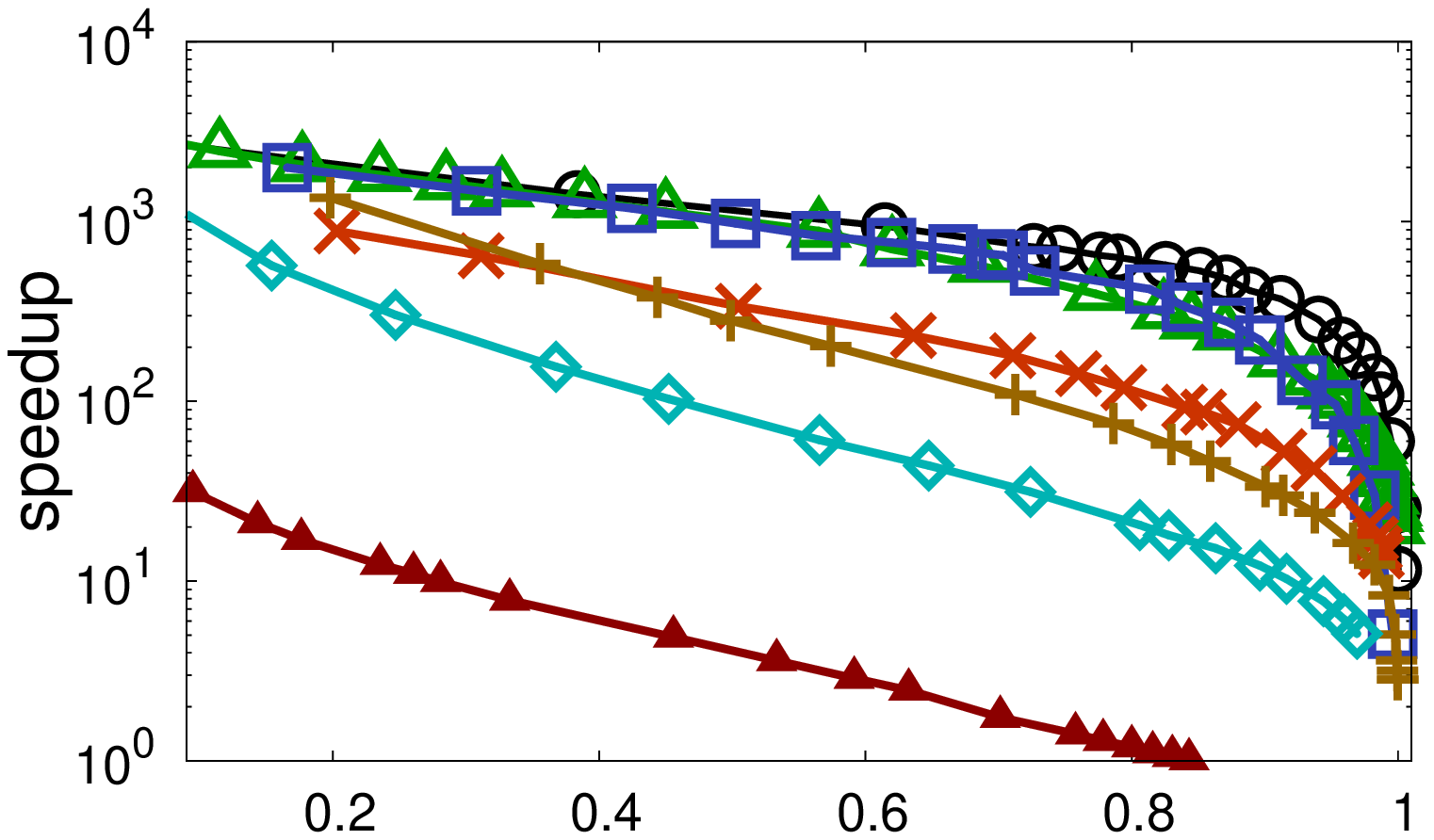}}
      \subfigure[\small Sift ]{
      \label{fig:exp_recall_sift} 
      \includegraphics[width=0.234\linewidth]{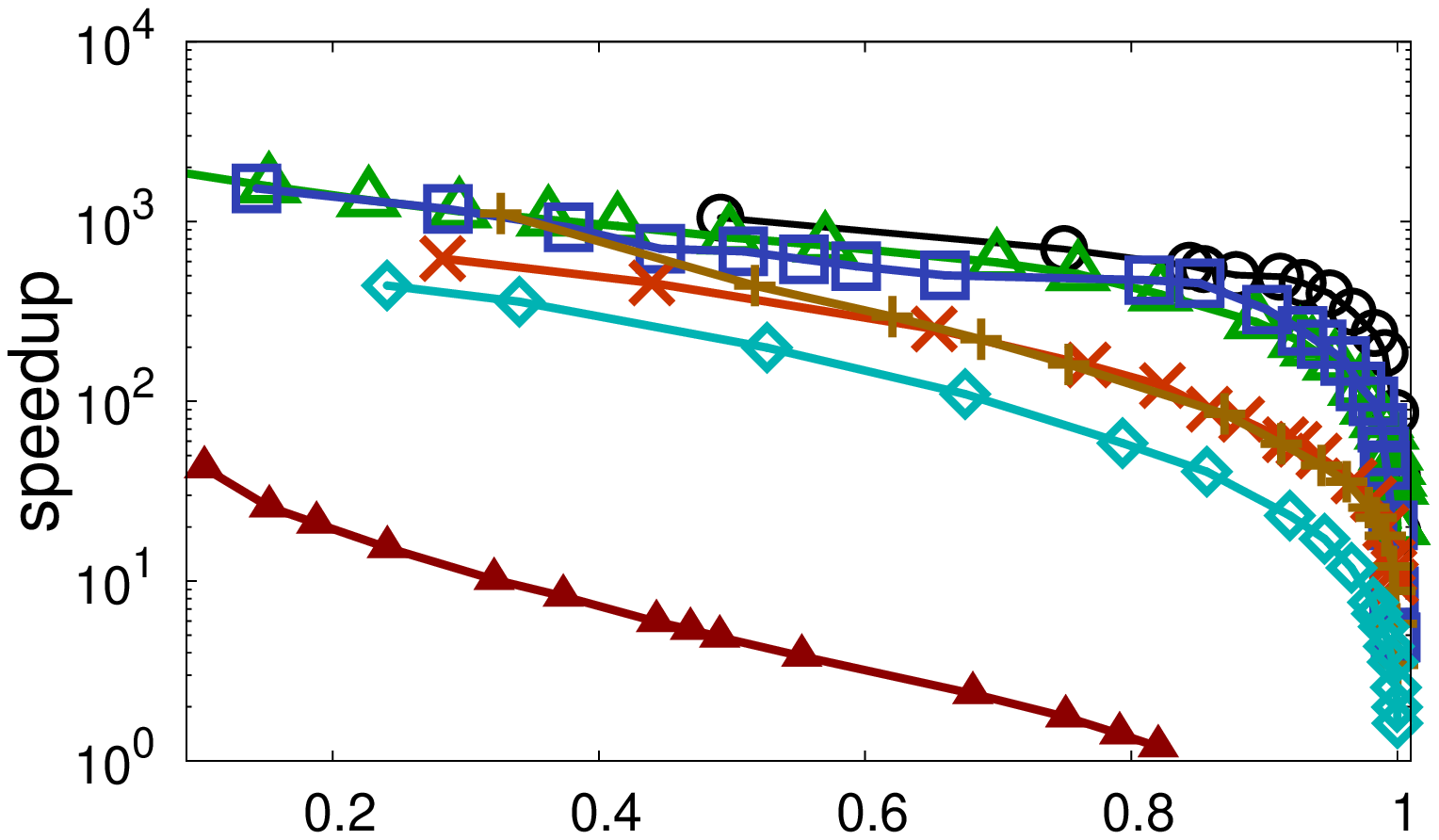}}
      \subfigure[\small Gauss ]{
      \label{fig:exp_recall_gauss} 
      \includegraphics[width=0.234\linewidth]{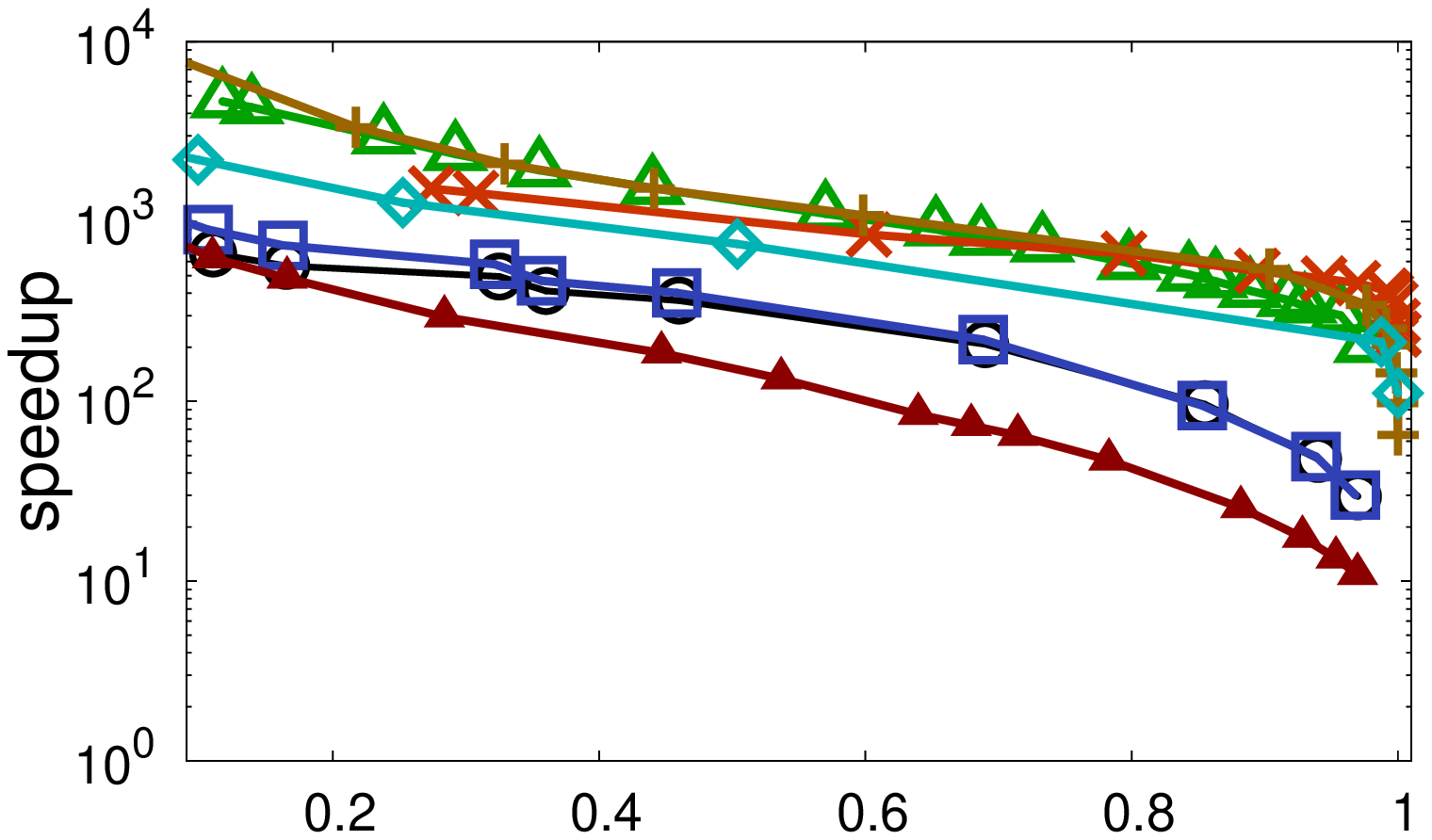}}
\end{minipage}%
\vspace{-3mm}
\caption{\small Speedup vs Recall on Different Datasets}
\label{fig:exp_final_recall}
\end{figure*}

\begin{figure*}[htb]
\centering%
\begin{minipage}[b]{0.8\linewidth}
\centering
\includegraphics[width=1.0\linewidth]{exp-fig/5.6.1-recall/final_title}%
\vspace{-3mm}
\end{minipage}
\begin{minipage}[t]{1.0\linewidth}
\centering
      \subfigure[\small Nusw]{
      \label{fig:exp_recall_nus} 
      \includegraphics[width=0.23\linewidth]{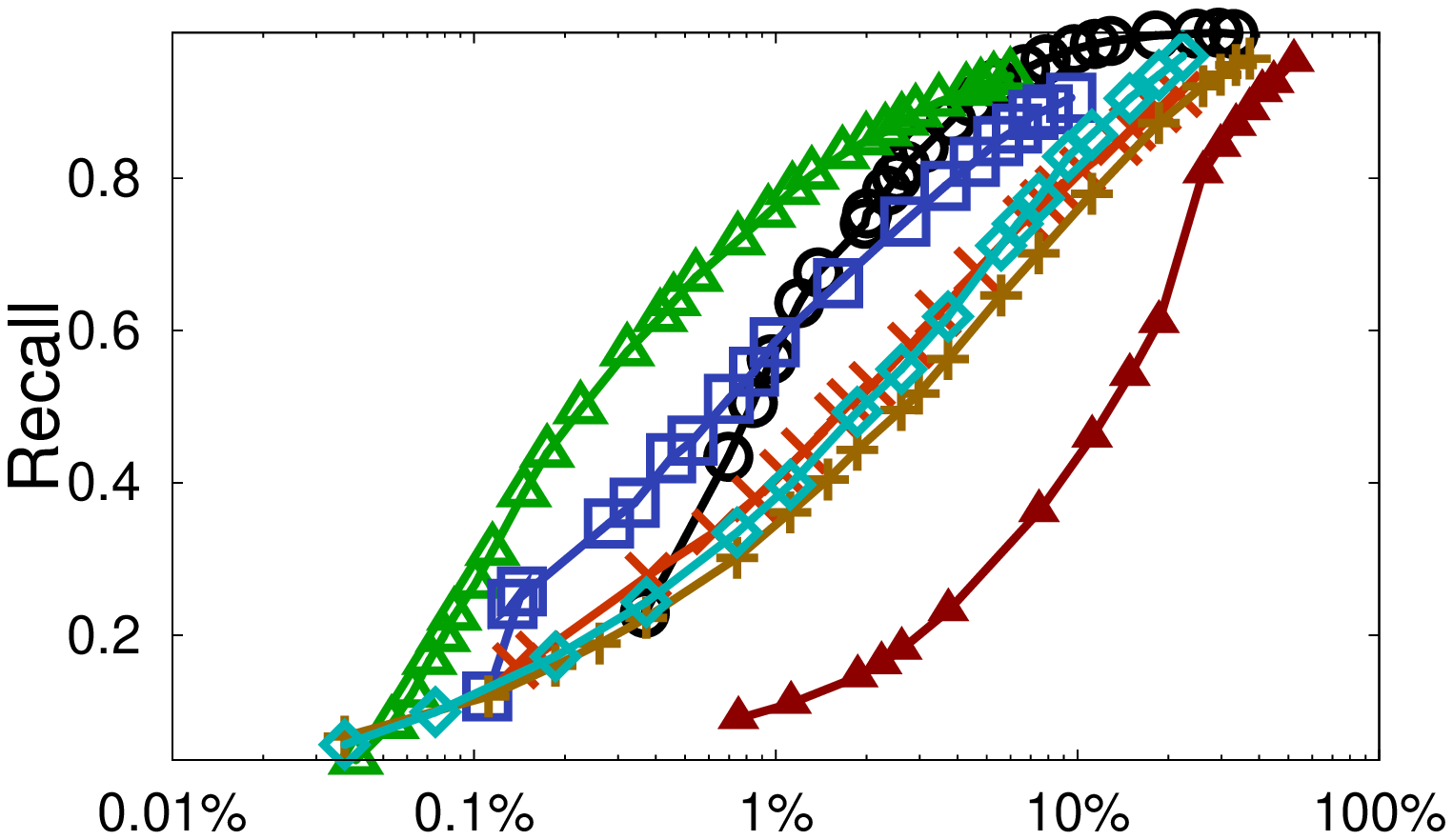}}
      \subfigure[\small Gist]{
      \label{fig:exp_recall_msong} 
      \includegraphics[width=0.23\linewidth]{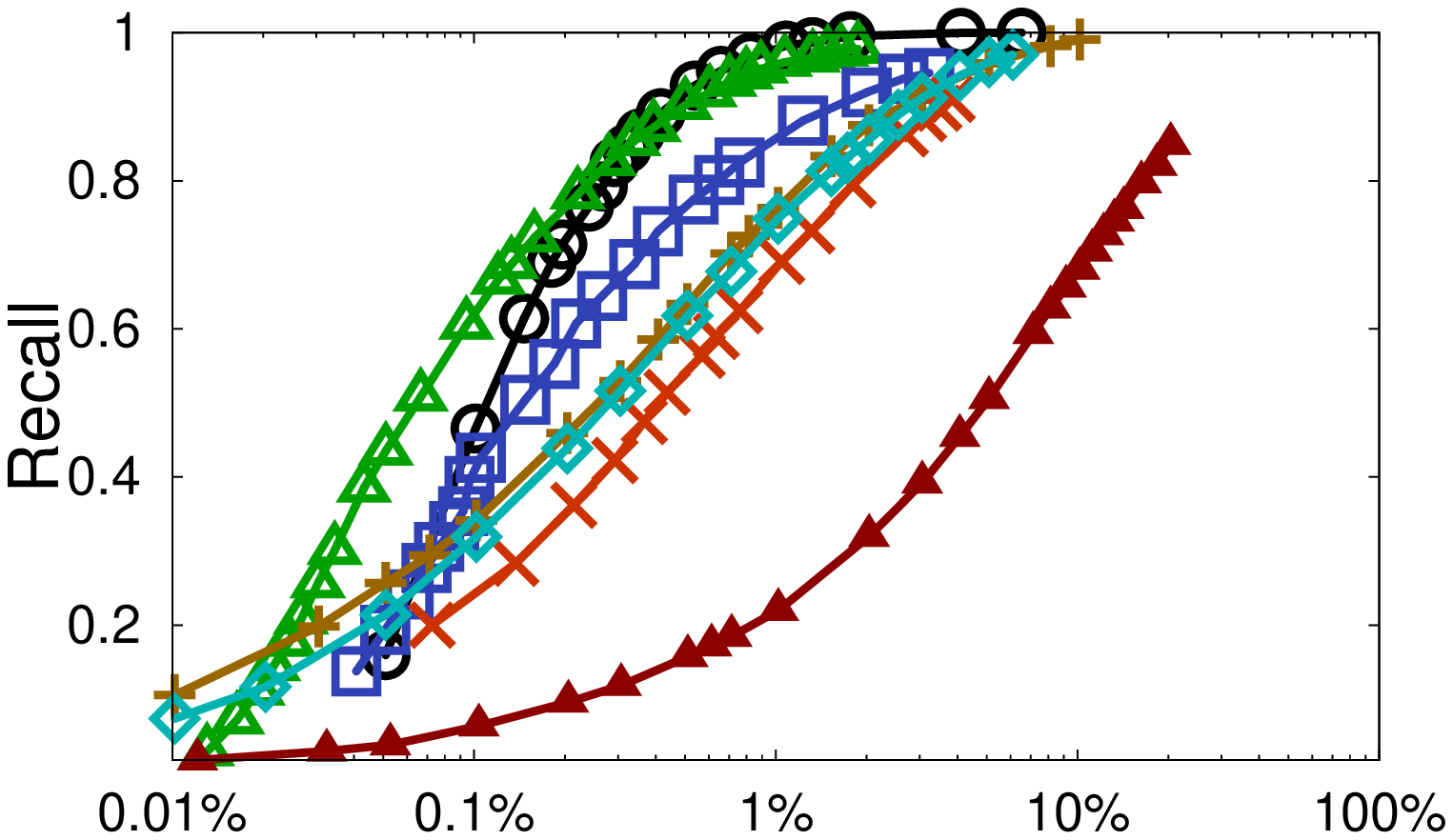}}
      \subfigure[\small Msong]{
      \label{fig:exp_recall_sift} 
      \includegraphics[width=0.23\linewidth]{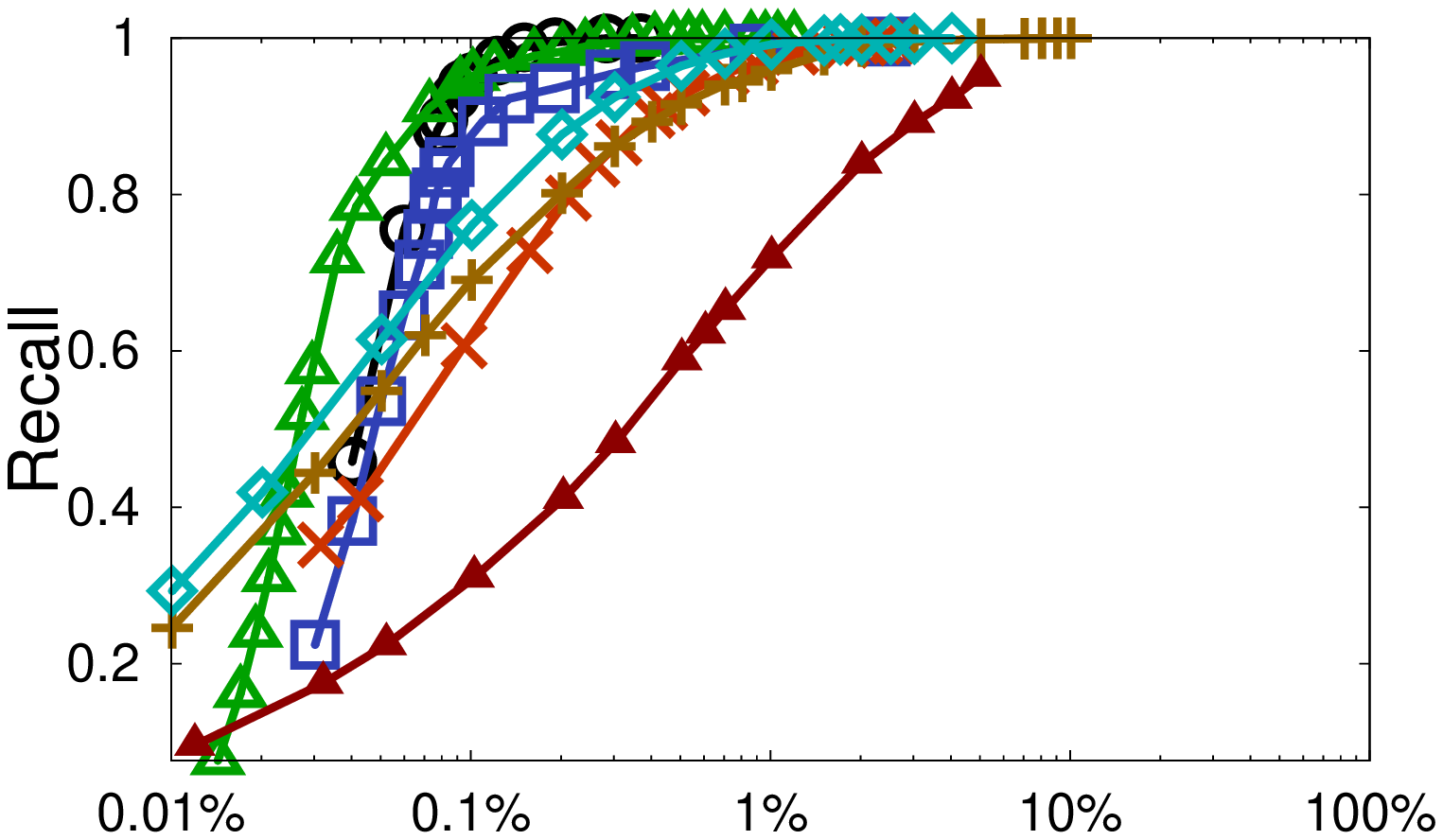}}
      \subfigure[\small Deep ]{
      \label{fig:exp_recall_deep} 
      \includegraphics[width=0.23\linewidth]{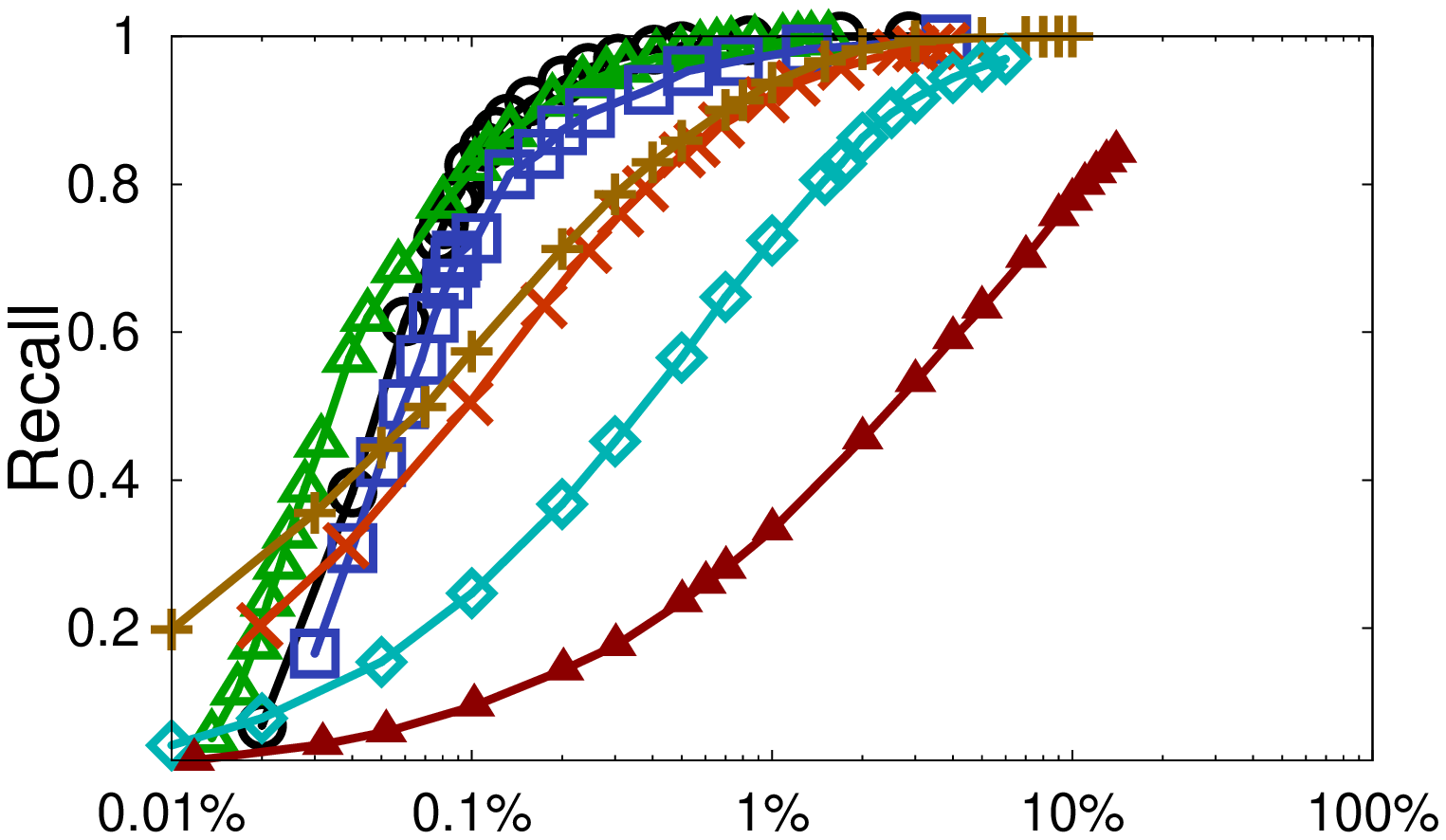}}
\end{minipage}%
\vspace{-3mm}
\caption{\small Recall vs Percentage of Data Points Accessed}
\label{fig:exp_final_recall_N}
\end{figure*}

\begin{figure*}[tbh]
\centering
\includegraphics[width=1.0\linewidth]{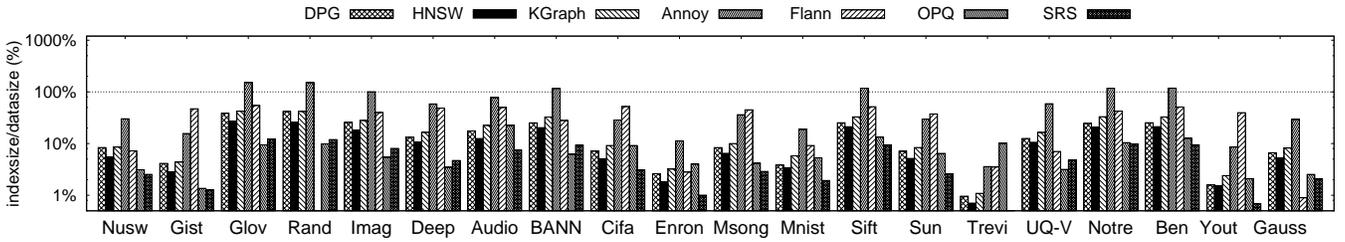}
\caption{\small The Ratio of Index Size and Data Size (\%) }
\vspace{-3mm}
\label{fig:exp_bar_indexsize}
\end{figure*}

\begin{figure*}[tbh]
\centering
\includegraphics[width=1.0\linewidth]{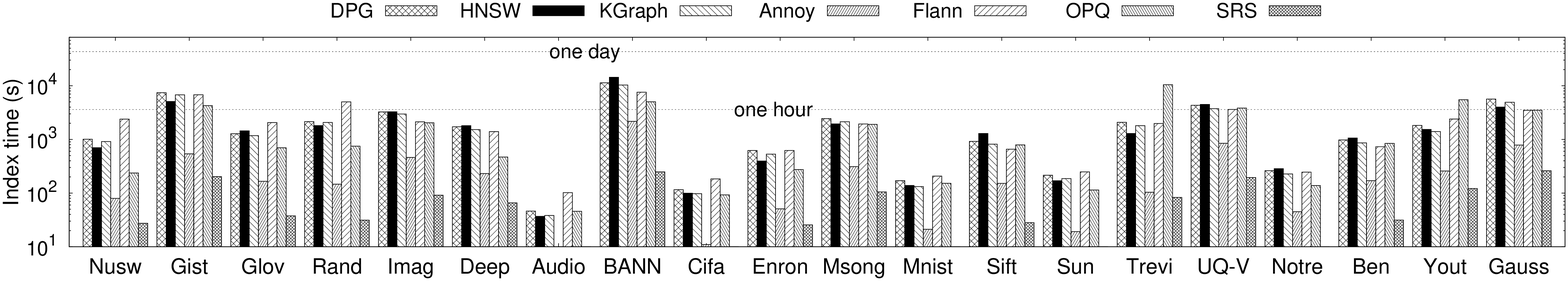}
\caption{\small Index Construction Time (seconds) }
\vspace{-3mm}
\label{fig:exp_bar_indextime}
\end{figure*}

\subsection{Second Round Evaluation}
\label{subsec:exp_final}
In the second round evaluation, we conduct comprehensive experiments on
\textit{seven} representative algorithms: \Algsrs{}, \AlgOPQ{}, \Algflann{}, \Algannoy{}, \Alghnsw{}, \Algkgraph{}, and \Algdpg{}.

\subsubsection{Search Quality and Time}
\label{subsubsec:exp_final_search}

In the first set of experiments, Figure~\ref{fig:exp_bar_speedup} reports the speedup of seven algorithms
when they reach the recall around $0.8$ on $20$ datasets.
Note that the speedup is set to $1.0$ if an algorithm cannot outperform the linear scan algorithm.
Among seven algorithms, \Algdpg{} and \Alghnsw{} have the best overall search performance and \Algkgraph{} follows.
It is shown that \Algdpg{} enhances the performance of \Algkgraph{}, especially on hard datasets: \Datanusw{}, \Datagist{},
\Dataglov{} and \Datarand{}. As reported thereafter, the improvement is also more significant on higher recall.
For instance, \Algdpg{} is ranked after \Algkgraph{} on three datasets under this setting (recall $0.8$),
but it eventually surpasses \Algkgraph{} on higher recall. Overall, \Algdpg{} and \Alghnsw{} have the best performance for different datasets.

\AlgOPQ{} can also achieve a decent speedup under all settings.
Not surprisingly, \Algsrs is slower than other competitors with a huge margin as it does not exploit the data distribution.
Similar observations are reported in Figure~\ref{fig:exp_bar_recall}, which depicts the recalls achieved by the algorithms
with speedup around $50$.

We evaluate the trade-off between search quality (Recall) and search time (Speedup and the percentage of data points accessed). The number of data points to be accessed ($N$) together with other parameters such as the number of trees (\Algannoy{}) and the search queue size (\Alghnsw{}, \Algkgraph{} and \Algdpg{}), are tuned to achieve various recalls with different search time (i.e., speedup).
Figure~\ref{fig:exp_final_recall} illustrates speedup of the algorithms on eight datasets with recall varying from $0$ to $1$. It further demonstrates the superior search performance of \Algdpg{} on high recall and hard datasets (Figure~\ref{fig:exp_final_recall}(a)-(d)). The overall performances of \Alghnsw{}, \Algkgraph{} and \Algannoy{} are also very competitive, followed by \Algflann{}.
It is shown that the performance of both \Algdpg{} and \Algkgraph{} are ranked lower than that of \Alghnsw{}, \Algannoy{}, \Algflann{} and \AlgOPQ{} in Figure~\ref{fig:exp_recall_gauss} where the data points are clustered.

In Figure~\ref{fig:exp_final_recall_N}, we evaluate the recalls of the algorithms against the percentage of data points accessed, i.e., whose distances to the query in the data space are calculated. As the search of proximity-based methods starts from random entrance points and then gradually approaches the results while other algorithms initiate their search from the most promising candidates, the search quality of these methods is outperformed by \Algannoy{}, \Algflann{} and even \AlgOPQ{} at the beginning stage. Nevertheless, their recall values rise up faster than other algorithms when more data points are accessed. Because of the usage of the hierarchical structure in \Alghnsw{}, \Alghnsw{} could approach the results more quickly.

\subsubsection{Searching with Different $K$}

\begin{figure*}[tbh]
\centering%
\begin{minipage}[b]{1.0\linewidth}
\centering
\includegraphics[width=0.8\linewidth]{exp-fig/5.6.1-recall/final_title}%
\end{minipage}
\vfill
\begin{minipage}[t]{1.0\linewidth}
\centering
\subfigure[Gist]{
\includegraphics[width=0.234\linewidth]{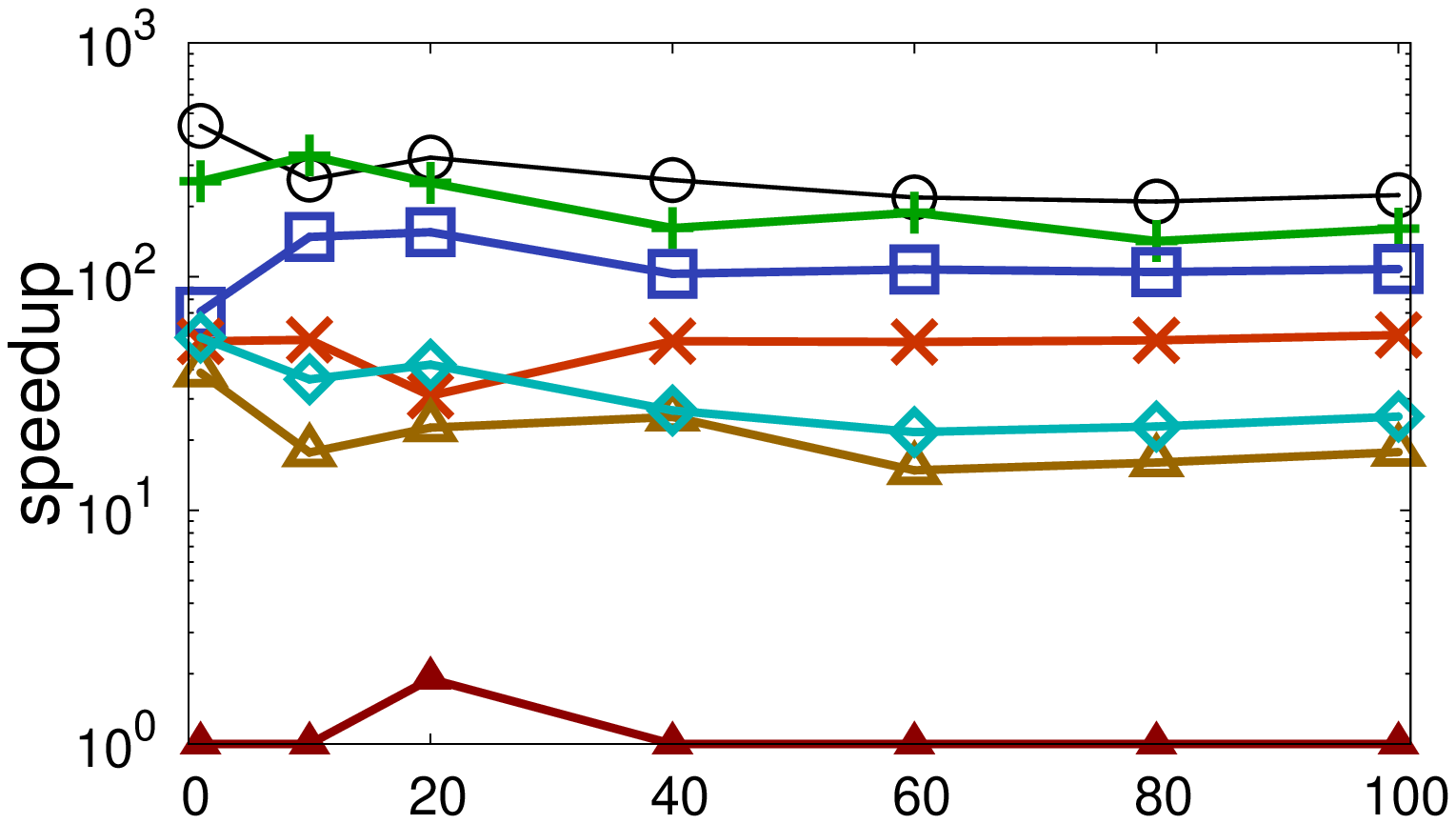}}
\subfigure[Sift]{
\includegraphics[width=0.234\linewidth]{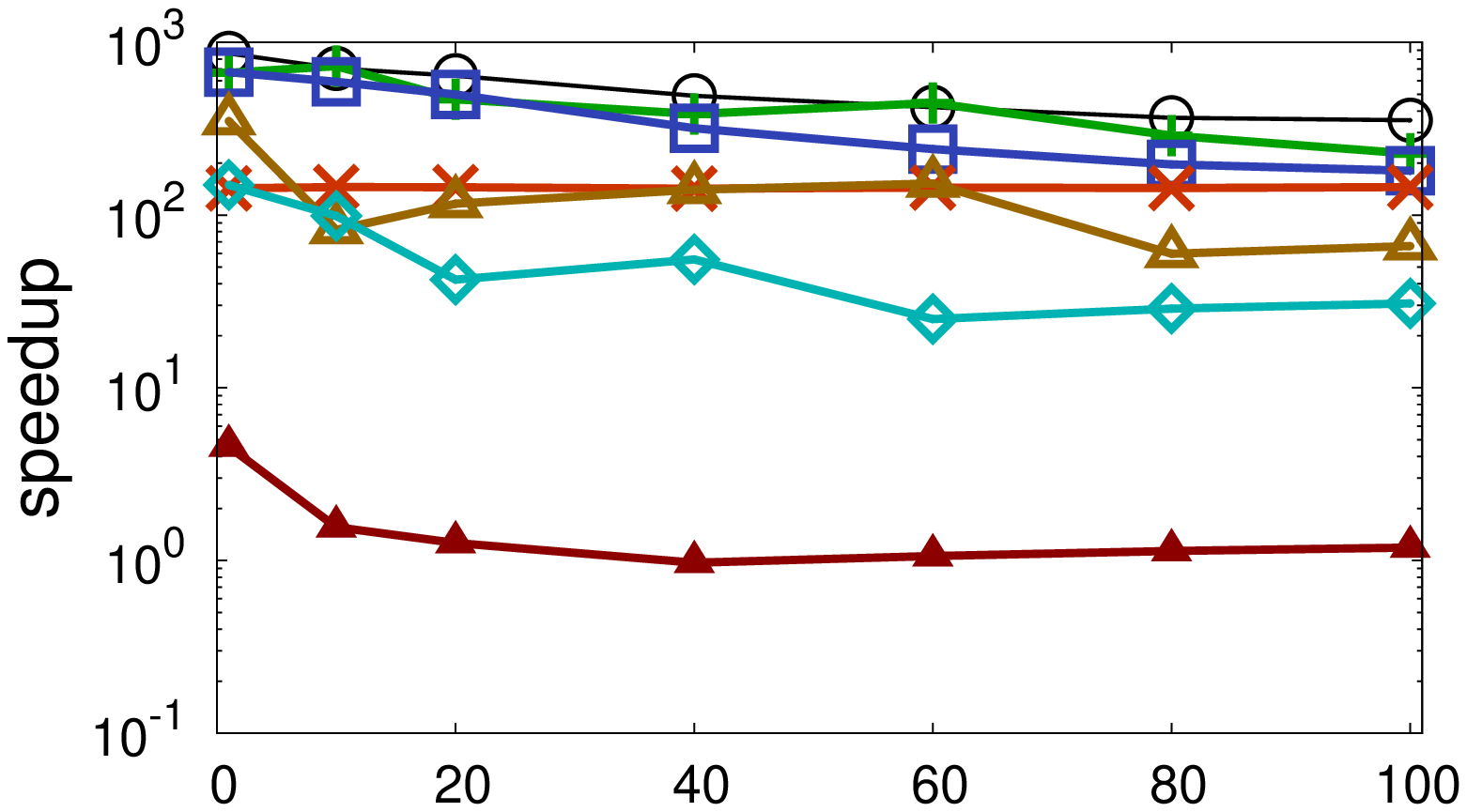}}
\subfigure[MNIST]{
\includegraphics[width=0.234\linewidth]{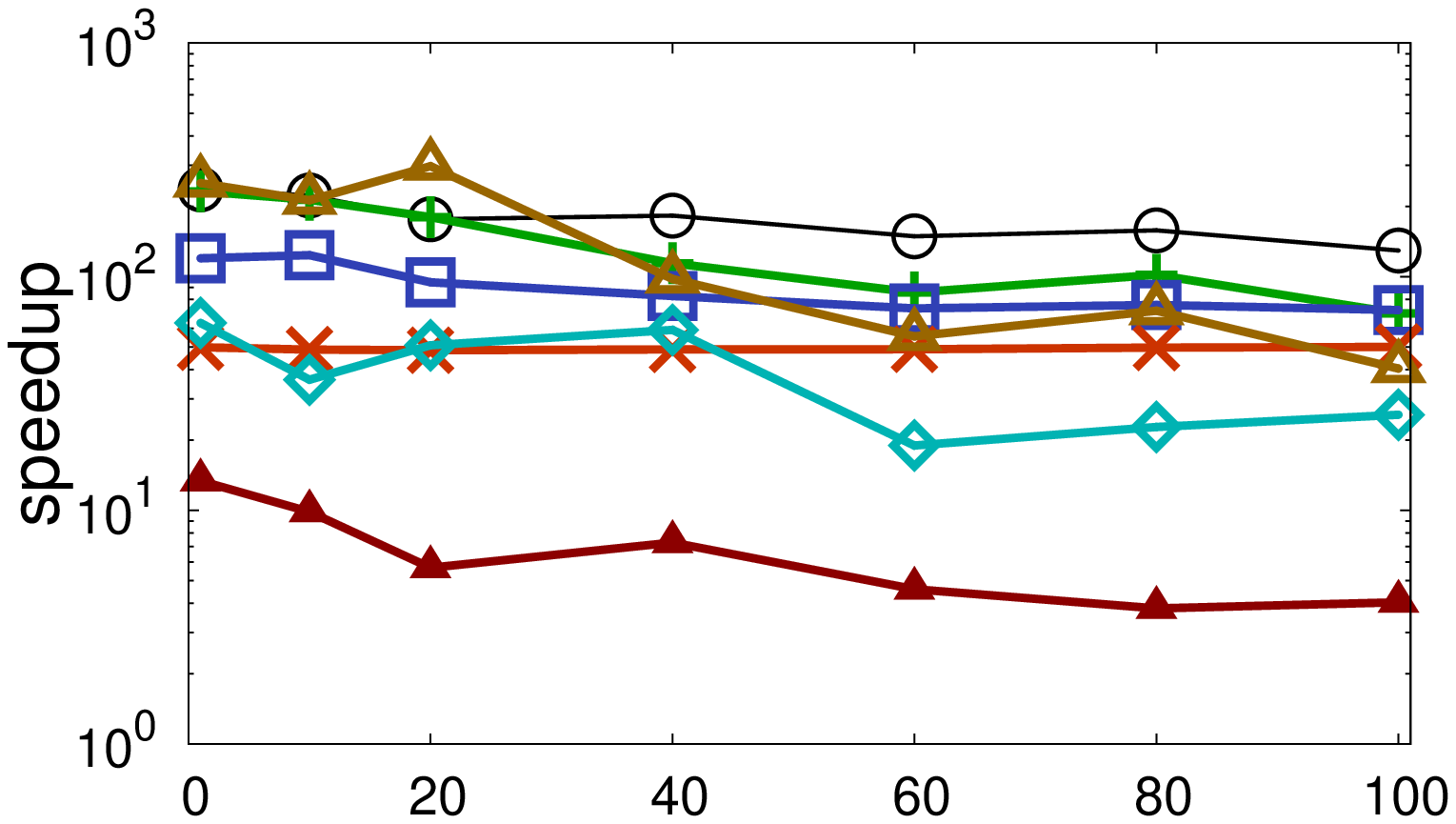}}
\subfigure[Deep]{
\includegraphics[width=0.234\linewidth]{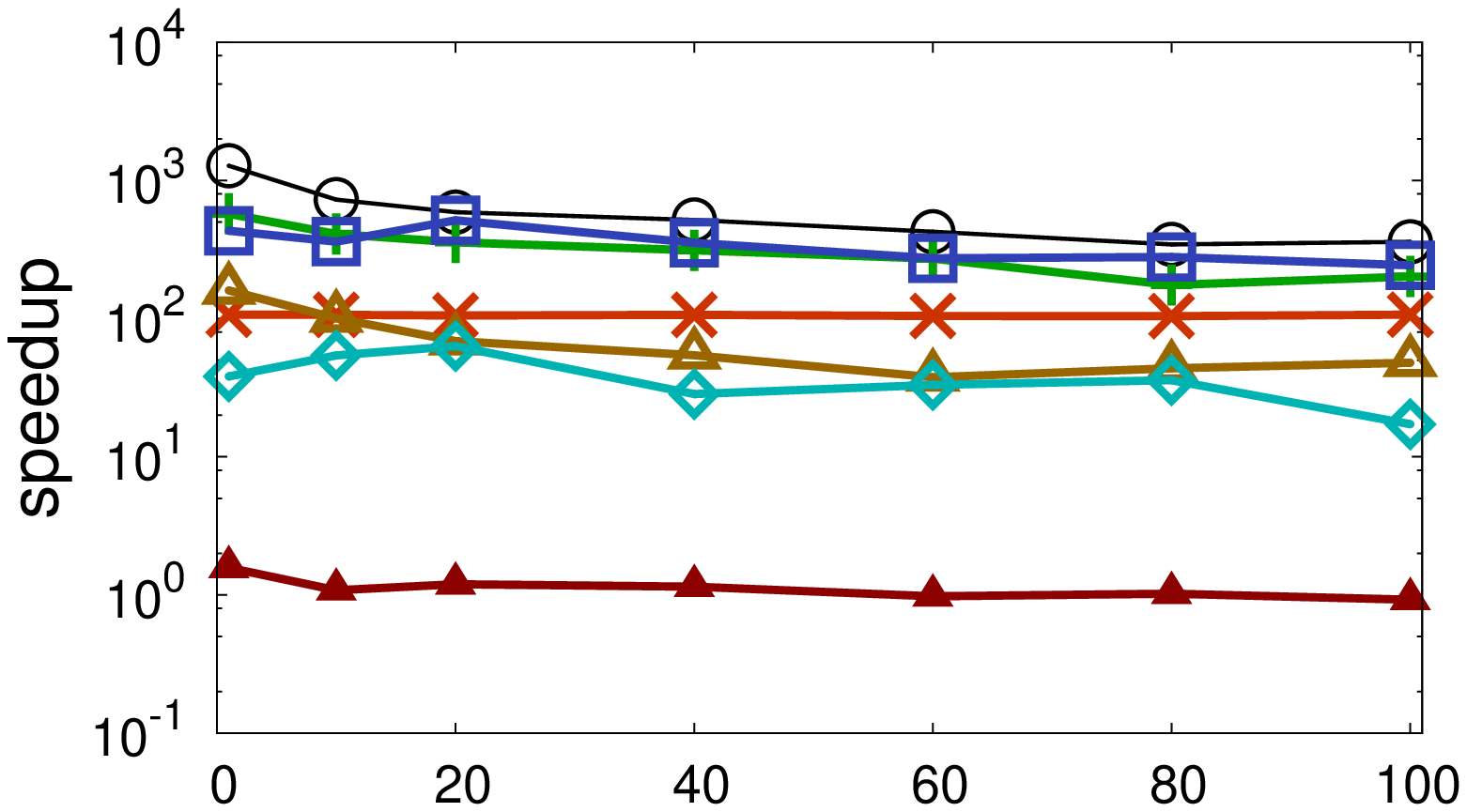}}
\end{minipage}
\caption{\small Speedup with recall of 0.8 with Diff $K$ }
\vspace{-3mm}
\label{fig:exp_K}
\end{figure*}

We evaluate the speedup at the recall of $0.8$ with different $K$. As shown in figure \ref{fig:exp_K}, \Algdpg{} and \Alghnsw{} almost has the best search performance for $K$ between 1 and 100, followed by \Algkgraph{} and \Algannoy{}. The similar ranking could be observed in different $K$.

\subsubsection{Index Size and Construction Time}
\label{subsubsec:exp_index}

In addition to search performance, we also evaluate the index size and construction time of seven algorithms on $20$ datasets.
Figure~\ref{fig:exp_bar_indexsize} reports ratio of the index size (exclude the data points) and the data size.
Except \Algannoy{}, the index sizes of all algorithms are smaller than the corresponding data sizes.

The index sizes of \Algdpg{}, \Algkgraph{}, \Alghnsw{} and \Algsrs{} are irrelevant to the dimensionality because
a fixed number of neighbor IDs and projections are kept for each data point
by neighborhood-based methods (\Algdpg{}, \Algkgraph{} and \Alghnsw{}) and \Algsrs{}, respectively.
Consequently, they have a relatively small ratio on data with high dimensionality (e.g., \Datatrevi{}).
Overall, \AlgOPQ{} and \Algsrs{} have the smallest index sizes, less than $5$\% among most of the datasets,
followed by \Alghnsw{}, \Algdpg{}, \Algkgraph{} and \Algflann{}.
It is shown that the rank of the index size of \Algflann{} varies dramatically over $20$ datasets
because it may choose three possible index structures.
\Algannoy{} needs to maintain a considerable number of trees for a good search quality, and hence has the largest index size.

Figure~\ref{fig:exp_bar_indextime} reports the index construction time of the algorithms on $20$ datasets.
\Algsrs{} has the best overall performance due to its simplicity. Then
\Algannoy{} ranks the second. The construction time of \AlgOPQ{} is related to the dimensionality because of the calculation of the sub-codewords (e.g., \Datatrevi{}).
\Alghnsw{}, \Algkgraph{} and \Algdpg{} have similar construction time.
Compared with \Algkgraph{}, \Algdpg{} doesn't spend large extra time for the graph diversification.
Nevertheless, they can still build the indexes within one hour for $16$ out of $20$ datasets.

\subsubsection{Scalability Evaluation}
\label{subsubsec:exp_scalability}

\begin{figure*}[hbt]
\centering%
\begin{minipage}[b]{1.0\linewidth}
\centering
\includegraphics[width=0.8\linewidth]{exp-fig/5.6.1-recall/final_title}%
\end{minipage}
\vfill
\begin{minipage}[t]{1.0\linewidth}
\centering
      \subfigure[\small Vaying $n$, \Databann{} ]{
      \label{fig:exp_final_scalability_sp_n} 
      \includegraphics[width=0.234\linewidth]{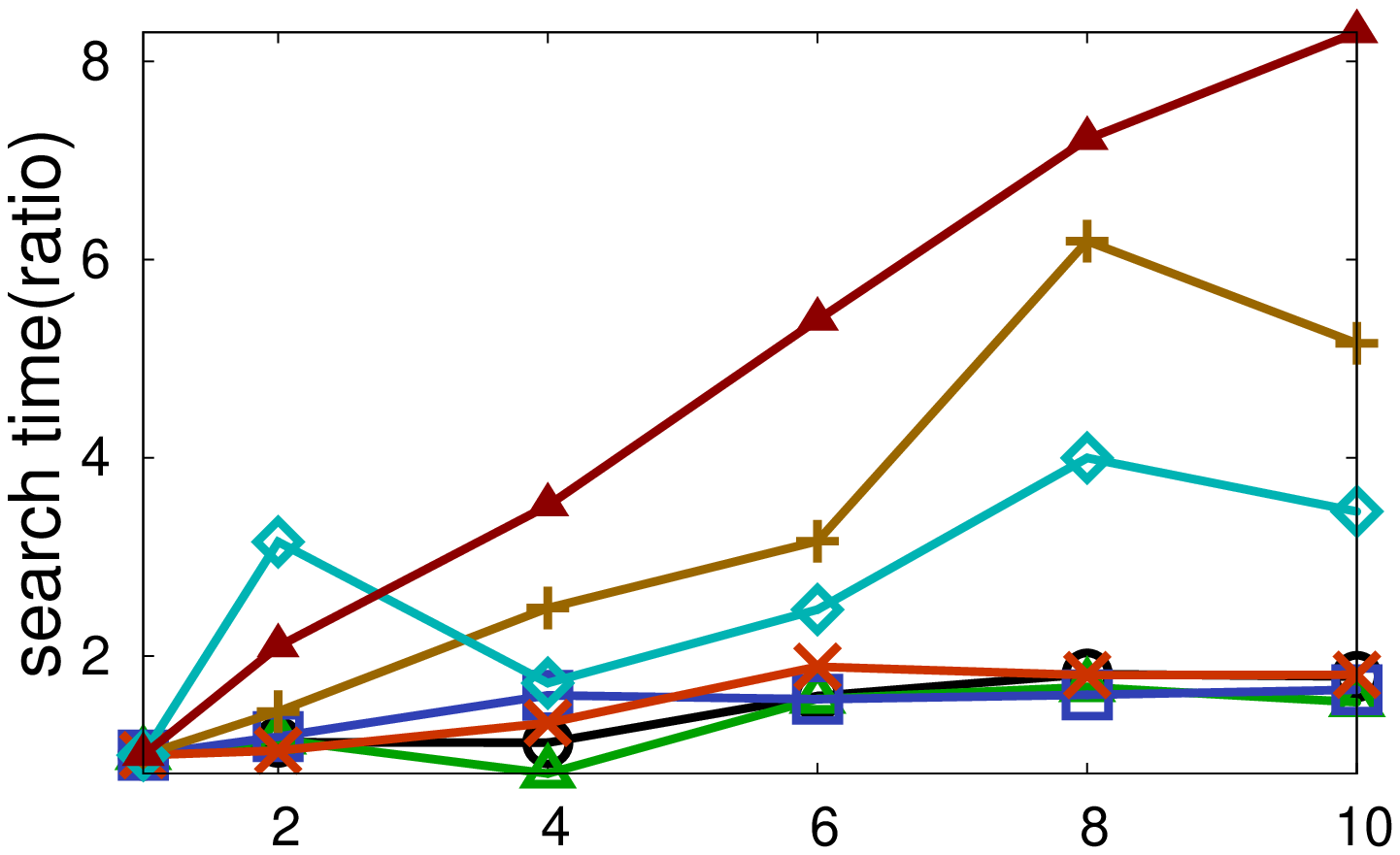}}
      \subfigure[\small Vaying $n$, \Databann{} ]{
      \label{fig:exp_final_scalability_size_n} 
      \includegraphics[width=0.234\linewidth]{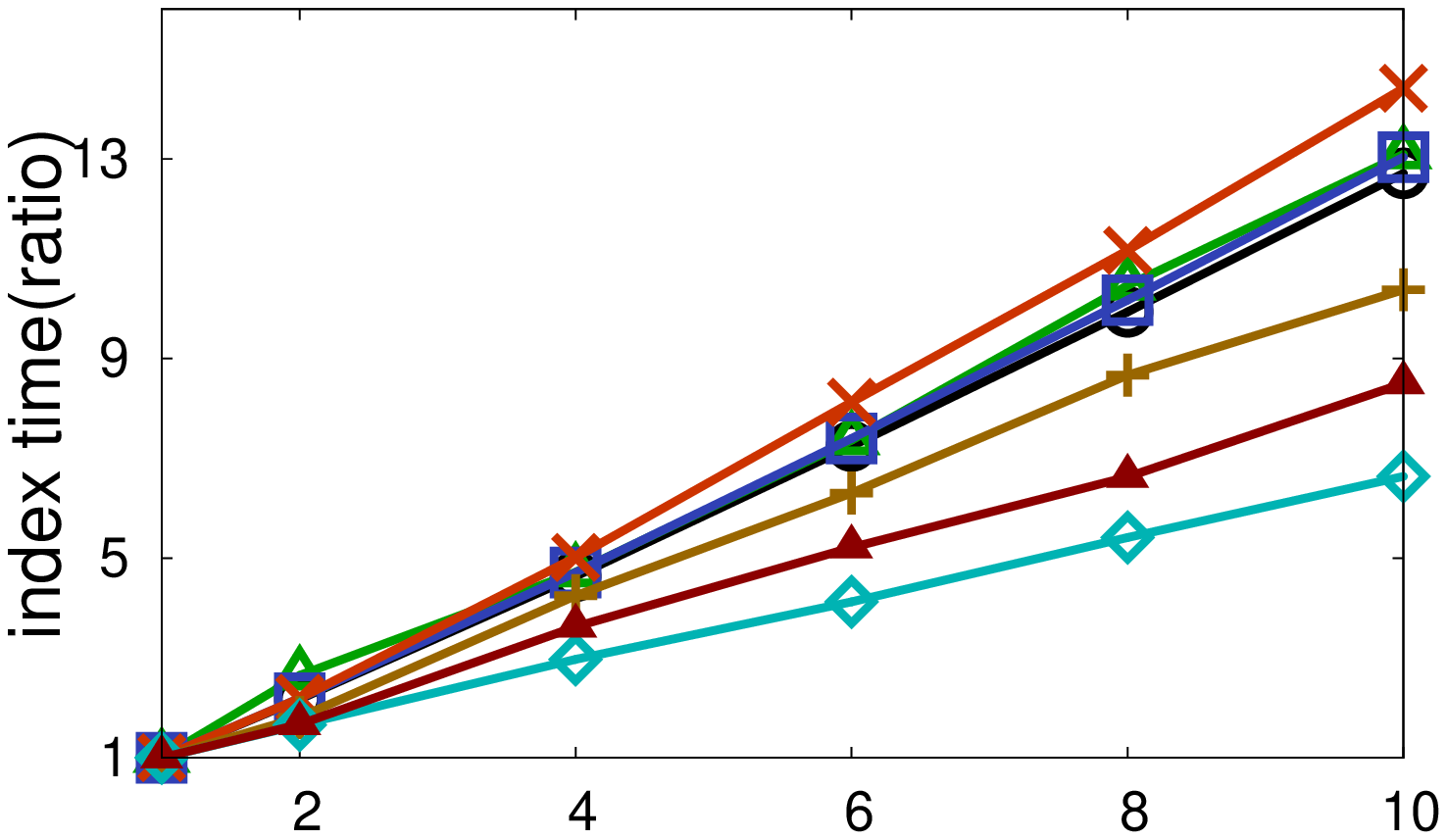}}
      \subfigure[\small Vaying $n$, \Databann{} ]{
      \label{fig:exp_final_scalability_time_n} 
      \includegraphics[width=0.234\linewidth]{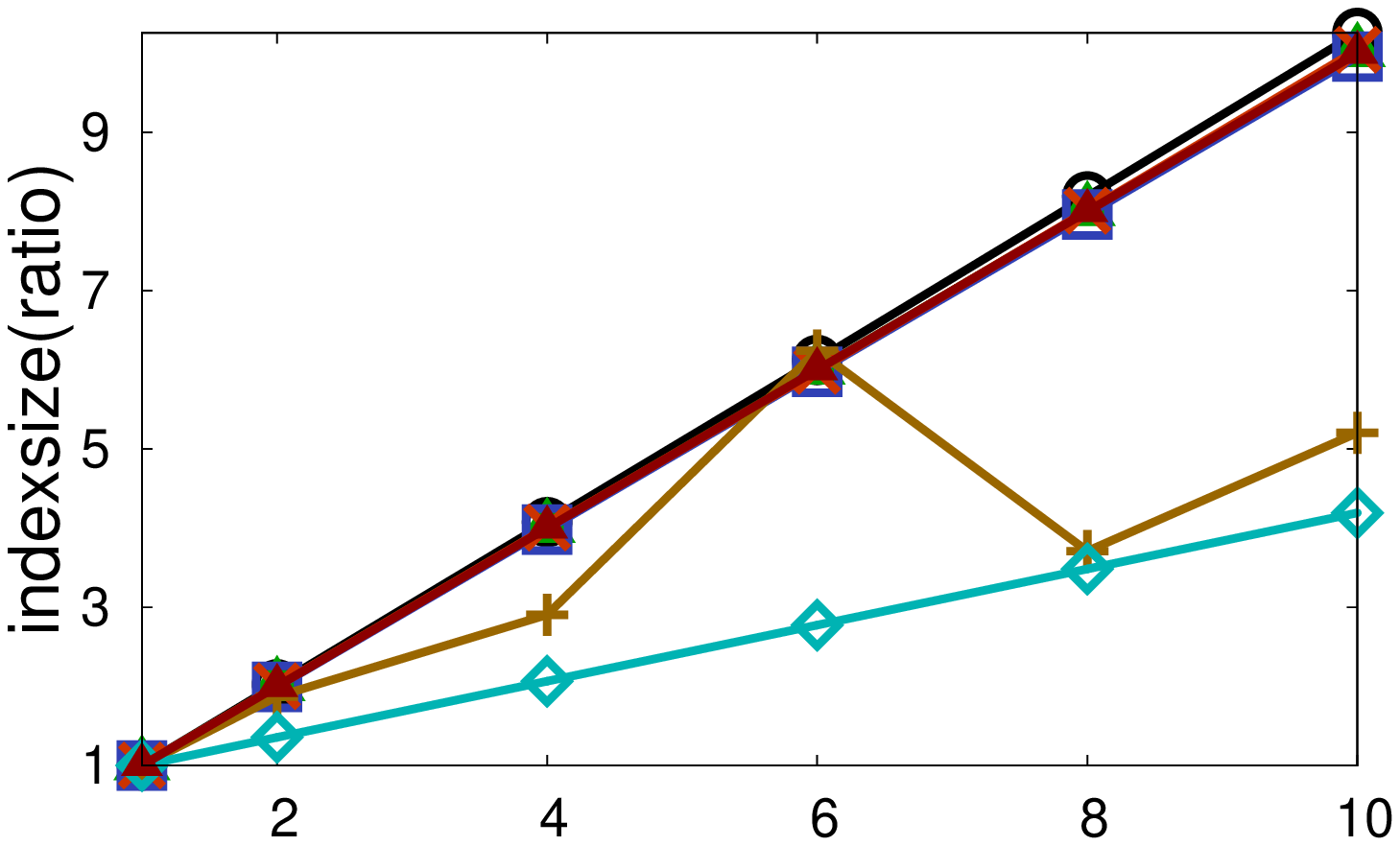}}
\end{minipage}
\begin{minipage}[t]{1.0\linewidth}
\centering
      \subfigure[{\small Varying $d$, \Datarand{}} ]{
      \label{fig:exp_final_scalability_sp_d} 
      \includegraphics[width=0.234\linewidth]{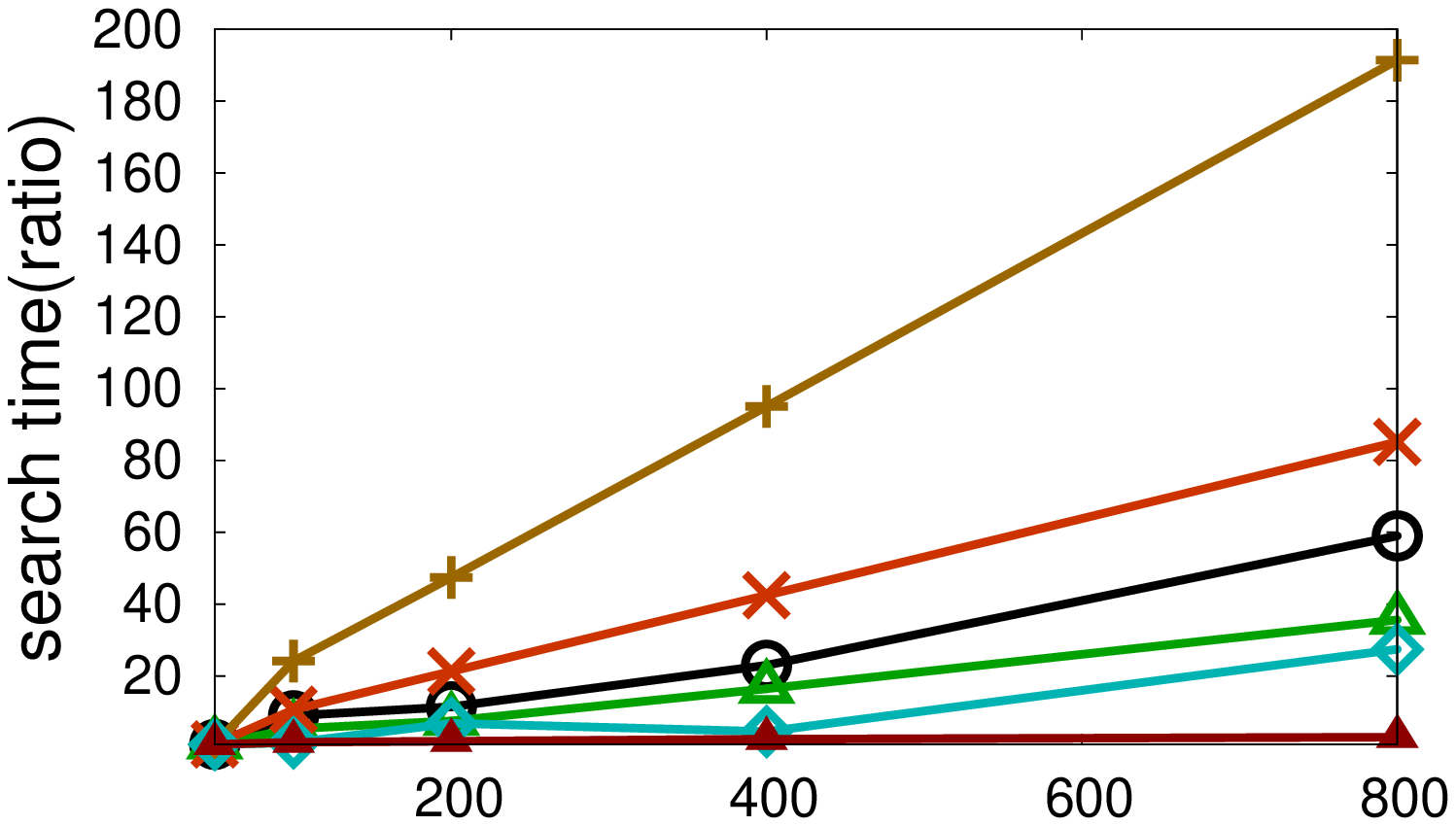}}
      \subfigure[{\small Varying $d$, \Datarand{}} ]{
      \label{fig:exp_final_scalability_size_d} 
      \includegraphics[width=0.234\linewidth]{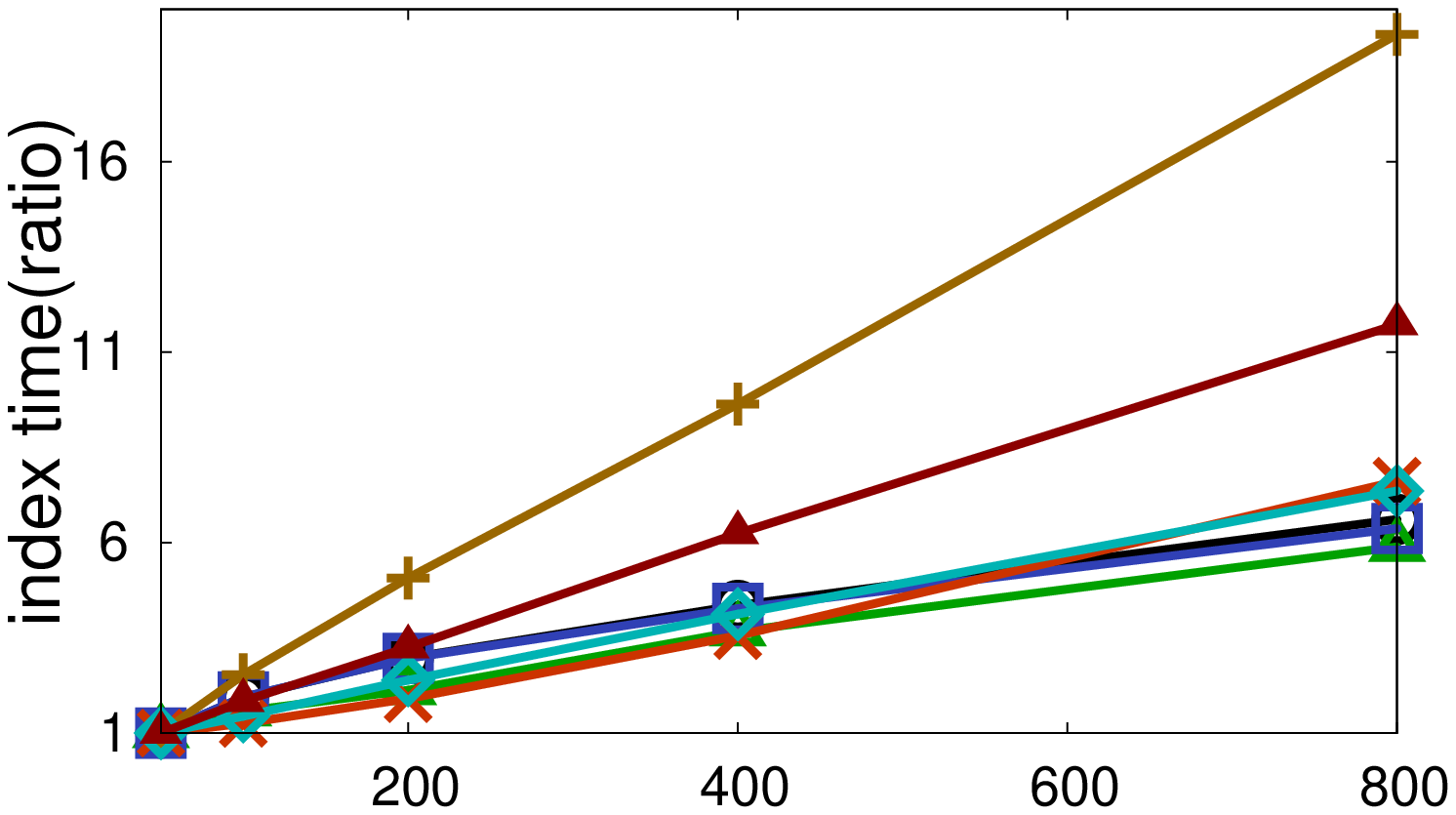}}
      \subfigure[{\small Varying $d$, \Datarand{}} ]{
      \label{fig:exp_final_scalability_time_d} 
      \includegraphics[width=0.234\linewidth]{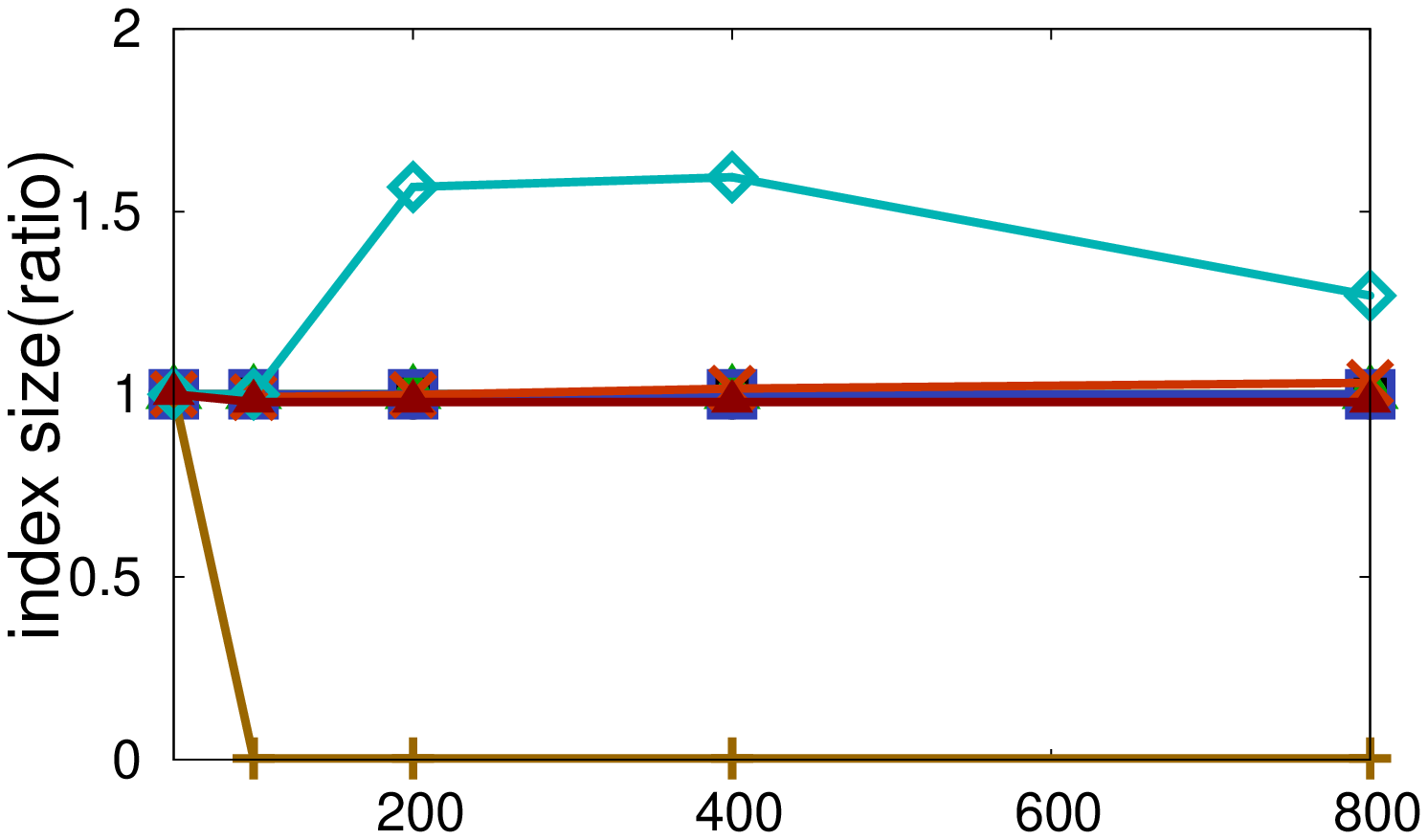}}
\end{minipage}%
\caption{\small Scalability vs data size ($n$) and dim ($d$)}
\vspace{-2mm}
\label{fig:exp_final_scalability}
\end{figure*}

In Figure~\ref{fig:exp_final_scalability}, we investigate the scalability of the
algorithms in terms of the search time,
index size and index construction time against the growth of the number of data points ($n$) and the dimensionality ($d$).
The target recall is set to $0.8$ for evaluation of $n$ and $d$, respectively.
Particularly, \Databann{} is deployed with the number of data points growing
from $1$M to $10$M.
Following the convention, we use random data, \Datarand{}, to evaluate the
impact of the dimensionality which varies from $50$ to $800$.
To better illustrate the scalability of the algorithms, we report the
\textit{growth ratio} of the algorithms against the increase of $n$ and $d$.
For instance, the index size growth ratio of \Algdpg{} with $n=4M$ is obtained
by its index size divided by the index size of \Algdpg{} with $n=1M$.

With the increase of the number of data points ($n$), \Algdpg{}, \Algkgraph{}, \Alghnsw{} and \Algannoy{} have the best search scalability while \Algsrs{} ranks the last.
On the other hand, \AlgOPQ{} has the best scalability over index size and construction time,
followed by \Algflann{}. It is noticed that the performance of \Algflann{} is rather unstable mainly because it chooses \Algflannkd{} when $n$ is $6$M and $10$M, and \Algflannhkm{} otherwise.


Regarding the growth of dimensionality, \Algflann{} has the worst overall performance
which simply goes for brute-force linear scan when $d \geq 100$.
As expected, \Algdpg{}, \Algkgraph{}, \Alghnsw{} and \Algsrs{} have the best index size
scalability since their index sizes are independent of $d$.
It is interesting that \Algsrs{} has the best search scalability, and its speedup even outperforms \Algdpg{} when $d \geq 2000$.
This is probably credited to its theoretical worse performance guarantee.
Note that we do not report the performance of \Algkgraph{} in Figure~\ref{fig:exp_final_scalability_sp_d}
because it is always outperformed by linear scan algorithm.
This implies that \Algkgraph{} is rather vulnerable to the high dimensionality, and hence justifies the importance of \Algdpg{}, which achieves much better search scalability towards the growth of dimensionality.

\subsubsection{Harder Query Workload}
\label{subsubsec:exp_query_sensitivity}

\begin{figure}[tbhp]
\centering
\includegraphics[width=.9\linewidth]{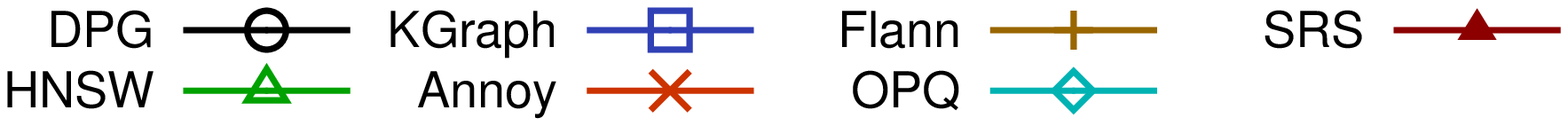}
\centering
\subfigure[\small Speedup with Recall of $0.8$  ]{
\includegraphics[width=.7\linewidth]{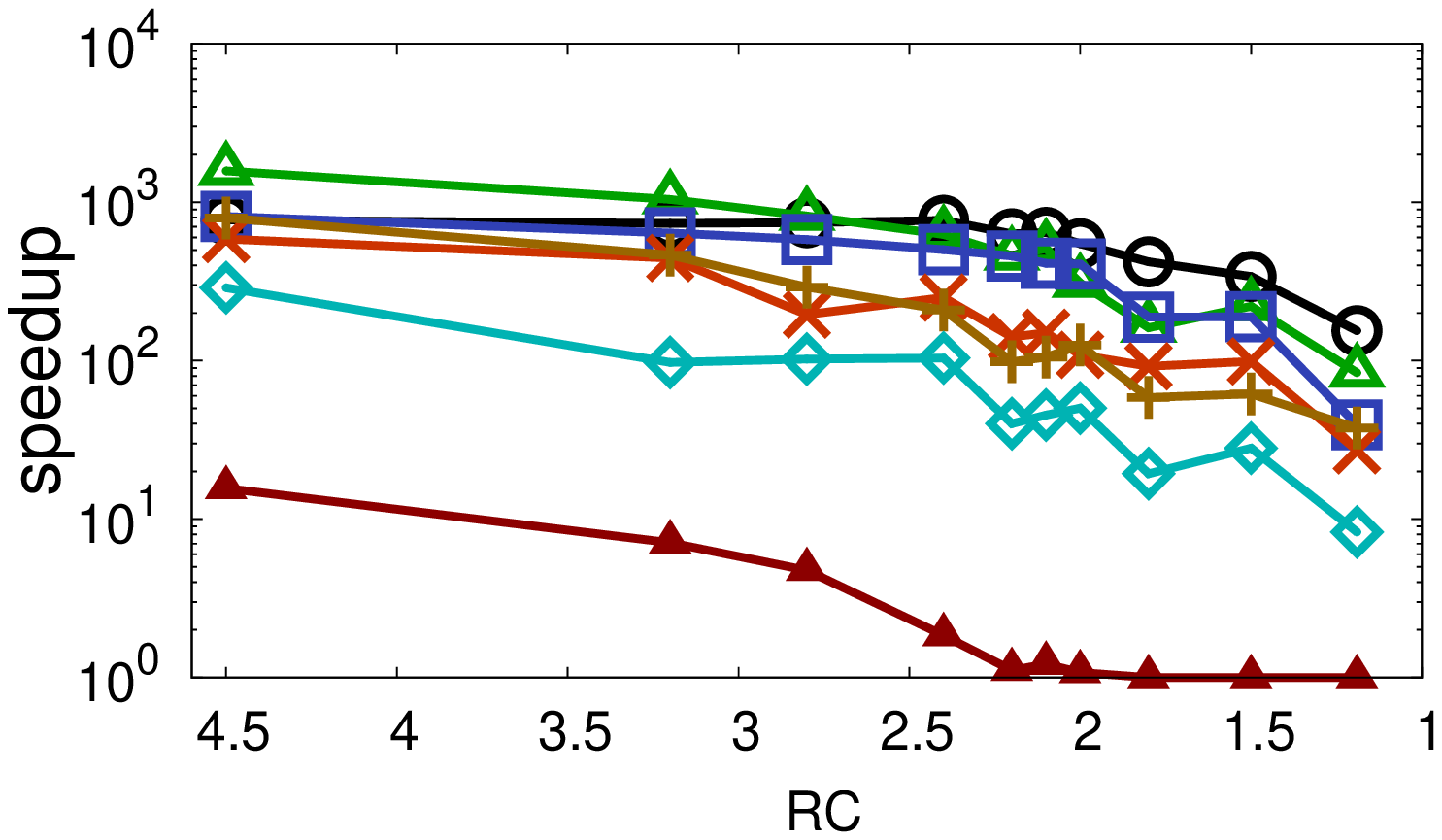}}
\caption{\small Queries with Different RC Values (\Datasift{})}
\vspace{-2mm}
\label{fig:exp_final_RC}
\end{figure}

We evaluate the algorithm performances when the distribution of the query
workload becomes different from that of the datasets. %
We control the generation of increasingly different query workloads by
perturbing the default queries by a fixed length $\delta$ in a random direction. By
increasingly large $\delta$ for the default queries on \Datasift{}, we obtain
query workloads whose RC values vary from $4.5$ (without perturbation) to $1.2$.
Intuitively, queries with smaller RC values are harder as the distance of a
random point in the dataset and that of the NN point becomes less
distinguishable. Figure~\ref{fig:exp_final_RC} shows the speedups of the
algorithms with recall at around $0.8$ on the harder query workloads
characterized by the RC values. The speedups of all the algorithms decrease with the increase of RC values.
\Alghnsw{} has the best performance on easy queries, followed by \Algdpg{}, \Algkgraph{}, \Algannoy{} and \Algflann{}.
Nevertheless, \Algdpg{} is the least affected and still achieves more than 100x
speedup (with a recall at least 0.8) on the hardest settings. This demonstrates
the robustness of \Algdpg{} against different query workloads.

\subsection{Summary}
\label{subsec:exp_summary}

Table~\ref{tab:summary} ranks the performances of the seven algorithms from various
perspectives including search performance, index size, index construction time,
and scalability. We also indicate that \Algsrs{} is the only one with
theoretical guarantee of searching quality, and it is very easy to tune the
parameters 
for search quality and search time. The tuning of
\Algannoy{} is also simple, simply varying the number of the trees. It is much
complicated to tune the parameters of \Algflann{}. Authors therefore developed the
auto-configure algorithm to handle this.

Below are some recommendations for users according to our comprehensive
evaluations. 

\begin{itemize}
\item When there are sufficient computing resources (both main memory and CPUs)
  for the off-line index construction, and sufficient main memory to hold the
  resulting index, \Algdpg{} and \Alghnsw{} are the best choices for ANNS on high dimensional data
  due to their outstanding search performance in terms of robustness to the
  datasets, result quality, search time and search scalability.

  We also recommend \Algannoy{}, due to its excellent search
  performance, and robustness to the datasets. Additionally, a nice property of
  \Algannoy{} is that, compared with proximity graph based approaches, it can provide better trade-off
  between search performance and index size/construction time. This is because, one can
  reduce the number of trees without hurting the search performance
  substantially.

  Note that \Algkgraph{} also provides overall excellent performance except on
  few datasets (e.g., the four hard datasets and \Datayout). We recommend
  \Algannoy{} instead of \Algkgraph{} as \Algdpg{} is an improved version of
  \Algkgraph{} with better performance, and \Algannoy{} performs best in
  the few cases where both \Algdpg{} and \Algkgraph{} do not perform as well
  (e.g., \Datayout{}, and \Datagauss{}).

\item To deal with large scale datasets (e.g., $1$ billion of data points) with
  moderate computing resources, \AlgOPQ{} and \Algsrs{} are good candidates due to
  their small index sizes and construction time. It is worthwhile to mention
  that, \Algsrs{} can easily handle the data points updates and have
  theoretical guarantee, which distinguish itself from other five algorithms.
\end{itemize}

\begin{table*}[htbp]
\small
\centering
\begin{tabular}{|c|c|c|c|c|c|c|c|c|c|}
\hline
\multirow{2}{*}{\textbf{Category}} & \multirow{2}{*}{\textbf{Search Performance}} & \multicolumn{2}{c|}{\textbf{Index}} & \multicolumn{2}{c|}{\textbf{Index Scalabiliity}} & \multicolumn{2}{c|}{\textbf{Search Scalabiliity}} & \multirow{2}{*}{\textbf{\begin{tabular}[c]{@{}c@{}}Theoretical\\  Guarantee\end{tabular}}} & \multirow{2}{*}{\textbf{\begin{tabular} [c]{@{}c@{}}Tuning\\ Difficulty\end{tabular}}} \\ \cline{3-8}
                                   &                                         & \textbf{Size}    & \textbf{Time}    & \textbf{Datasize}         & \textbf{Dim}         & \textbf{Datasize}          & \textbf{Dim}         &                                                                                            &                                                                                       \\ \hline
DPG                                & \textbf{1st}                            & 4th              & 7th              & =4th                      & \textbf{=1st}        & \textbf{=1st}              & 5th                  & No                                                                                         & Medium                                                                             \\ \hline
HNSW                               & \textbf{1st}                            & 3rd              & 5th              & =4th                      & 4th                  & \textbf{=1st}               & 4th                  & No                                                                                         & Medium                                                                               \\ \hline
KGraph                             & 3rd                                     & 5th              & 6th              & =4th                      & \textbf{=1st}        & \textbf{=1st}              & 7th                  & No                                                                                         & Medium                                                                                \\ \hline
Annoy                              & 4th                                     & 7th              & 2nd              & 7th                       & 3rd                  & 6th                        & =2nd         & No                                                                                         & \textbf{Easy}                                                                         \\ \hline
FLANN                              & 5th                                     & 6th              & 4th              & =2nd                      & 7th                  & \textbf{=1st}              & 6th                  & No                                                                                         & Hard                                                                                  \\ \hline
OPQ                                & 6th                                     & 2nd              & 3rd              & \textbf{1st}              & =5th                 & 5th                        & =2nd                  & No                                                                                         & Medium                                                                                \\ \hline
SRS                                & 7th                                     & \textbf{1st}     & \textbf{1st}     & =2nd                      & =5th                 & 7th                        & \textbf{1st}         & \textbf{Yes}                                                                               & \textbf{Easy}                                                                         \\ \hline
\end{tabular}
\caption{Ranking of the Algorithms Under Different Criteria}
\label{tab:summary}
\end{table*}





\section{Further Analyses}
\label{sec:analysis}

 In this section, we analyze the  most competitive algorithms in our
 evaluations, grouped by category, in order to understand their strength and
 weakness.

\subsection{Space Partition-based Approach}
\label{subsec:anaylis_partition}

Our comprehensive experiments show that \Algannoy{}, \Algflann{} and \AlgOPQ{}
have the best performance among the space partitioning-based methods. Note that
\Algflann{} chooses \Algflannhkm{} in most of the datasets. Therefore, all
three algorithms are based on \Algkmeans{} space partitioning.%
%

We identify that a key factor for the effectiveness of \Algkmeans{}-style space
partitioning is that the large number of clusters, typically $\Theta(n)$. Note
that we cannot \emph{directly} apply \Algkmeans with $k = \Theta(n)$ because
\begin{inparaenum}[(i)]
\item the index construction time complexity of \Algkmeans{} is linear to $k$, and
\item the time complexity to identify the partition where the query is located
  takes $\Theta(n)$ time.
\end{inparaenum}
Both \AlgOPQ{} and \Algflannhkm{}/\Algannoy{} achieve this goal indirectly by
using the ideas of subspace partitioning and recursion, respectively.

\begin{figure}[thb]
  \centering
  \subfigure[\small Partition Quality]{
    \label{fig:exp_ana_quality} 
    \includegraphics[width=0.48\linewidth]{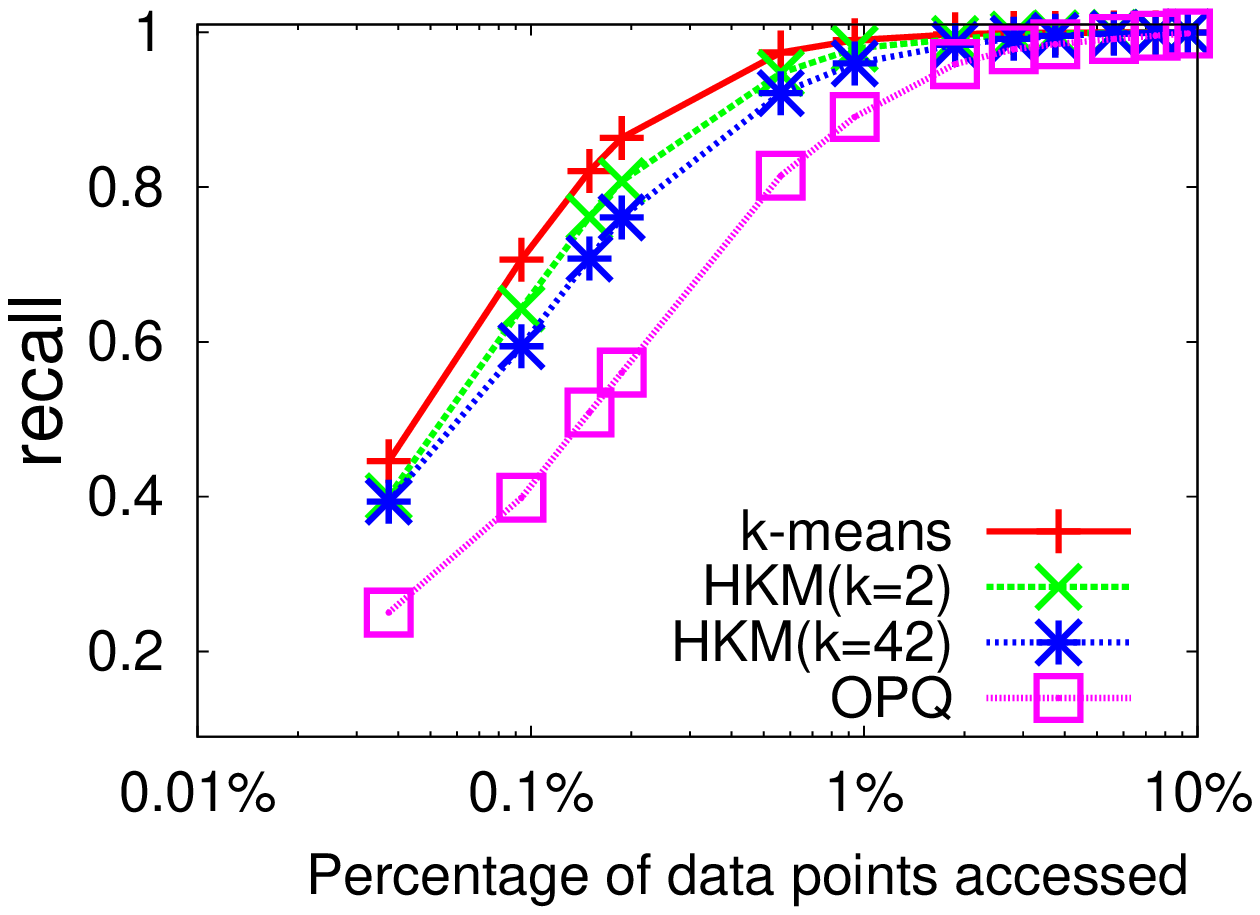}}
  \subfigure[{\small Multiple HKM Trees} ]{
    \label{fig:exp_ana_multiple} 
    \includegraphics[width=0.48\linewidth]{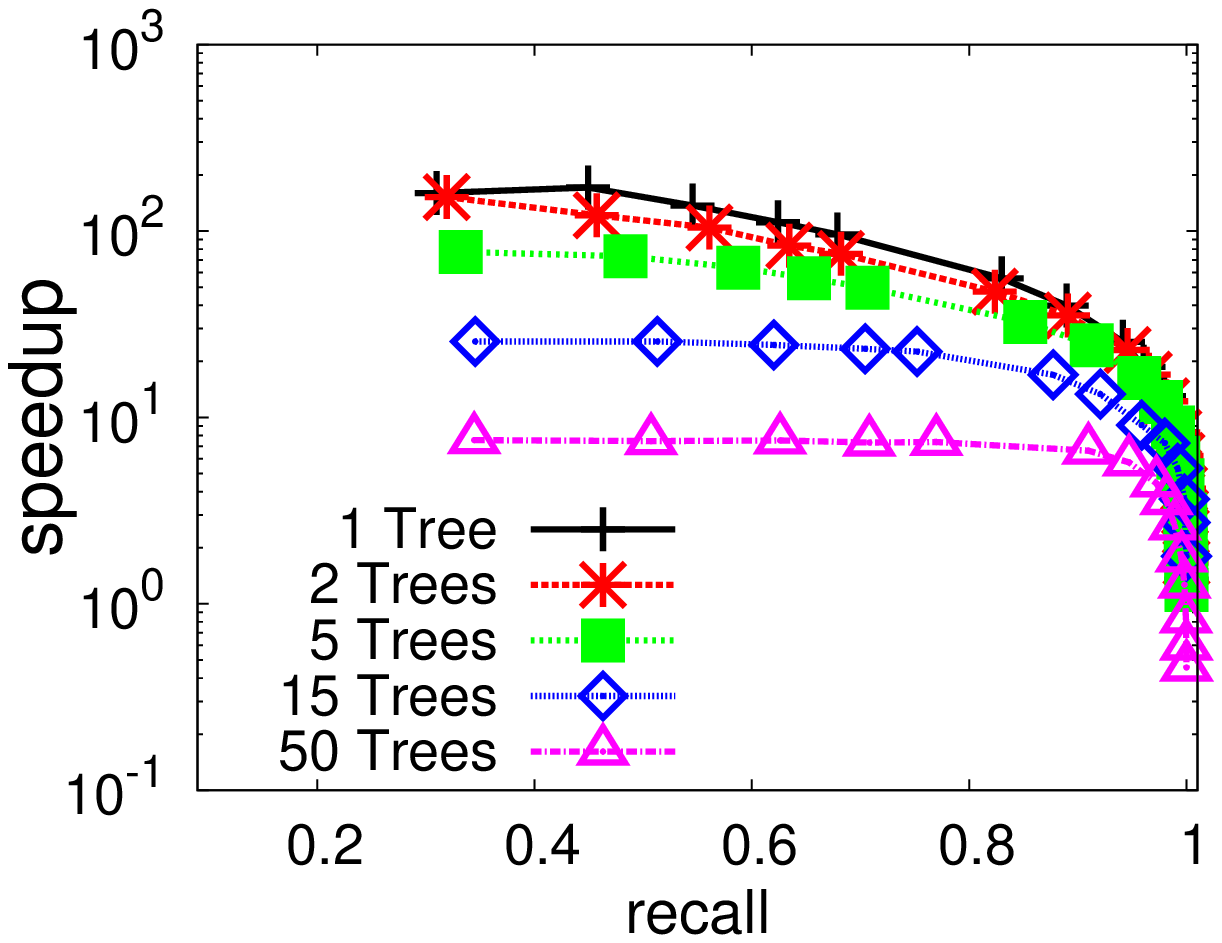}}
  \vspace{-4mm}
  \caption{\small Analyses of Space Partitioning-based Methods}
  \label{fig:exp_ana_partition}
\end{figure}

We perform experiments to understand which idea is more effective. We consider
the goal of achieving \Algkmeans{}-style space partitioning with approximately
the same number of partitions. %
Specifically, we consider the following choices:
\begin{inparaenum}[(i)]
\item Use \Algkmeans{} directly with $k = 18,611$.
\item Use \AlgOPQ{} with $2$ subspaces and each has 256 clusters. The number of effective partitions
  (i.e., non-empty partitions) is also $18,611$.
\item Use \Algflannhkm with branching factor $L=2$ and $L=42$, respectively. We
  also modify the stopping condition so that the resulting trees have $18,000$
  and $17,898$ partitions, respectively.
\end{inparaenum}
Figure~\ref{fig:exp_ana_quality} reports the recalls of the above choices on
\Dataaudio{} against the percentage of data points accessed. %
Partitions are accessed ordered by the distances of their centers to the
query. %
We can see that \AlgOPQ{}-based partition has the worst performance, followed by
(modified) \Algflannhkm{} with $L = 42$, and then $L = 2$. \Algkmeans{} has the
best performance, although the performance differences between the latter three
are not significant. %
Therefore, our analysis suggests that hierarchical \Algkmeans{}-based
partitioning is the most promising direction so far.


Our second analysis is to investigate whether we can further boost the search
performance by using multiple hierarchical \Algkmeans{} trees. Note that
\Algannoy already uses multiple trees and it significantly outperforms a single
hierarchical \Algkmeans{} tree in \Algflannhkm on most of the datasets. %
It is natural to try to enhance the performance of \Algflannhkm in a similar
way.
We set up an experiment to construct multiple \Algflannhkm{} trees.
In order to build different trees, we perform \Algkmeans{} clustering on a set
of random samples of the input data points. Figure~\ref{fig:exp_ana_multiple}
shows the resulting speedup vs recall where we use up to 50 trees. %
We can see that it is not cost-effective to apply multiple trees for
\Algflannhkm{} on \Dataaudio{}, mainly because the trees obtained are still
similar to each other, and hence the advantage of multiple trees cannot offset
the extra indexing and searching overheads. %
Note that \Algannoy{} does not suffer from this problem because $2$-means
partition with limited number of samples and iterations naturally provides
diverse partitions.

\subsection{Neighborhood-based Approach}
\label{subsec:anaysis_nb}

\begin{figure}
  \centering
  \subfigure[\small Min \# of hops to any 20-NNs (\Datayout{})]{
    \label{fig:ana-minhop-youtube}
    \includegraphics[width=.9\linewidth]{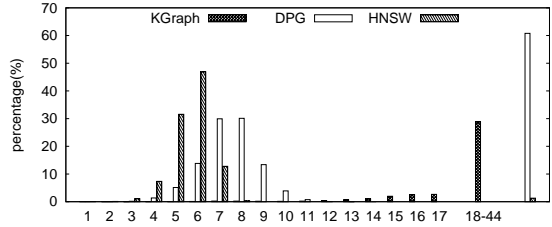}
  }
  \vspace{-4mm}
  \subfigure[\small Min \# of hops to any 20-NNs (\Datagist{}) ]{
    \label{fig:ana-minhop-gist}
    \includegraphics[width=.9\linewidth]{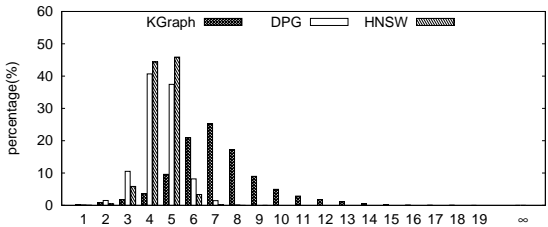}
  }
  \vspace{-1mm}
  \caption{\small minHops Distributions of \Algkgraph{} and \Algdpg{}}
  \label{fig:ana-minhop}
\end{figure}


Our first analysis is to understand why \Algkgraph{}, \Algdpg{} and \Alghnsw{} work very
well (esp.~attaining a very high recall) in most of the datasets.
%
%
Our preliminary analysis indicates that this is because
\begin{inparaenum}[(i)]
\item the \kNN{} points of a query are typically \emph{closely connected} in the
  neighborhood graph, and
\item most points are \emph{well connected} to at least one of the \kNN{} points of a
  query.
\end{inparaenum}
(ii) means there is a high empirical probability that one of the $p$ entry points
selected randomly by the search algorithm can reach one of the \kNN{} points,
and (i) ensures that most of the \kNN{} points can be returned. %
By well connected, we mean there are many paths from an entry point to one of
the \kNN{} point, hence there is a large probability that the ``hills'' on one of
the path is low enough so that the search algorithm won't stuck in the local
minima.

We also investigate why \Algkgraph{} does not work well on some datasets and why
\Algdpg{} and \Alghnsw{} works much better. \Algkgraph{} does not work on \Datayout{} and
\Datagauss{} mainly because both datasets have many well-separated clusters.
Hence, the index of \Algkgraph{} has many disconnected components. Thus,
unless one of the entrance points used by its search algorithm is located in the
same cluster as the query results, there is no or little chance for \Algkgraph{}
to find any near point. %
On the other hand, mainly due to the diversification step and the use of the reverse
edges in \Algdpg{}, there are edges linking points from different clusters,
hence resulting in much improved recalls.
Similarly, in \Alghnsw{}, the edges are also well linked.

For example, we perform the experiment where we use the NN of the query as the entrance point of the
search on \Datayout{}. \Algkgraph{} then achieves 100\%
recall. In addition, we plot the distribution of the minimum number of hops
(minHops) for a data point to reach
any of the \kNN{} points of a query\footnote{The figures are very similar for
  all the tested queries.} for the indexes of \Algkgraph{} and \Algdpg{} on
\Datayout{} and \Datagist{} in Figure~\ref{fig:ana-minhop}. We can observe that
\begin{itemize}
\item For \Algkgraph, there are a large percentage of data points that cannot
  reach any \kNN{} points (i.e., those corresponding to $\infty$ hops) on
  \Datayout{} (60.38\%), while the percentage is low on \Datagist{} (0.04\%).
\item The percentages of the  $\infty$ hops are much lower for \Algdpg{} (1.28\% on \Datayout{} and
  0.005\% on \Datagist{}).
\item There is no $\infty$ hops for \Alghnsw{} on both datasets.
\item \Algdpg{} and \Alghnsw{} have much more points with small minHops than \Algkgraph{}, which
  contributes to making it easier to reach one of the \kNN{} points. Moreover,
  on \Datayout{}, \Alghnsw{} has the most points with small minHops over three
  algorithms, which results in a better performance as shown in
  Figure~\ref{fig:exp_final_recall}(g).
\end{itemize}




\subsection{Comparisions with Prior Benchmarks}
\label{sec:comp-with-prior}

We have verified that the performance results obtained in our evaluation
generally match prior evaluation results, and we can satisfactorily explain most
of the few discrepancies.

\myparagraph{Comparison with \annbenchmark's Results.} While the curves in both
evaluations have similar shapes, the relative orders of the best performing
methods are different. This is mainly due to the fact that we turned off all
hardware-specific optimizations in the implementations of the methods.
Specifically, we disabled distance computation using SIMD and multi-threading in
\Algkgraph{}, \texttt{-ffast-math} compiler option in \Algannoy{},
multi-threading in \Algflann{}, and distance computation using SIMD,
multi-threading, prefetching technique and \texttt{-Ofast} compiler option in methods implemented in
the NonMetricSpaceLib, i.e., \Algsw{}, \AlgNAPP, \Algvptree{} and \Alghnsw{}). In addition, we disabled the optimized search implementation used in \Alghnsw{}.
We confirm that the results resemble \annbenchmark{}'s more when these
optimizations are turned on.

Disabling these hardware-specific optimizations allows us to gain more insights
into the actual power of the algorithms. In fact, the optimizations can be
easily added to the implementations of all the algorithms.

\myparagraph{\Algkgraph{} and \Algsw{}.}
\Algkgraph{} was ranked very low in the \annbenchmark{}
study~\cite{misc:url/ann-benchmark} possibly due to an error in the
test\footnote{Our evaluation agrees with \\
\url{http://www.kgraph.org/index.php?n=Main.Benchmark}}.

\myparagraph{\Algannoy{}.} %
We note that the performance of latest version of \Algannoy{} (based on
randomized hierarchical $2$-means trees) is noticeably better than its earlier
versions (based on multiple heuristic RP-trees), and this may affect prior
evaluation results.



\section{CONCLUSION AND FUTURE WORK}
\label{sec:con}
NNS is an fundamental problem with both significant theoretical values and
empowering a diverse range of applications.
%
It is widely believed that there is no practically competitive algorithm to
answer exact NN queries in sublinear time with linear sized index. A natural
question is whether we can have an algorithm that \emph{empirically} returns
\emph{most of} the \kNN{} points in a \emph{robust} fashion by building an index
of size $O(n)$ and by accessing at most $\alpha n$ data points, where $\alpha$
is a small constant (such as 1\%).

In this paper, we evaluate many state-of-the-art algorithms proposed in
different research areas and by practitioners in a comprehensive manner. We
analyze their performances and give practical recommendations.


Due to various constraints, the study in this paper is inevitably limited. In
our future work, we will
\begin{inparaenum}[(i)]
\item use larger datasets (e.g., $100$M+ points);
\item consider high dimensional sparse data;
\item use more complete, including exhaustive method, to tune the algorithms;
\item consider other distance metrics, as in~\cite{DBLP:journals/pvldb/NaidanBN15}.
\end{inparaenum}

Finally, our understanding of high dimensional real data is still vastly
inadequate. This is manifested in many heuristics with no reasonable
justification, yet working very well in \emph{real} datasets.
We hope that this study opens up more questions that call for innovative
solutions by the entire community.



{
\bibliographystyle{abbrv}
\bibliography{ref}
}

\newpage
\newpage
\appendix
\section{Parameter Setting}
\label{app:parameter}
In this section, we carefully tune all algorithms to achieve a good search performance with reasonable index space and construction time overheads. By default, we use the number of data points verified (i.e., computing the exact distance to the query), denoted by $N$, to achieve the trade-off between the search quality and search speed unless specially mentioned.

\subsection{SRS}

We test SRS in two versions: External-Memory version and In-Memory version. In this paper, we do not use the early termination test and stop the searching when it has accessed enough points. Meanwhile, the approximation ratio $c$ is set to 4 and the page size for external memory version is $4096$. For the sake of fairness, the success probability of all data-independent methods is set to $1/2 - 1/e $. Therefore, there are two parameters $T'$(the maximum number of data points accessed in the query processing), $m$ (the number of dimension of projected space) for SRS algorithm. We change the number of the accessed data points to tune the trade-off between the search quality and search speed under a certain $m$.

\myparagraph{External-Memory Version}

\begin{figure}[tbh]
\begin{minipage}[t]{1.0\linewidth}
\centering
\subfigure[\small Audio ]{
      \label{fig:exp_srs_ex_audio} 
      \includegraphics[width=0.48\linewidth]{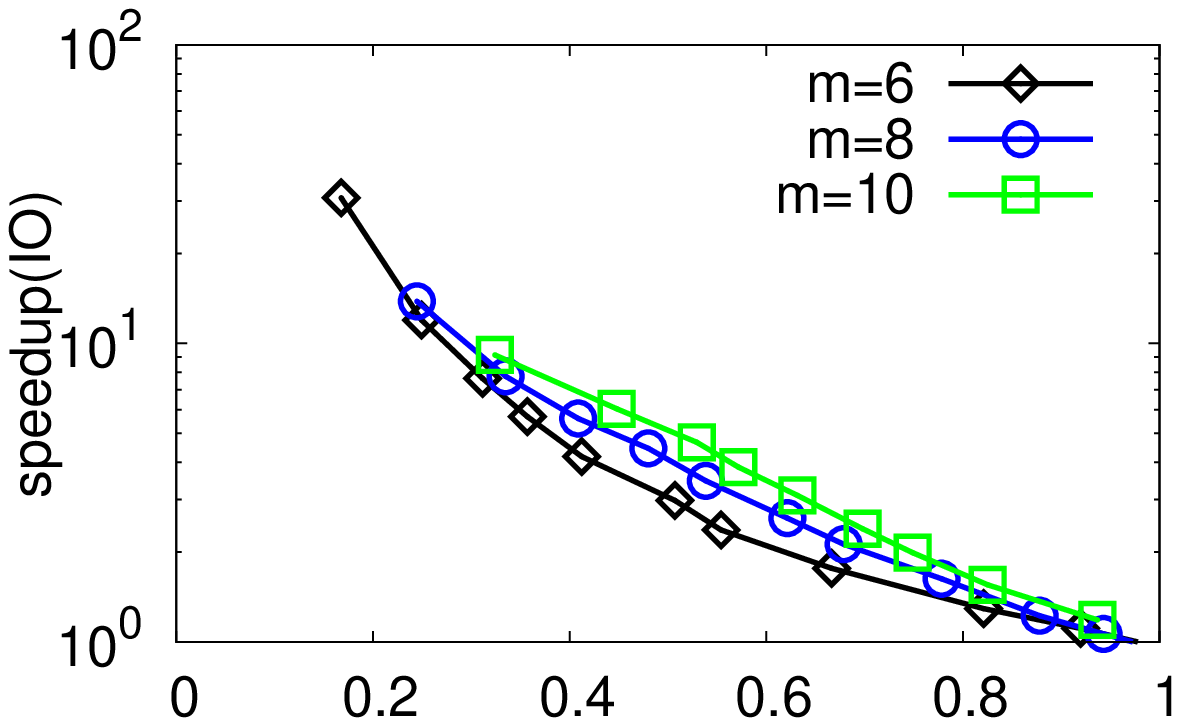}}
\subfigure[\small Sift ]{
      \label{fig:exp_srs_ex_sift} 
      \includegraphics[width=0.48\linewidth]{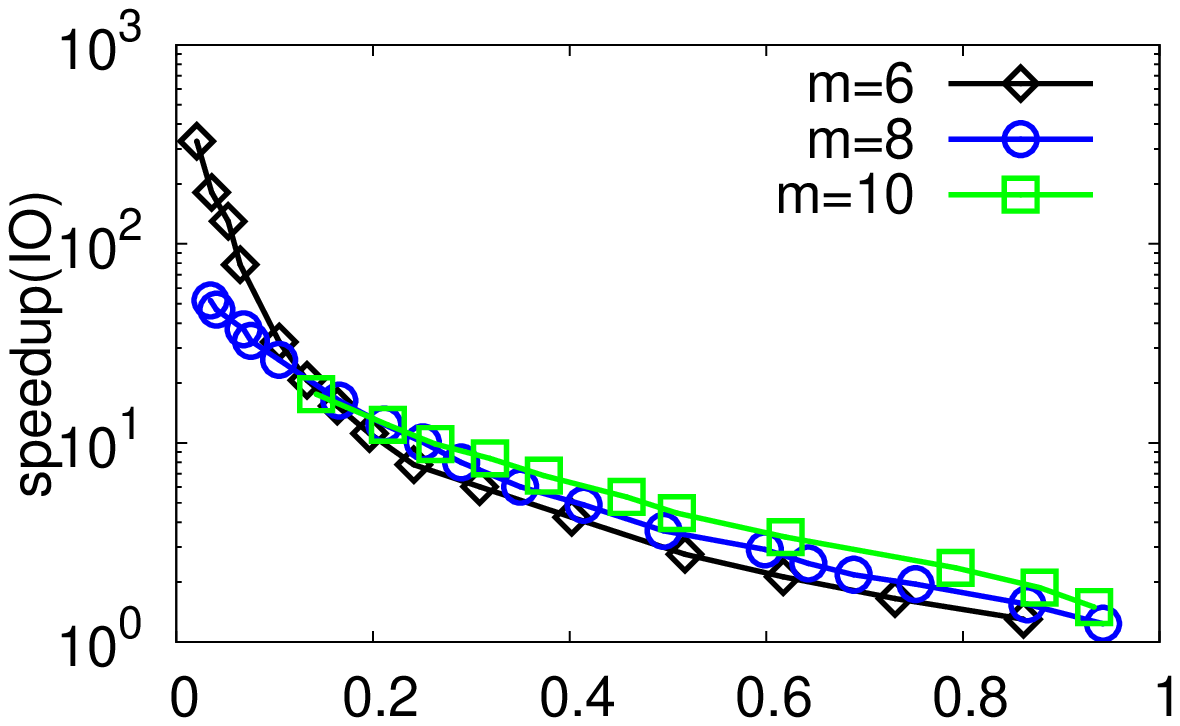}}
\end{minipage}%
\vspace{-4mm}
\caption{\small IO Speedup vs Recall for Diff $m$ (Ex-Memory SRS)}
\label{fig:exp_srs_ex}
\end{figure}

Figure \ref{fig:exp_srs_ex} plots the changes of the search performance with different projection dimensions. As the increase of the projection dimensionality $m$, SRS could achieve a better performance.

\myparagraph{In-Memory version}

\begin{figure}[tbh]
\begin{minipage}[t]{1.0\linewidth}
\centering
\subfigure[\small Audio ]{
      \label{fig:exp_srs_in_audio} 
      \includegraphics[width=0.48\linewidth]{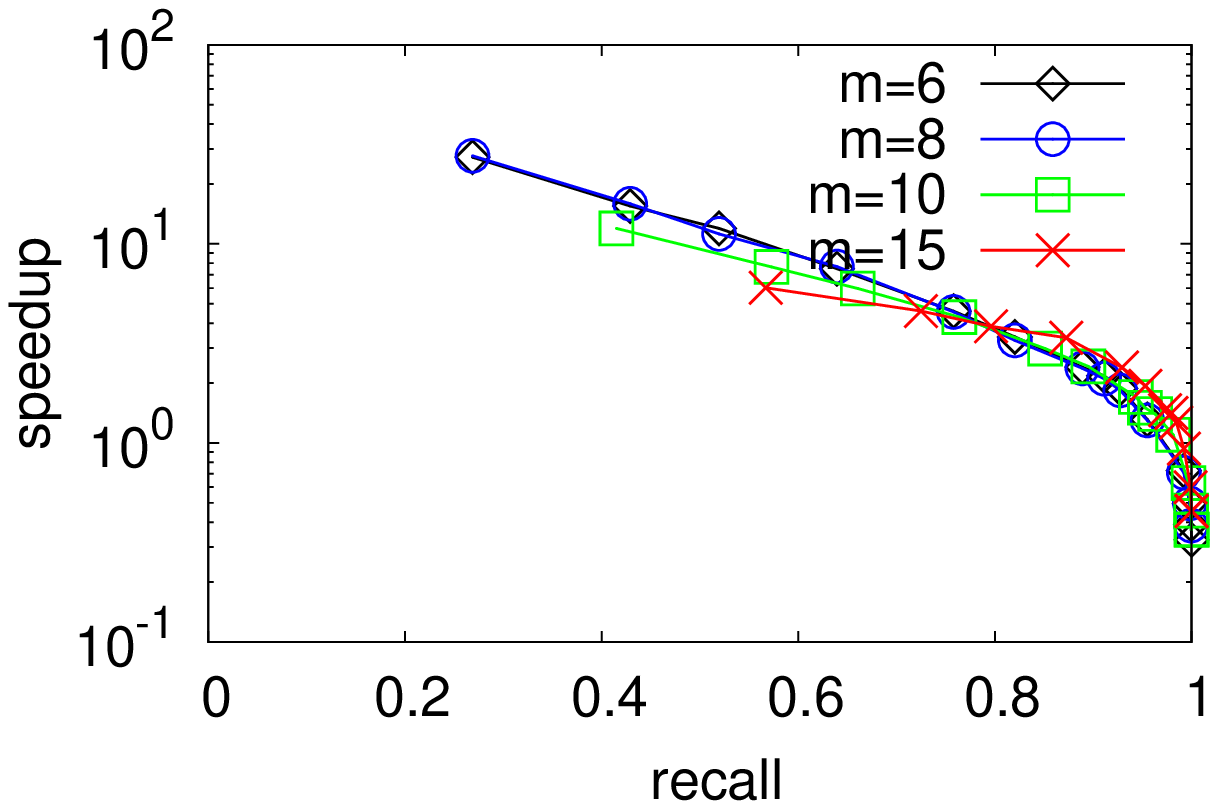}}
\subfigure[\small Sift ]{
      \label{fig:exp_srs_in_sift} 
      \includegraphics[width=0.48\linewidth]{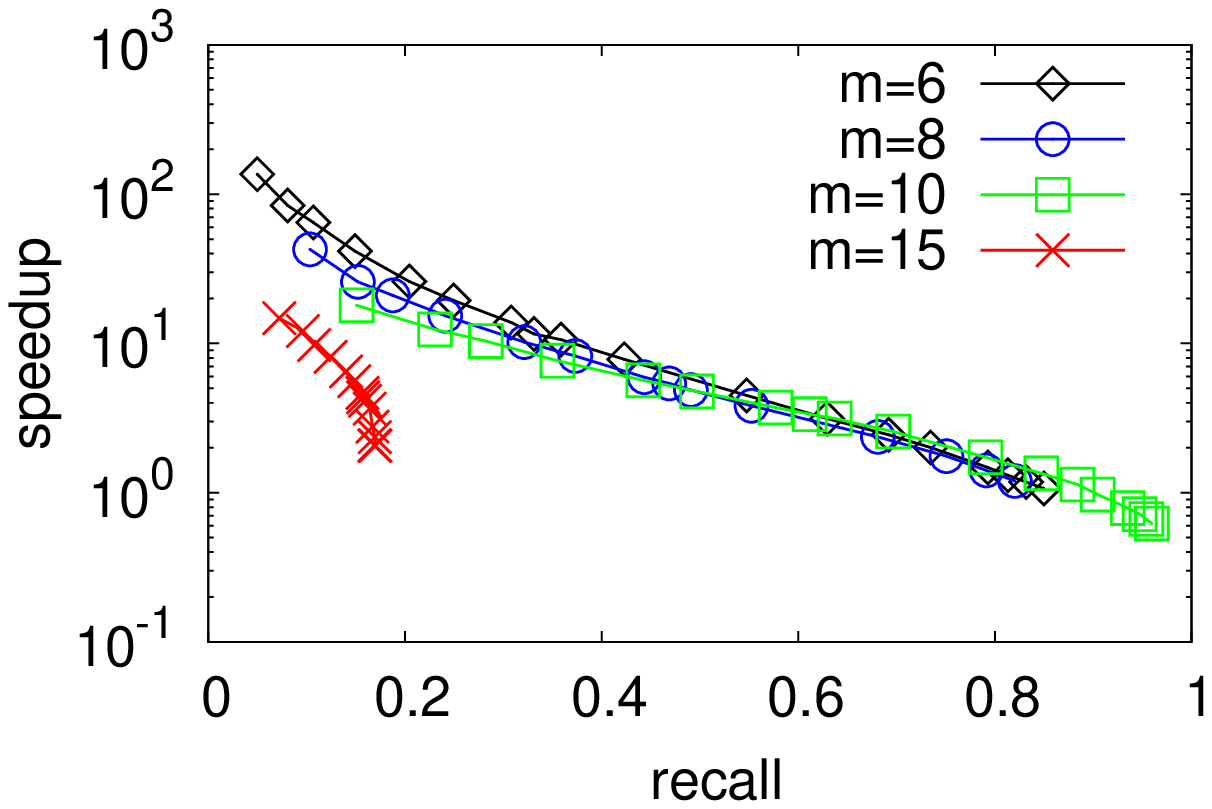}}
\subfigure[\small Mnist ]{
      \label{fig:exp_srs_in_MNIST} 
      \includegraphics[width=0.48\linewidth]{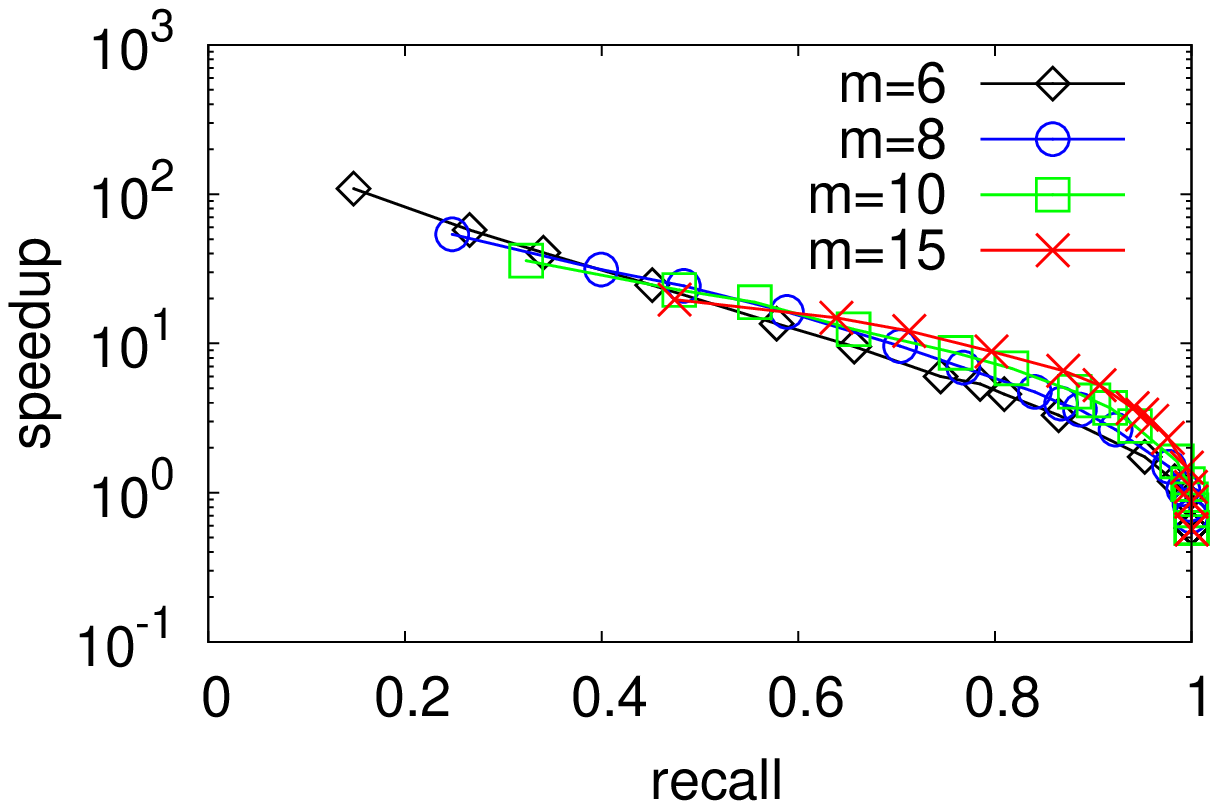}}
\subfigure[\small Gist ]{
      \label{fig:exp_srs_in_gist} 
      \includegraphics[width=0.48\linewidth]{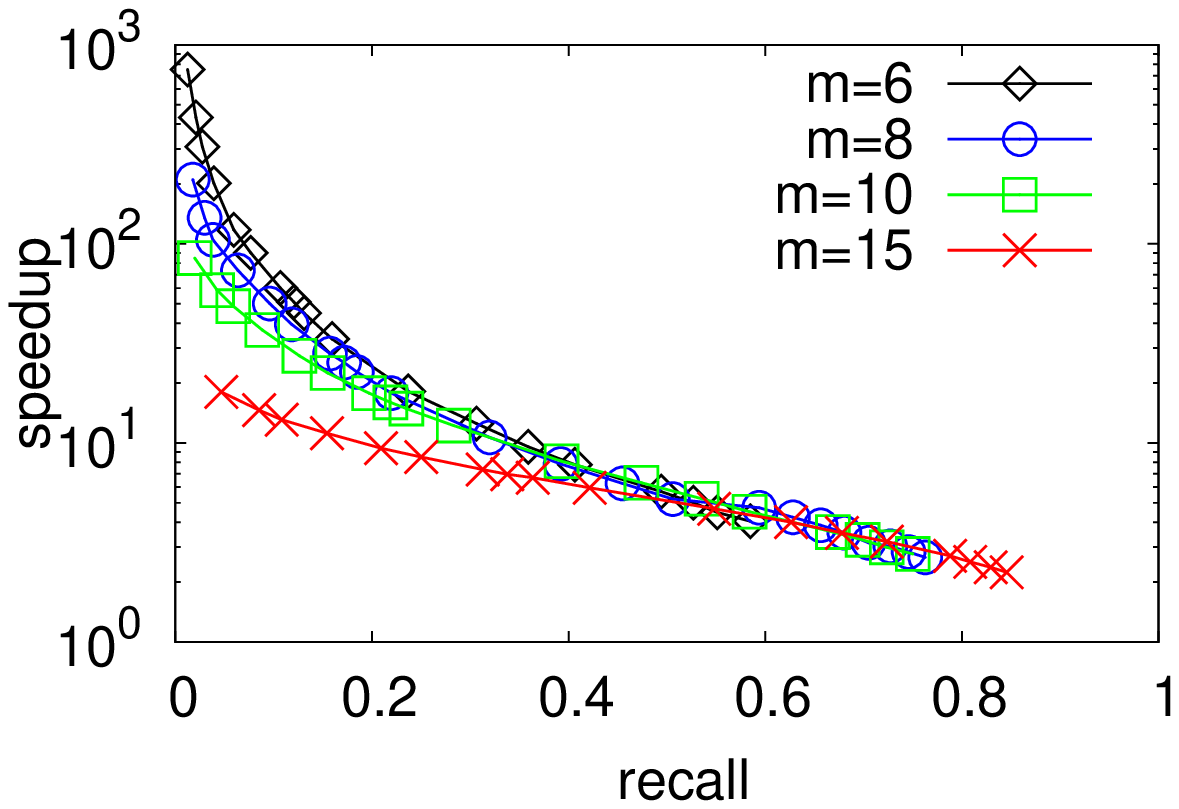}}
\end{minipage}%
\vspace{-4mm}
\caption{\small Speedup vs Recall for Diff $m$ (In-Memory SRS)}
\label{fig:exp_srs_in}
\end{figure}

For In-Memory version, we compare the speedup using the ratio of the brute-force time and the search time. From figure \ref{fig:exp_srs_in}, we can see as the increase of the value of $m$, higher speedup could be achieved for high recall while the search speed would be more faster when one requires a relatively moderate value of recall. Considering various aspects, values of $m$ from 8 to 10 provide a good trade-off.

\subsection{QALSH}

Because QALSH didn't release the source code of In-Memory version, we only test the performance of External-Memory version. We use the default setting for QALSH and tune the value of $c$(approximation ratio) to obtain different search performance.

\begin{figure}[tbh]
\begin{minipage}[t]{1.0\linewidth}
\centering
\subfigure[\small Audio ]{
      \label{fig:exp_srs_in_sift} 
      \includegraphics[width=0.48\linewidth]{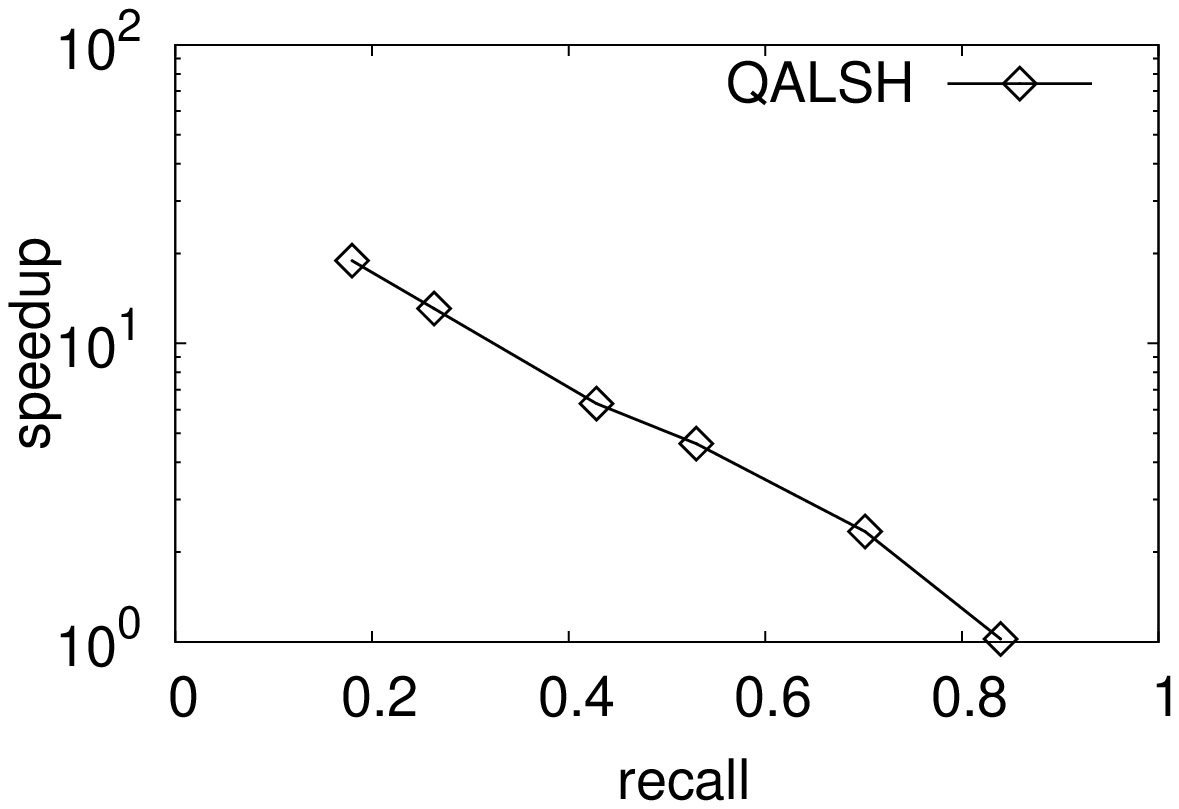}}
\subfigure[\small Sift ]{
      \label{fig:exp_srs_in_sift} 
      \includegraphics[width=0.48\linewidth]{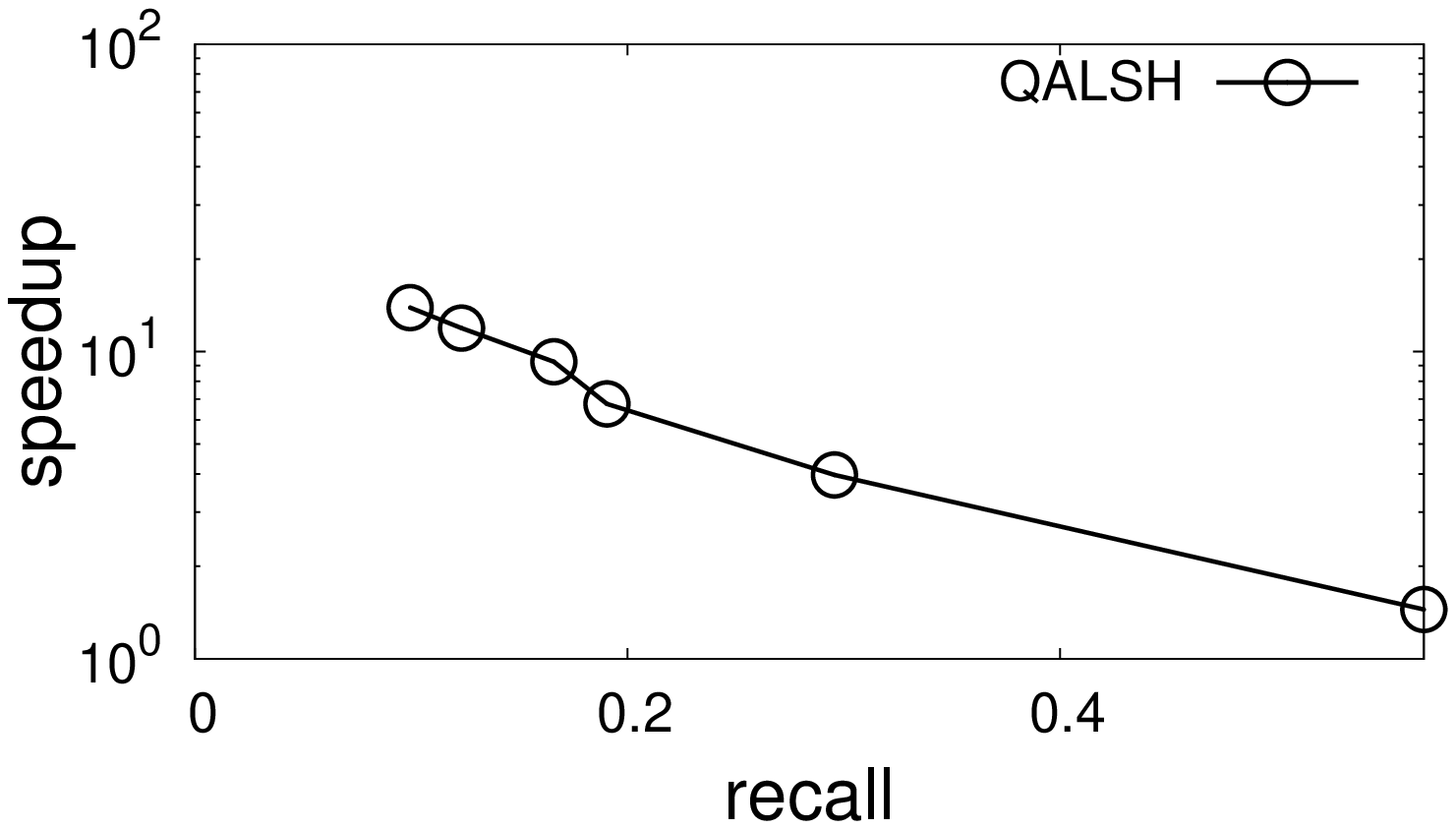}}
\end{minipage}%
\vspace{-4mm}
\caption{\small IO Speedup vs Recall for QALSH}
\label{fig:exp_srs_qalsh}
\end{figure}

%
%
%

\subsection{Scalable Graph Hashing}

In SGH, we use the default setting recommended by the author. For kernel feature construction, we use Gaussian kernel and take 300 randomly sampled  points as kernel bases. Hence, we compare the search accuracies with different hashcode length $b$. For most of the datasets, $b=8$ always obtains the worst search performance compared with larger hashcodes and the best speedup could be achieved with 128 bits.

\begin{figure}[tbh]
\begin{minipage}[t]{1.0\linewidth}
\centering
\subfigure[\small Audio ]{
      \label{fig:exp_SGH_in_audio} 
      \includegraphics[width=0.48\linewidth]{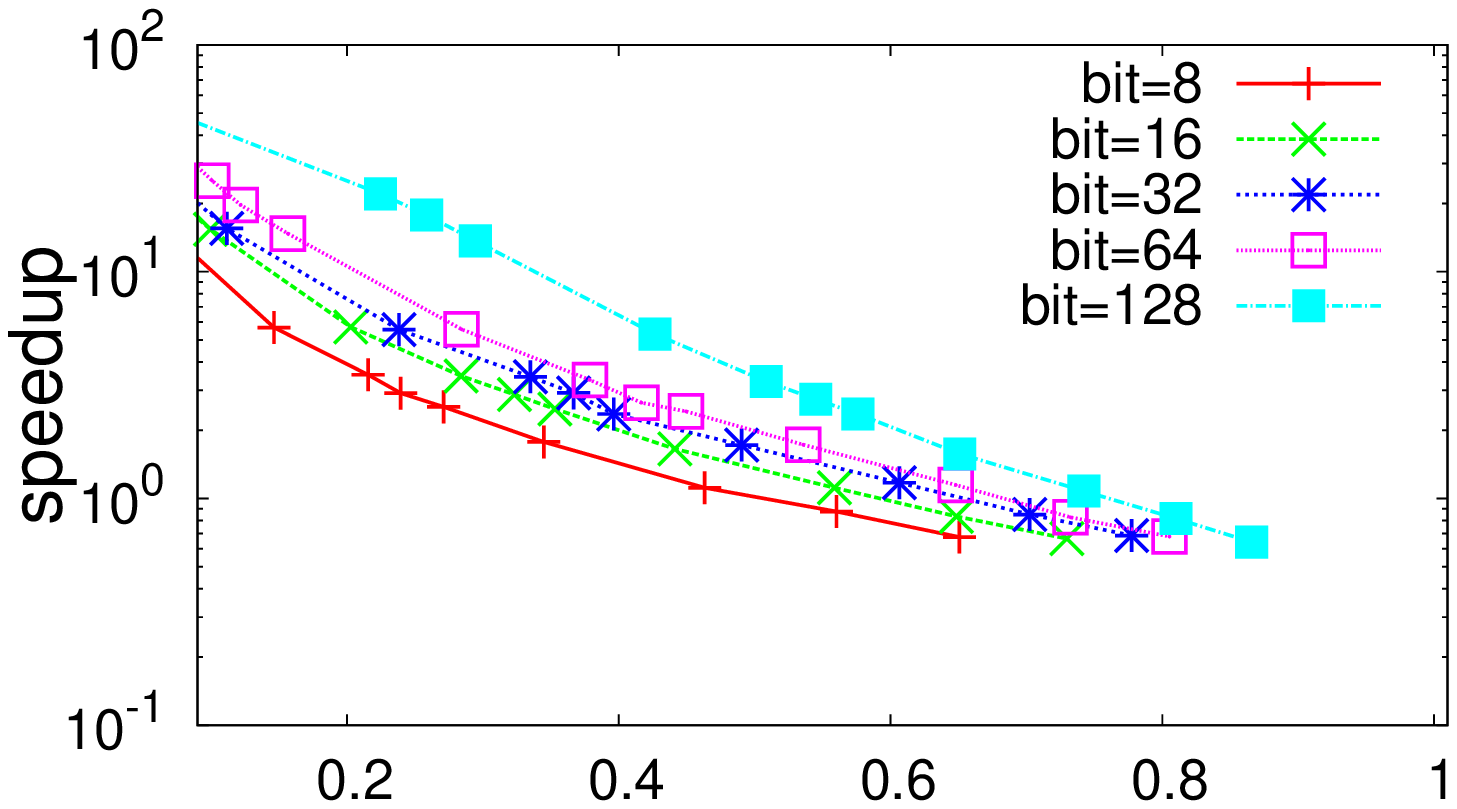}}
\subfigure[\small Cifa ]{
      \label{fig:exp_SGH_in_cifar10} 
      \includegraphics[width=0.48\linewidth]{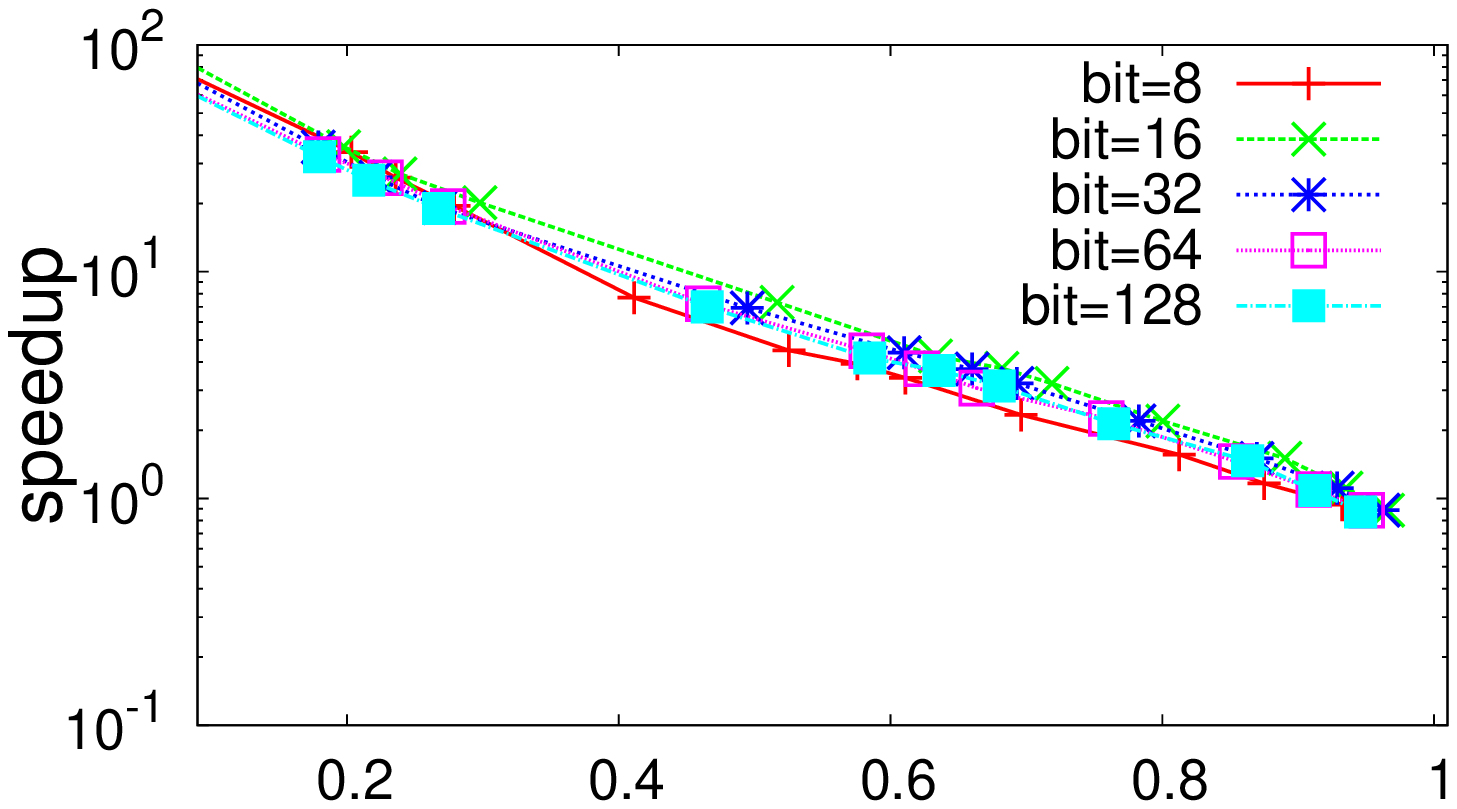}}
\subfigure[\small Deep ]{
      \label{fig:exp_SGH_in_deep} 
      \includegraphics[width=0.48\linewidth]{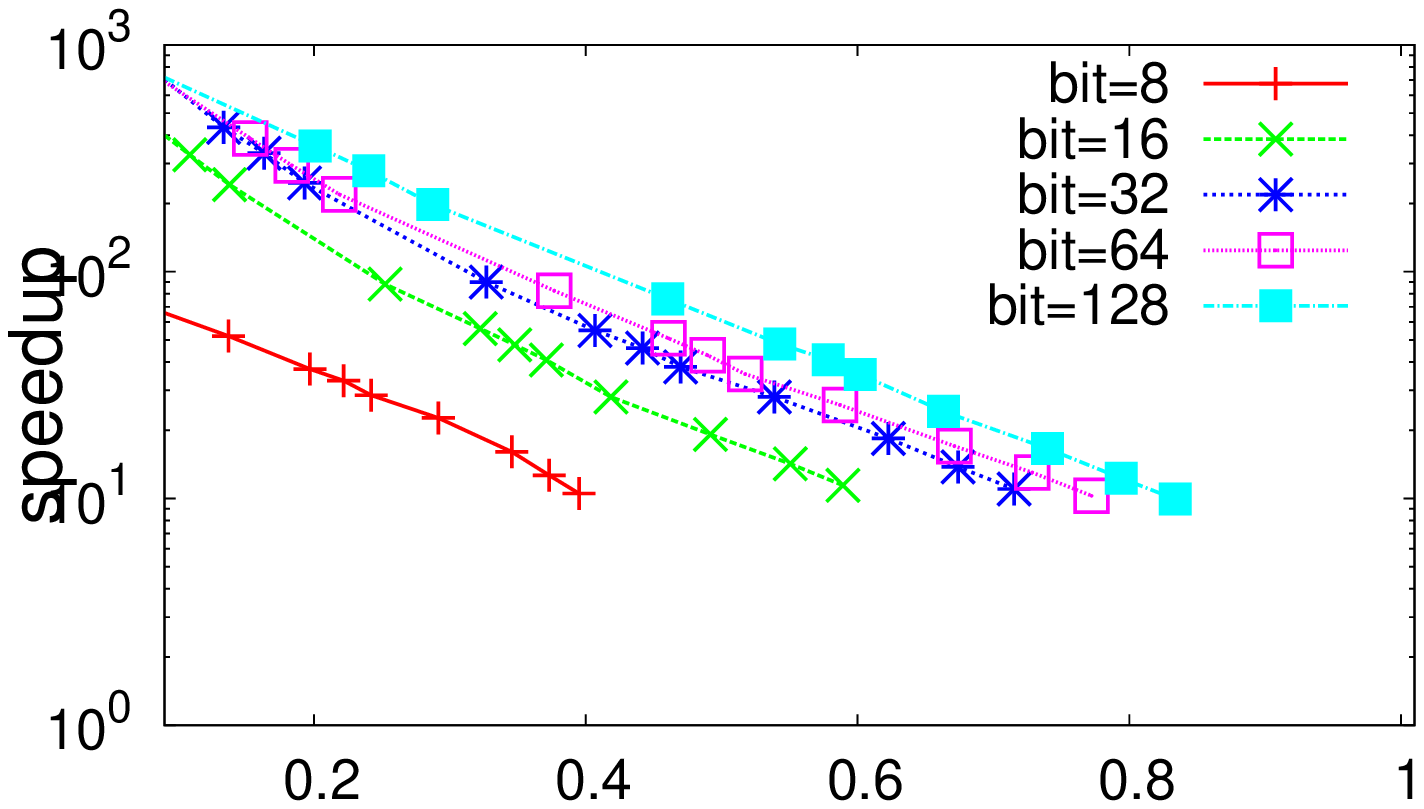}}
\subfigure[\small Yout ]{
      \label{fig:exp_SGH_in_youtube} 
      \includegraphics[width=0.48\linewidth]{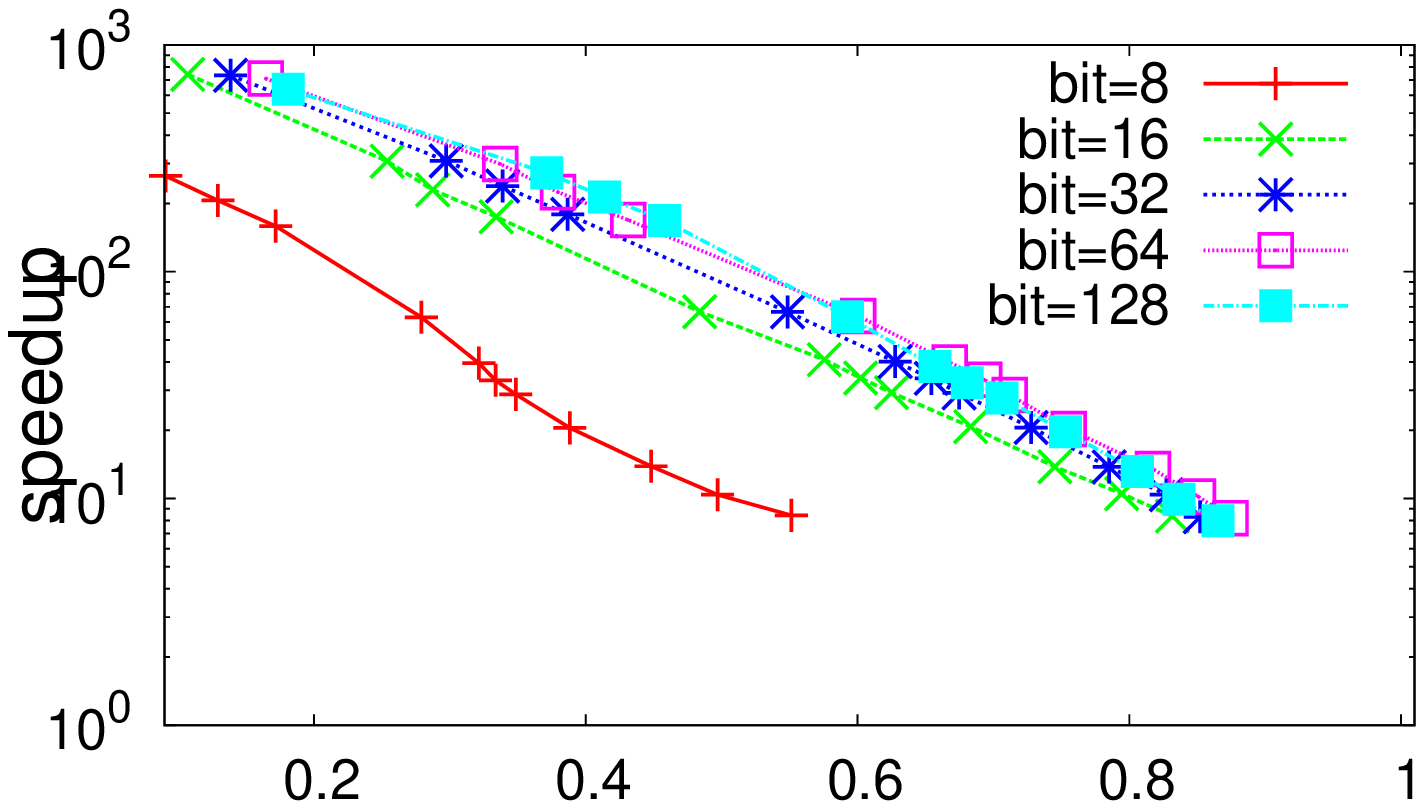}}
\end{minipage}%
\vspace{-4mm}
\caption{\small Speedup vs Recall for Diff $b$ (\textbf{SGH})}
\label{fig:exp_SGH}
\end{figure}

\subsection{Anchor Graph Hashing}
To run 1-AGH and 2-AGH, we have to fix three parameters: $m$ (the number of anchor), $s$ (number of nearest anchors need to be considered for each point) and hash code length $b$. We focus $m$ to 300 and $s$=5.

Following are the comparisons for 1-AGH and 2-AGH. We first show the search performance for a single-layer AGH and two-layer AGH with different length of hashcode. For most datasets, it seems using longer hashcode could obtain the much higher performance. Hence, we only compare the performance for $b=64$ and $b=128$ on both layer AGH. We can observe that 2-AGH provides superior performance compared to 1-AGH for majority of datasets.

\begin{figure}[tbh]
\centering
\subfigure[\small Sift for 1-AGH ]{
      \label{fig:exp_AGH_in_audio} 
      \includegraphics[width=0.48\linewidth]{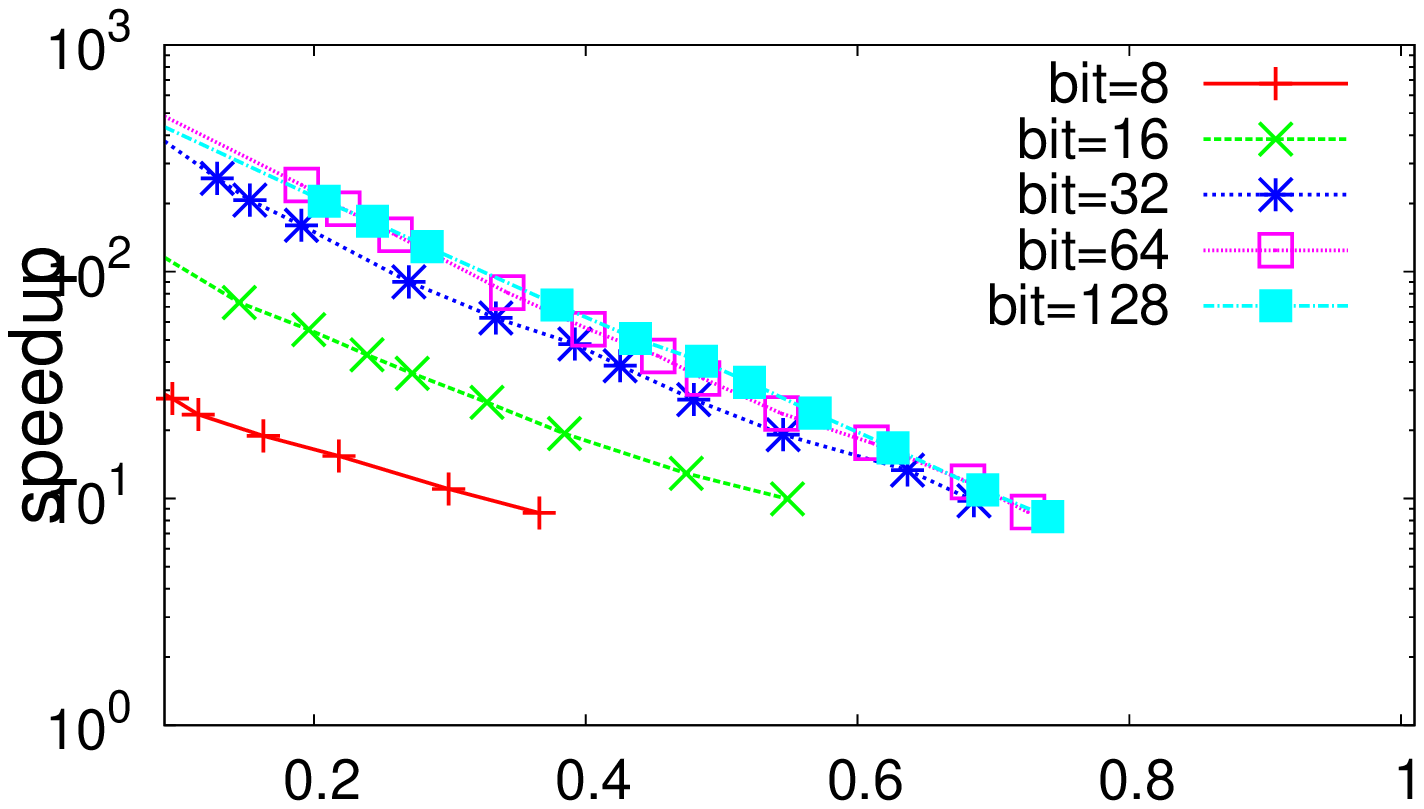}}
\subfigure[\small Yout for 1-AGH ]{
      \label{fig:exp_AGH_in_cifar10} 
      \includegraphics[width=0.48\linewidth]{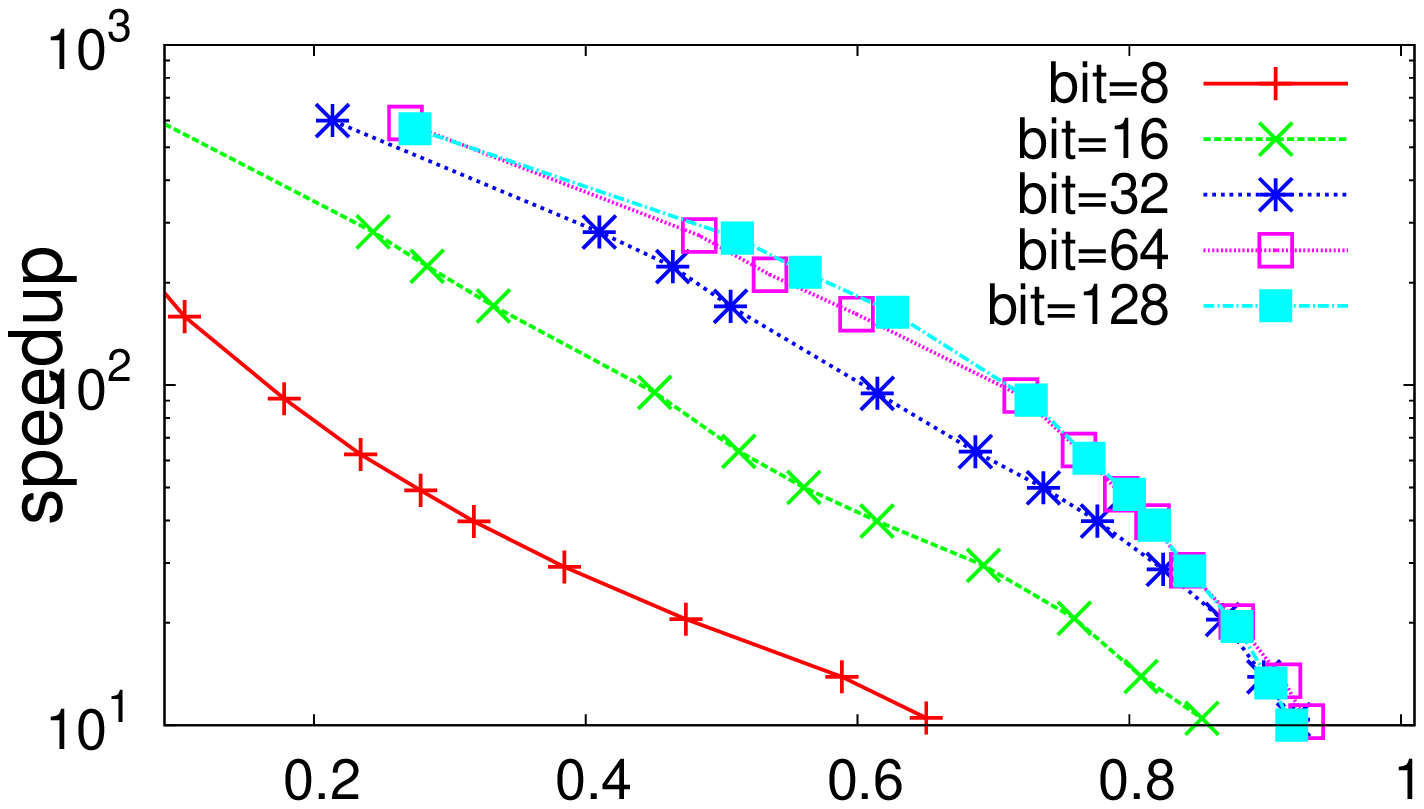}}
\subfigure[\small Sift for 2-AGH ]{
      \label{fig:exp_AGH_in_cifar10} 
      \includegraphics[width=0.48\linewidth]{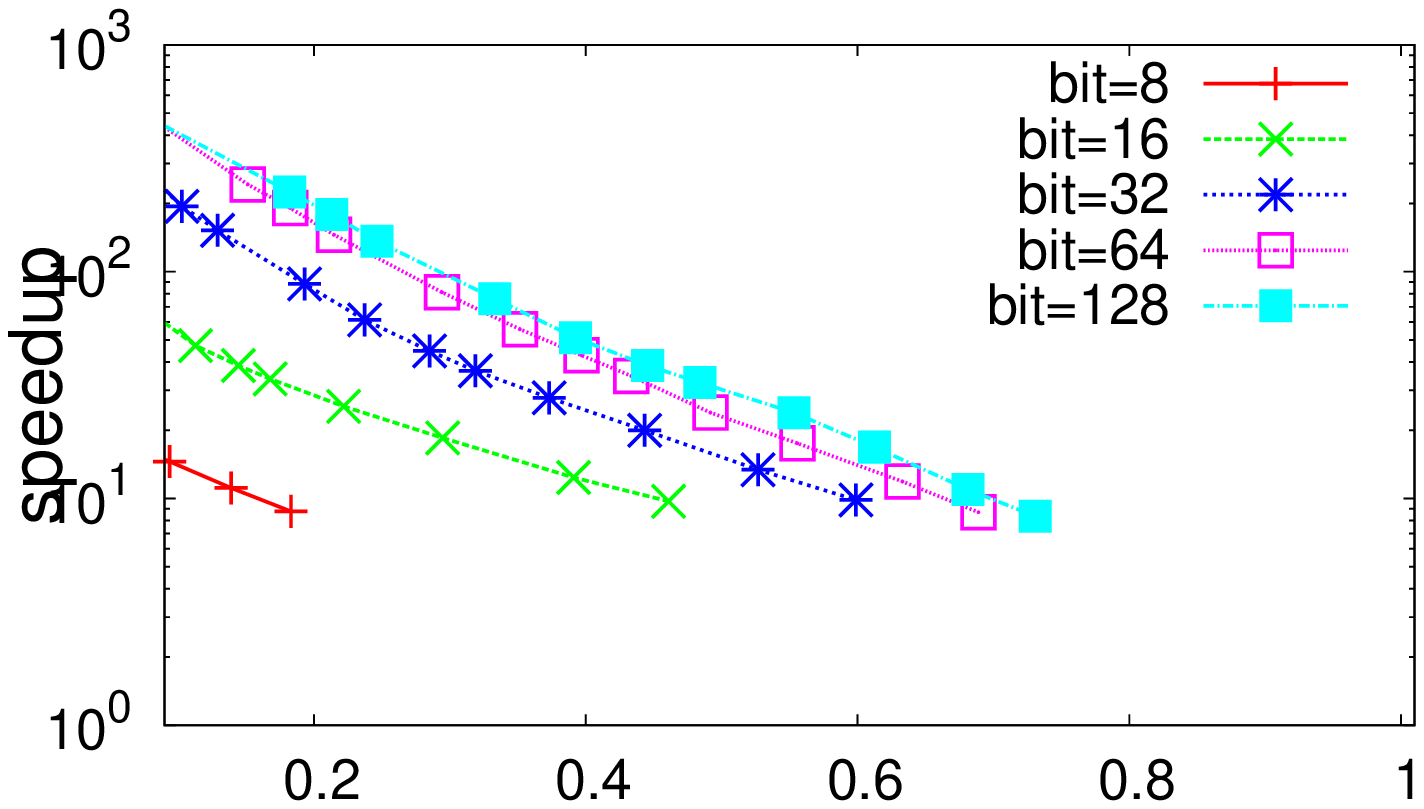}}
\subfigure[\small Yout for 2-AGH ]{
      \label{fig:exp_AGH_in_cifar10} 
      \includegraphics[width=0.48\linewidth]{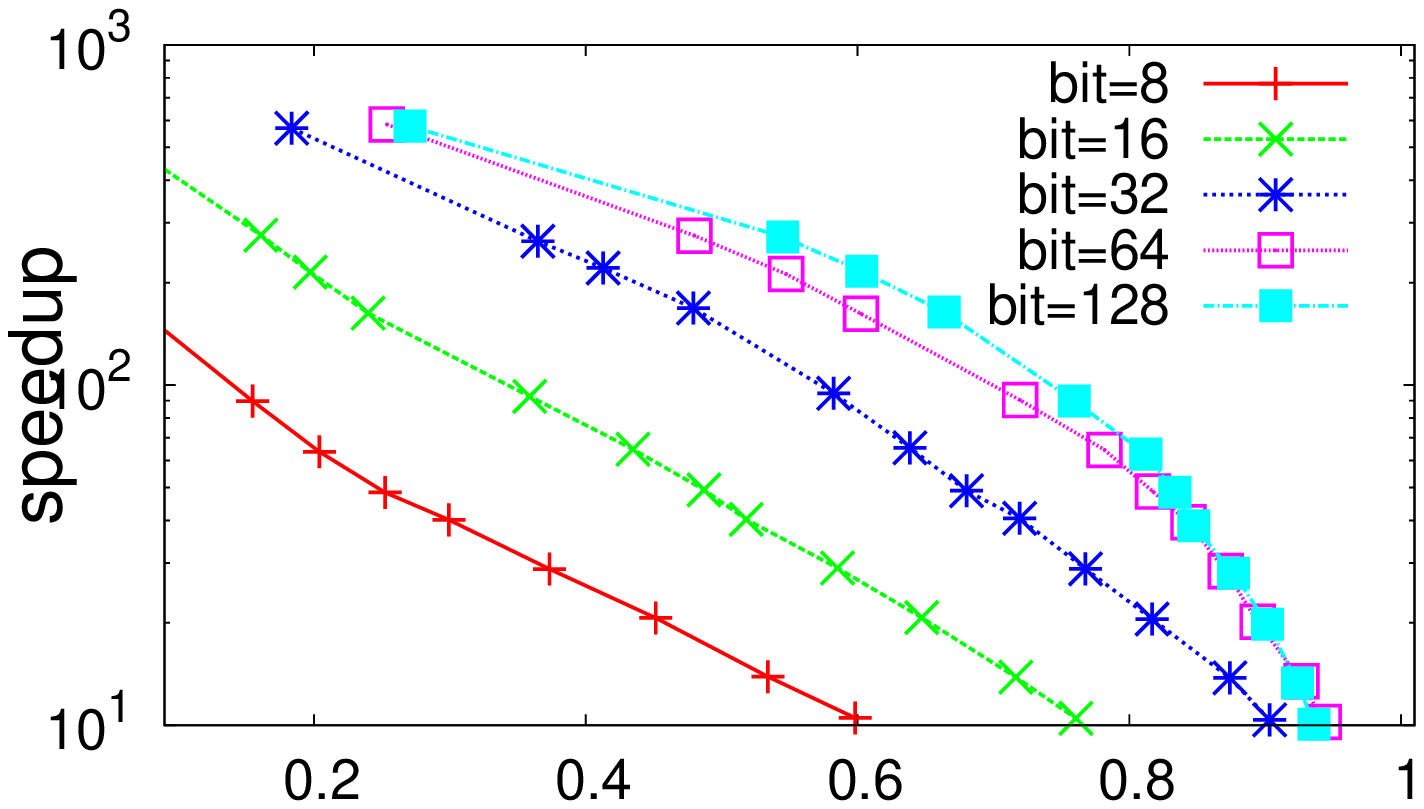}}
\vspace{-4mm}
\caption{\small Speedup vs Recall for Diff $b$ (\textbf{AGH})}
\label{fig:exp_AGH}
\end{figure}

\begin{figure}[tbh]
\centering
\subfigure[\small Sift ]{
      \label{fig:exp_SGH_in_deep} 
      \includegraphics[width=0.48\linewidth]{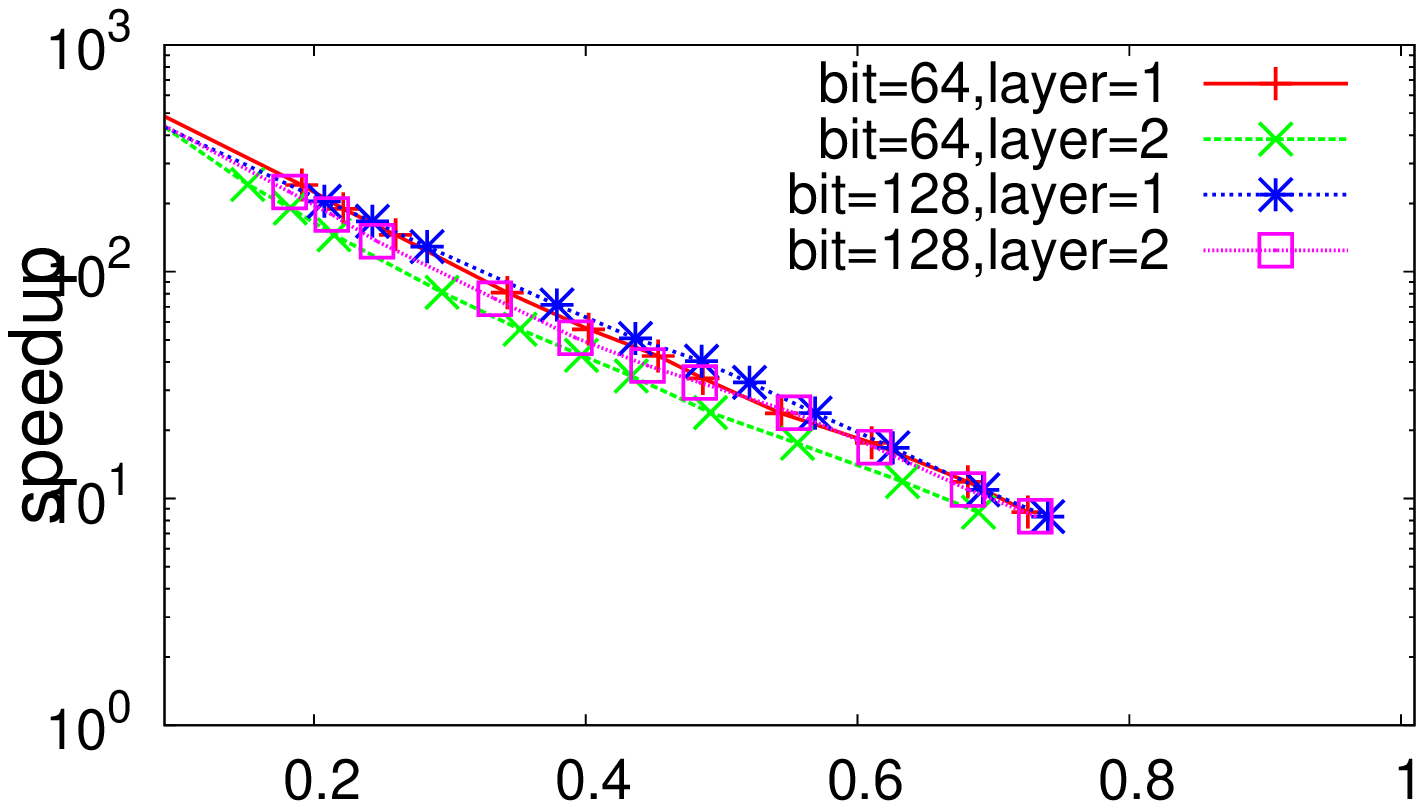}}
\subfigure[\small Yout ]{
      \label{fig:exp_SGH_in_deep} 
      \includegraphics[width=0.48\linewidth]{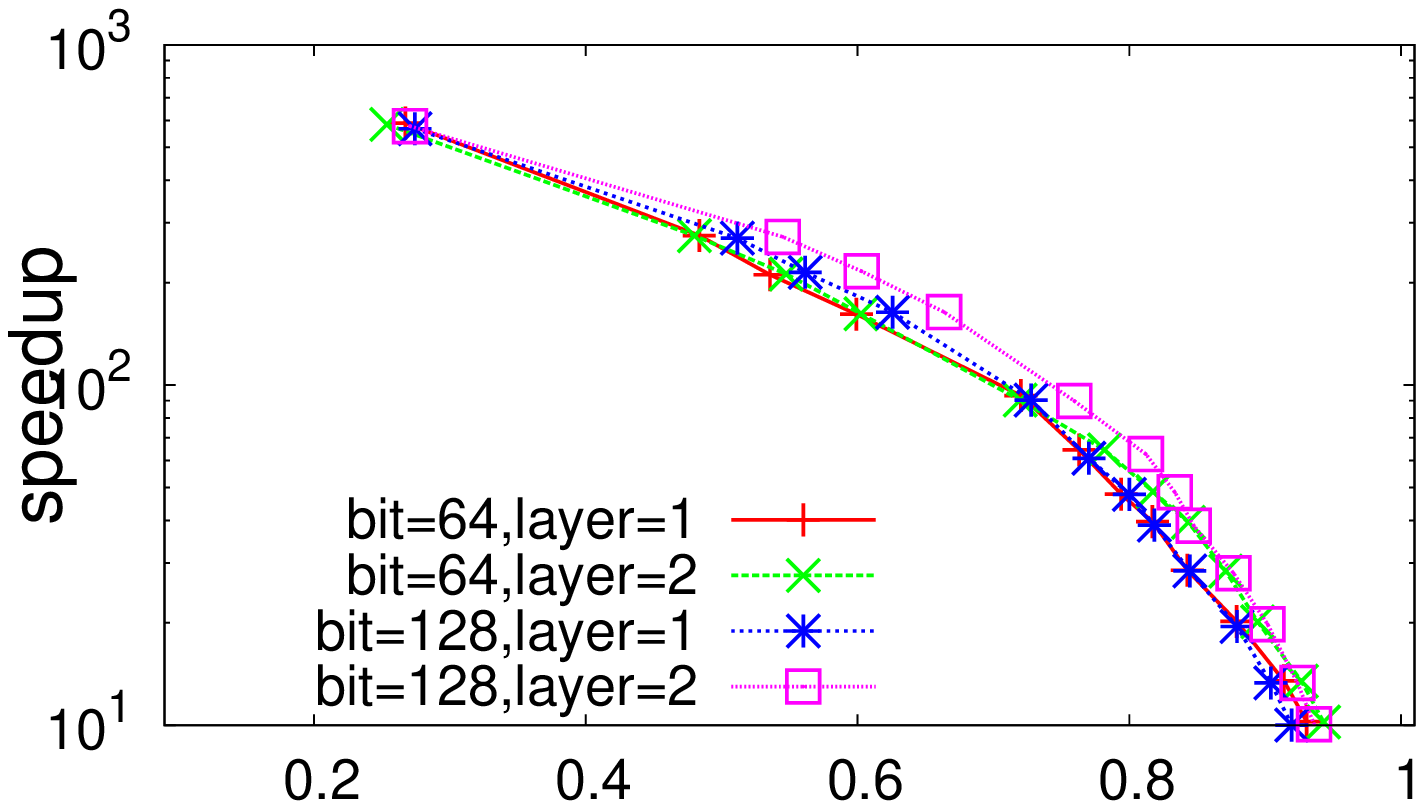}}
\vspace{-4mm}
\caption{\small Speedup vs Recall for \textbf{1-AGH} and \textbf{2-AGH}}
\label{fig:exp_AGH}
\end{figure}

\subsection{Neighbor-Sensitive Hashing}

we set the number of pivots $m$ to $4b$ where $b$ is the length of hash code and used k-means++ to generate the pivots as described by the authors. The value of $\eta$ was set to 1.9 times of the average distance from a pivot to its closest pivot. We tune the value of $b$ to get the best performance.

\begin{figure}[tbh]
\begin{minipage}[t]{1.0\linewidth}
\centering
\subfigure[\small Audio ]{
      \label{fig:exp_NSH_in_audio} 
      \includegraphics[width=0.48\linewidth]{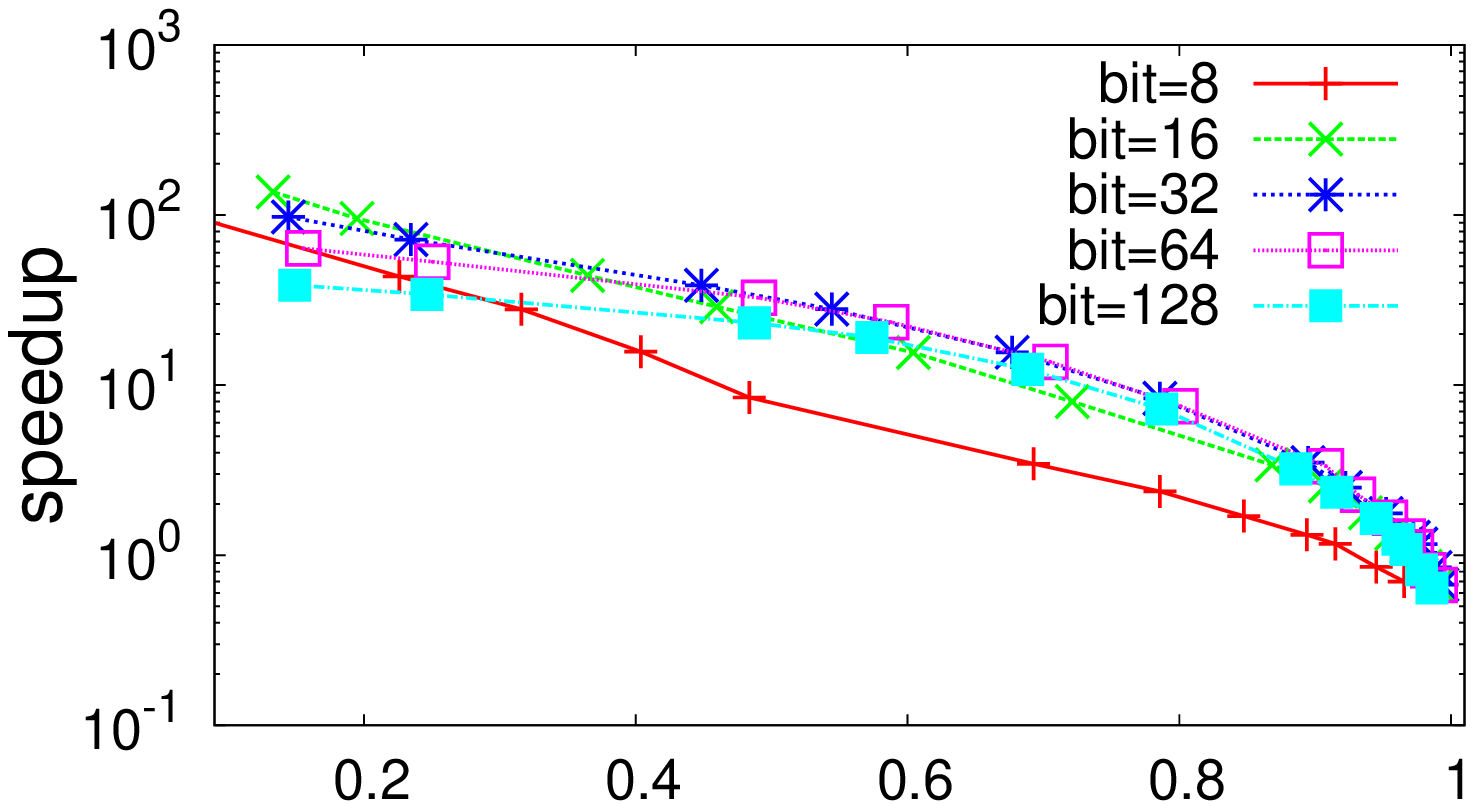}}
\subfigure[\small Sift ]{
      \label{fig:exp_NSH_in_sift} 
      \includegraphics[width=0.48\linewidth]{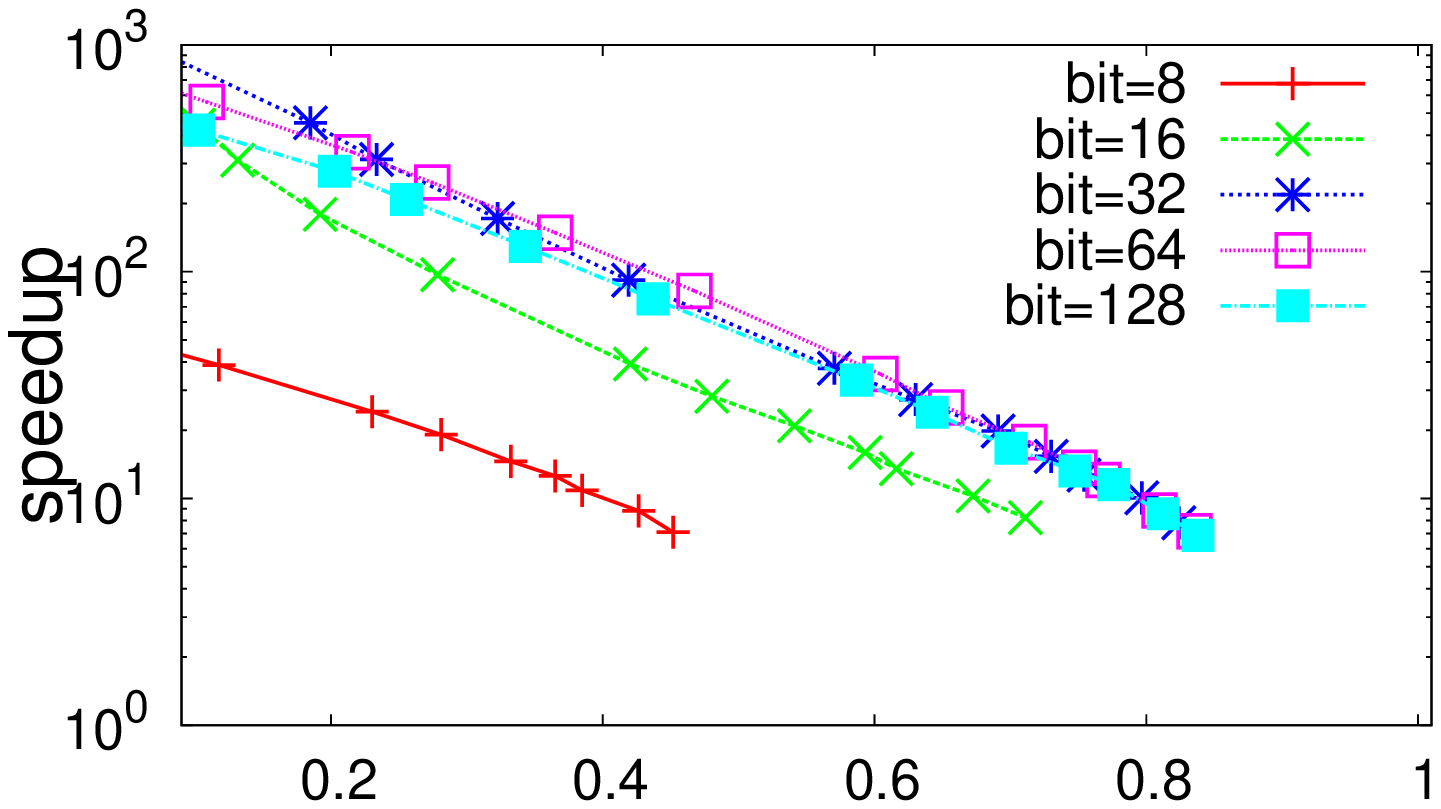}}
\subfigure[\small Sun ]{
      \label{fig:exp_NSH_in_sun} 
      \includegraphics[width=0.48\linewidth]{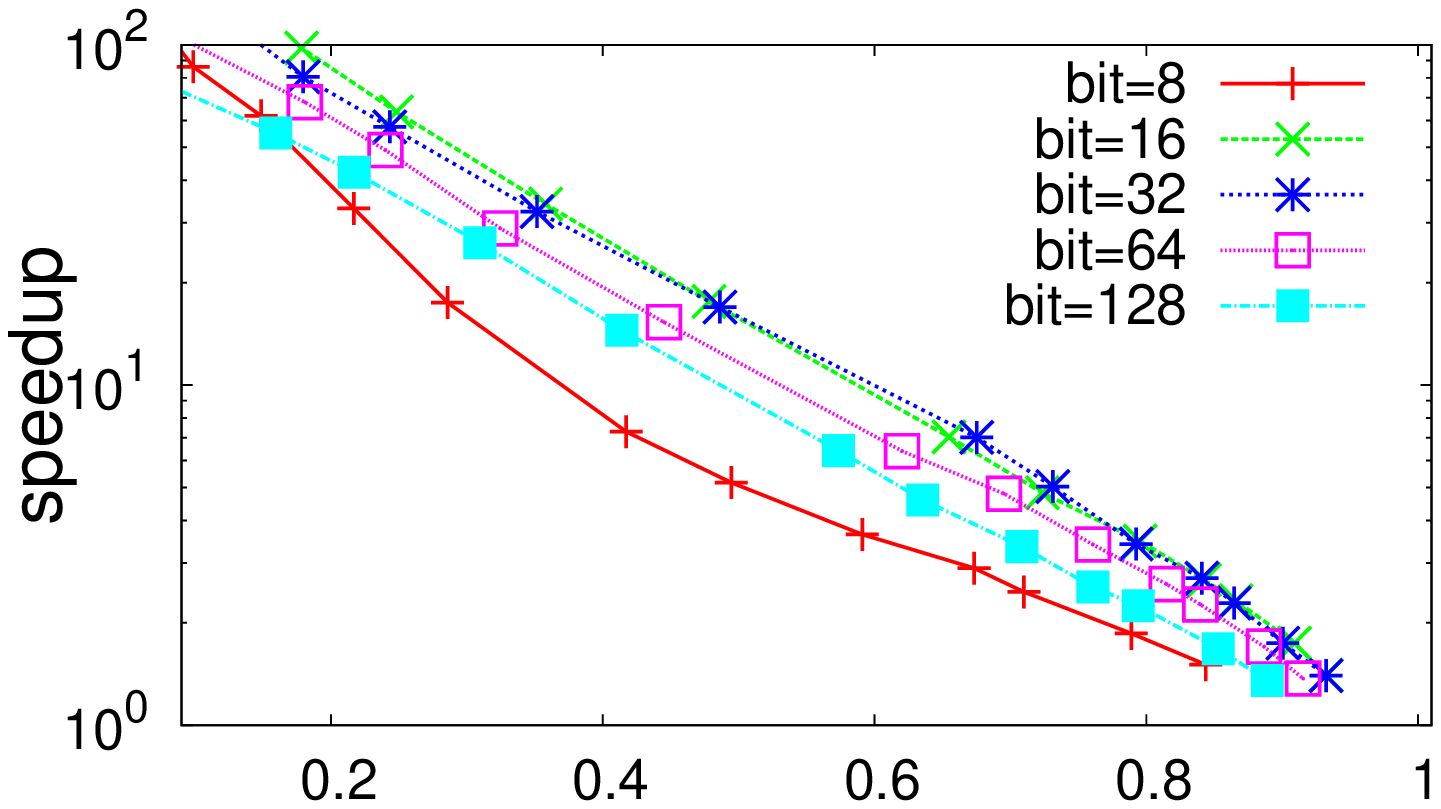}}
\subfigure[\small Trevi ]{
      \label{fig:exp_NSH_in_trevi} 
      \includegraphics[width=0.48\linewidth]{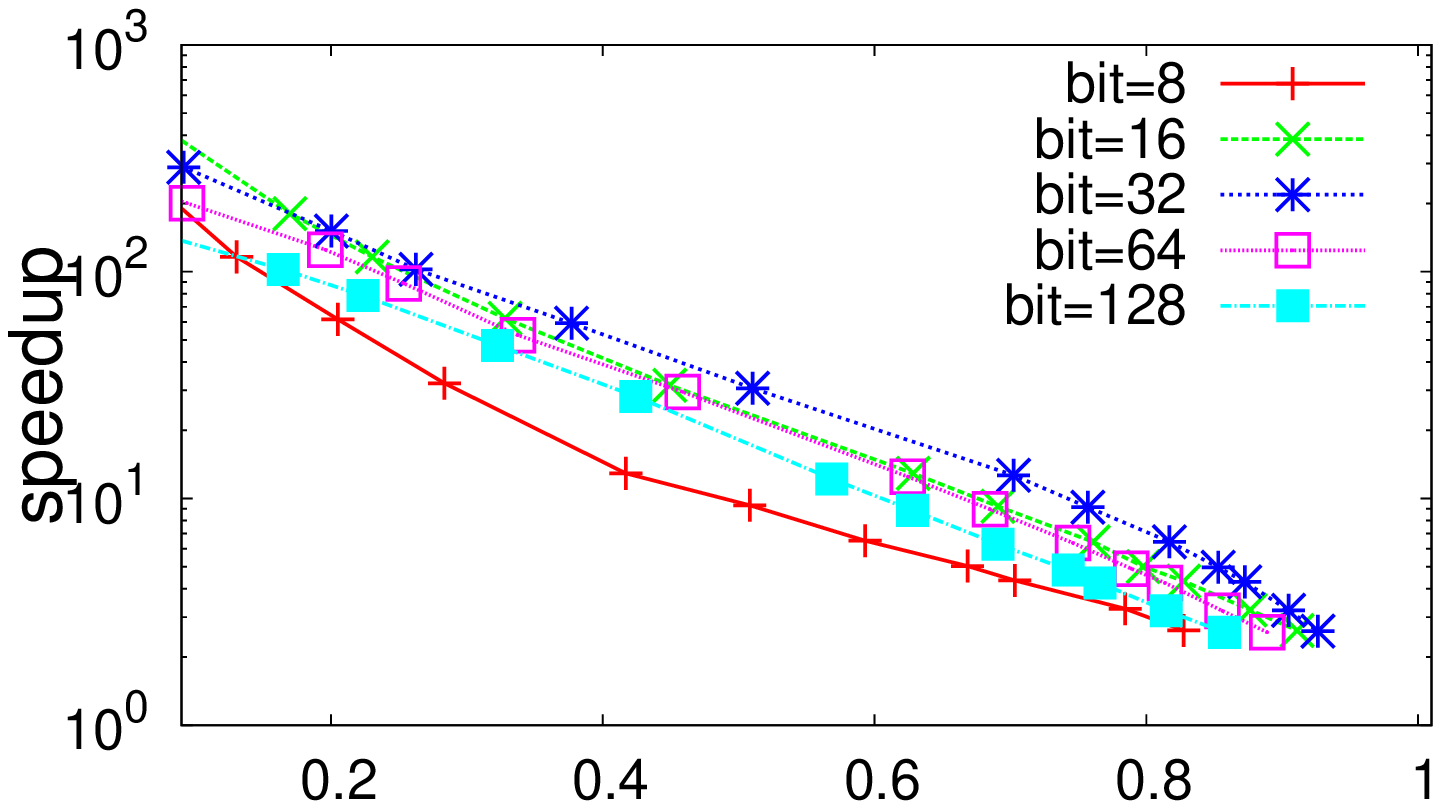}}
\end{minipage}%
\vspace{-4mm}
\caption{\small Speedup vs Recall for Diff $b$ (\textbf{NSH})}
\label{fig:exp_NSH}
\end{figure}

\subsection{NAPP}

Tunning NAPP involves selection of three parameters: $PP$ (the total number of pivots), $P_{i}$ (the number of indexed pivots) and $P_{s}$(the number of the shared pivots with the query).
According to the experiments in \cite{DBLP:journals/pvldb/NaidanBN15}, the values of $PP$ between 500 and 2000 provide a good trade-off. The large value of $PP$ will take long construction time. We will tune the value of $PP$ from 500 to 2000. The author also recommend the value of $P_{i}$ to be 32. We will change $P_{s}$ to get different search performance.

\begin{figure}[tbh]
\begin{minipage}[t]{1.0\linewidth}
\centering
\subfigure[\small Audio ]{
      \label{fig:exp_KDTree_in_audio} 
      \includegraphics[width=0.48\linewidth]{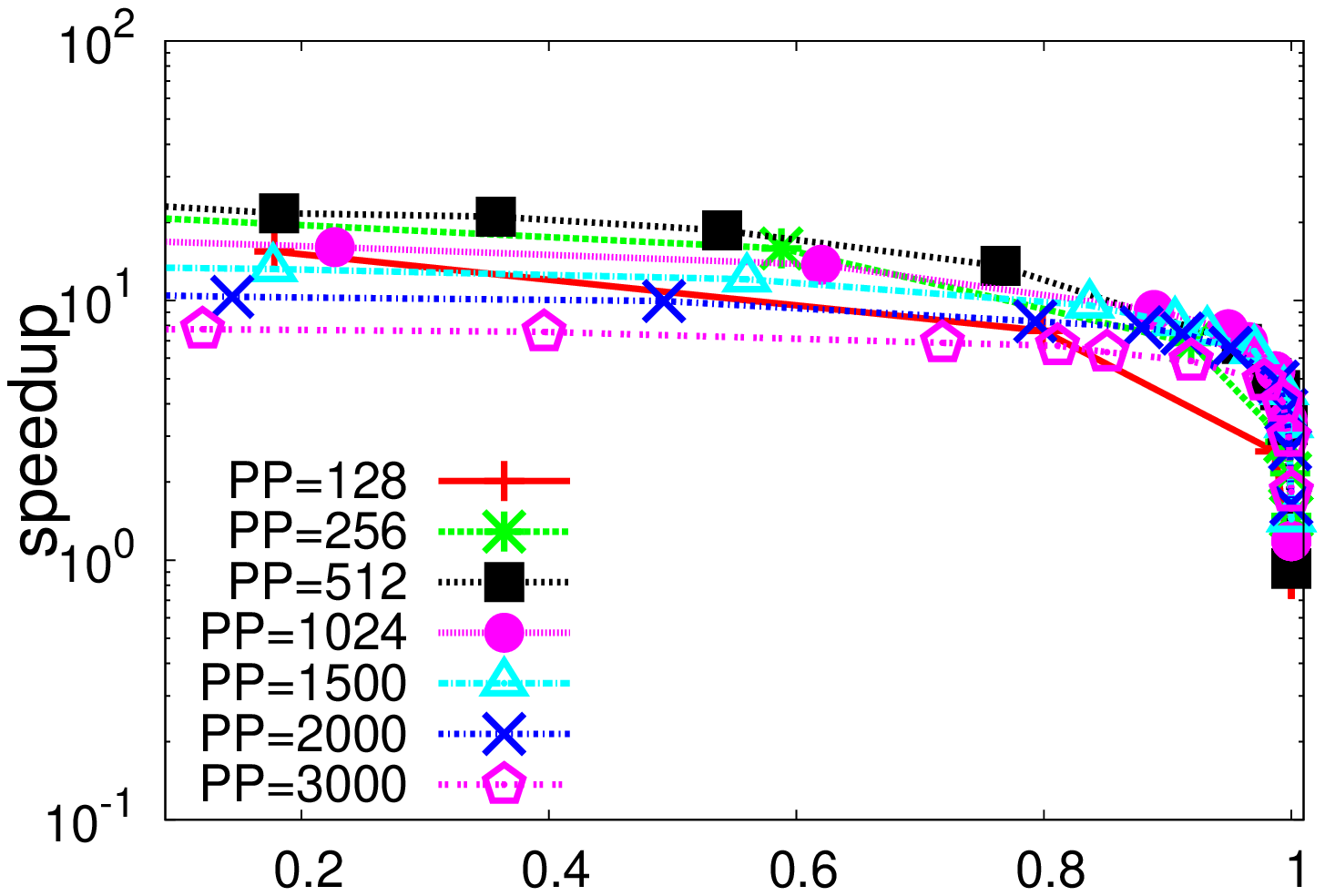}}
\subfigure[\small Sift ]{
      \label{fig:exp_KDTree_in_deep} 
      \includegraphics[width=0.48\linewidth]{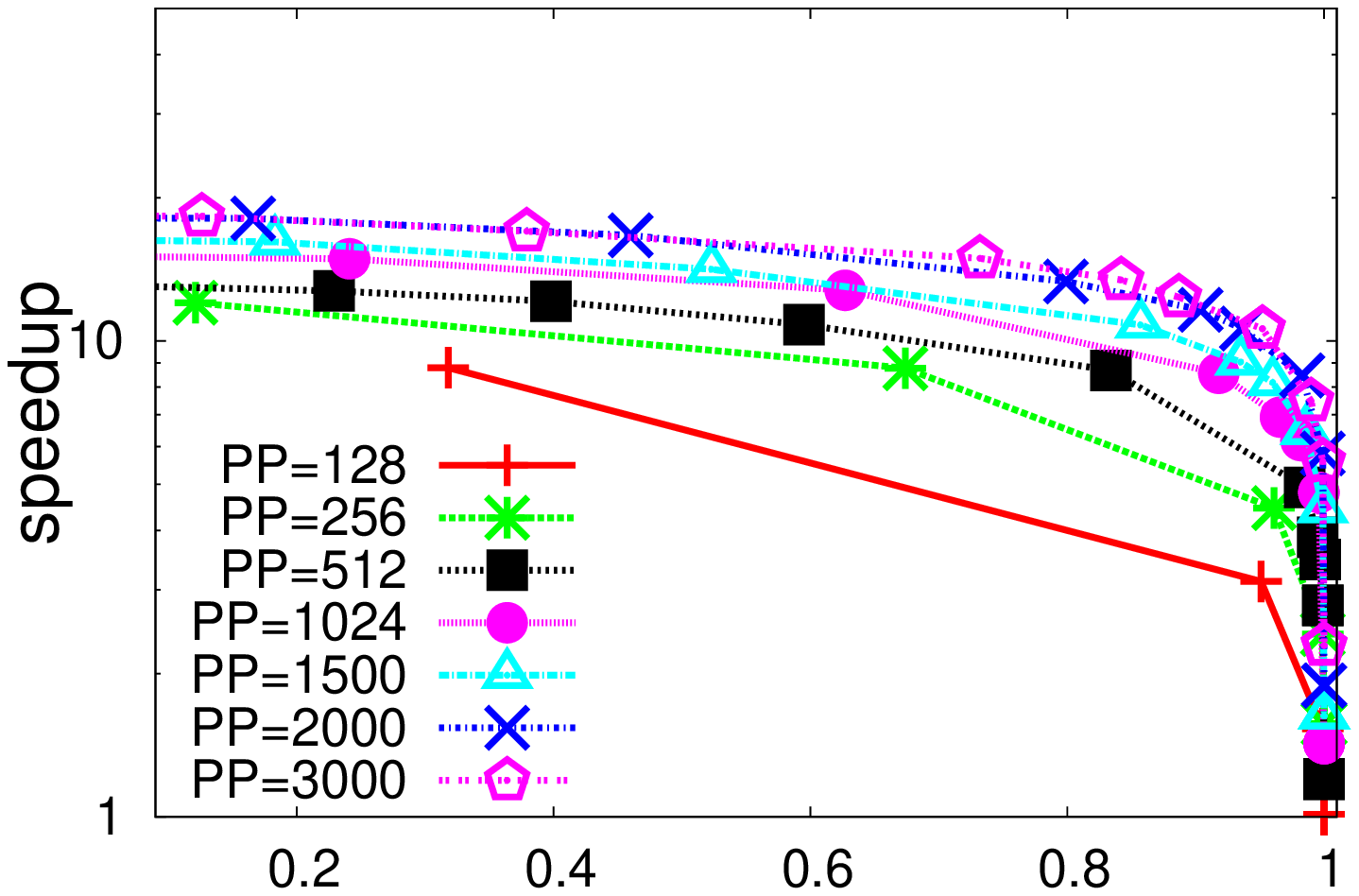}}
\subfigure[\small Deep ]{
      \label{fig:exp_KDTree_in_sit} 
      \includegraphics[width=0.48\linewidth]{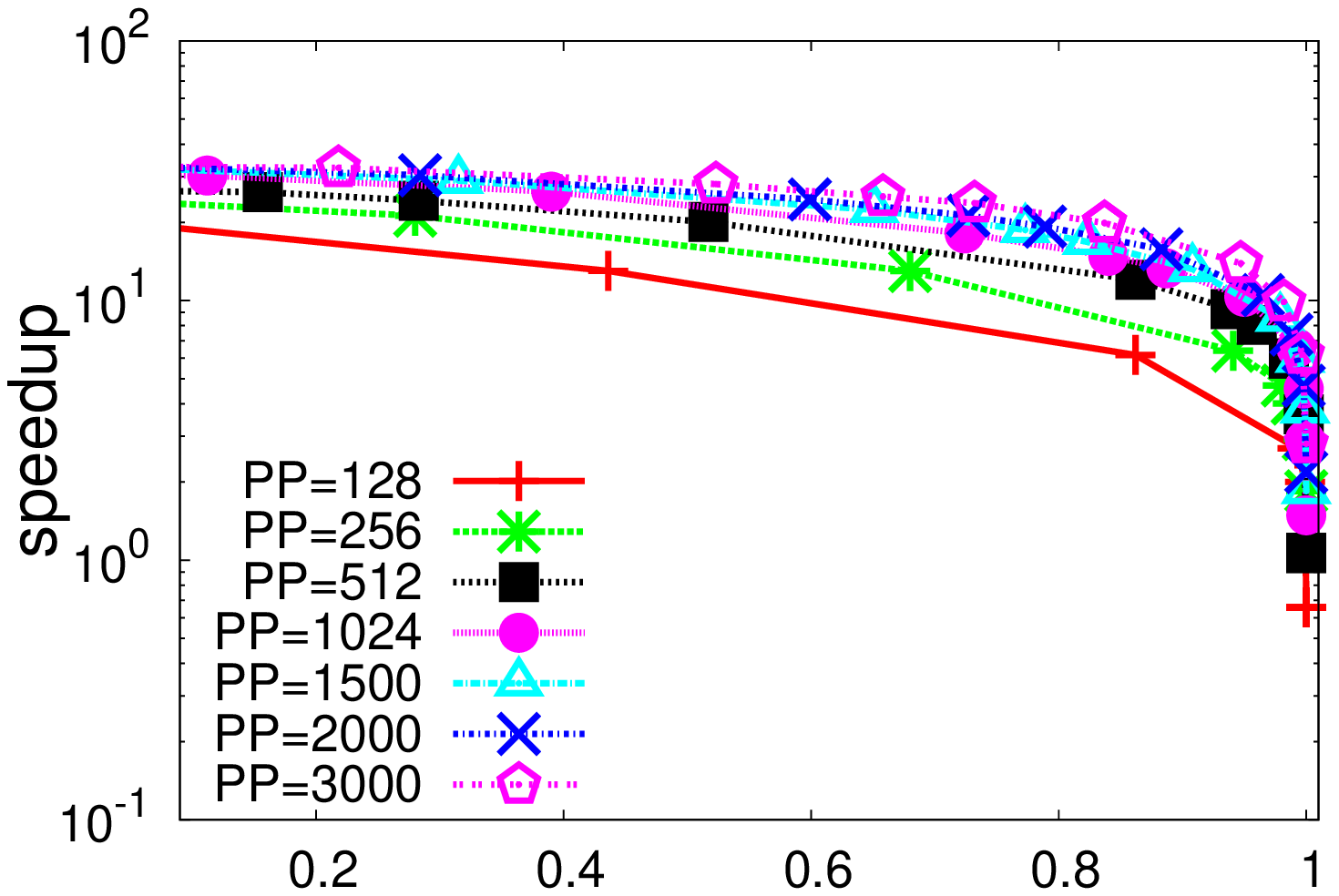}}
\subfigure[\small Glove ]{
      \label{fig:exp_KDTree_in_gist} 
      \includegraphics[width=0.48\linewidth]{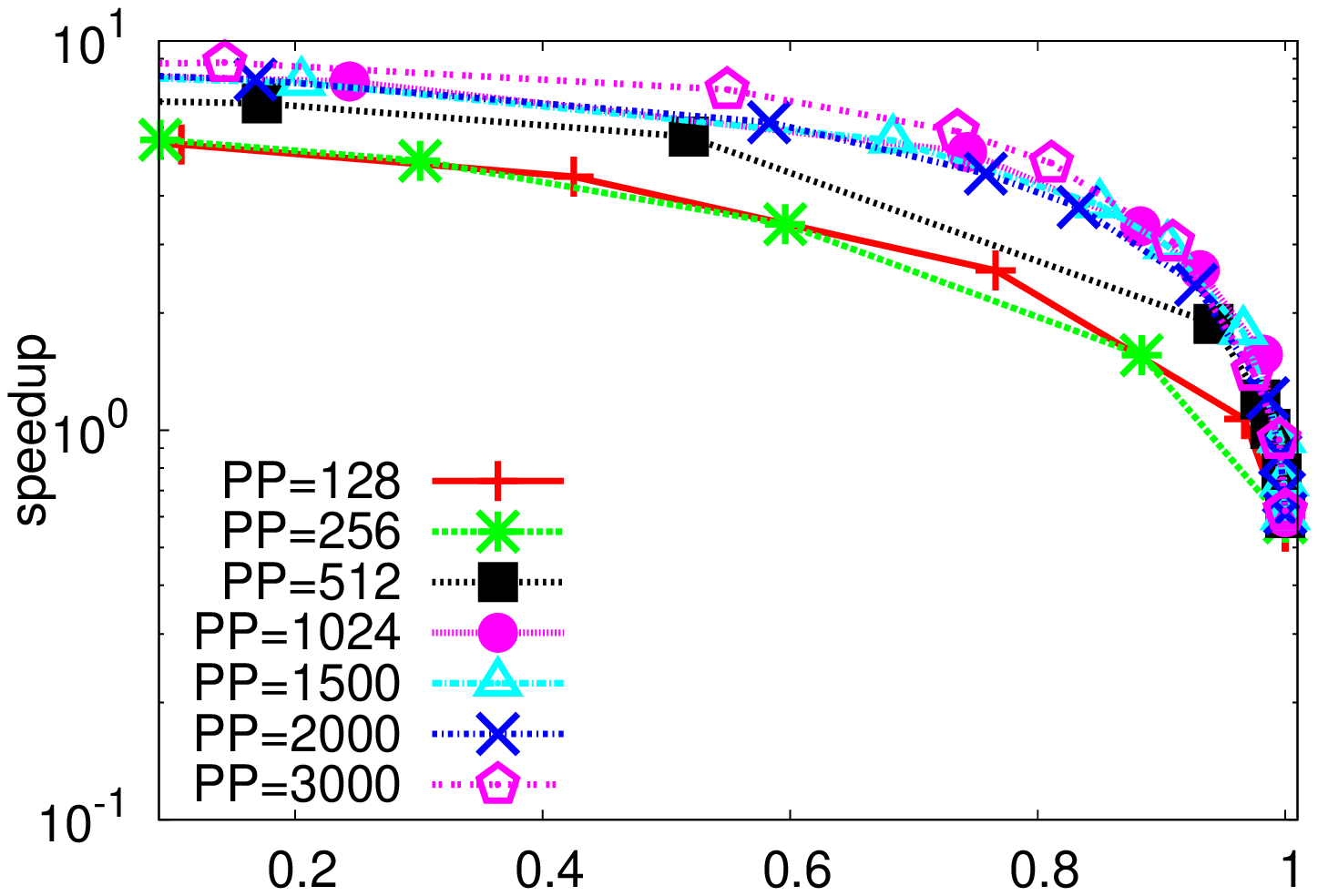}}
\end{minipage}%
\vspace{-4mm}
\caption{\small Speedup vs Recall for Diff $m$ (\textbf{NAPP})}
\label{fig:exp_NAPP}
\end{figure}

\subsection{Selective Hashing}

The experimentation was performed with the default parameter settings provided by the authors. The total number of buckets of per table is 9973. The number of radii $\mathcal{G}$ is set to 20. We change the number of the retrieved points $T$  and the approximation ratio $c$ to get different recalls.

\begin{figure}[tbh]
\begin{minipage}[t]{1.0\linewidth}
\centering
\subfigure[\small Sift]{
      \label{fig:exp_NSH_in_sift_speedup} 
      \includegraphics[width=0.48\linewidth]{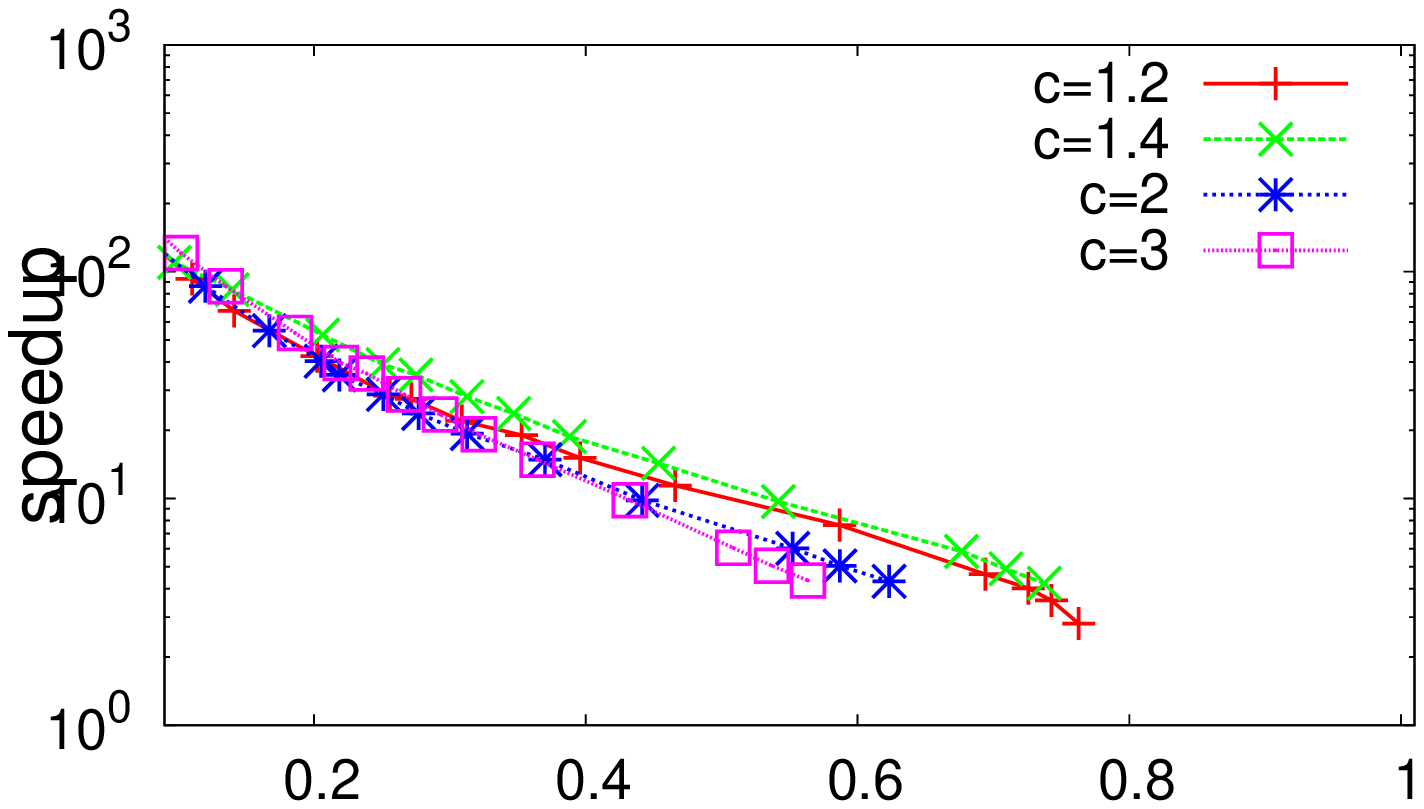}}
\subfigure[\small Mnist ]{
      \label{fig:exp_NSH_in_MNIST} 
      \includegraphics[width=0.48\linewidth]{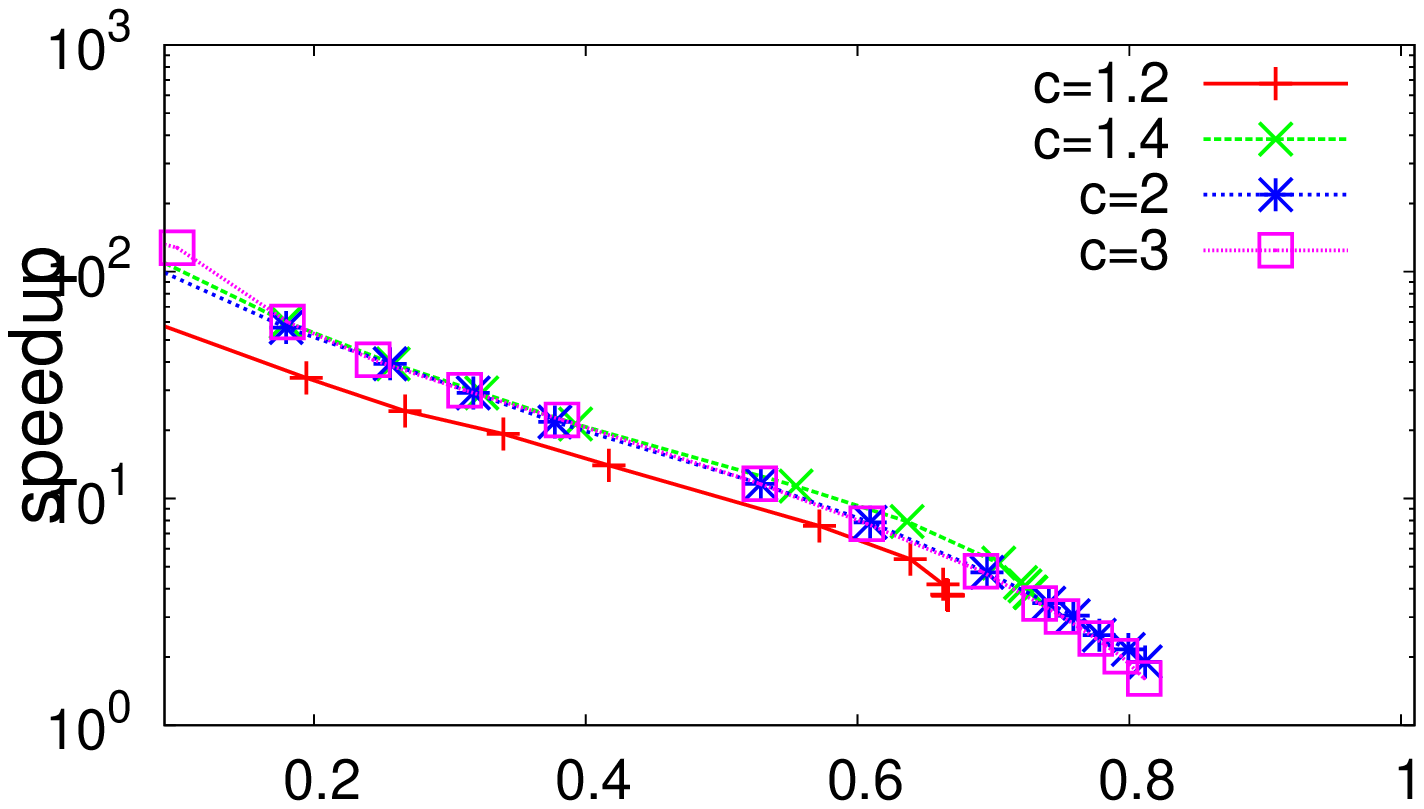}}
\end{minipage}%
\vspace{-4mm}
\caption{\small Speedup vs Recall for Diff $c$ (\textbf{SH})}
\label{fig:exp_SH}
\end{figure}

\subsection{OPQ}

An optimized product quantizer with $M$ subspaces and $k$ sub-codeword in each is used to build inverted multi-index. The product quantizer is optimized using the non-parametric solution initialized by the natural order. The OPQ is generated using $M=2$ and evaluates at most $T$ data points to obtain different search trade-off. For most of the datasets, $k=10$ would achieve a good performance while $k=8$ would be better for the datasets with small data points.

\begin{figure}[tbh]
\begin{minipage}[t]{1.0\linewidth}
\centering
\subfigure[\small Audio ]{
      \label{fig:exp_OPQ_in_audio} 
      \includegraphics[width=0.48\linewidth]{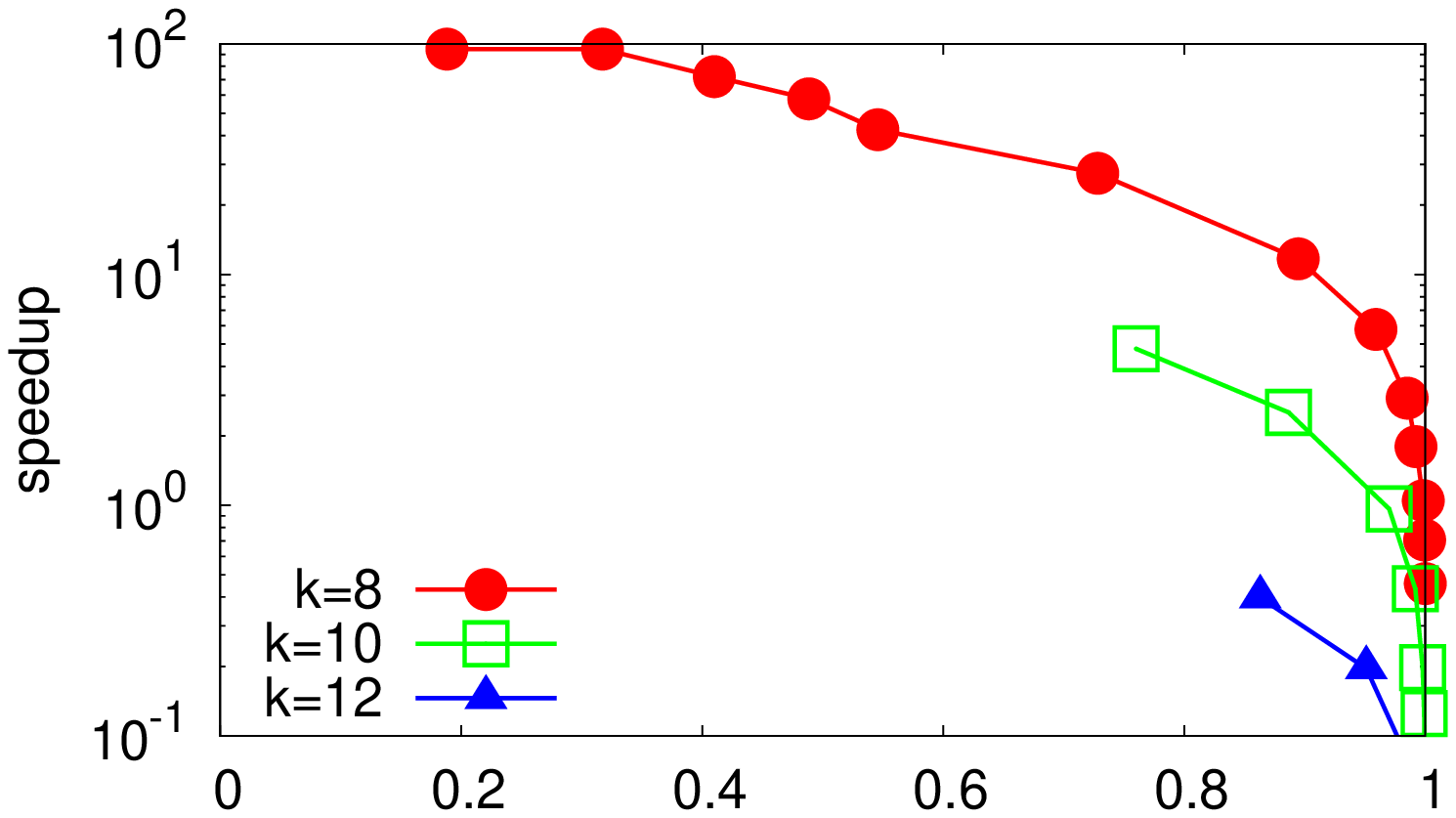}}
\subfigure[\small Sift ]{
      \label{fig:exp_OPQ_in_sift} 
      \includegraphics[width=0.48\linewidth]{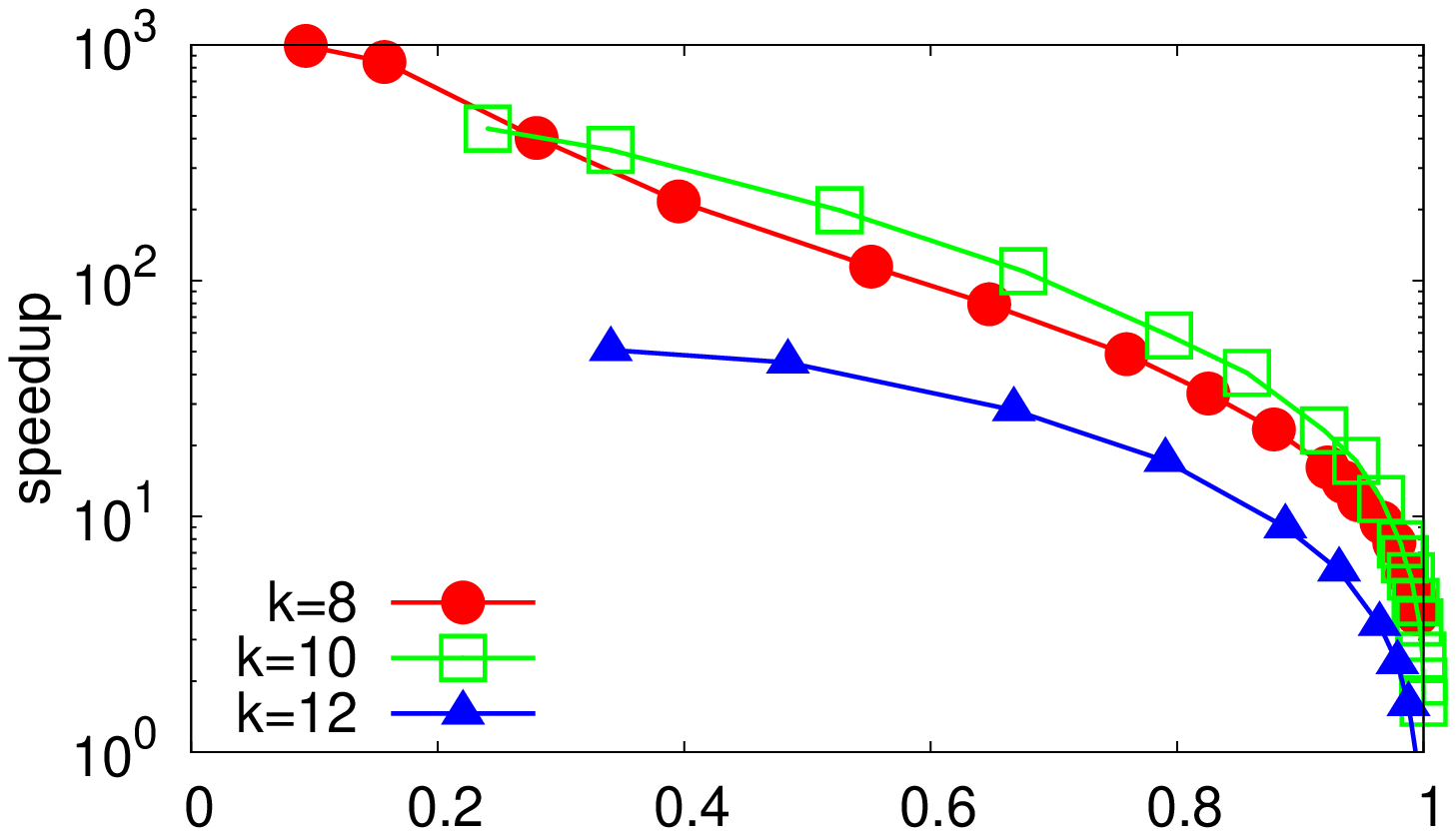}}
\subfigure[\small Gist ]{
      \label{fig:exp_OPQ_in_gist} 
      \includegraphics[width=0.48\linewidth]{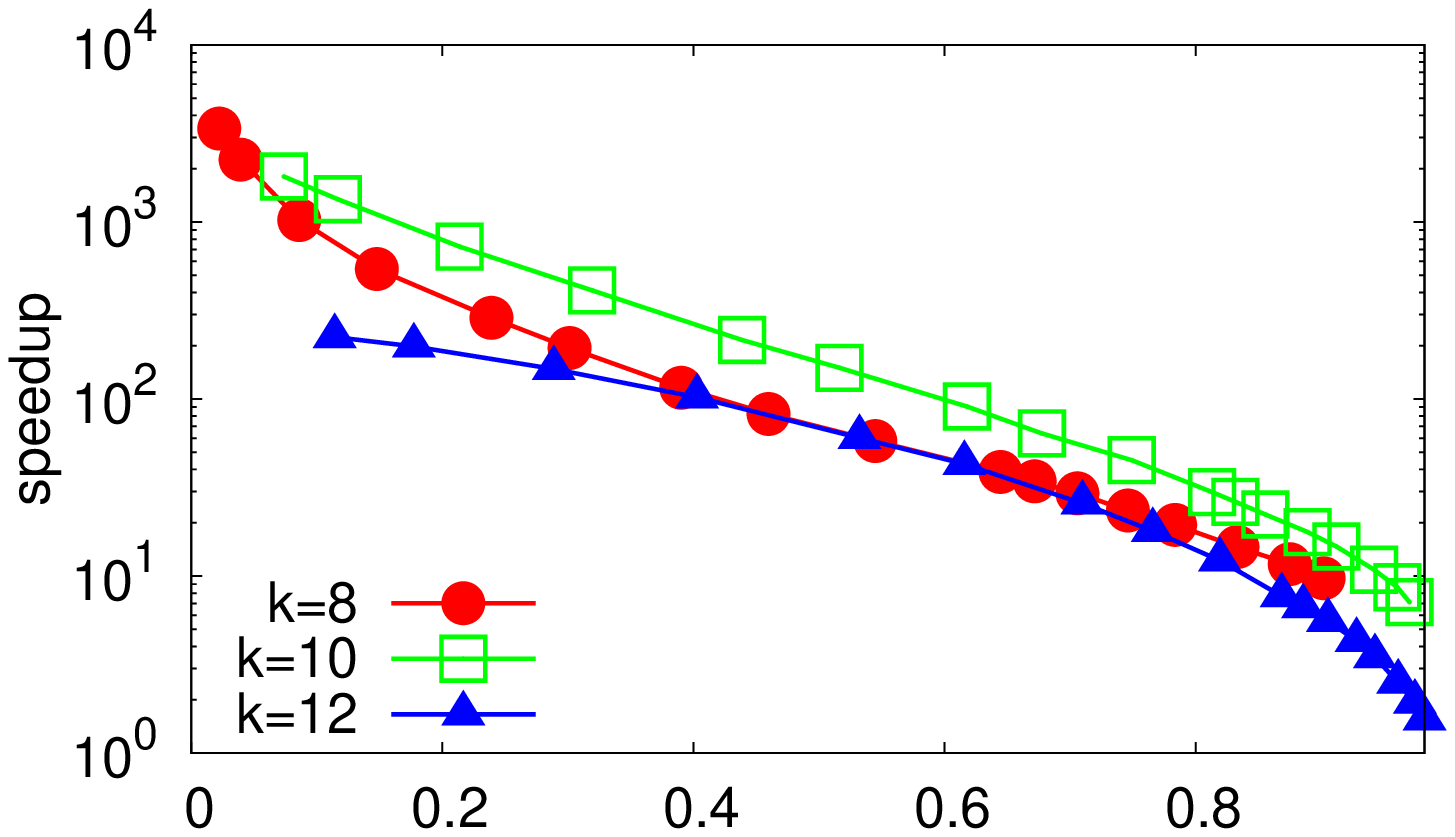}}
\subfigure[\small Nusw ]{
      \label{fig:exp_OPQ_in_nuswide} 
      \includegraphics[width=0.48\linewidth]{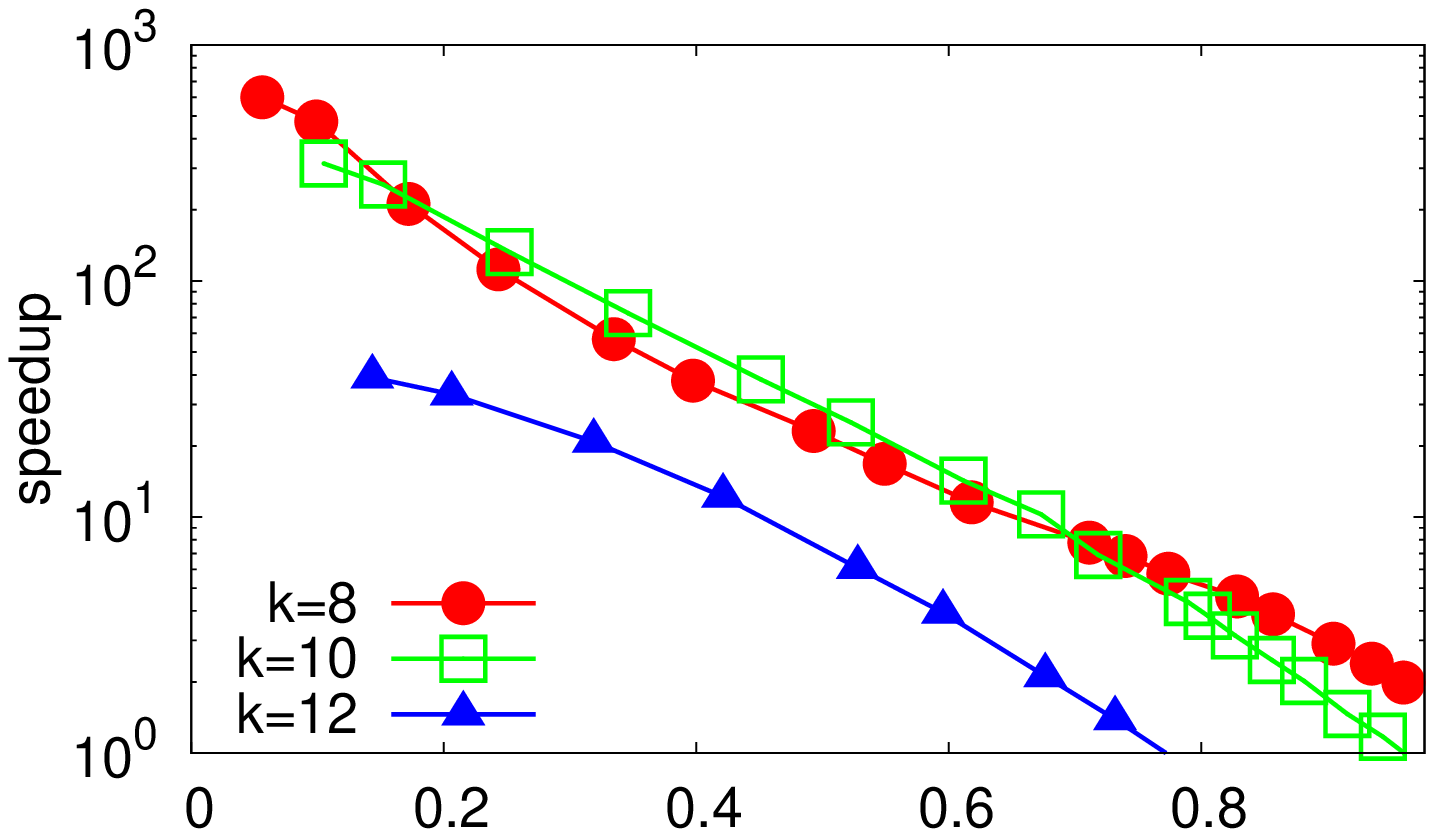}}
\end{minipage}%
\vspace{-4mm}
\caption{\small Speedup vs Recall for Diff $k$ (\textbf{OPQ})}
\label{fig:exp_OPQ}
\end{figure}

\subsection{VP-tree}

In NonMetricSpaceLibrary, VP-tree uses a simple polynomial pruner to generate the optimal parameters $\alpha_{left}$ and $\alpha_{right}$. We use auto-tuning procedure to produce different search performance by given the input parameter $target$ (which is the expected recall) and $b$ (the maximum number of points in leaf nodes).

\begin{figure}[tbh]
\begin{minipage}[t]{1.0\linewidth}
\centering
\subfigure[\small imageNet]{
      \label{fig:exp_VPTree_in_imageNet} 
      \includegraphics[width=0.48\linewidth]{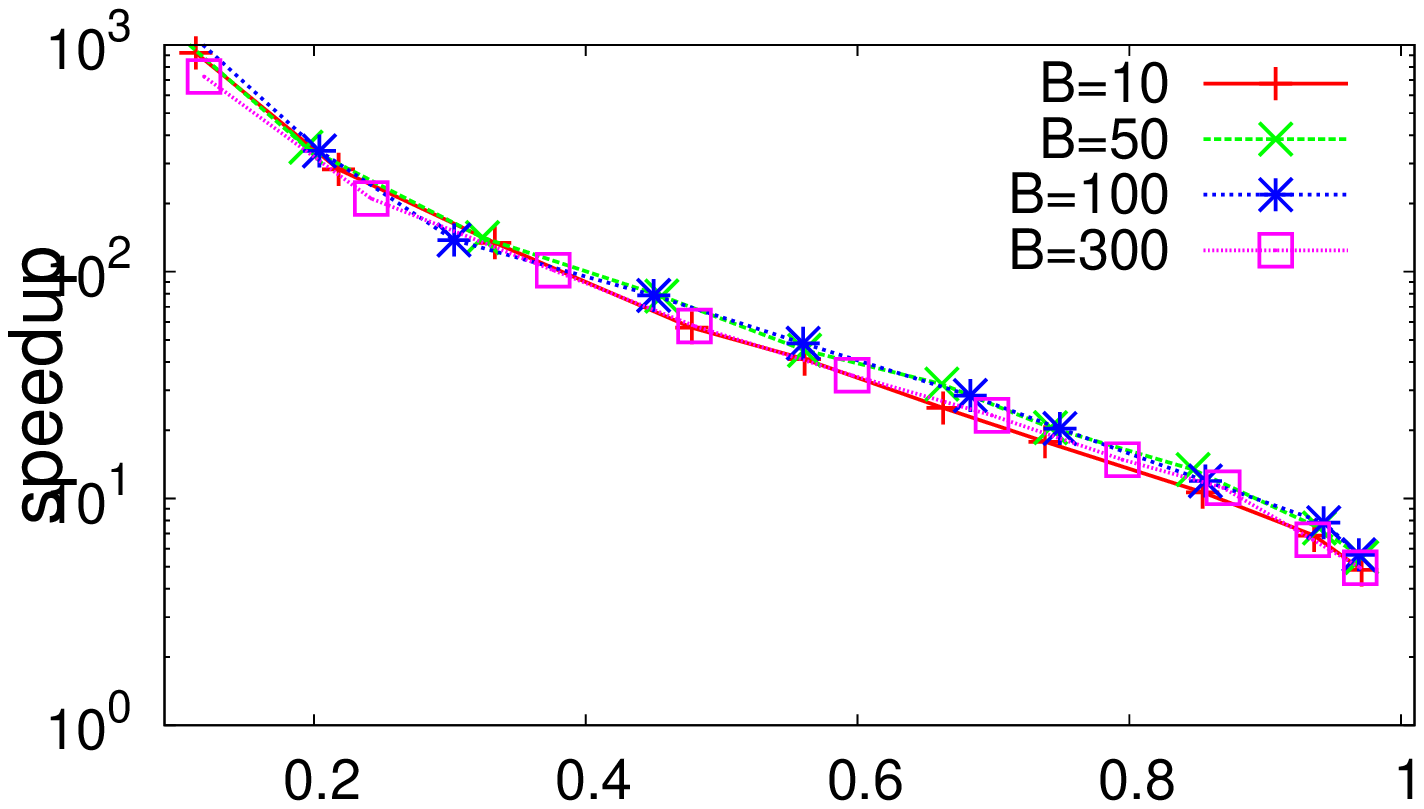}}
\subfigure[\small Sift ]{
      \label{fig:exp_VPTree_in_sift} 
      \includegraphics[width=0.48\linewidth]{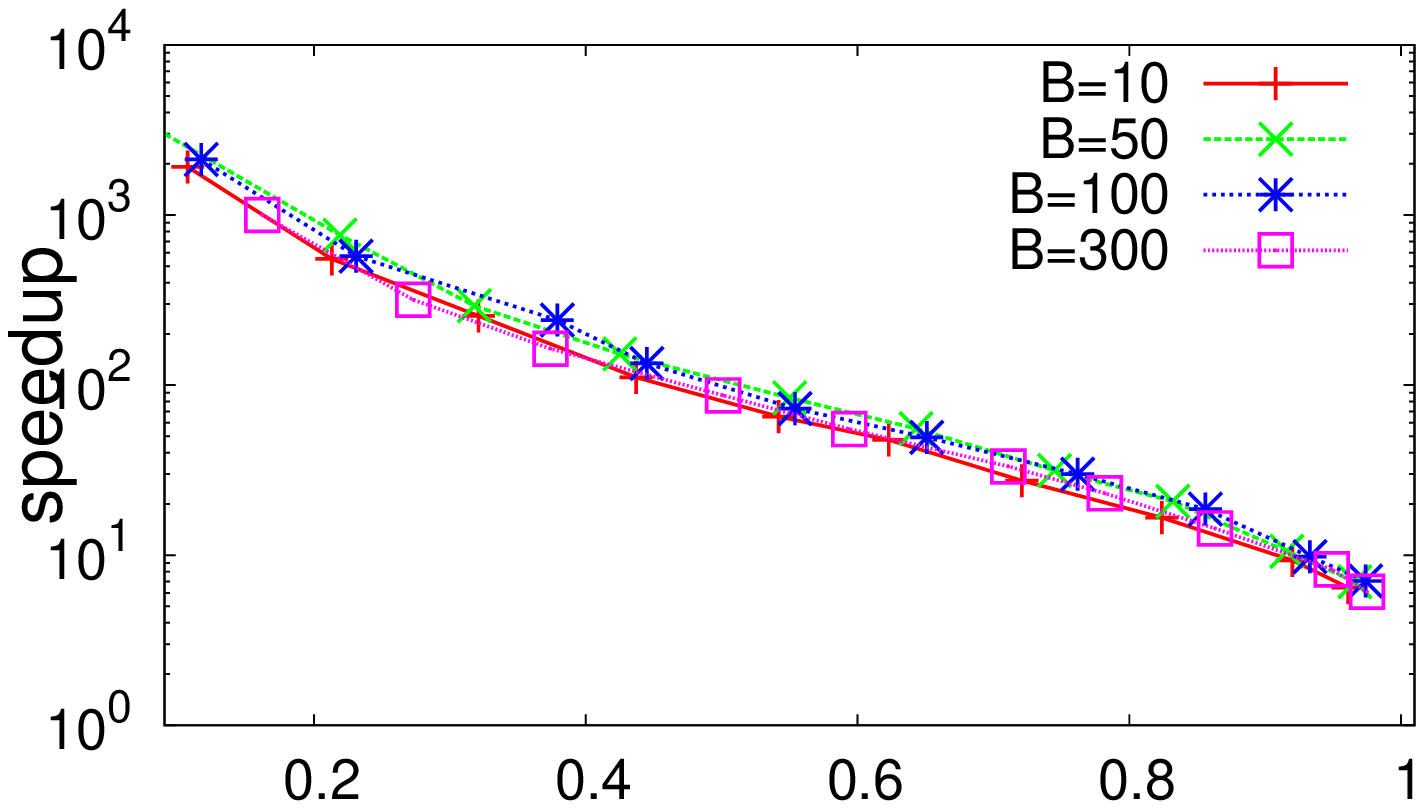}}
\subfigure[\small Deep ]{
      \label{fig:exp_VPTree_in_deep} 
      \includegraphics[width=0.48\linewidth]{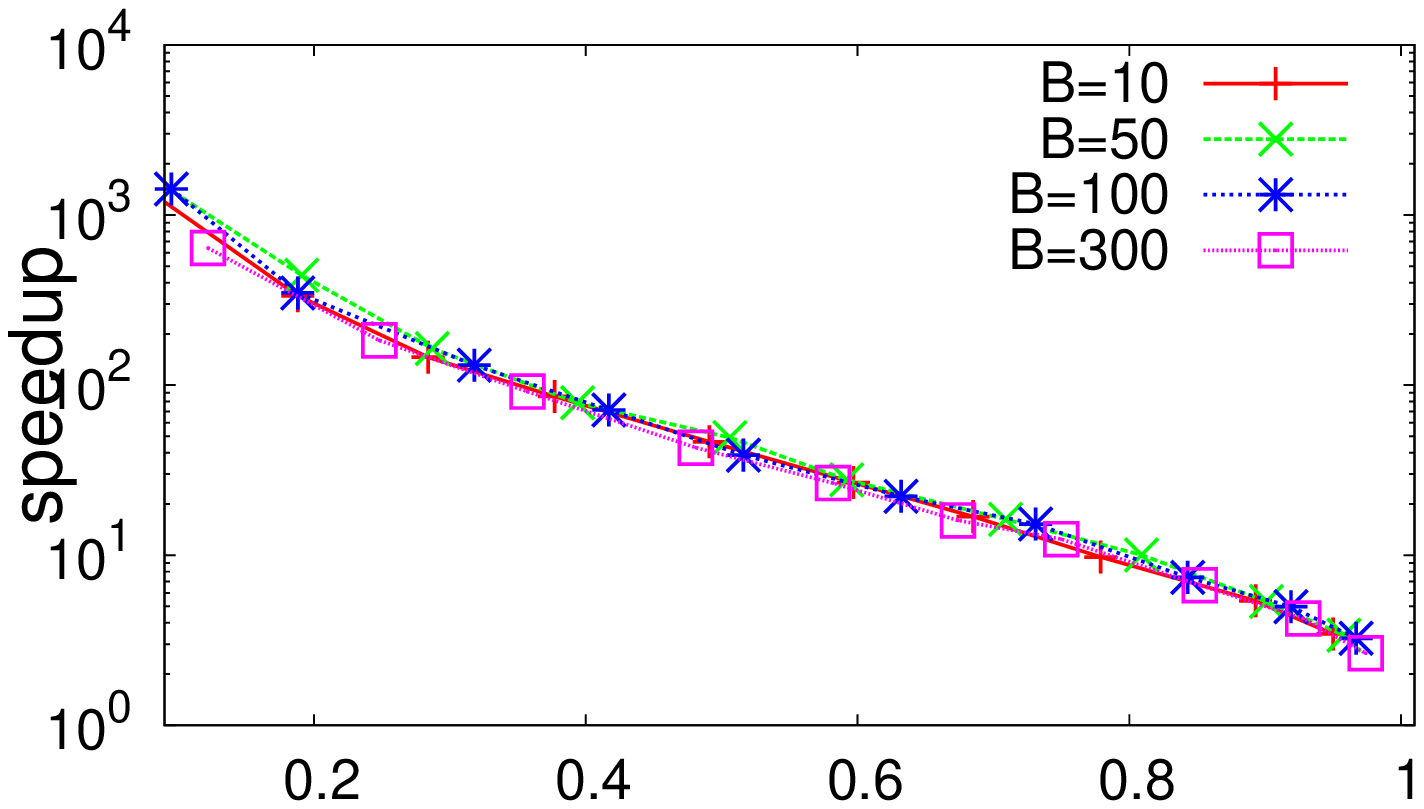}}
\subfigure[\small Glove ]{
      \label{fig:exp_VPTree_in_glove} 
      \includegraphics[width=0.48\linewidth]{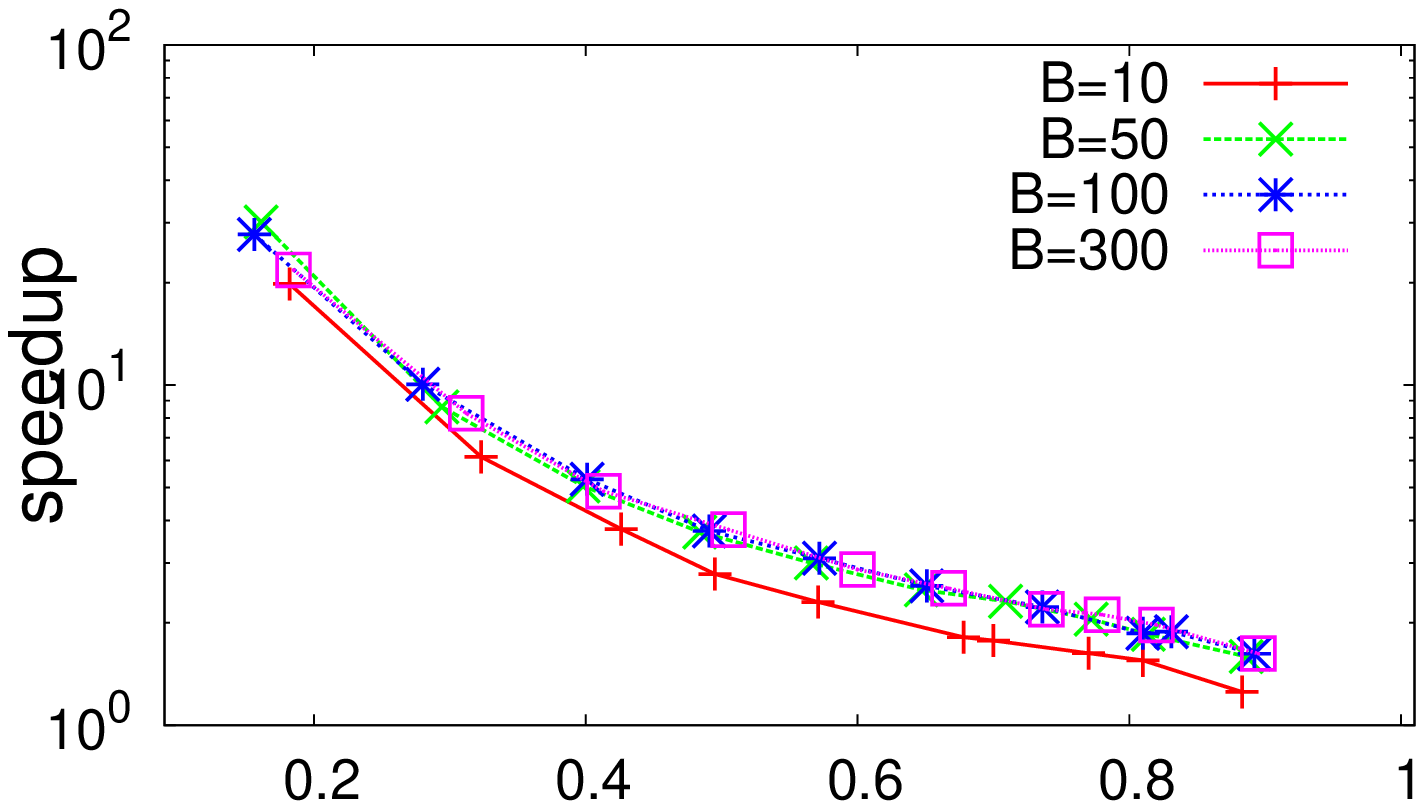}}
\end{minipage}%
\vspace{-4mm}
\caption{\small Speedup vs Recall for Diff $b$ (\textbf{VP-tree}) }
\label{fig:exp_VPTree}
\end{figure}

\subsection{Annoy}

Annoy only involves one parameter: trees number $f$. Table \ref{tab:annoy_indextime} and \ref{tab:annoy_indexsize} shows the index sizes and the construction time complexities are linear to $f$. The search performance could be significantly improved by using multiple Annoy trees while the growth rate changes slowly when $f$ is larger than $50$ for most of the datasets.Considering the search performance and index performance comprehensively, we build $50$ trees for all the datasets.

\begin{table}[htb]
  \small
  \centering
  \begin{tabular}{|c|r|r|r|r|r|}
    \hline \textbf{Name} & \textbf{$1$Tree} & \textbf{$10$Trees} & \textbf{$50$Trees} & \textbf{$100$Trees} & \textbf{$200$Trees} \\
    \hline
    \hline Audio  & 0.04                      & 0.5         & 2.3          & 4.5         & 9.4        \\
    \hline Yout   & 2.3                       & 22.1        & 112.6        & 217         & 442         \\
    \hline Sift   & 1.8                       & 17          & 85           & 128         & 352         \\
    \hline Gauss  & 8.2                       & 77          & 384          & 608         & 1538        \\
    \hline
  \end{tabular}                                                                                                                    \\
  \vspace{-2mm}
  \caption{\small construction time using different trees} 
  \label{tab:annoy_indextime}
\end{table}

\begin{table}[htb]
  \small
  \centering
  \begin{tabular}{|c|r|r|r|r|r|}
    \hline \textbf{Name} & \textbf{$1$Tree} & \textbf{$10$Trees} & \textbf{$50$Trees} & \textbf{$100$Trees} & \textbf{$200$Trees} \\
    \hline
    \hline Audio                & 40.5                      & 45.9        & 70           & 100         & 160.1        \\
    \hline Yout                 & 2347                      & 2382        & 2539         & 2735        & 3128        \\
    \hline Sift                 & 512                       & 612         & 1058         & 1615        & 2730         \\
    \hline Gauss                & 3960                      & 4161        & 5055         & 6171        & 8407        \\
    \hline
  \end{tabular}                                                                                                                    \\
  \vspace{-2mm}
  \caption{\small index size using different trees} 
  \label{tab:annoy_indexsize}
\end{table}

\begin{figure}[tbh]
\begin{minipage}[t]{1.0\linewidth}
\centering
\subfigure[\small Audio ]{
      \label{fig:exp_Annoy_in_audio} 
      \includegraphics[width=0.48\linewidth]{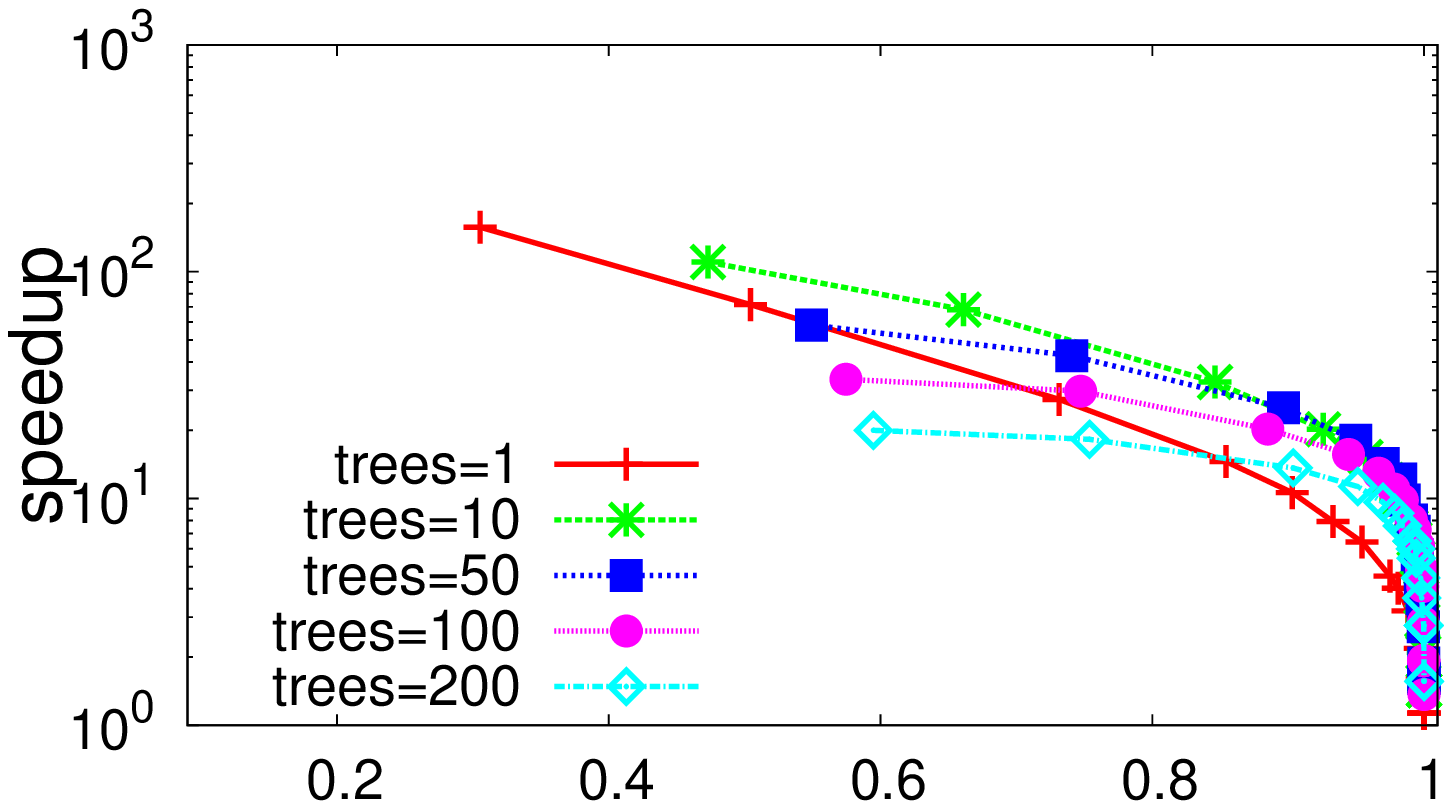}}
\subfigure[\small Sift ]{
      \label{fig:exp_Annoy_in_sift} 
      \includegraphics[width=0.48\linewidth]{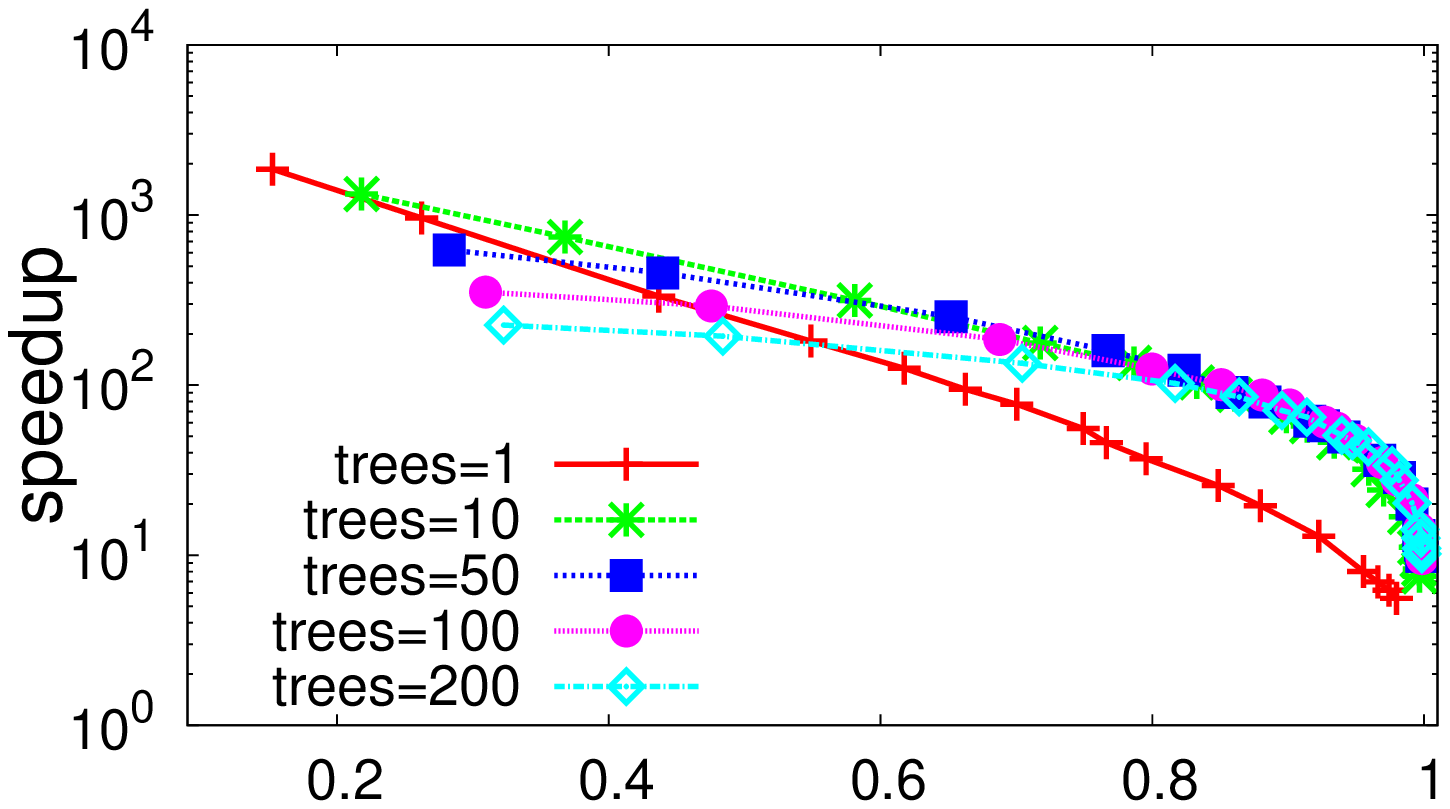}}
\subfigure[\small Yout ]{
      \label{fig:exp_Annoy_in_youtube} 
      \includegraphics[width=0.48\linewidth]{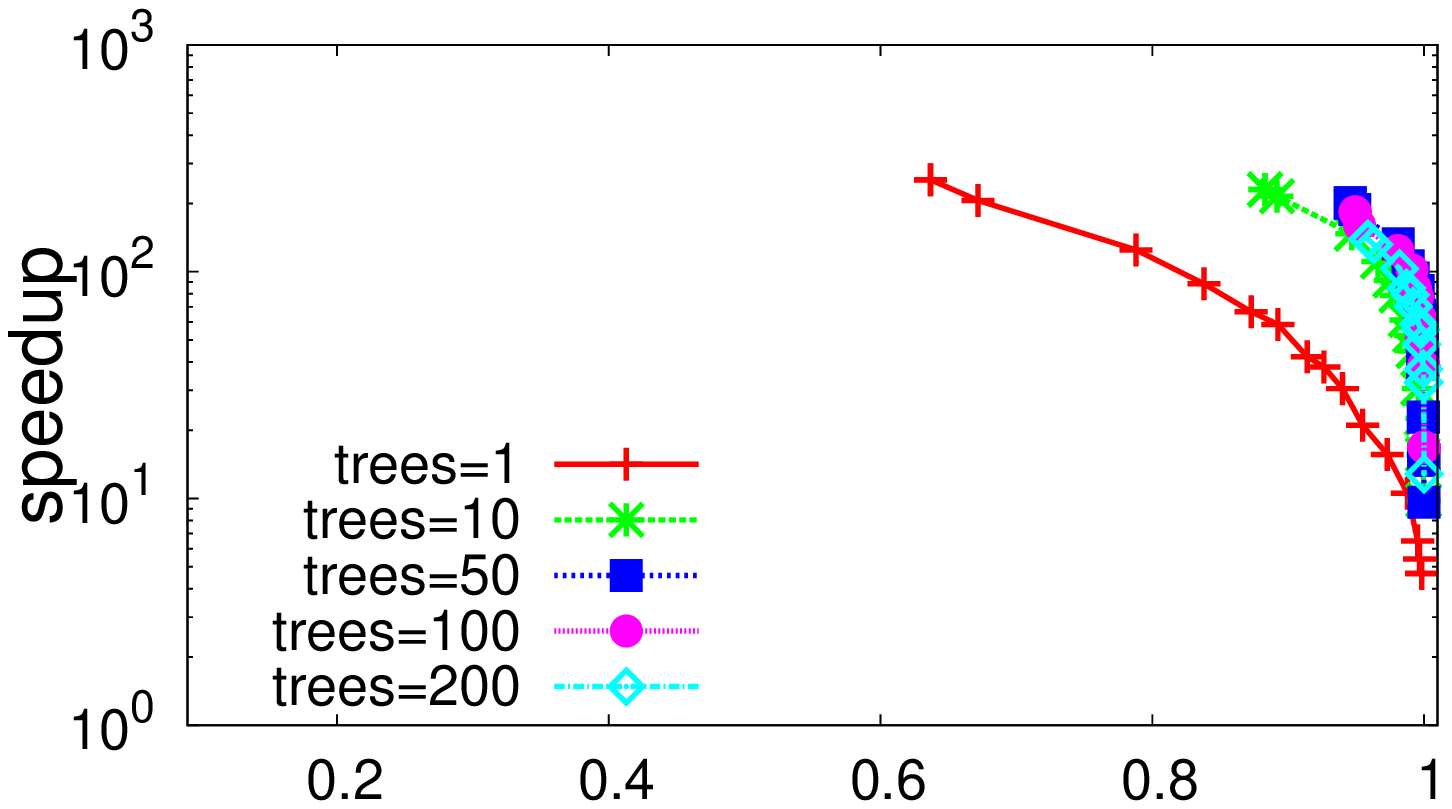}}
\subfigure[\small Gauss ]{
      \label{fig:exp_Annoy_in_gauss} 
      \includegraphics[width=0.48\linewidth]{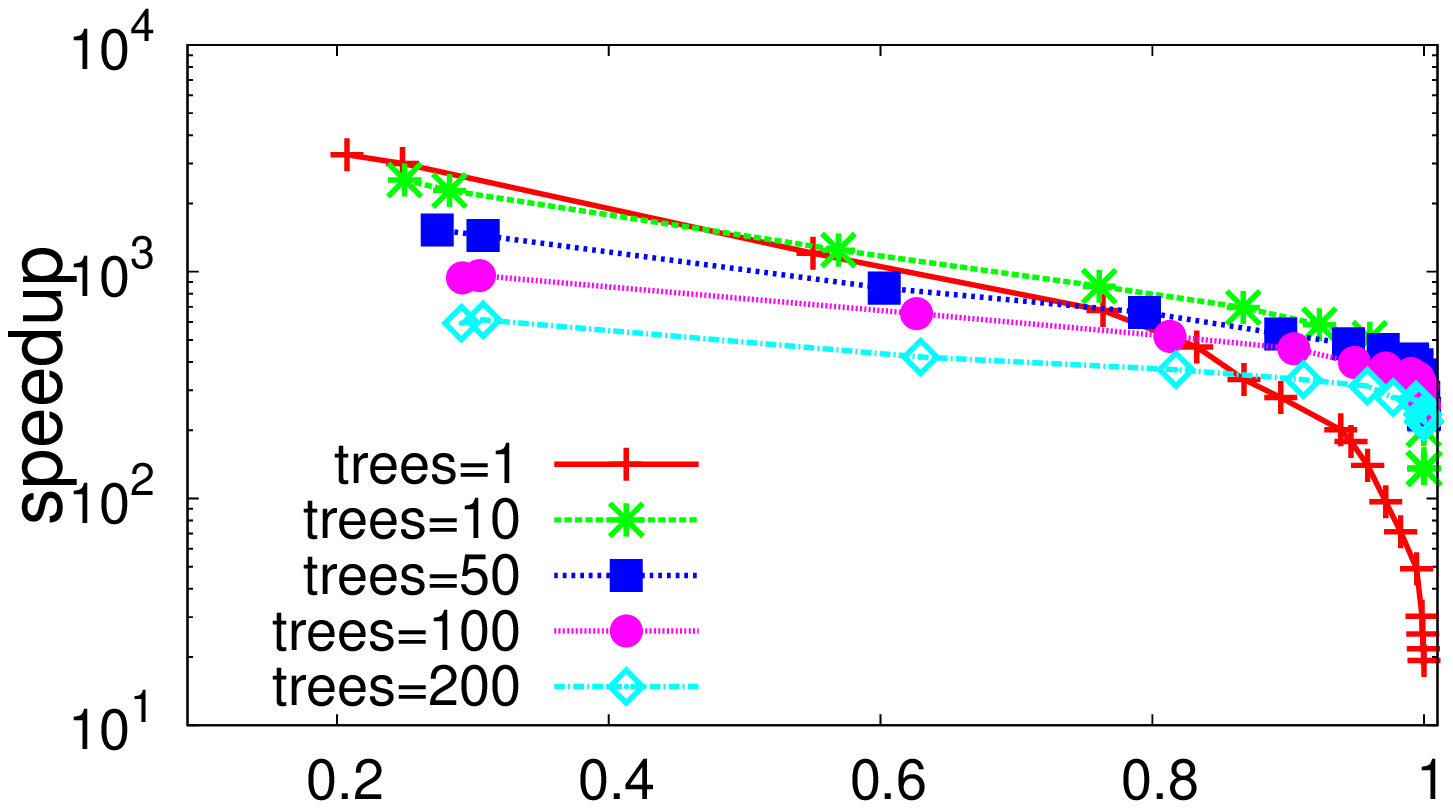}}
\end{minipage}%
\vspace{-4mm}
\caption{\small Speedup vs Recall for Diff $f$ (\textbf{Annoy})}
\label{fig:exp_SpH_in}
\end{figure}

\subsection{HKMeans}

We randomly select the init centers in the k-means clustering. According to the recommendations from the source code of \textbf{Flann}, we apply different combinations of iteration times $iter$ and branching size $b$ to generate the recall-speedup trade-off.

\begin{figure}[tbh]
\begin{minipage}[t]{1.0\linewidth}
\centering
\subfigure[\small Sift with branches=64 ]{
      \label{fig:exp_HKMeans_in_sift} 
      \includegraphics[width=0.48\linewidth]{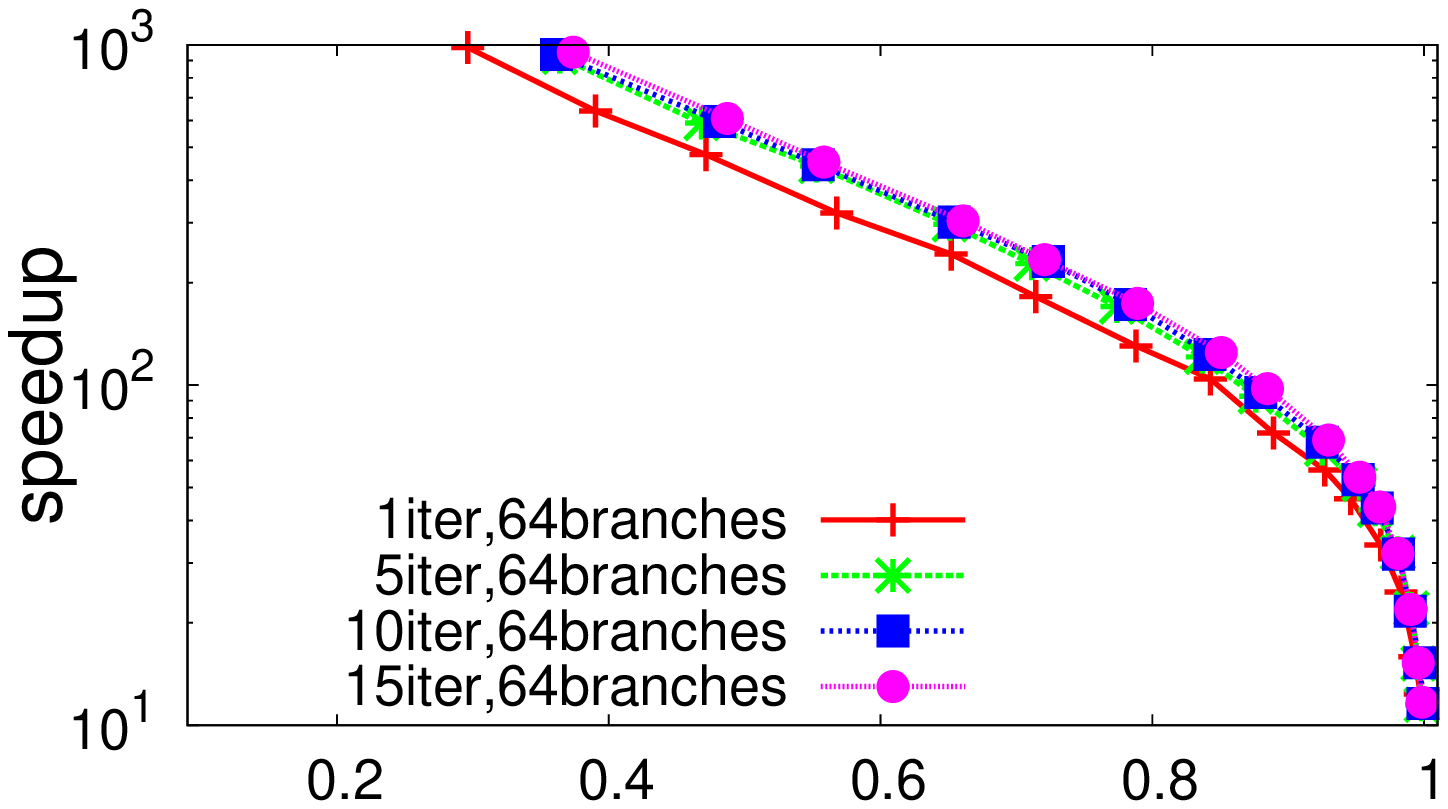}}
\subfigure[\small sift with branches=128 ]{
      \label{fig:exp_HKMeans_in_sift} 
      \includegraphics[width=0.48\linewidth]{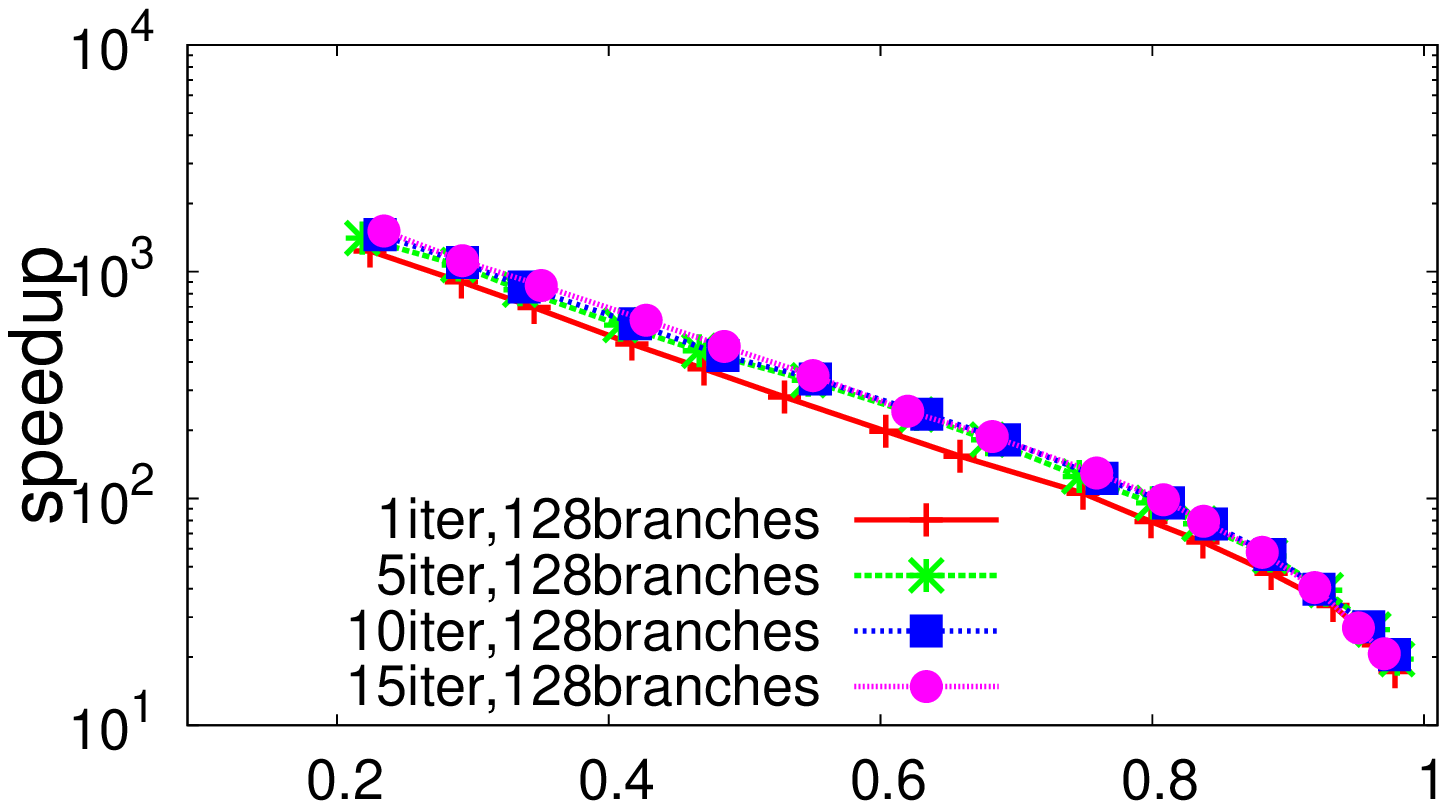}}
\end{minipage}%
\vspace{-4mm}
\caption{\small Speedup vs Recall for Diff $iter$ (\textbf{Flann-HKM})}
\label{fig:exp_SpH_in}
\end{figure}

\begin{figure}[tbh]
\begin{minipage}[t]{1.0\linewidth}
\centering
\subfigure[\small Sift]{
      \label{fig:exp_HKMeans_in_sift} 
      \includegraphics[width=0.48\linewidth]{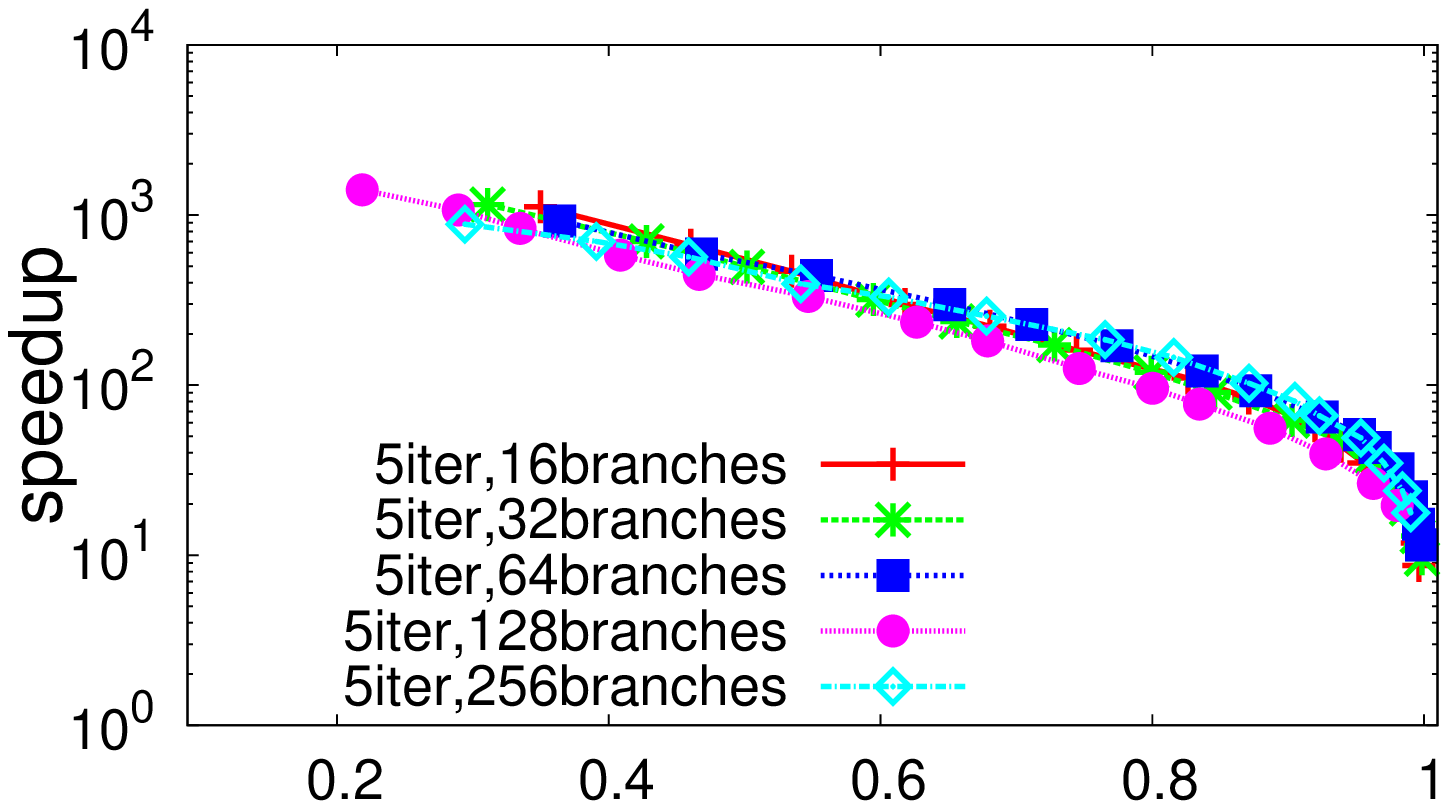}}
\subfigure[\small UQ\_V]{
      \label{fig:exp_HKMeans_in_uqvideo} 
      \includegraphics[width=0.48\linewidth]{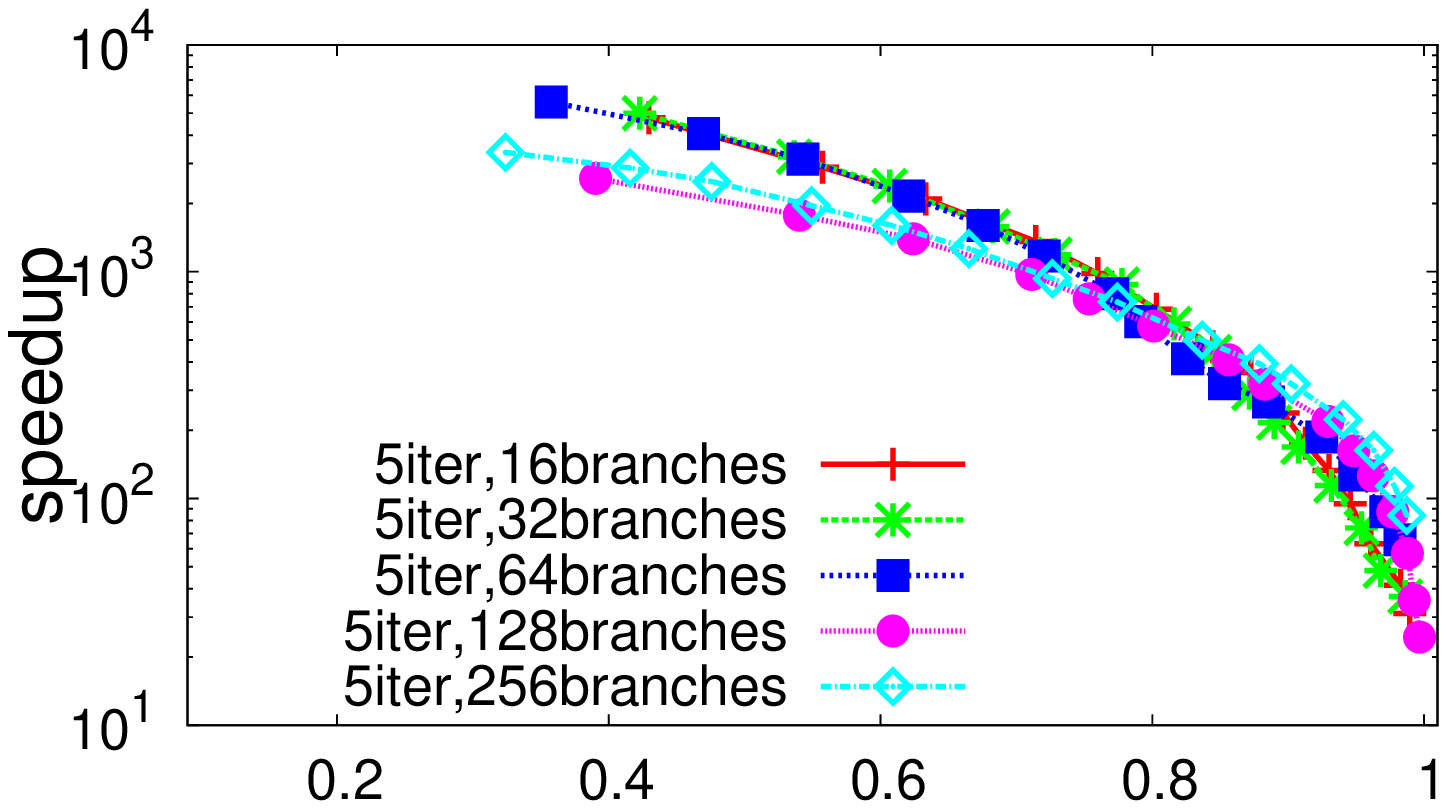}}
\end{minipage}
\vspace{-4mm}
\caption{\small Speedup vs Recall for Diff $b$ (\textbf{Flann-HKM})}
\label{fig:exp_HKMeans}
\end{figure}

\subsection{KDTree}

Except the parameter $t$ which is the number of Randomized kdtrees, we use the default setting provided by the source code. The search performance would be improved as the growth of the number of trees. While the speedup doesn't show considerable increase when $t$ is larger than 8.

\begin{figure}[tbh]
\begin{minipage}[t]{1.0\linewidth}
\centering
\subfigure[\small Audio ]{
      \label{fig:exp_KDTree_in_audio} 
      \includegraphics[width=0.48\linewidth]{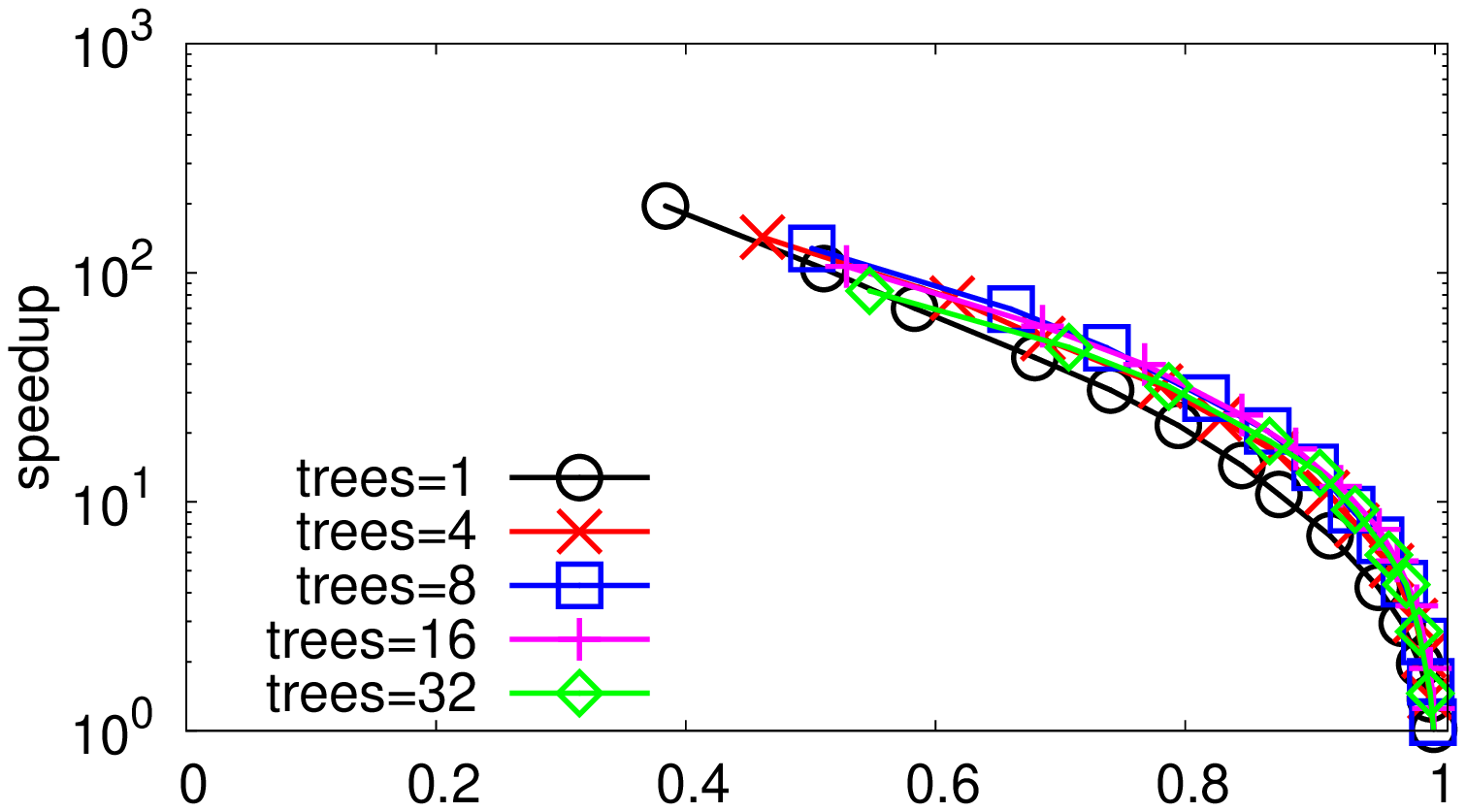}}
\subfigure[\small Deep ]{
      \label{fig:exp_KDTree_in_deep} 
      \includegraphics[width=0.48\linewidth]{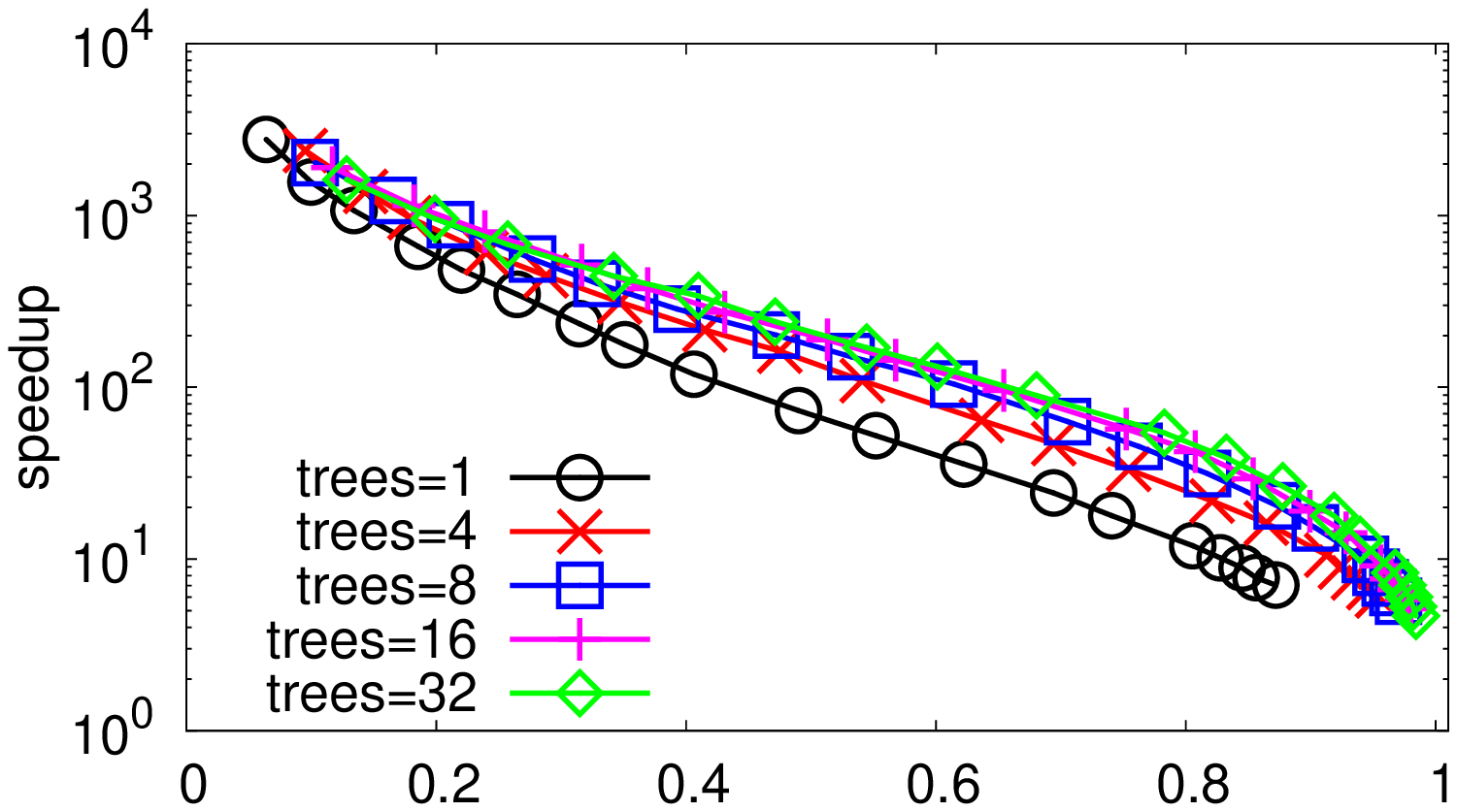}}
\subfigure[\small Sift ]{
      \label{fig:exp_KDTree_in_sit} 
      \includegraphics[width=0.48\linewidth]{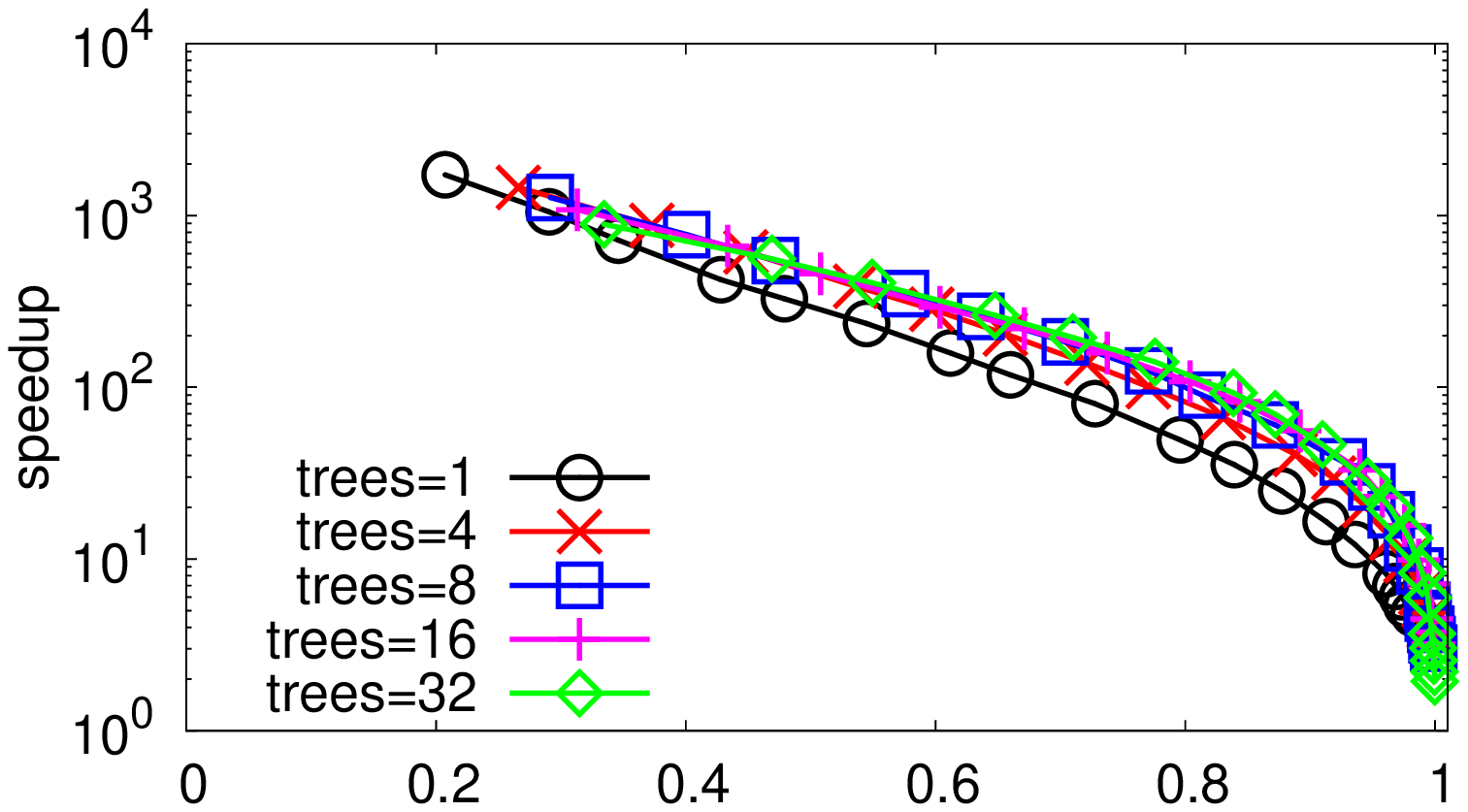}}
\subfigure[\small Gist ]{
      \label{fig:exp_KDTree_in_gist} 
      \includegraphics[width=0.48\linewidth]{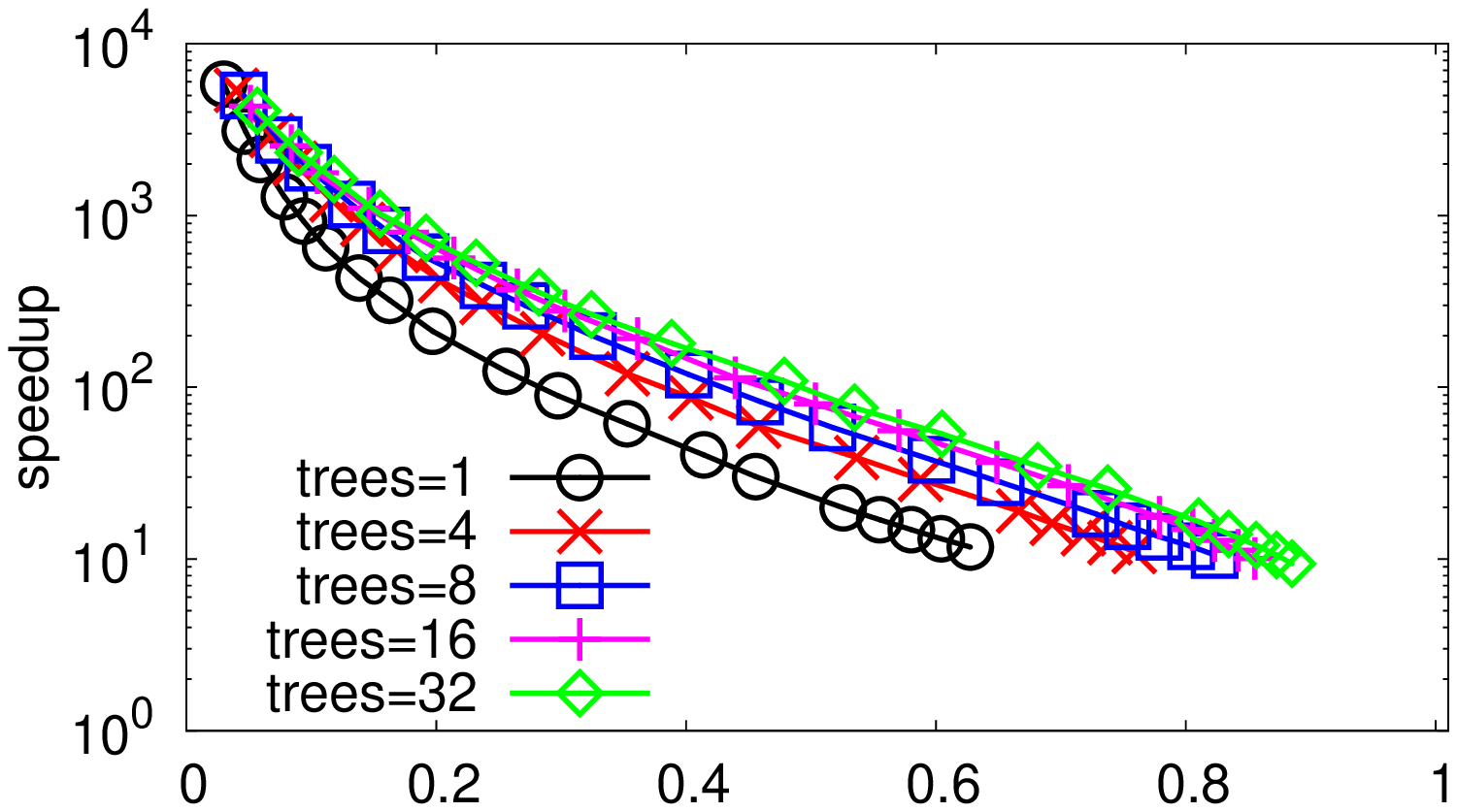}}
\end{minipage}%
\vspace{-4mm}
\caption{\small Speedup vs Recall for Diff $t$ (\textbf{Flann-KD})}
\label{fig:exp_SpH_in}
\end{figure}

\subsection{Flann}

Flann defines the cost as a combination of the search time, tree build time and the tree memory overhead.

We used the default search precisions (90 percent) and several combinations of the tradeoff factors $wb$ and $wm$. For the build time weight, $wb$, we used three different possible values: 0 representing the case where we don't care about the tree build time, 1 for the case where the tree build time and search time have the same importance and 0.01 representing the case where we care mainly about the search time but we also want to avoid a large build time. Similarly, the memory weight was chosen to be 0 for the case where the memory usage is not a concern, 100 representing the case where the memory use is the dominant concern and 1 as a middle ground between the two cases.

When we pay more attention to the size of the memory use, the search speedup is very low and almost declines to that of linear scan. For the datasets who have the medium data size or the system with enough memory, the $wm$ of 0 would provide a good search performance. Due to the large margin between large $wb$
and small $wb$ for the search performance, we select $0.01$ for $wb$ in this paper.
\begin{figure}[tbh]
\begin{minipage}[t]{1.0\linewidth}
\centering
\subfigure[\small Sift]{
      \label{fig:exp_Flann_in_sift} 
      \includegraphics[width=0.48\linewidth]{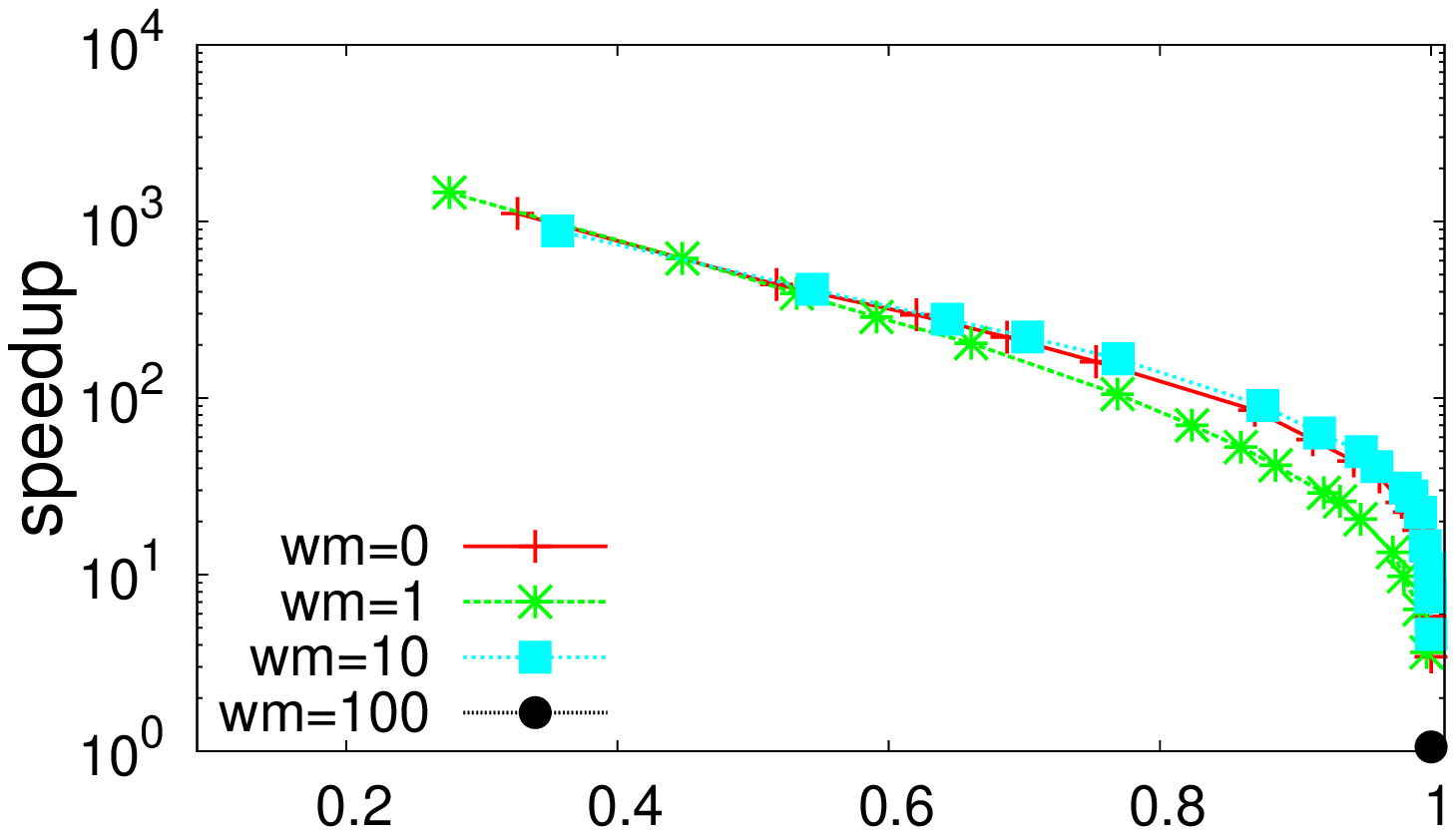}}
\subfigure[\small Mnist ]{
      \label{fig:exp_Flann_in_MNIST} 
      \includegraphics[width=0.48\linewidth]{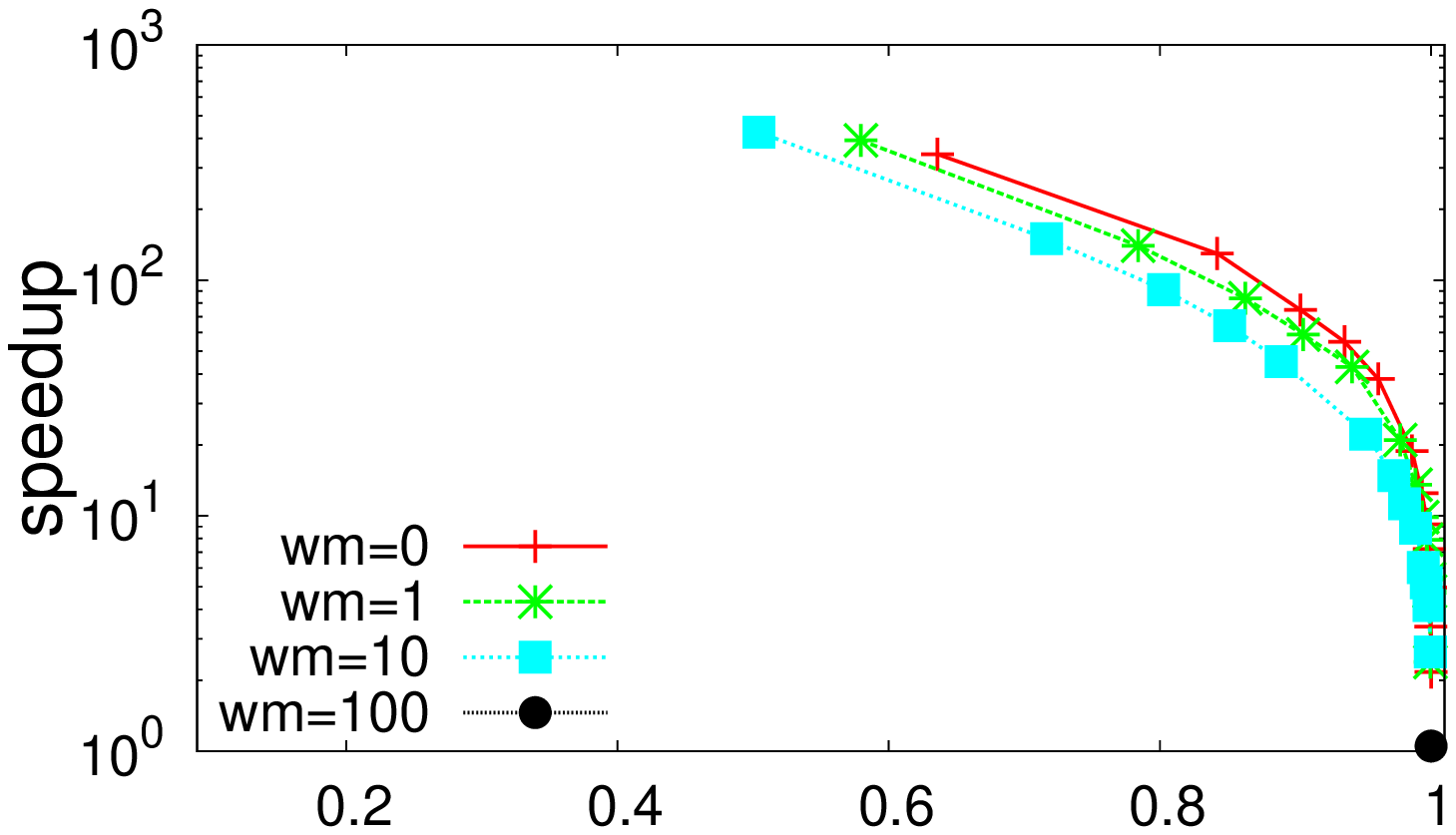}}
\end{minipage}%
\vspace{-4mm}
\caption{\small Speedup vs Recall for Diff $wm$ (\textbf{Flann})}
\label{fig:exp_SpH_in}
\end{figure}

\begin{figure}[tbh]
\begin{minipage}[t]{1.0\linewidth}
\centering
\subfigure[\small Sift ]{
      \label{fig:exp_Flann_in_sift} 
      \includegraphics[width=0.48\linewidth]{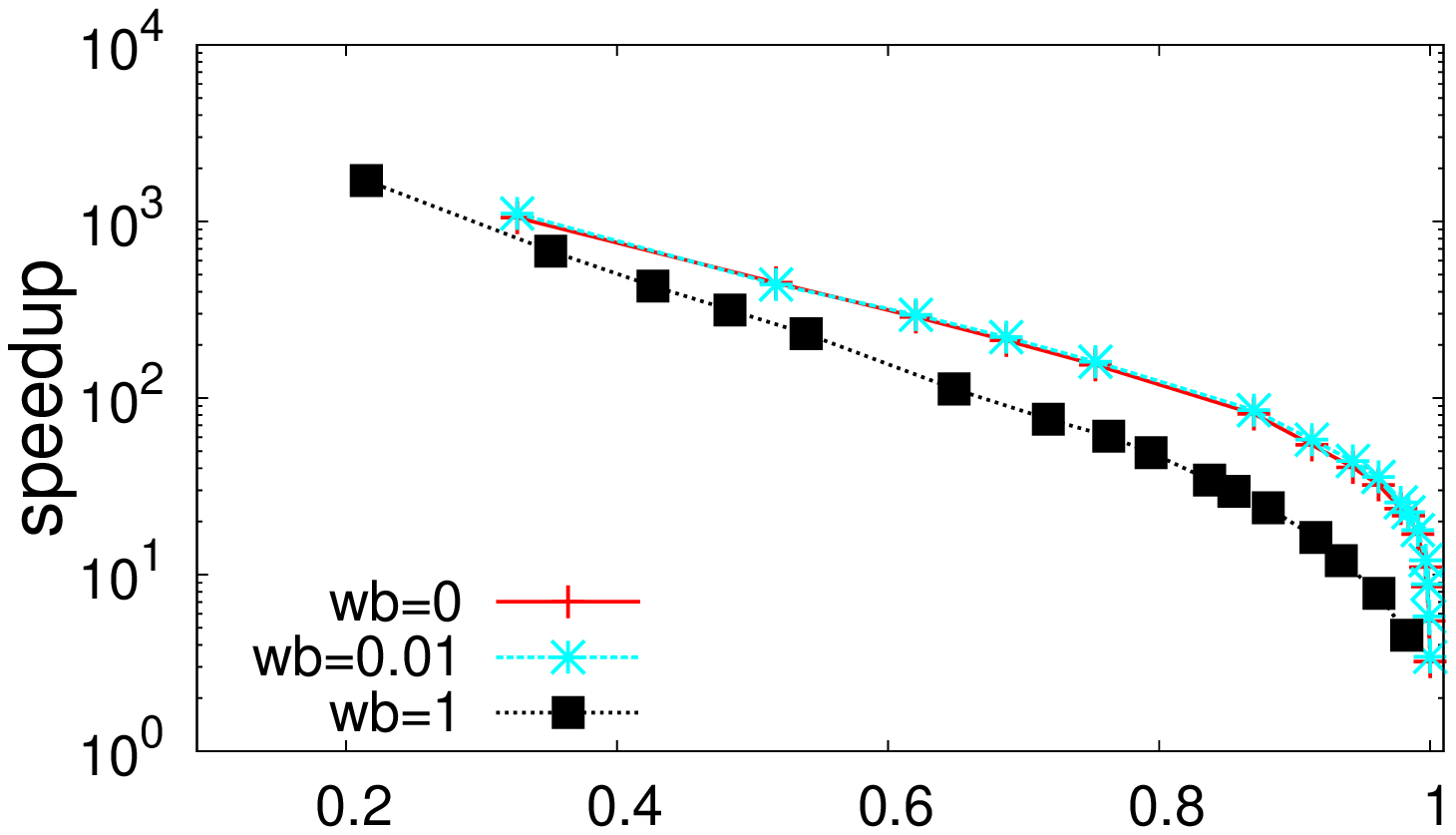}}
\subfigure[\small Mnist ]{
      \label{fig:exp_Flann_in_MNIST} 
      \includegraphics[width=0.48\linewidth]{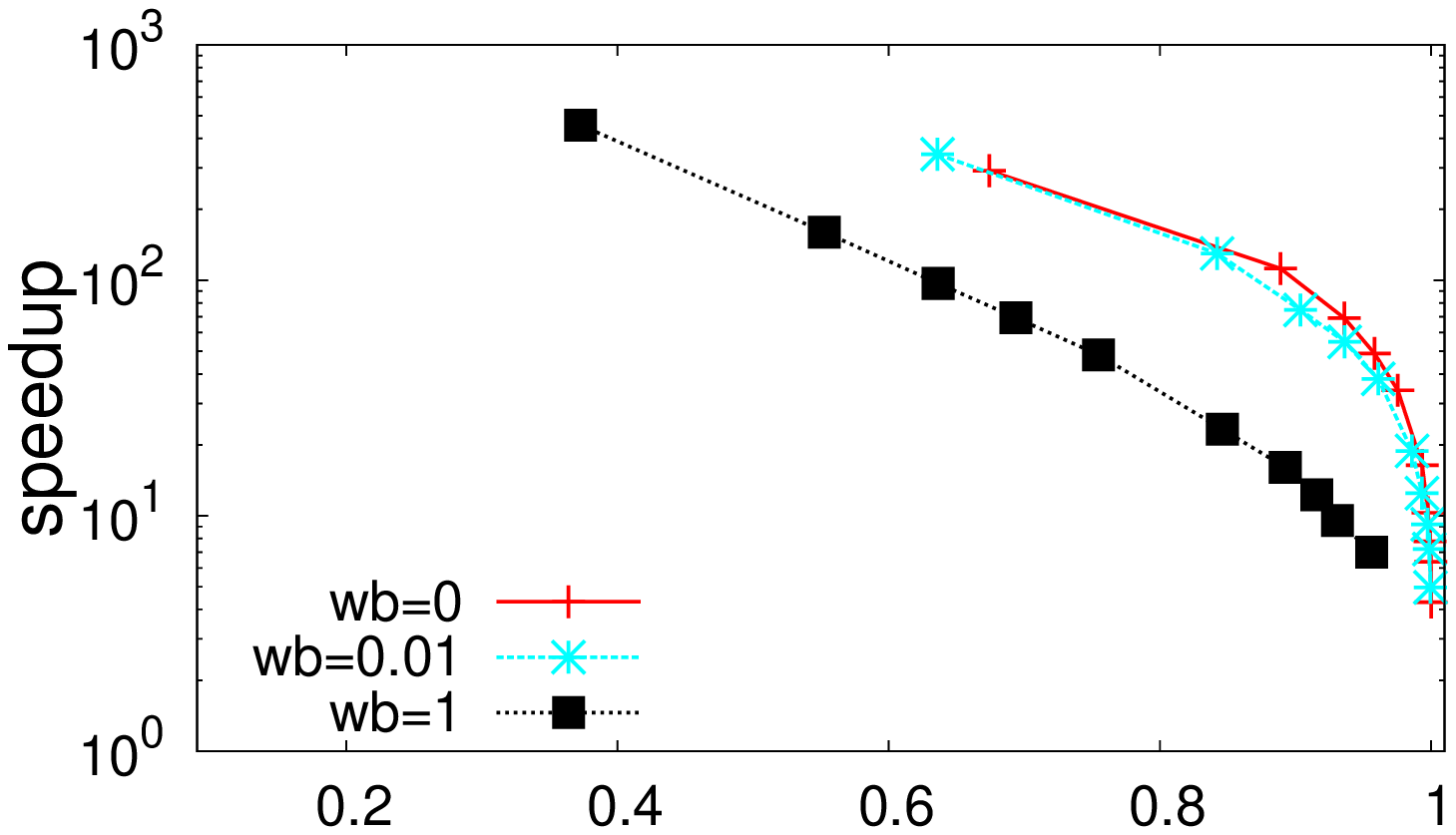}}
\end{minipage}%
\vspace{-4mm}
\caption{\small Speedup vs Recall for Diff $wb$ (\textbf{Flann})}
\label{fig:exp_SpH_in}
\end{figure}

\subsection{Small World}

\begin{figure}[tbh]
\begin{minipage}[t]{1.0\linewidth}
\centering
\subfigure[\small Ben ]{
      \label{fig:exp_SW_in_sift} 
      \includegraphics[width=0.48\linewidth]{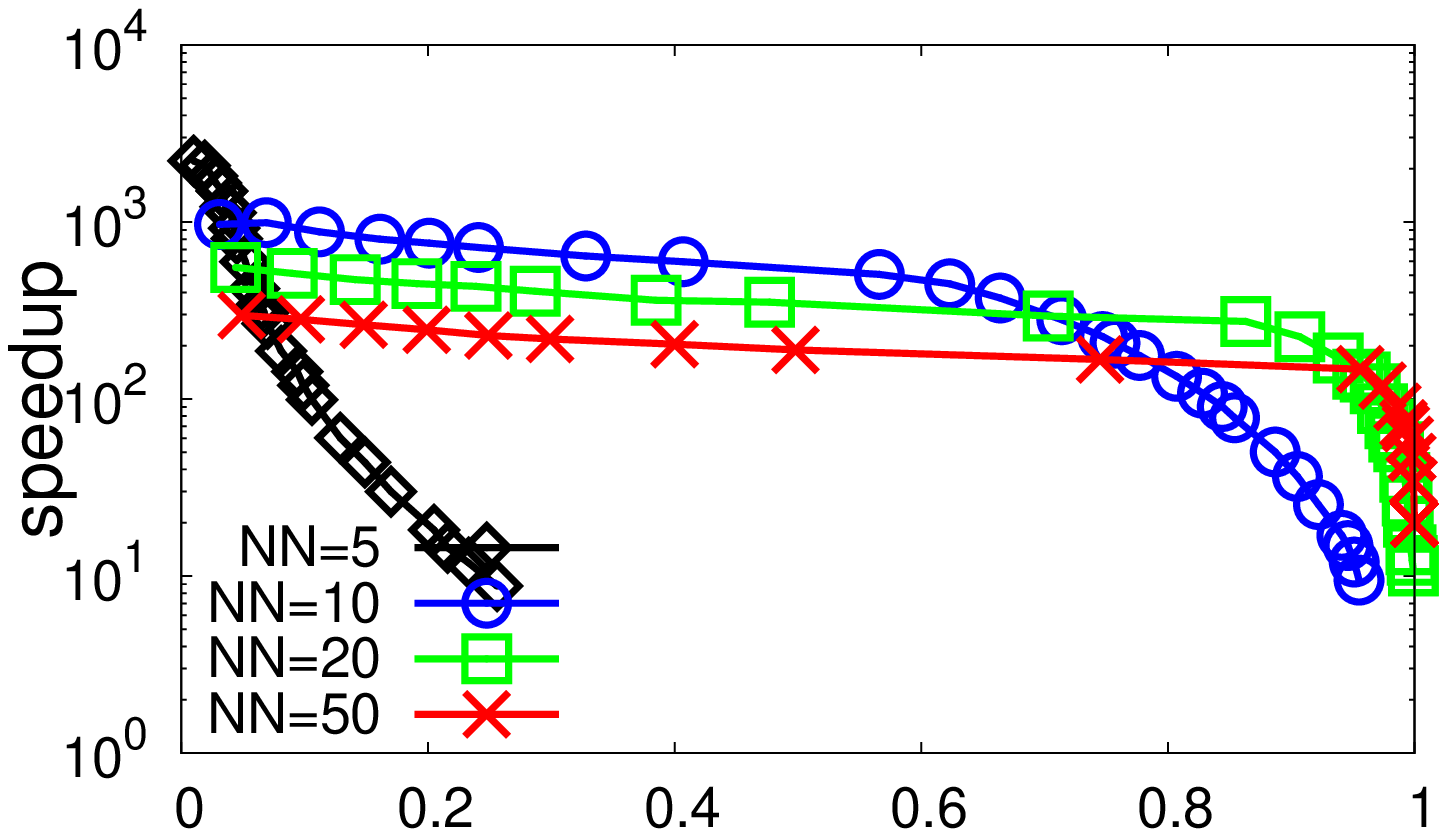}}
\subfigure[\small Deep ]{
      \label{fig:exp_SW_in_deep} 
      \includegraphics[width=0.48\linewidth]{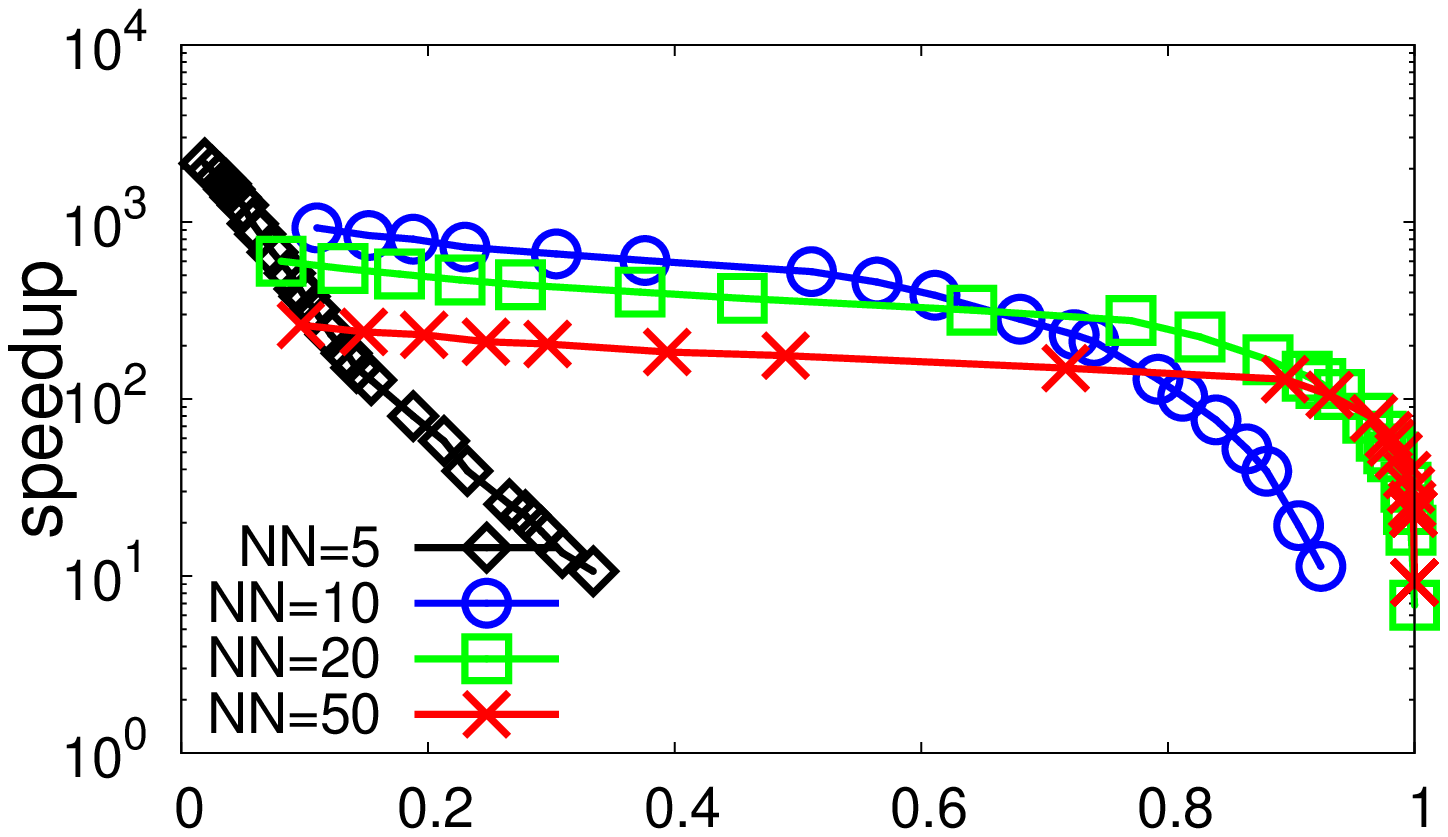}}
\subfigure[\small MSong ]{
      \label{fig:exp_SW_in_msong} 
      \includegraphics[width=0.48\linewidth]{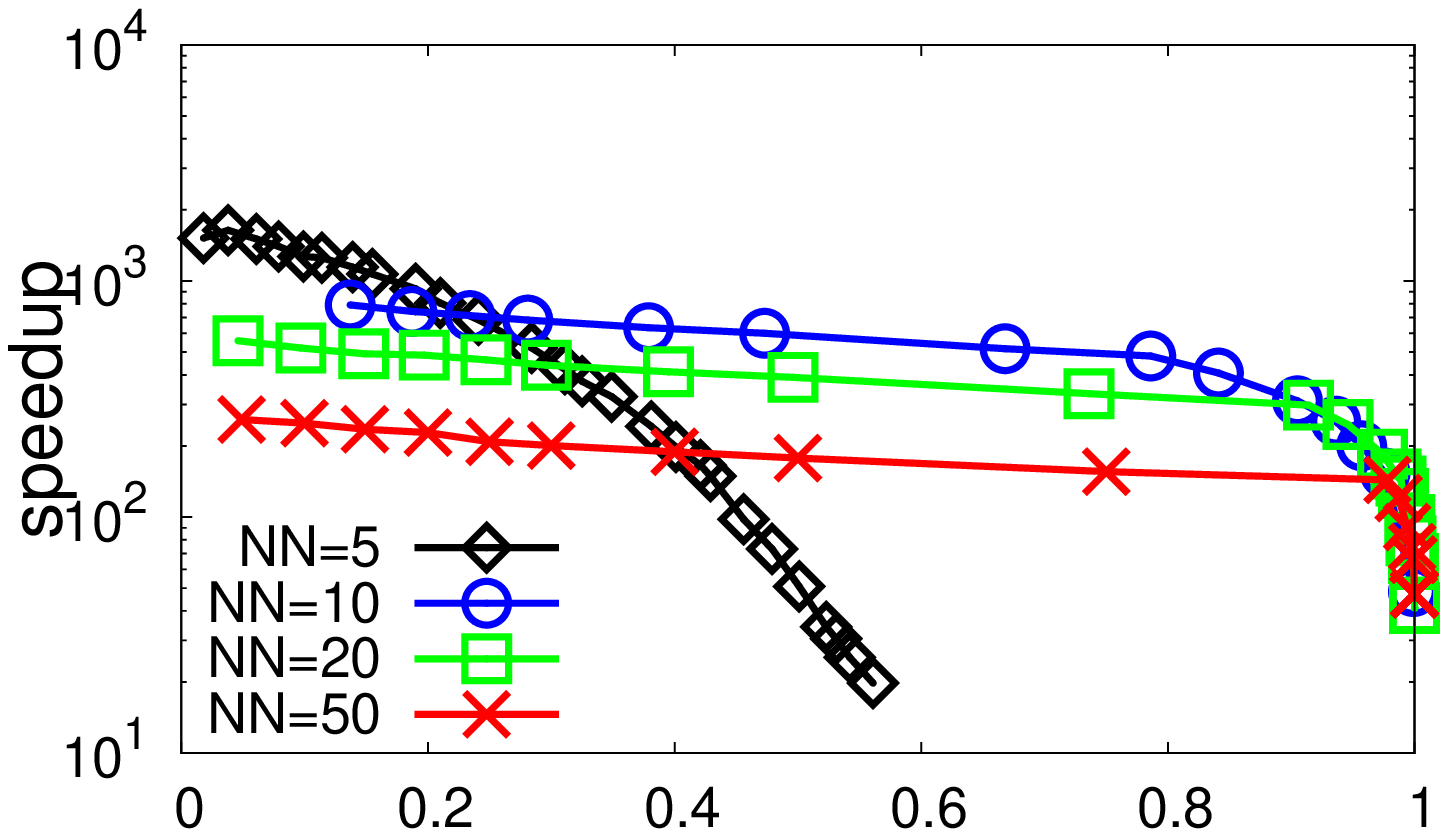}}
\subfigure[\small Glove ]{
      \label{fig:exp_SW_in_glove} 
      \includegraphics[width=0.48\linewidth]{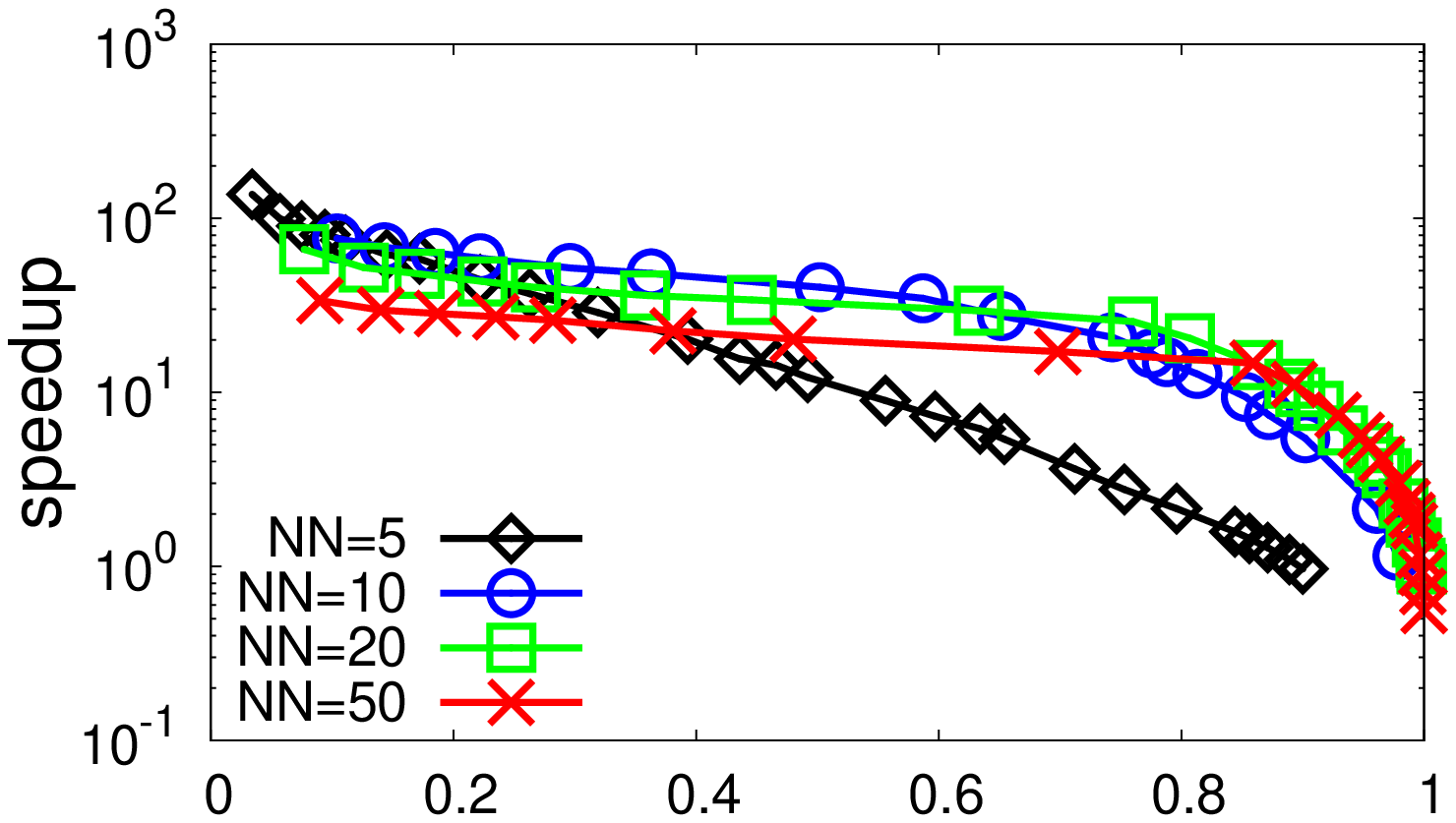}}
\end{minipage}%
\vspace{-4mm}
\caption{\small Speedup vs Recall for Diff $NN$(\textbf{SW}) }
\label{fig:exp_SW_NN}
\end{figure}

Small World involves the parameter NN( NN closest points are found during index constructing period). $S$ (number of entries using in searching procedure) is a searching parameter and we tune the value of $S$ to obtain the trade-off between the search speed and quality. The construction algorithm is computationally expensive and the time is roughly linear to the value of $NN$. Figure \ref{fig:exp_SW_NN} shows the speedup with different $NN$. The small value of $NN$ could provide a good search performance for low recall but decrease for high recall. For most of datasets, the algorithm could provide a good search tradeoff when $NN$ is 10 or 20.

\subsection{Hierarchical Navigable Small World}

\begin{figure}[tbh]
\centering
\subfigure[\small Sift ]{
      \label{fig:exp_HNSW_in_audio} 
      \includegraphics[width=0.48\linewidth]{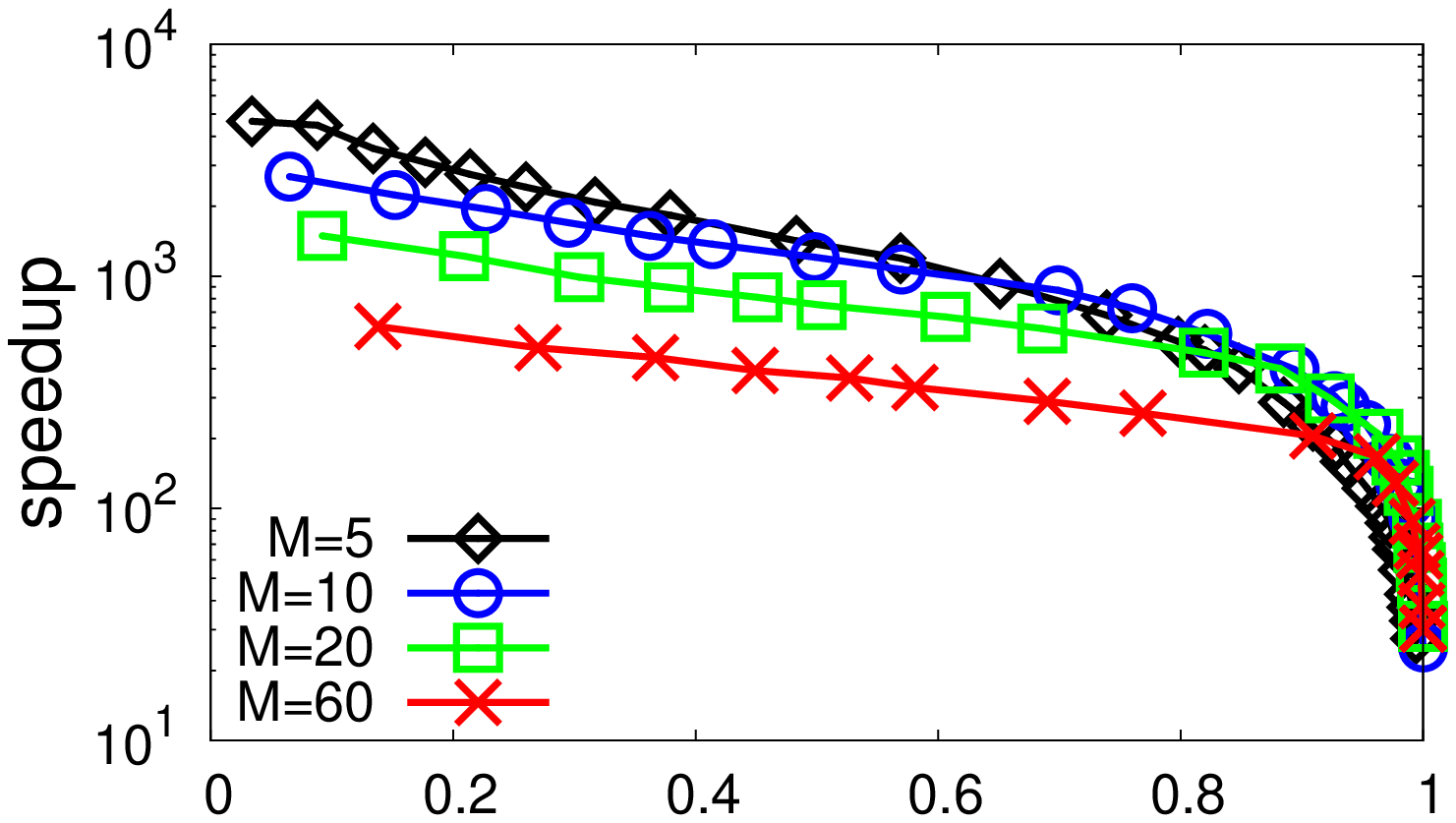}}
\subfigure[\small Deep]{
      \label{fig:exp_HNSW_in_deep} 
      \includegraphics[width=0.48\linewidth]{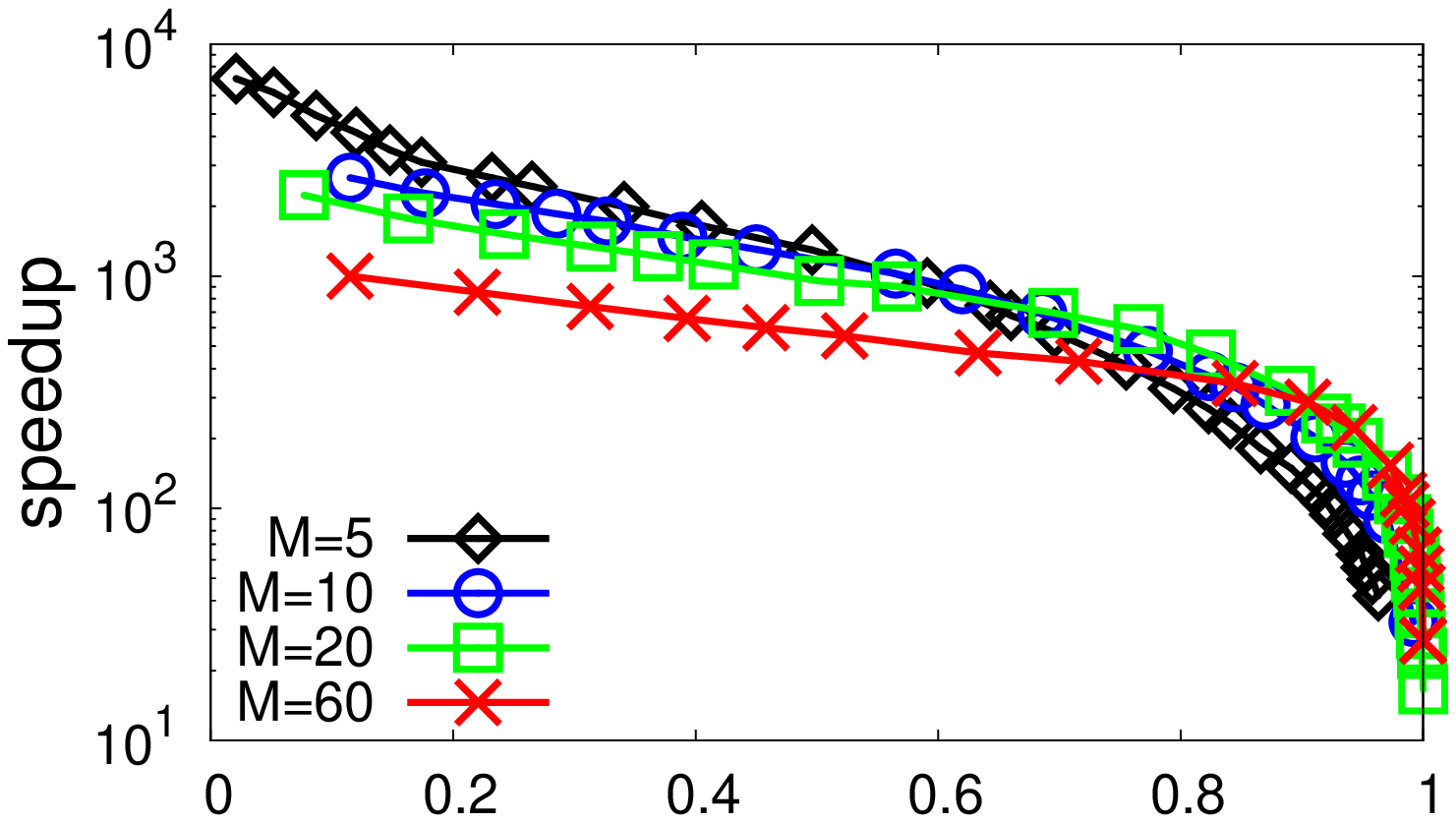}}
\subfigure[\small MNIST]{
      \label{fig:exp_HNSW_in_mnist} 
      \includegraphics[width=0.48\linewidth]{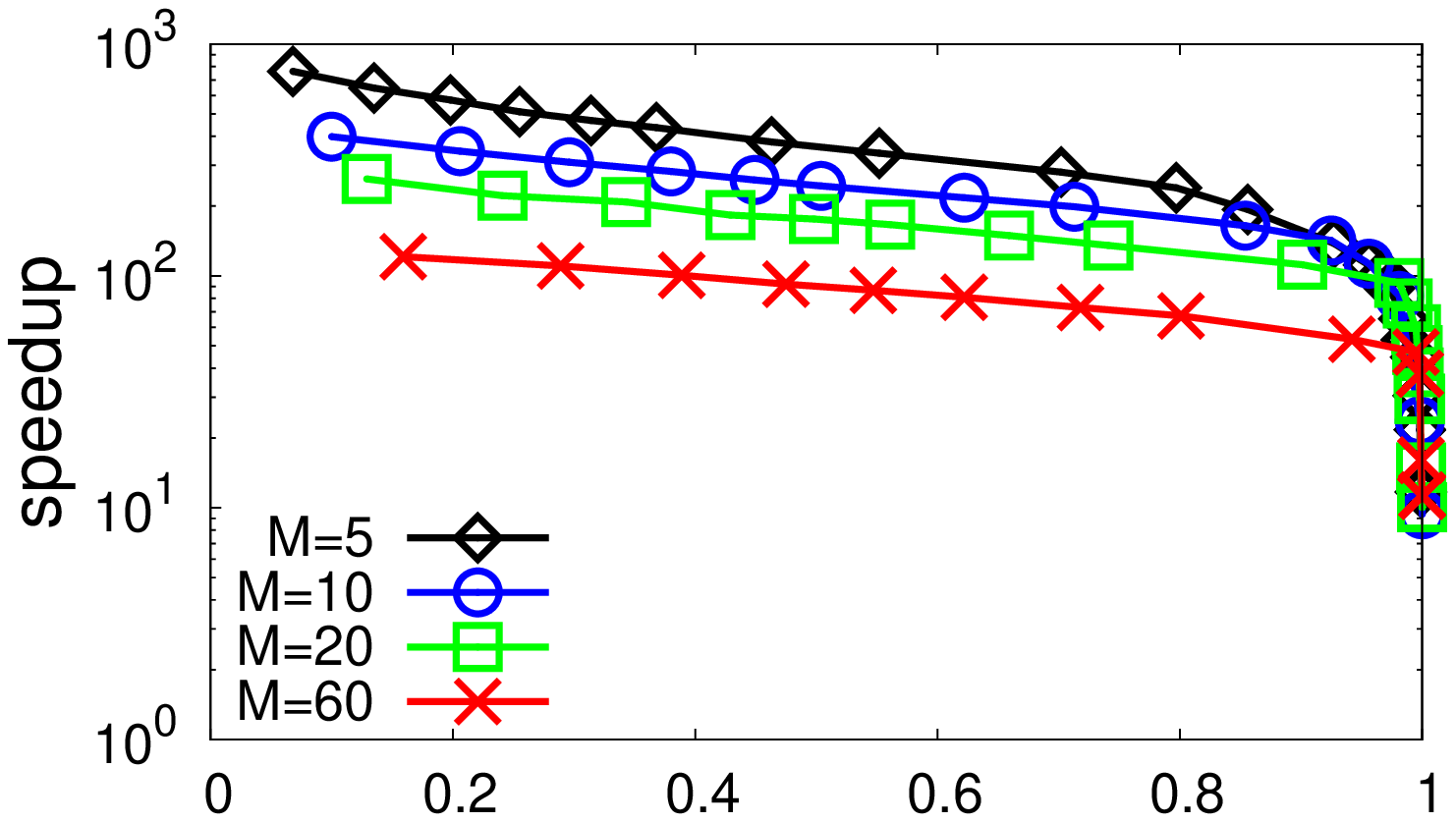}}
\subfigure[\small Ben]{
      \label{fig:exp_HNSW_in_ben} 
      \includegraphics[width=0.48\linewidth]{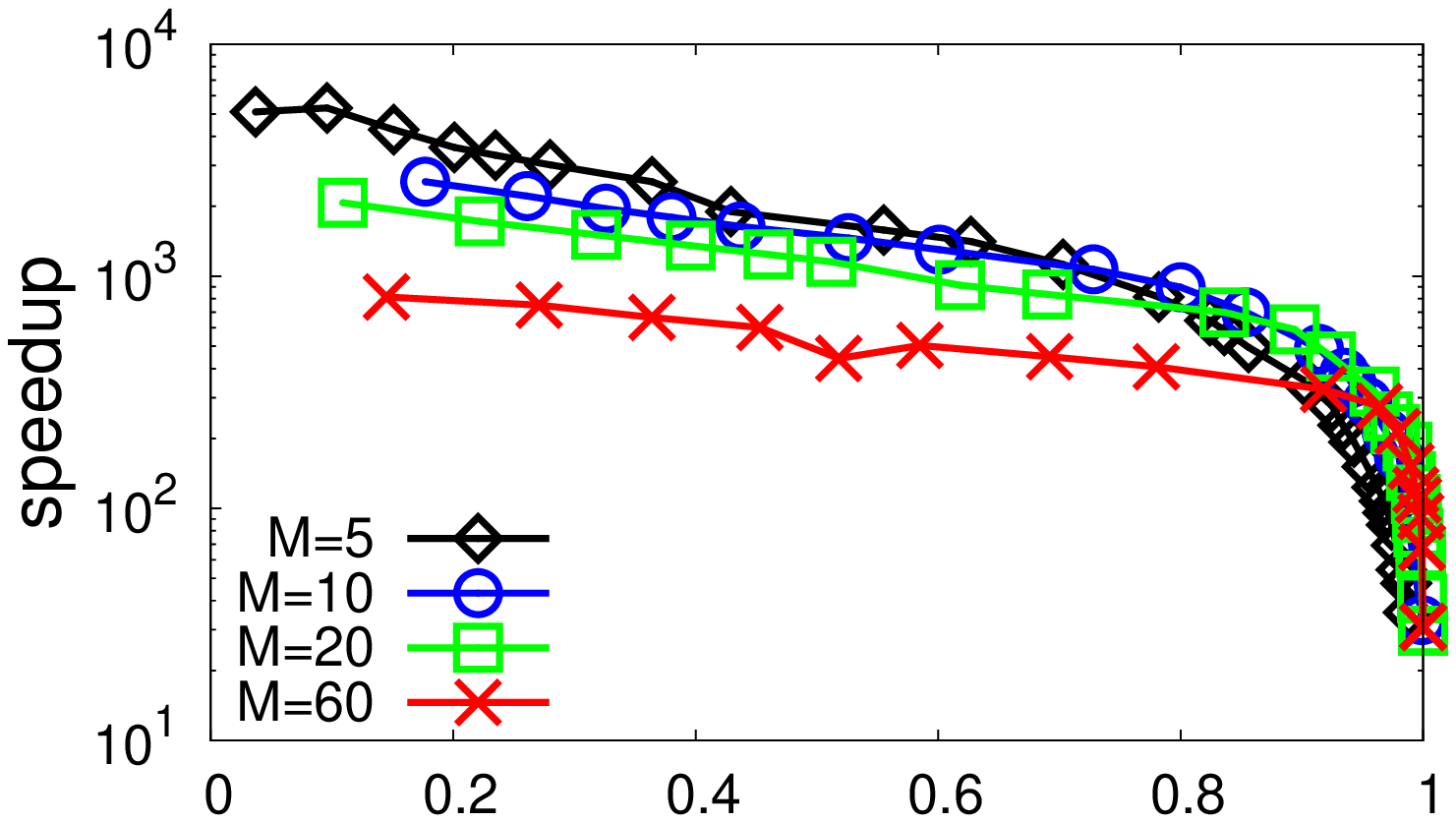}}
\vspace{-4mm}
\caption{\small Speedup vs Recall for Diff $M$(\textbf{HNSW}) }
\label{fig:exp_HNSW}
\end{figure}

Two main parameters are adopted to get the tradeoff between the index complexity and search performance: $M$ indicates the size of the potential neighbors in some layers for indexing phase, and $efConstruction$ is used to controlled the quality of the neighbors during indexing. We use the default value of $efConstruction$, which is set to 200. $efSearch$ is a parameter similar to $efConstruction$ to control the search quality. We change the value of $efSearch$ to get the tradeoff between the search speed and quality. Figure \ref{fig:exp_HNSW} show the search performance of different $M$. Similar to SW, the small value of $M$ could provide a good search performance for low recall but decrease for high recall. We set $M=10$ as a default value.

\subsection{Rank Cover Tree}

In RCT, there are parameters: $\Delta$($1/\Delta$ is the sample rate sampled from the lower level, which could determine the height $h$ of the tree), $p$(the maximum number of parents for each node) and coverage parameter $\omega$.Acording to the the recommendations from the authors, we select $h=4$ to build the rank cover tree.

\begin{figure}[tbh]
\begin{minipage}[t]{1.0\linewidth}
\centering
\subfigure[\small Sift for Diff $p$]{
      \label{fig:exp_Flann_in_sift} 
      \includegraphics[width=0.48\linewidth]{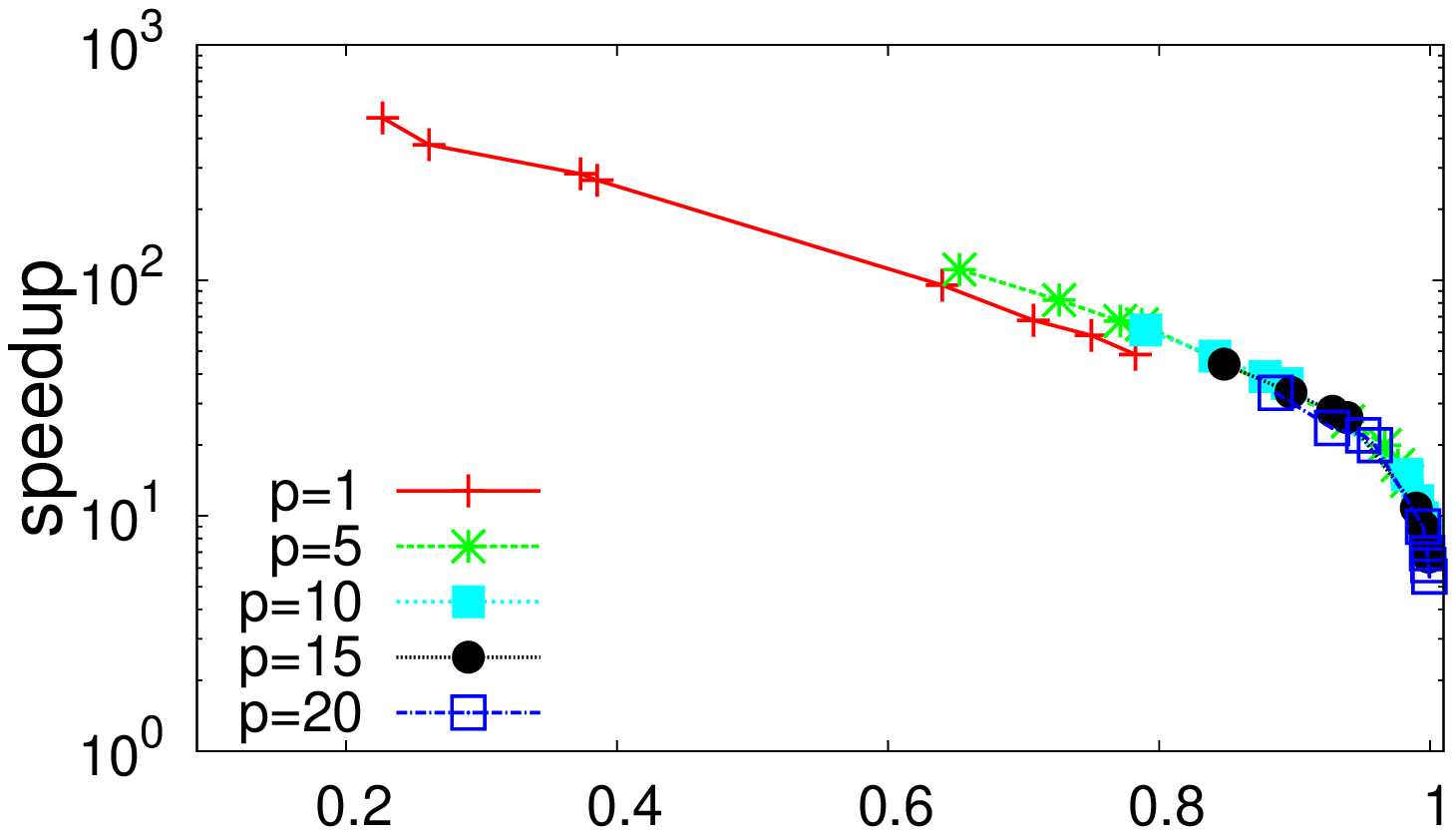}}
\subfigure[\small Nusw for Diff $p$ ]{
      \label{fig:exp_Flann_in_nuswide} 
      \includegraphics[width=0.48\linewidth]{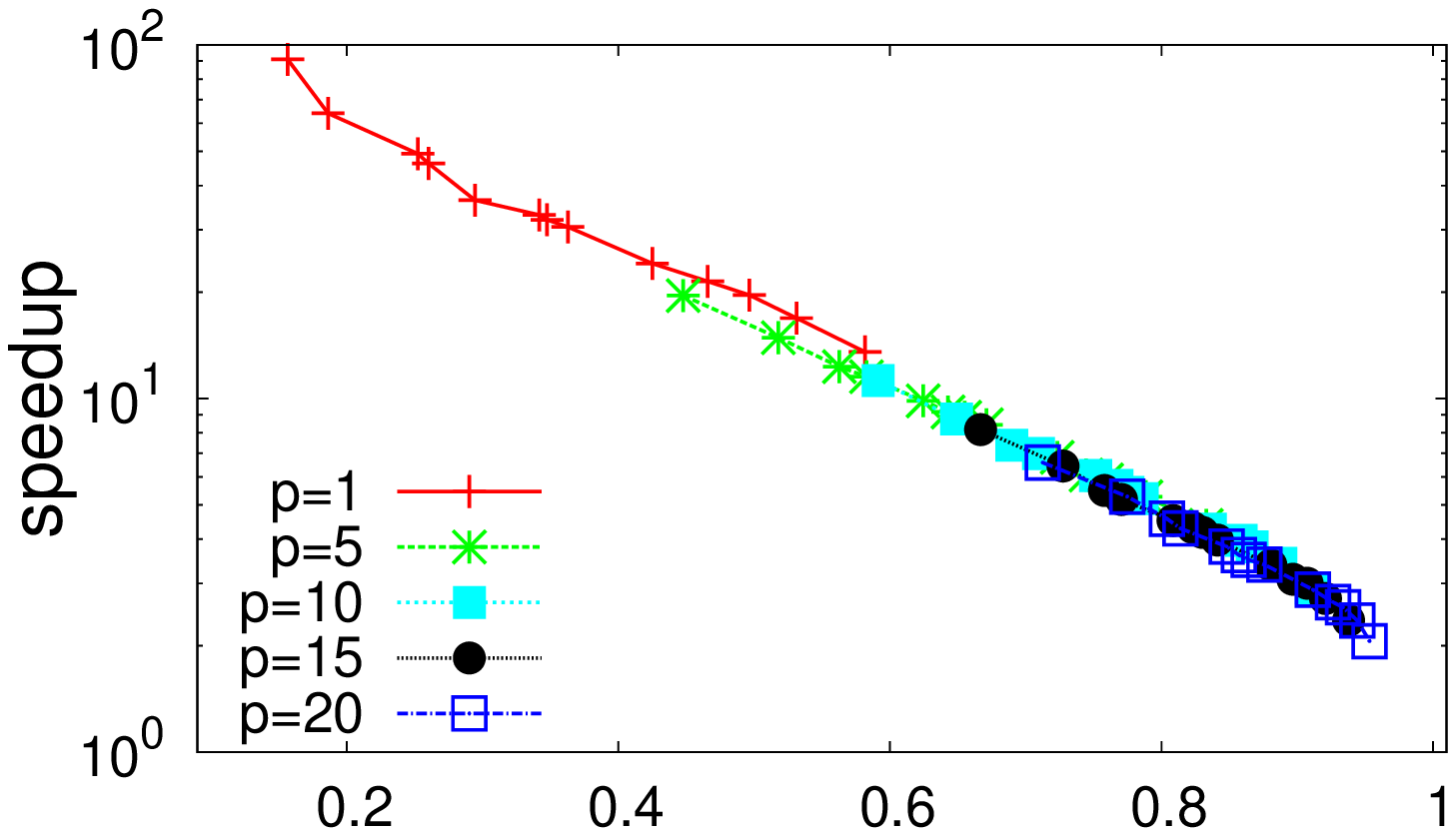}}
\subfigure[\small Sift for Diff $\omega$]{
      \label{fig:exp_Flann_in_sift} 
      \includegraphics[width=0.48\linewidth]{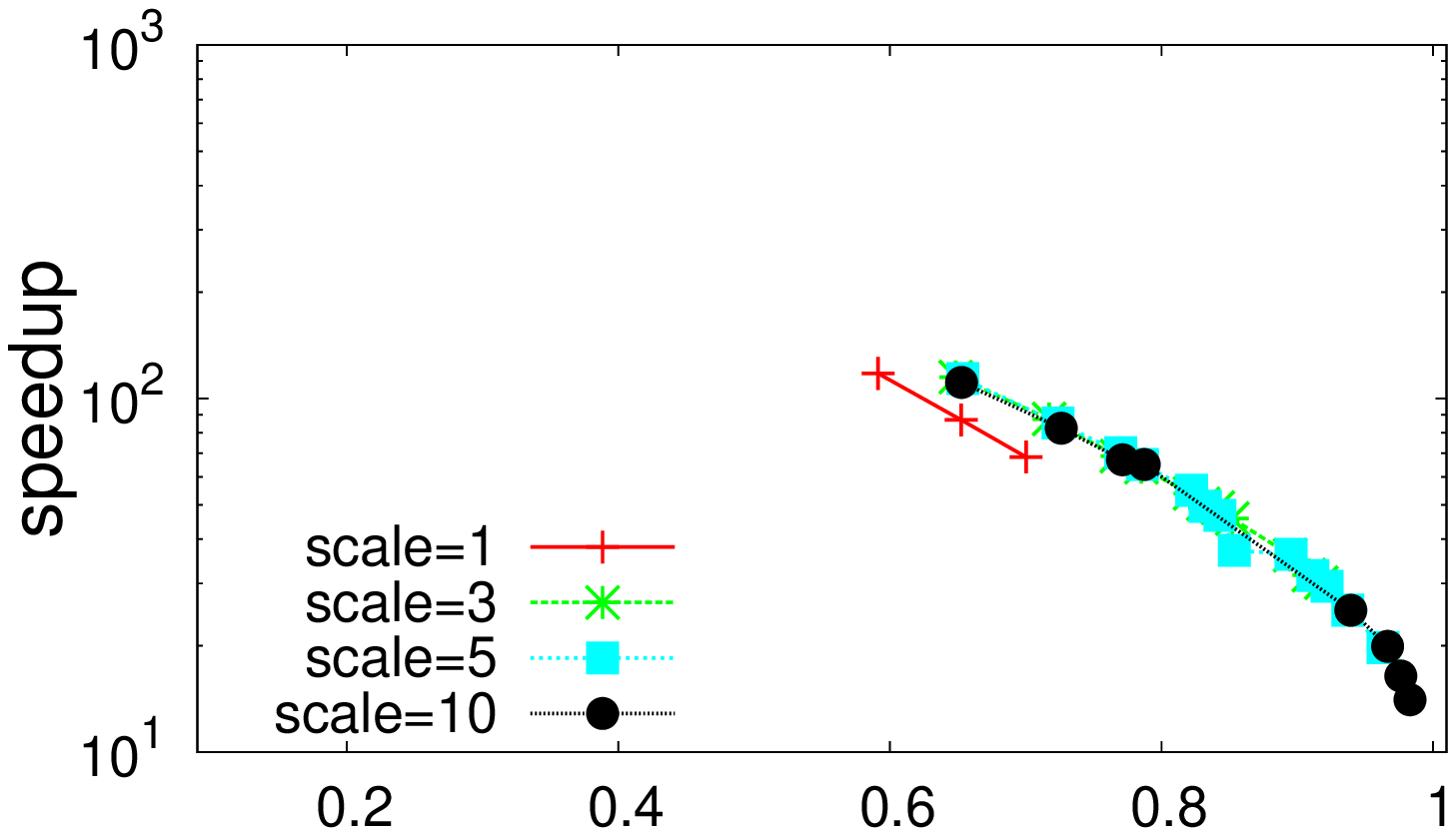}}
\subfigure[\small Nusw for Diff $\omega$]{
      \label{fig:exp_Flann_in_nuswide} 
      \includegraphics[width=0.48\linewidth]{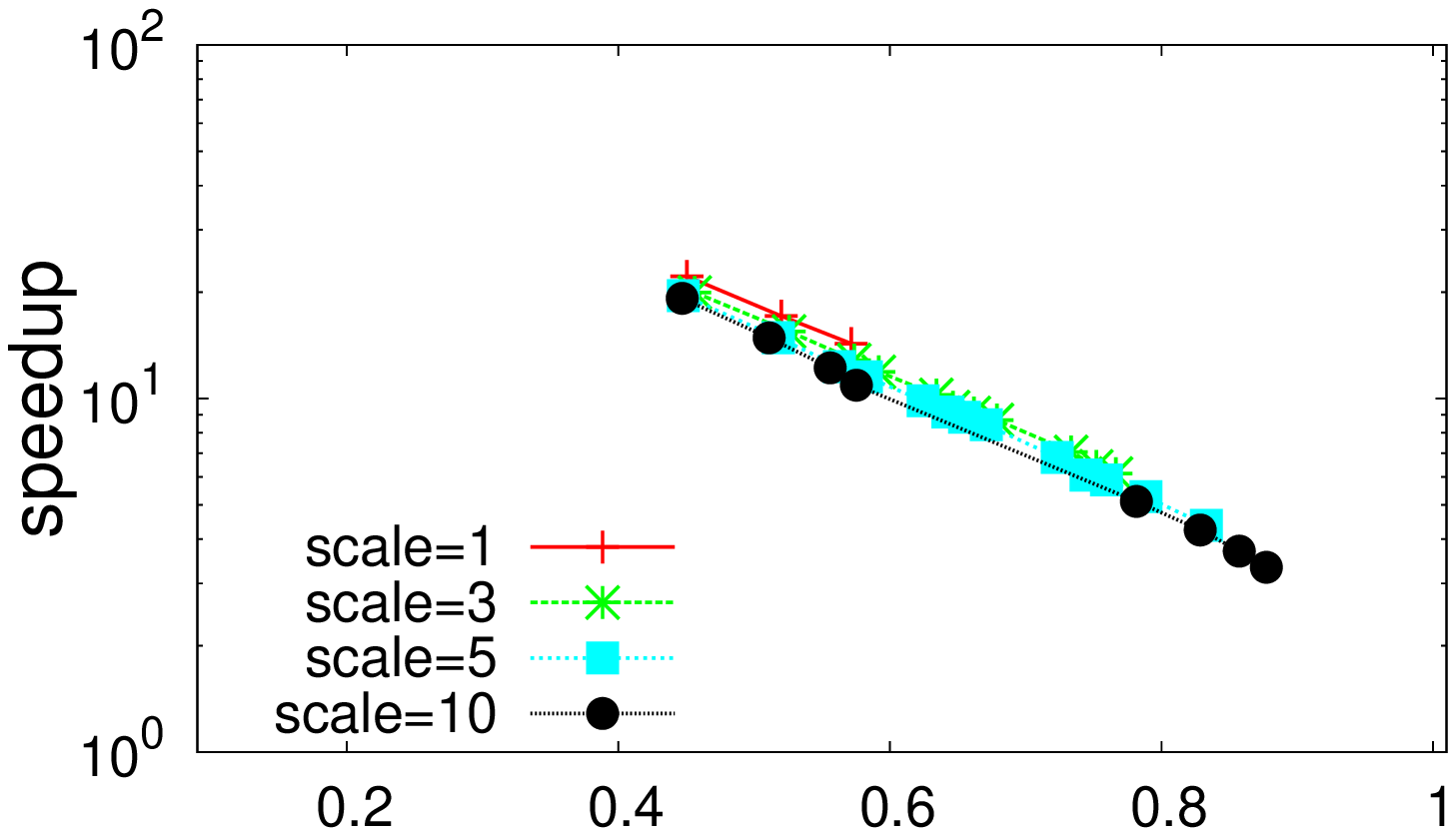}}
\end{minipage}%
\vspace{-4mm}
\caption{\small Speedup vs Recall (\textbf{RCT})}
\label{fig:exp_RCT}
\end{figure}

\subsection{KGraph}

\begin{figure}[tbh]
\begin{minipage}[t]{1.0\linewidth}
\centering
\subfigure[\small Audio ]{
      \label{fig:exp_SW_in_audio} 
      \includegraphics[width=0.48\linewidth]{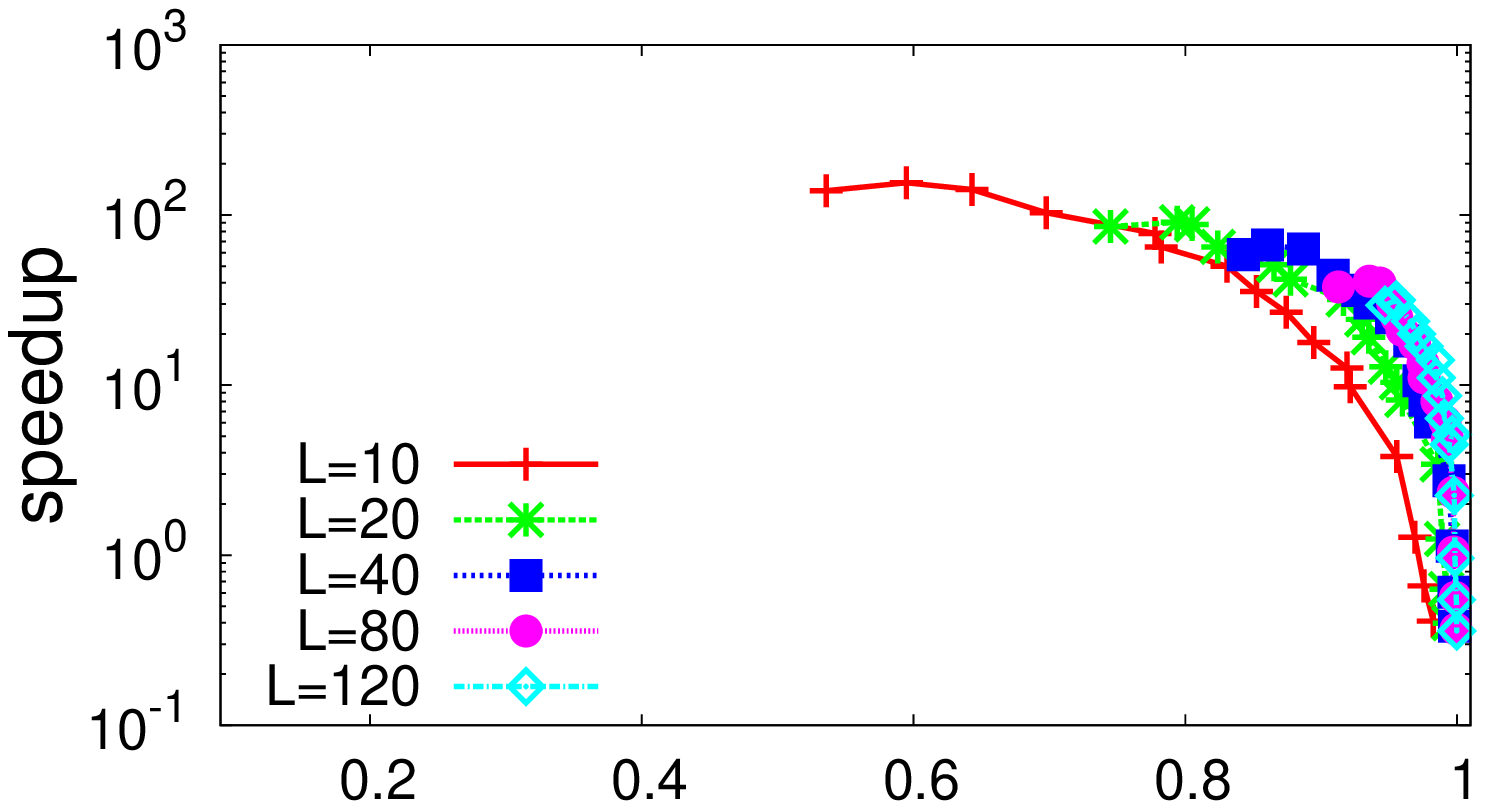}}
\subfigure[\small Sift ]{
      \label{fig:exp_SW_in_deep} 
      \includegraphics[width=0.48\linewidth]{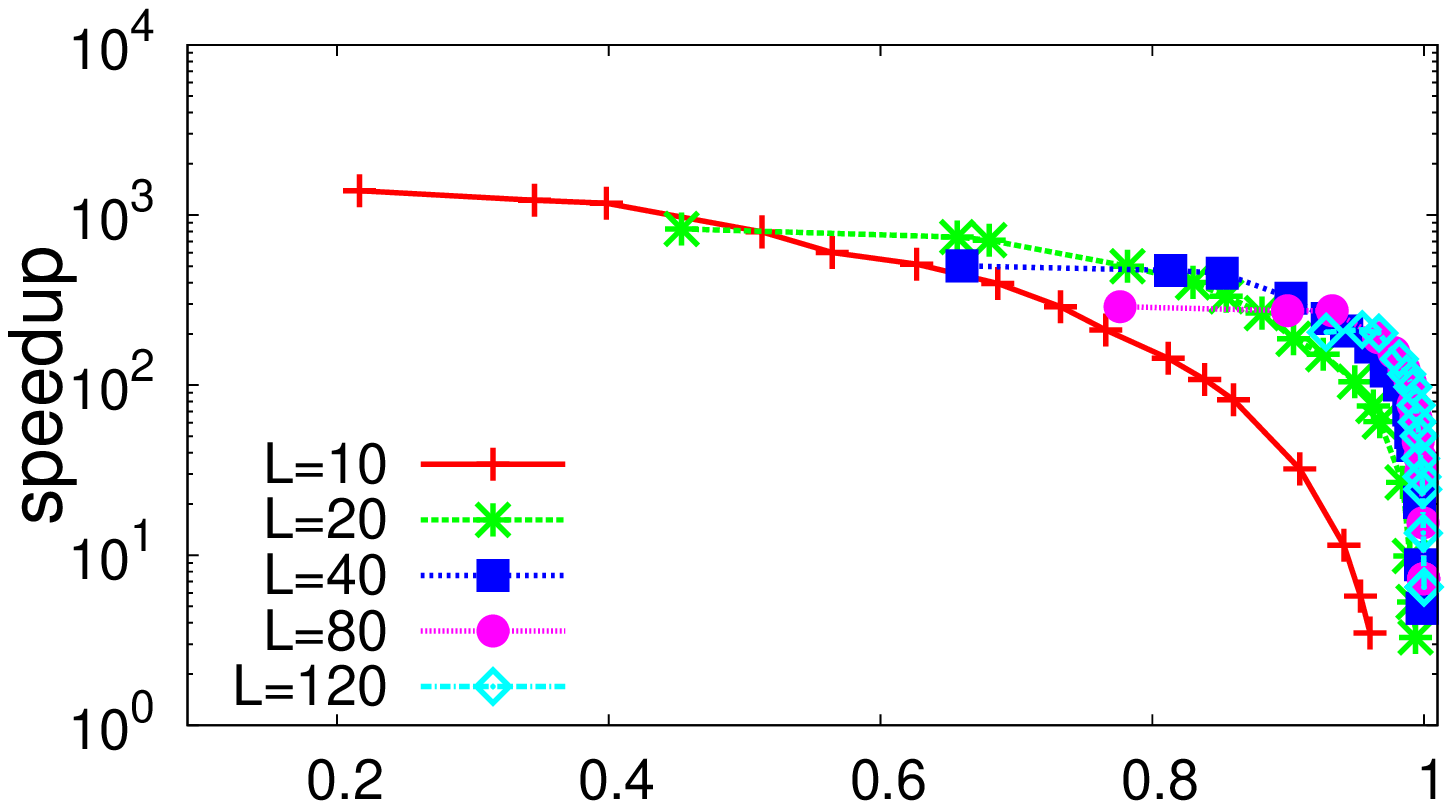}}
\subfigure[\small Mnist ]{
      \label{fig:exp_SW_in_sift} 
      \includegraphics[width=0.48\linewidth]{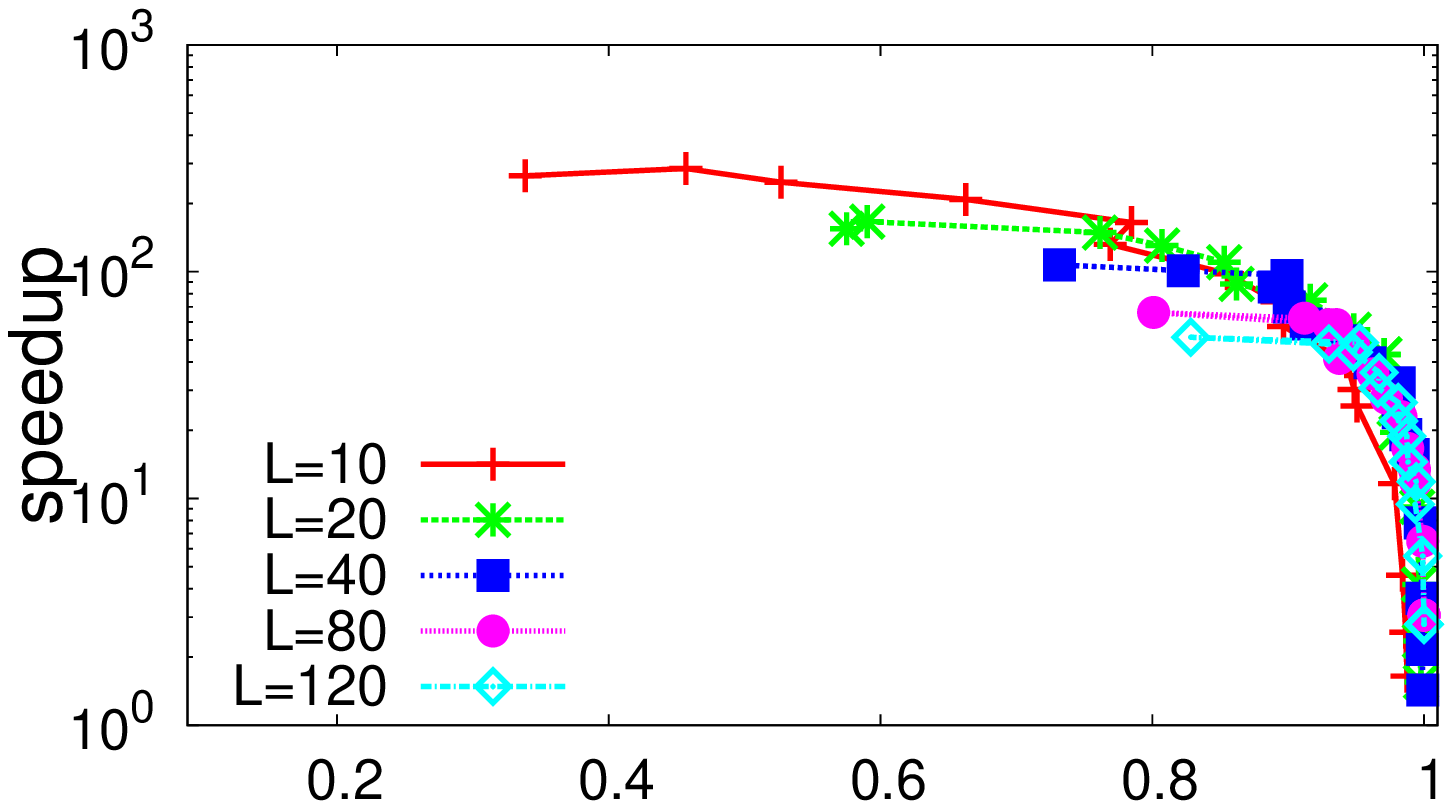}}
\subfigure[\small Notre ]{
      \label{fig:exp_SW_in_notre} 
      \includegraphics[width=0.48\linewidth]{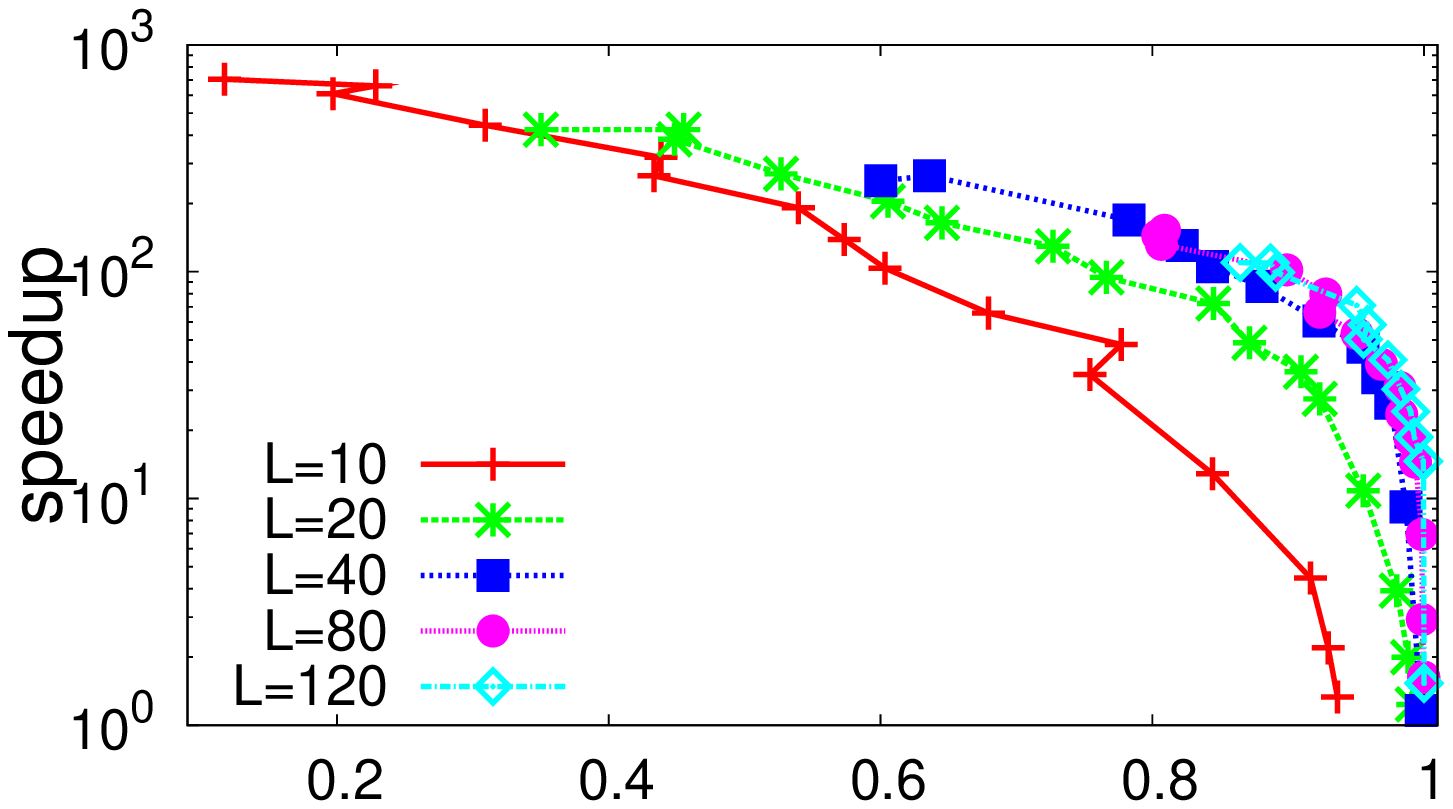}}
\end{minipage}%
\vspace{-4mm}
\caption{\small Speedup vs Recall for Diff $L$ (\textbf{KGraph}) }
\label{fig:exp_KGraph}
\end{figure}

K-NN Graph involves three parameters: $IK$ (the number of most similar objects connected with each vertex), sample rate $\rho$, termination threshold $\zeta$ and $P$ (initial entries count to start the search). The meaning of $\zeta$ is the loss in recall tolerable with early termination. We use a default termination threshold of 0.002. As reported by the author, the recall grows slowly beyond $\rho=0.5$. Here we study the impact of $IK$ and $\rho$ on performance. From figure \ref{fig:exp_KGraph}, we see that $K>40$ is need for most of the datasets.

\subsection{DPG}
\label{app:dpg}

The parameter tuning of \Algdpg{} is similar to that of \Algkgraph{}. In the experiments, \Algdpg{} has the same setting with \Algkgraph{} except that we use $\kappa = \frac{K}{2} = 20$ so that the index size of \Algdpg{} is the same as that of \Algkgraph{} in the worst case.

\begin{figure}[tbh]
\centering
\subfigure[\small Sift ]{
      \label{fig:exp_SW_in_audio} 
      \includegraphics[width=0.48\linewidth]{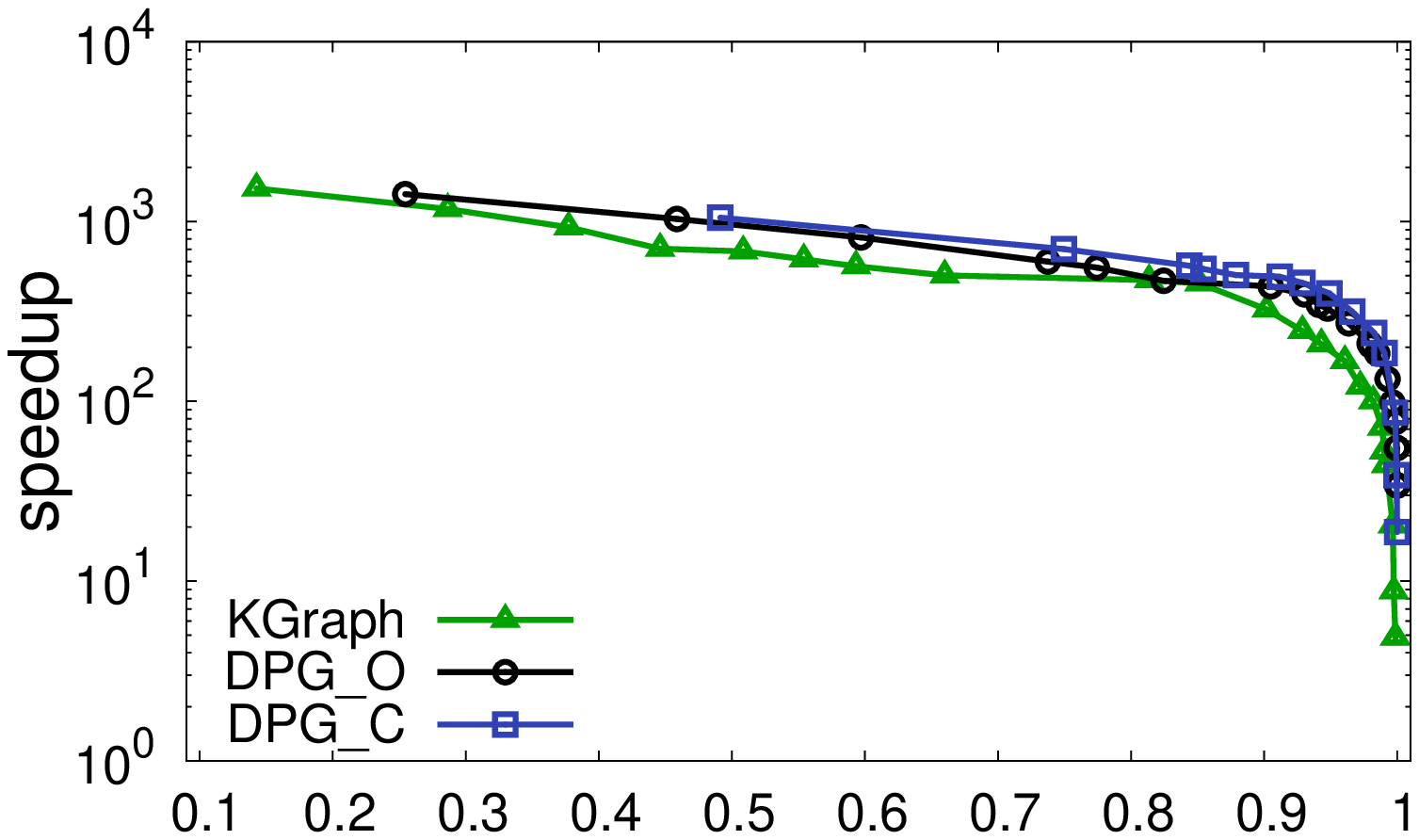}}
\subfigure[\small Gist ]{
      \label{fig:exp_SW_in_gist} 
      \includegraphics[width=0.48\linewidth]{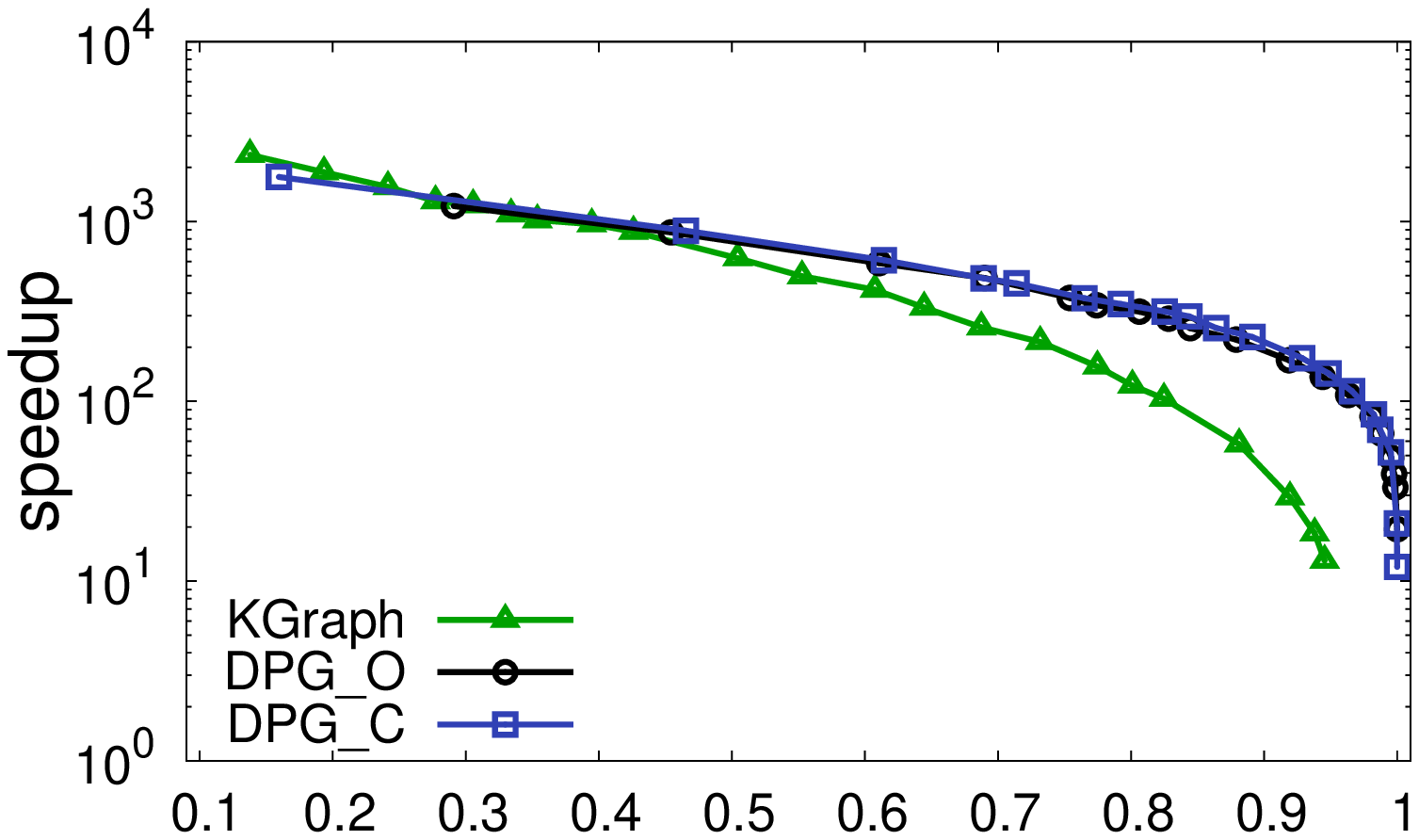}}
\subfigure[\small Imag ]{
      \label{fig:exp_SW_in_gist} 
      \includegraphics[width=0.48\linewidth]{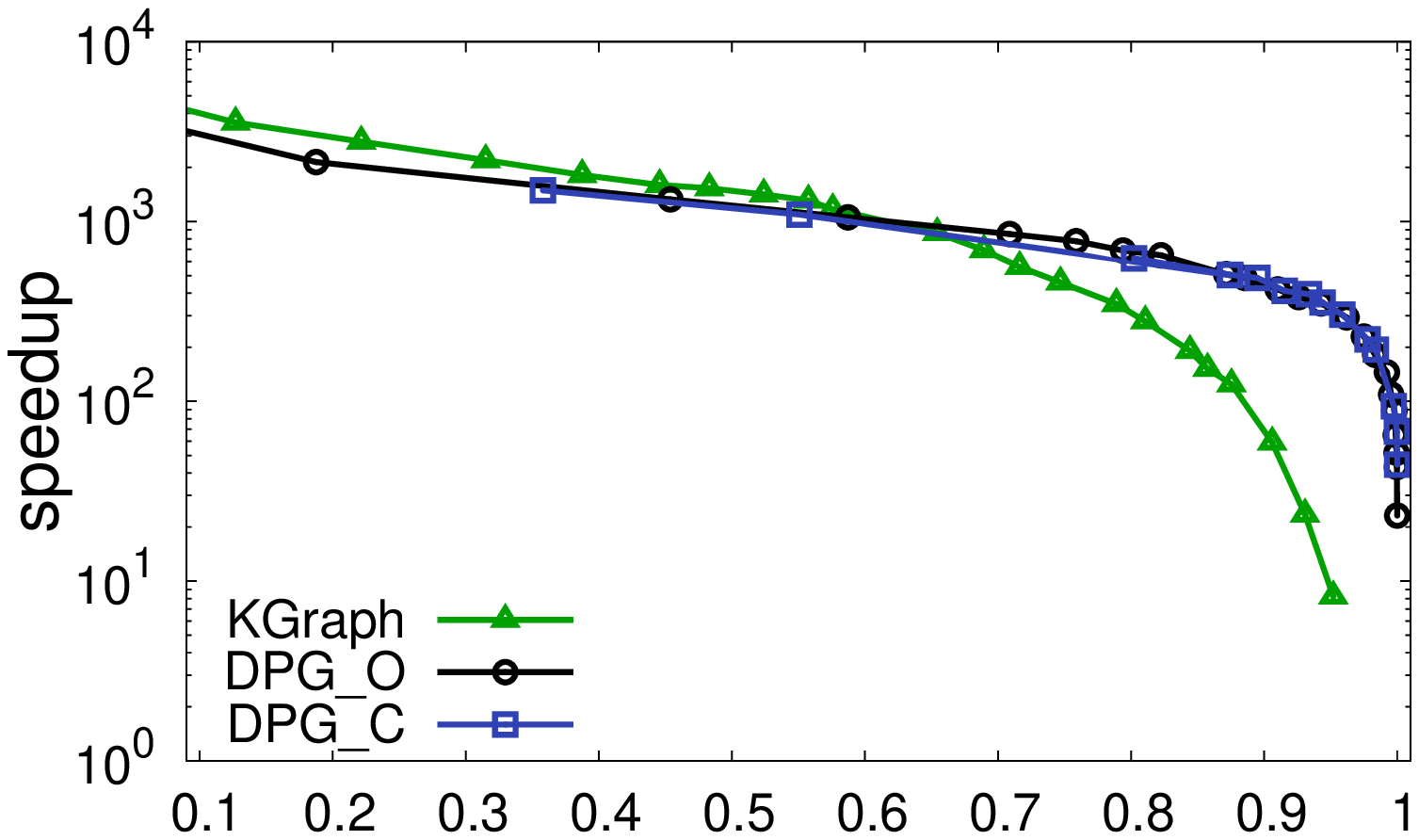}}
\subfigure[\small Notre ]{
      \label{fig:exp_SW_in_gist} 
      \includegraphics[width=0.48\linewidth]{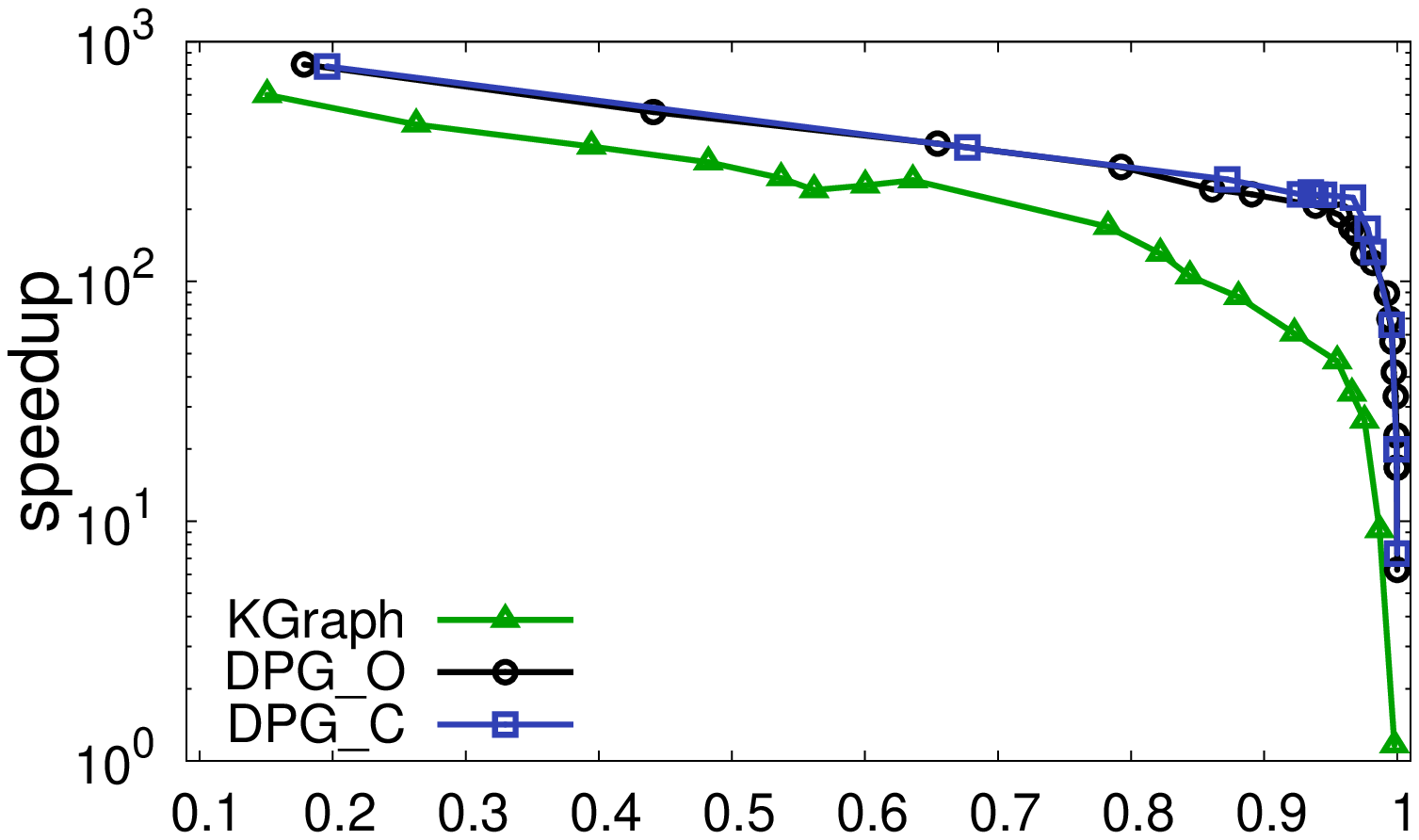}}
\subfigure[\small Ben ]{
      \label{fig:exp_SW_in_gist} 
      \includegraphics[width=0.48\linewidth]{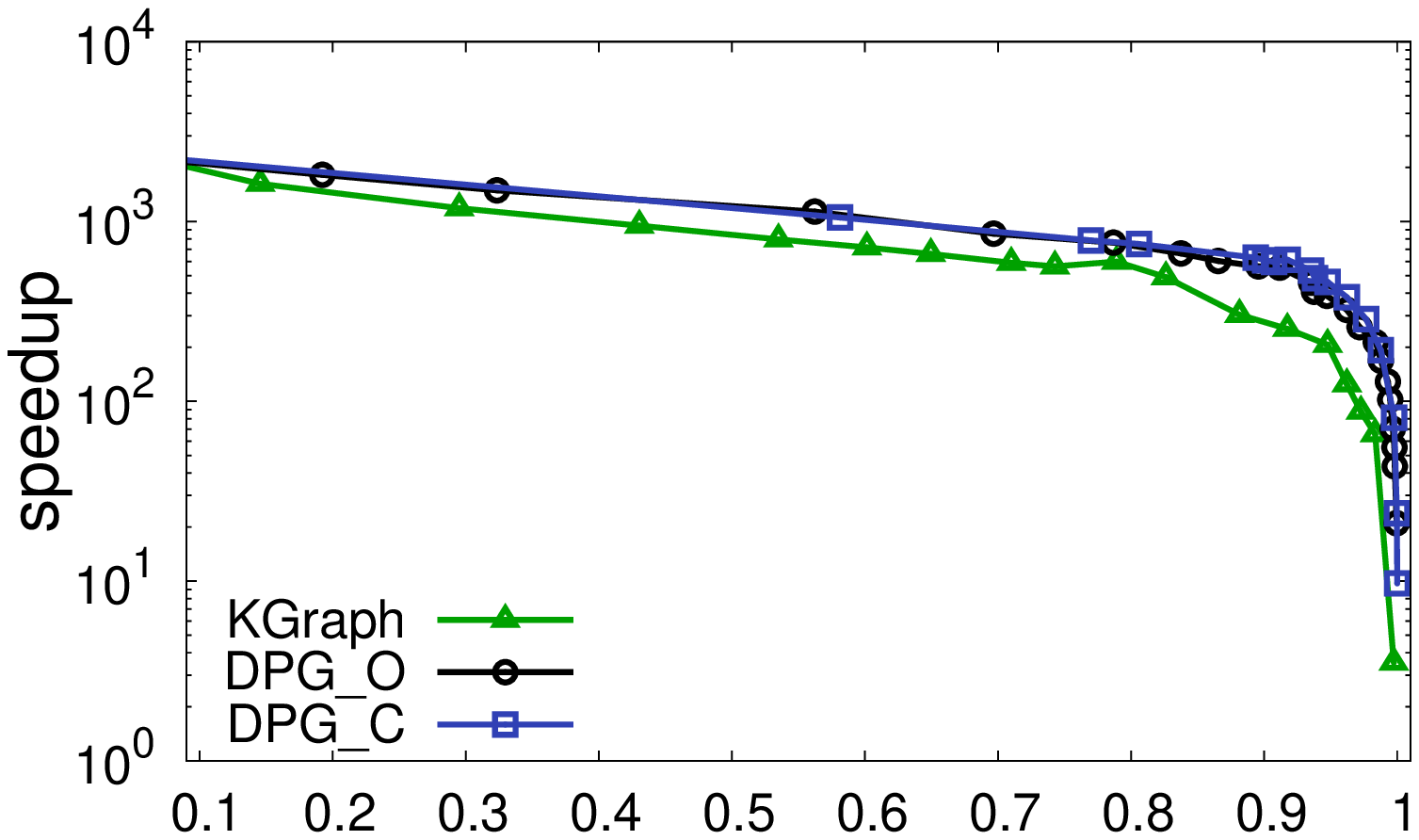}}
\subfigure[\small Nus ]{
      \label{fig:exp_SW_in_gist} 
      \includegraphics[width=0.48\linewidth]{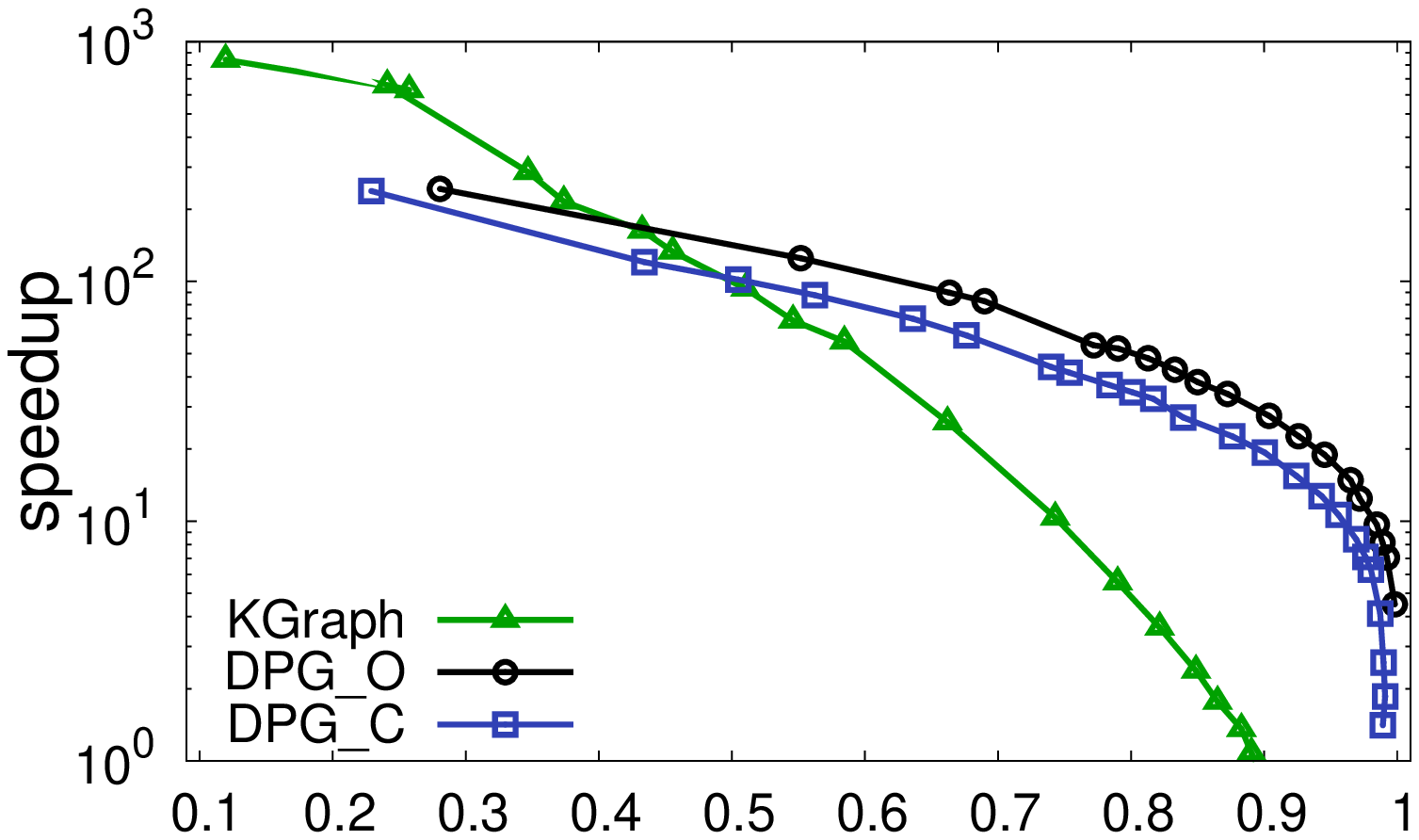}}
\vspace{-4mm}
\caption{\small Speedup vs Recall between counting-based DPG and angular-based DPG }
\label{fig:exp_DPG_C}
\end{figure}

Figure \ref{fig:exp_DPG_C} shows that using count based diversification (i.e., DPG\_C ) achieves similar search
performance as using angular similarity based diversification (i.e., DPG\_O).


\begin{figure*}[tbh]
\centering
\includegraphics[width=0.9\linewidth]{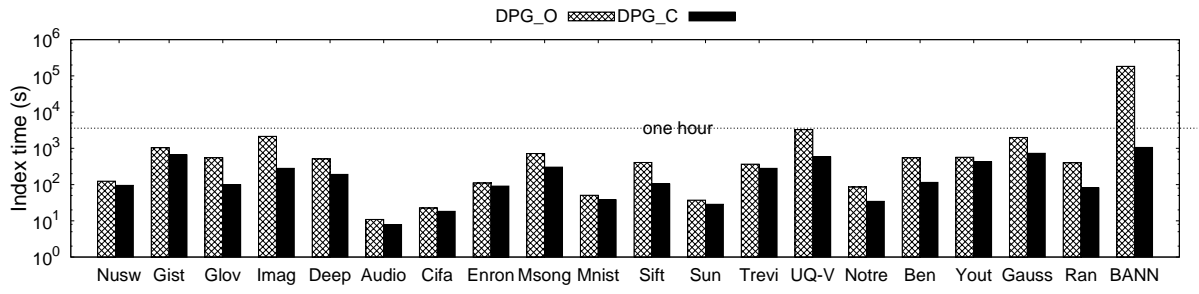}
\caption{\small diversification time between DPG\_C and DPG\_O}
\vspace{-3mm}
\label{fig:exp_DPG_Divtime}
\end{figure*}

%
Figure \ref{fig:exp_DPG_Divtime} shows the comparisons of the diversification time between counting-based DPG and angular-based DPG.  DPG\_C spends more less time than DPG\_O. The improvements are especially significant for the dataset with large data points.

\subsection{Default setting}

Below are the \emph{default} settings of the key parameters of the algorithms in
the second round evaluation in Section~\ref{subsec:exp_final}.
\begin{itemize}
\item \Algsrs. The number of projections ($m$) is set to $8$.
\item \AlgOPQ. The number of subspaces is $2$, and each subspace can have $2^{10}$
  codewords (i.e., cluster centers) by default.
\item \Algannoy. The number of the \Algannoy{} trees, $m$, is set to $50$.
\item \Algflann. We let the algorithm tune its own parameters.
\item \Alghnsw. The number of the connections for each point, $M$, is set to $10$.
\item \Algkgraph. By default, we use $K=40$ for the \KNN{} graph index. 
\item \Algdpg. We use $\kappa = \frac{K}{2} = 20$ so that the index size of
  \Algdpg{} is the same as that of \Algkgraph{} in the worst case.
\end{itemize}

%


\section{Supplement for the Second Round Evaluation}
\label{app:final}

\begin{figure*}[htb]
\centering%
\begin{minipage}[b]{0.8\linewidth}
\centering
\includegraphics[width=1.0\linewidth]{exp-fig/5.6.1-recall/final_title}%
\vspace{-3mm}
\end{minipage}
\begin{minipage}[t]{1.0\linewidth}
\centering
      \subfigure[\small audio]{
      \label{fig:exp_recall_nus} 
      \includegraphics[width=0.23\linewidth]{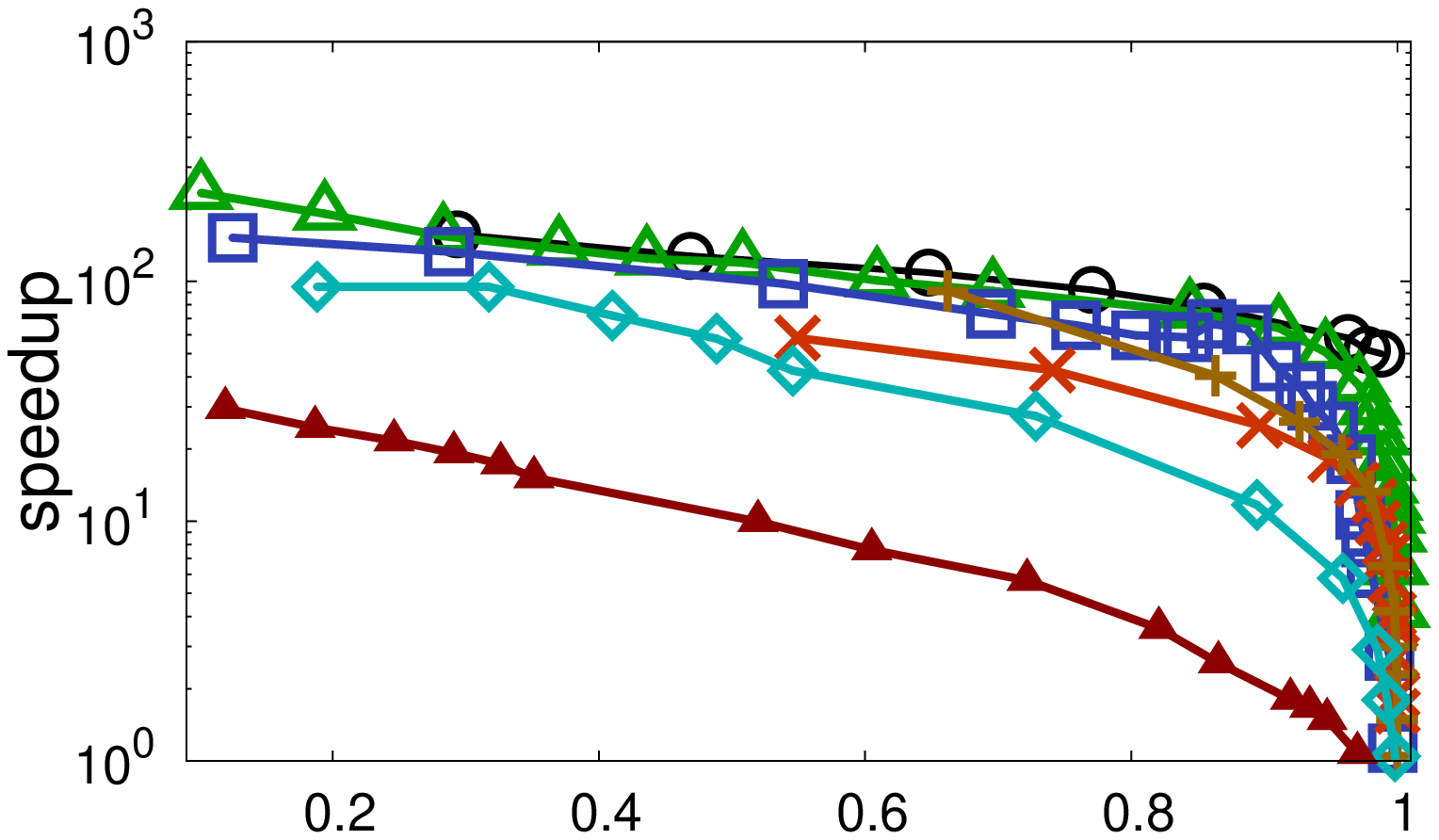}}
      \subfigure[\small cifar]{
      \label{fig:exp_recall_msong} 
      \includegraphics[width=0.23\linewidth]{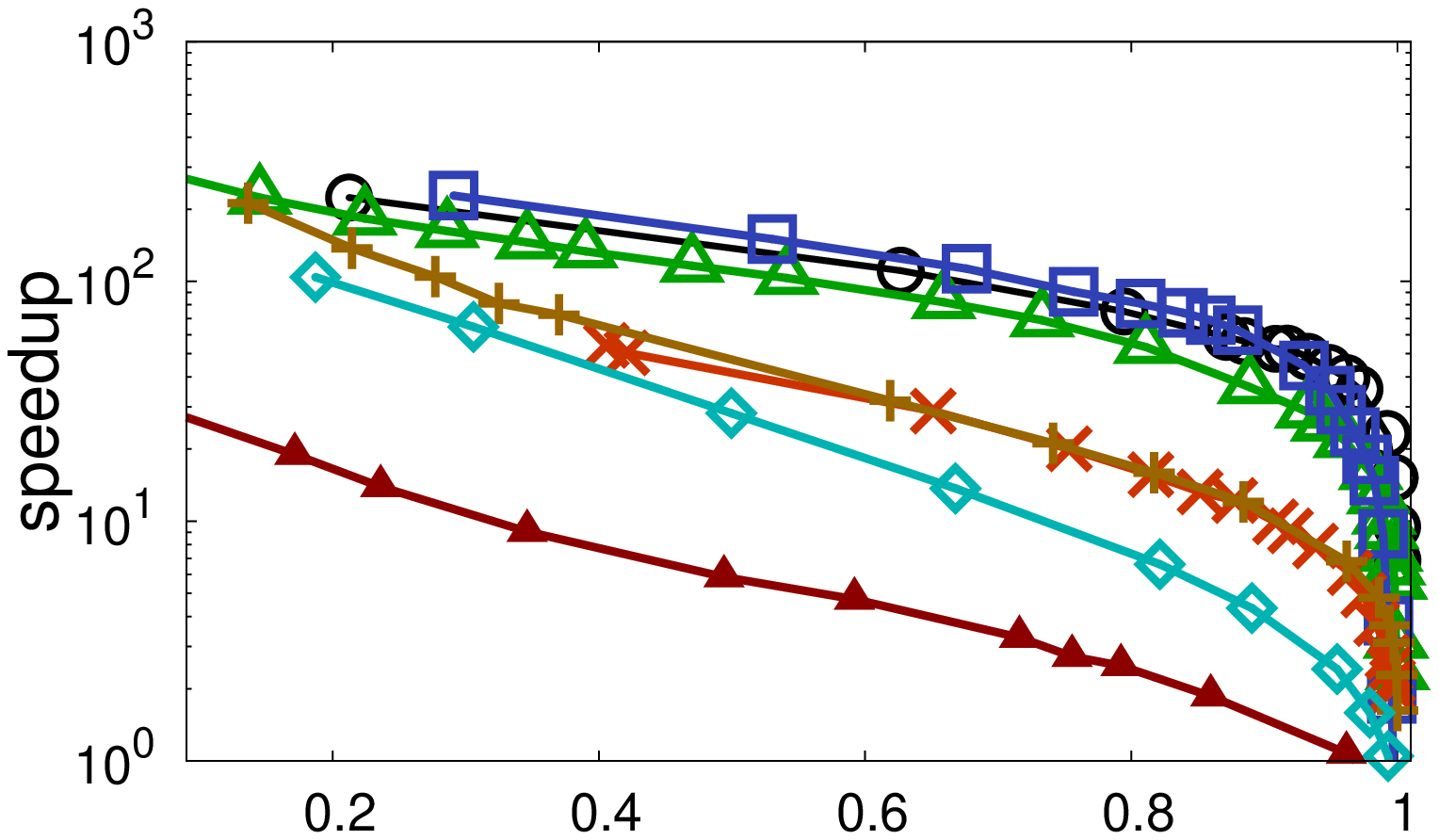}}
      \subfigure[\small enron]{
      \label{fig:exp_recall_sift} 
      \includegraphics[width=0.23\linewidth]{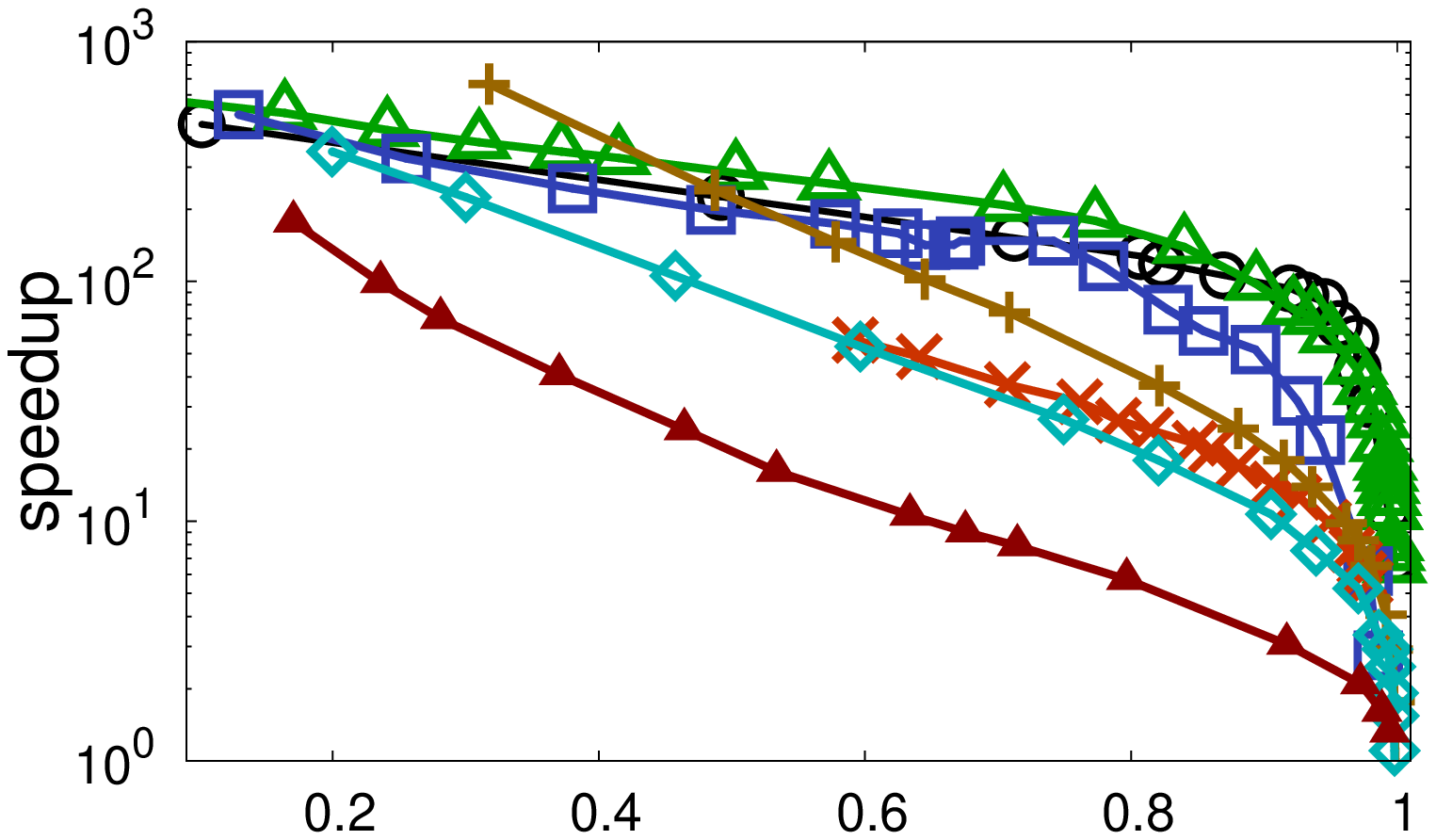}}
      \subfigure[\small imag ]{
      \label{fig:exp_recall_deep} 
      \includegraphics[width=0.23\linewidth]{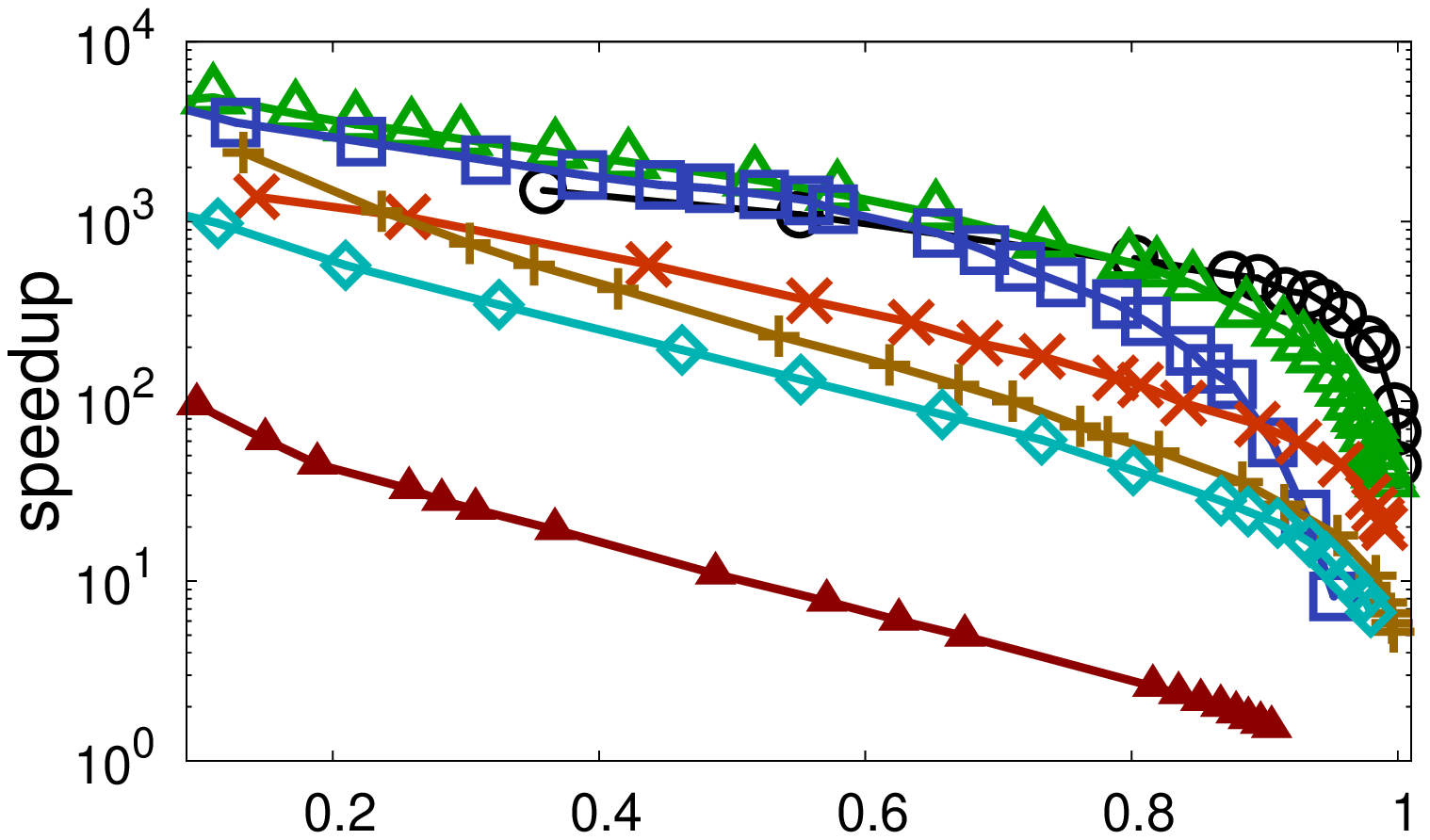}}
\end{minipage}%
\vspace{-1mm}
\begin{minipage}[t]{1.0\linewidth}
\centering
      \subfigure[\small Mnist]{
      \label{fig:exp_recall_nus} 
      \includegraphics[width=0.23\linewidth]{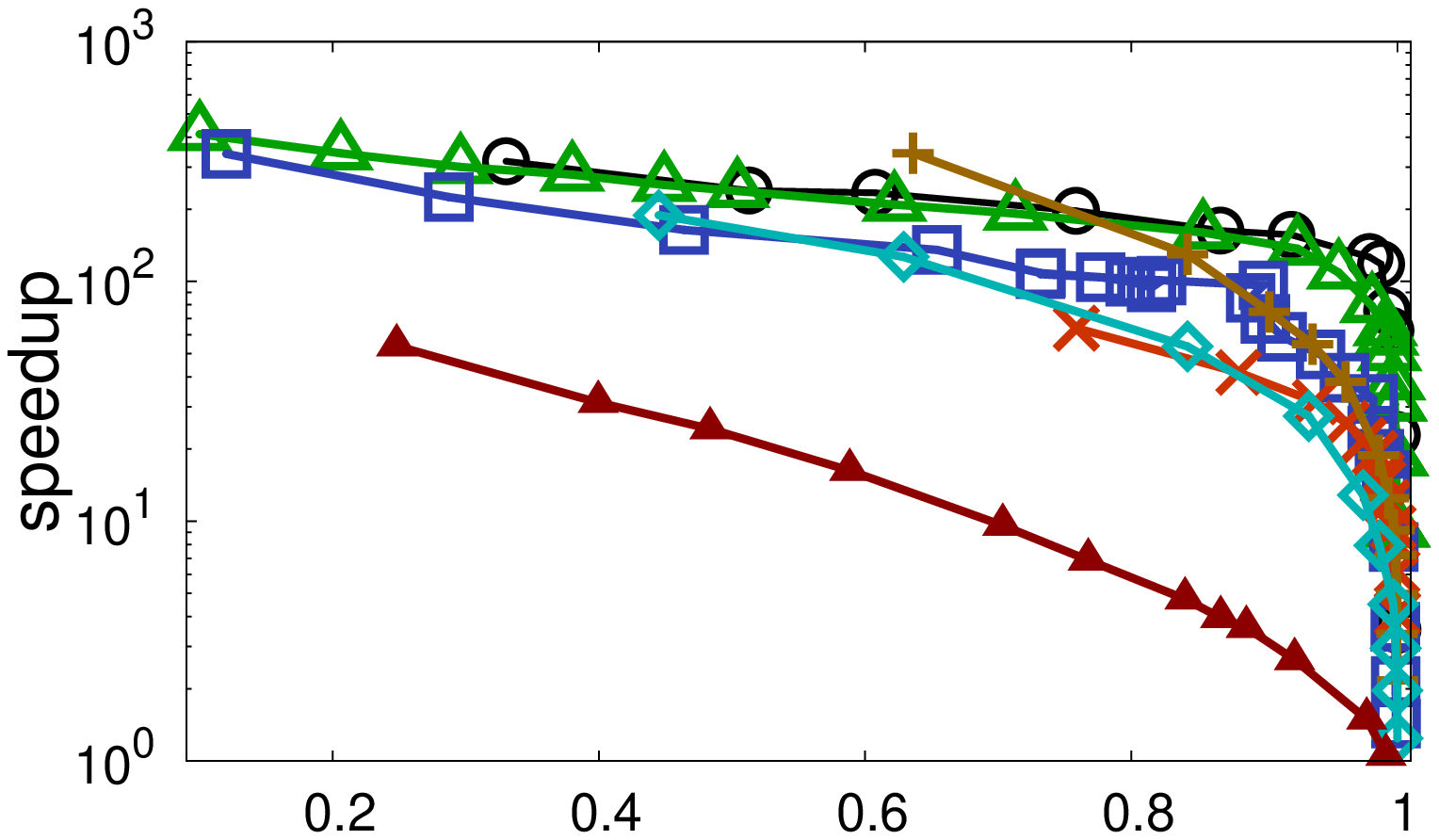}}
      \subfigure[\small Notre]{
      \label{fig:exp_recall_msong} 
      \includegraphics[width=0.23\linewidth]{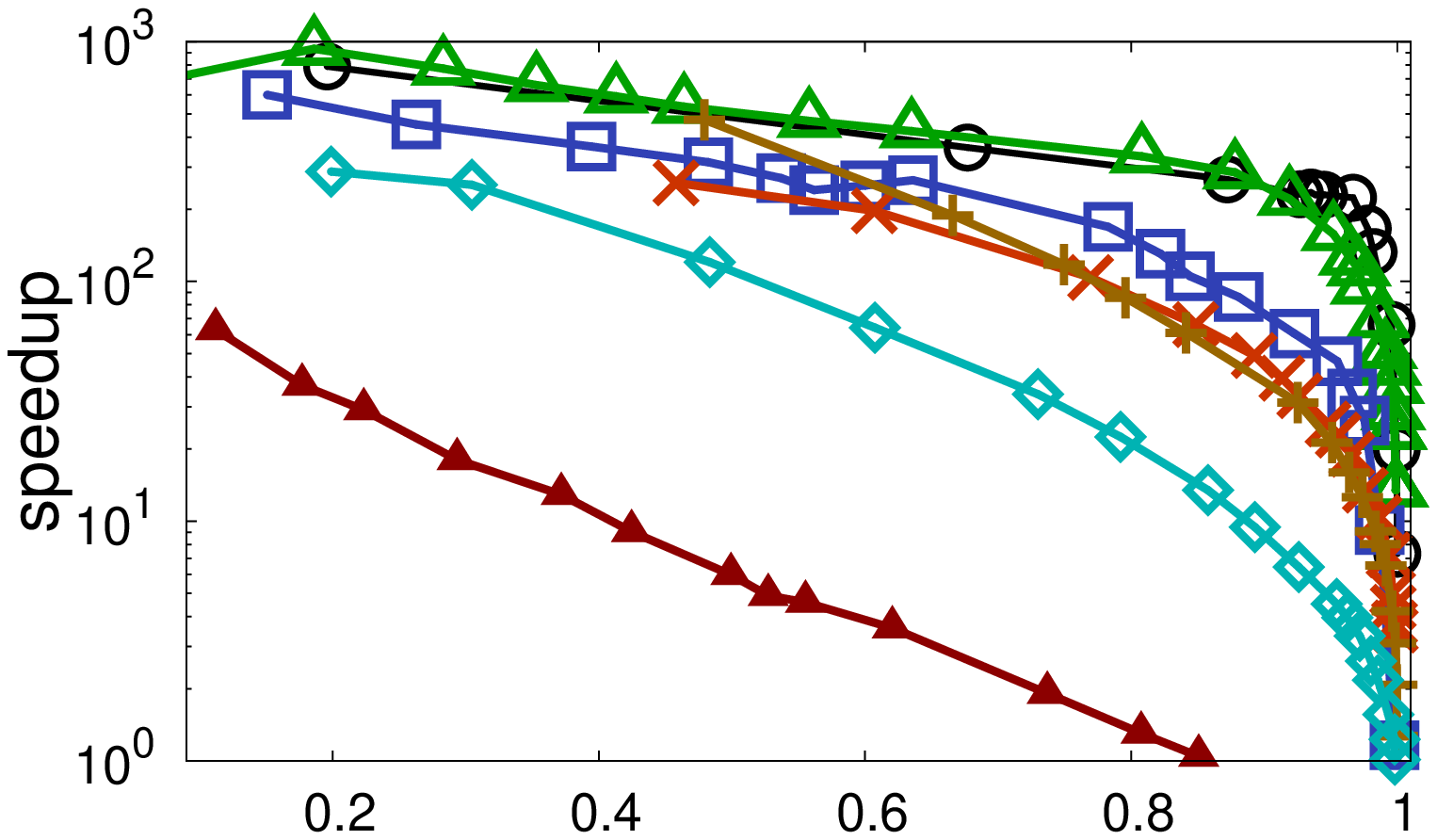}}
      \subfigure[\small Sun]{
      \label{fig:exp_recall_sift} 
      \includegraphics[width=0.23\linewidth]{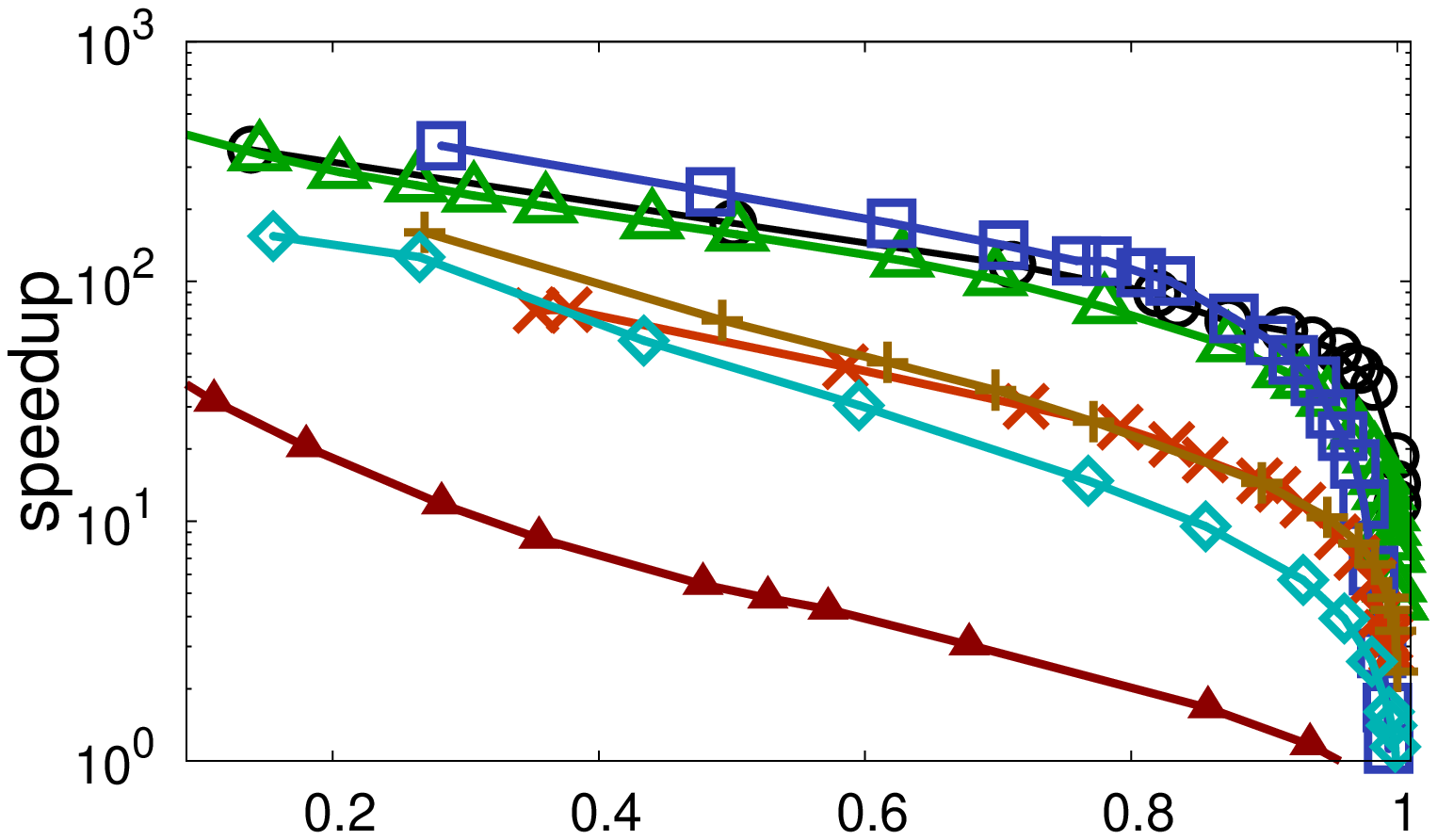}}
      \subfigure[\small Trevi ]{
      \label{fig:exp_recall_deep} 
      \includegraphics[width=0.23\linewidth]{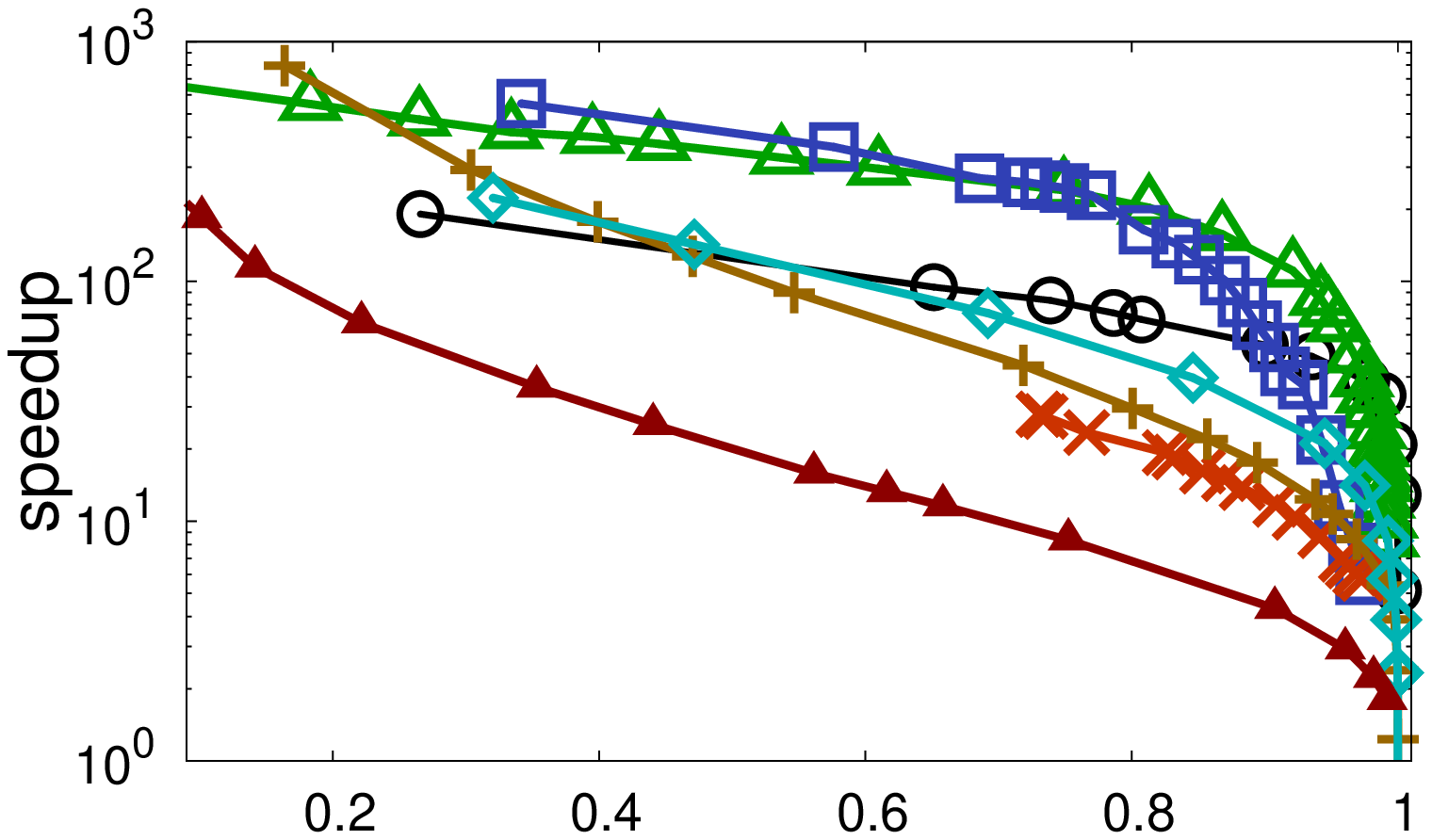}}
\end{minipage}%
\vspace{-1mm}
\begin{minipage}[t]{1.0\linewidth}
\centering
      \subfigure[\small Ben]{
      \label{fig:exp_recall_nus} 
      \includegraphics[width=0.23\linewidth]{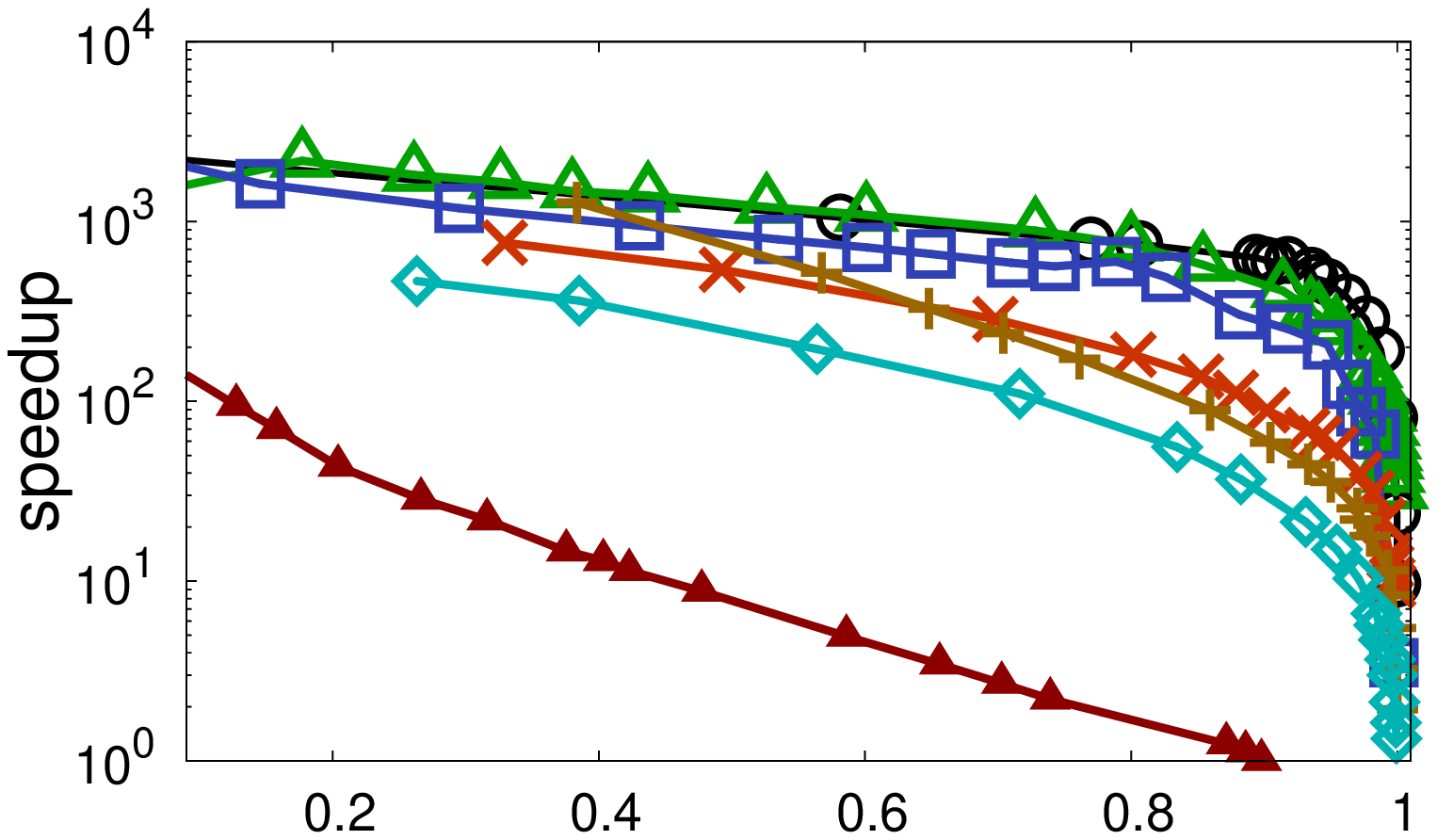}}
      \subfigure[\small UQ-V]{
      \label{fig:exp_recall_msong} 
      \includegraphics[width=0.23\linewidth]{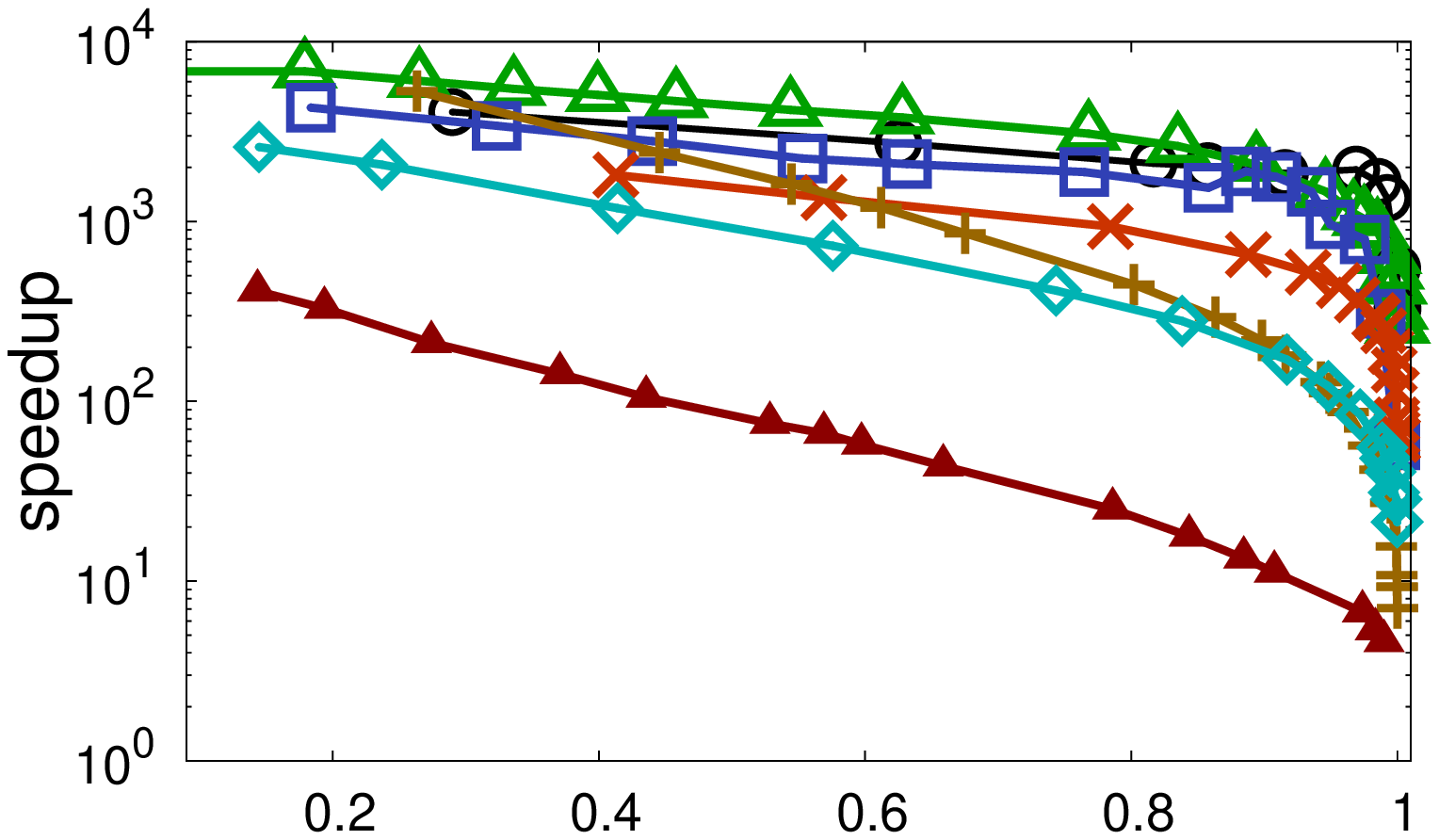}}
      \subfigure[\small Yout]{
      \label{fig:exp_recall_sift} 
      \includegraphics[width=0.23\linewidth]{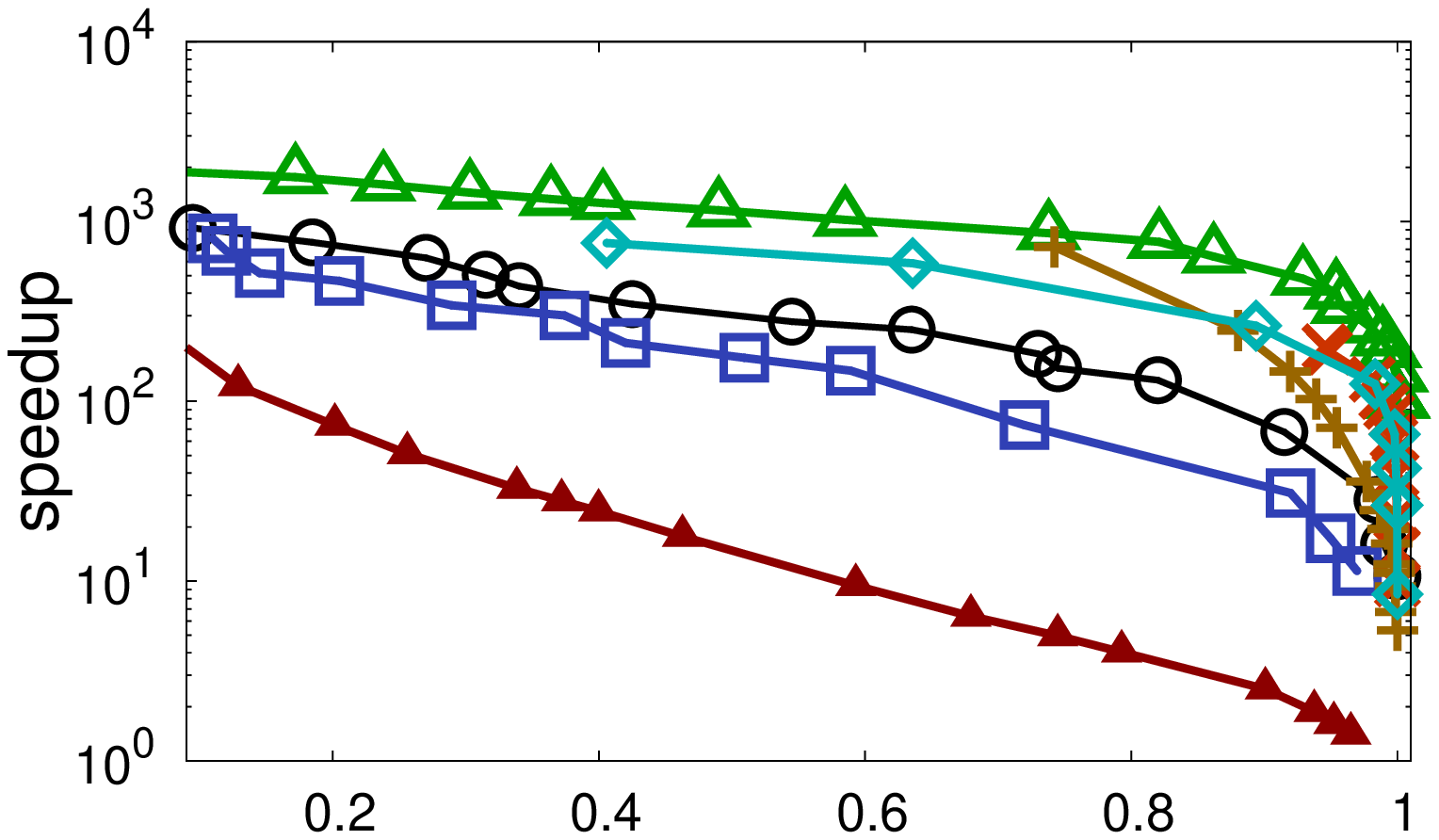}}
      \subfigure[\small BANN ]{
      \label{fig:exp_recall_deep} 
      \includegraphics[width=0.23\linewidth]{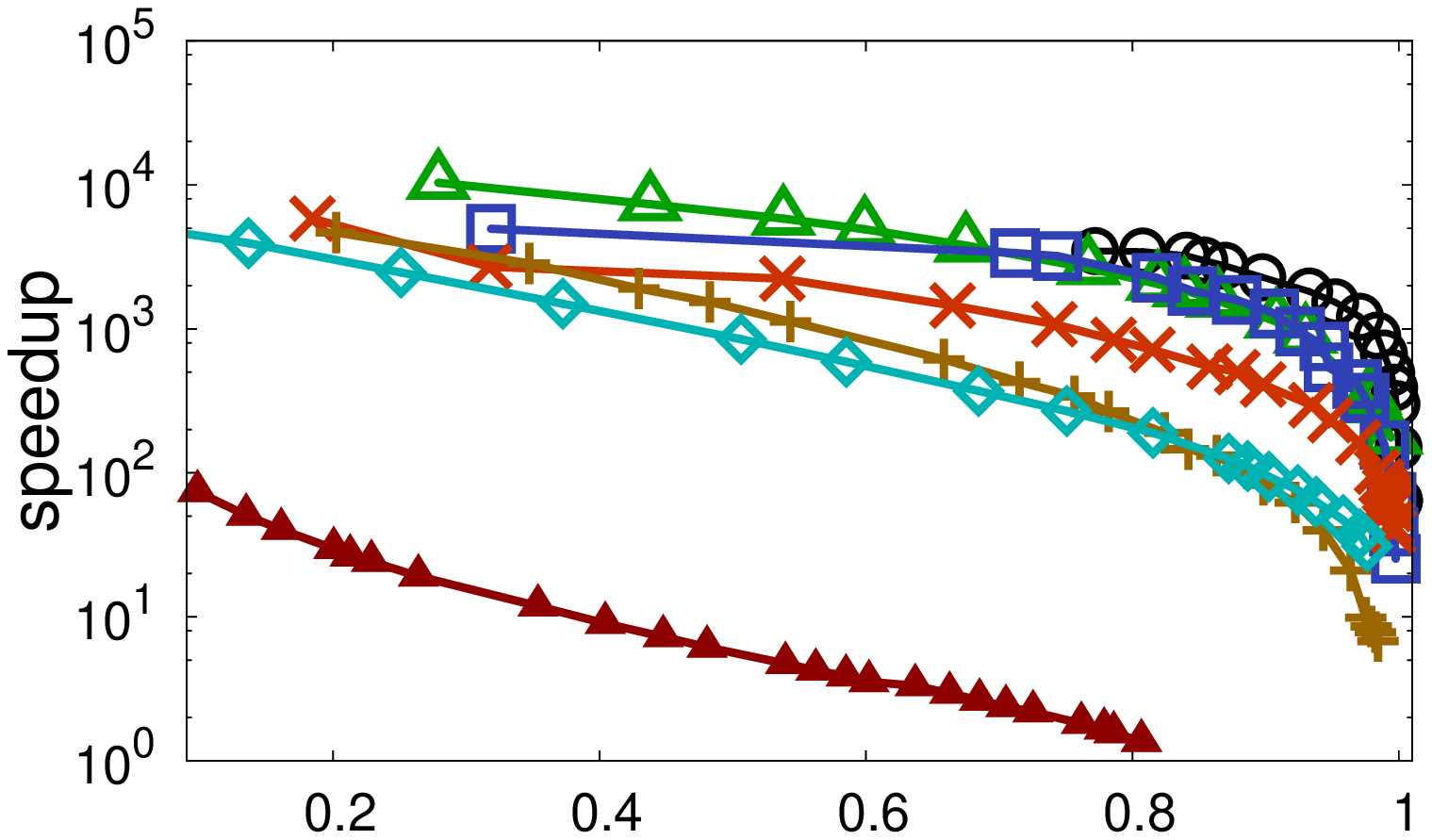}}
\end{minipage}%
\vspace{-1mm}
\caption{\small Speedup vs Recall}
\label{fig:app_final_speedup_recall}
\end{figure*}

\begin{figure*}[htb]
\centering%
\begin{minipage}[b]{0.8\linewidth}
\centering
\includegraphics[width=1.0\linewidth]{exp-fig/5.6.1-recall/final_title}%
\vspace{-1mm}
\end{minipage}
\begin{minipage}[t]{1.0\linewidth}
\centering
      \subfigure[\small audio]{
      \label{fig:exp_recall_nus} 
      \includegraphics[width=0.23\linewidth]{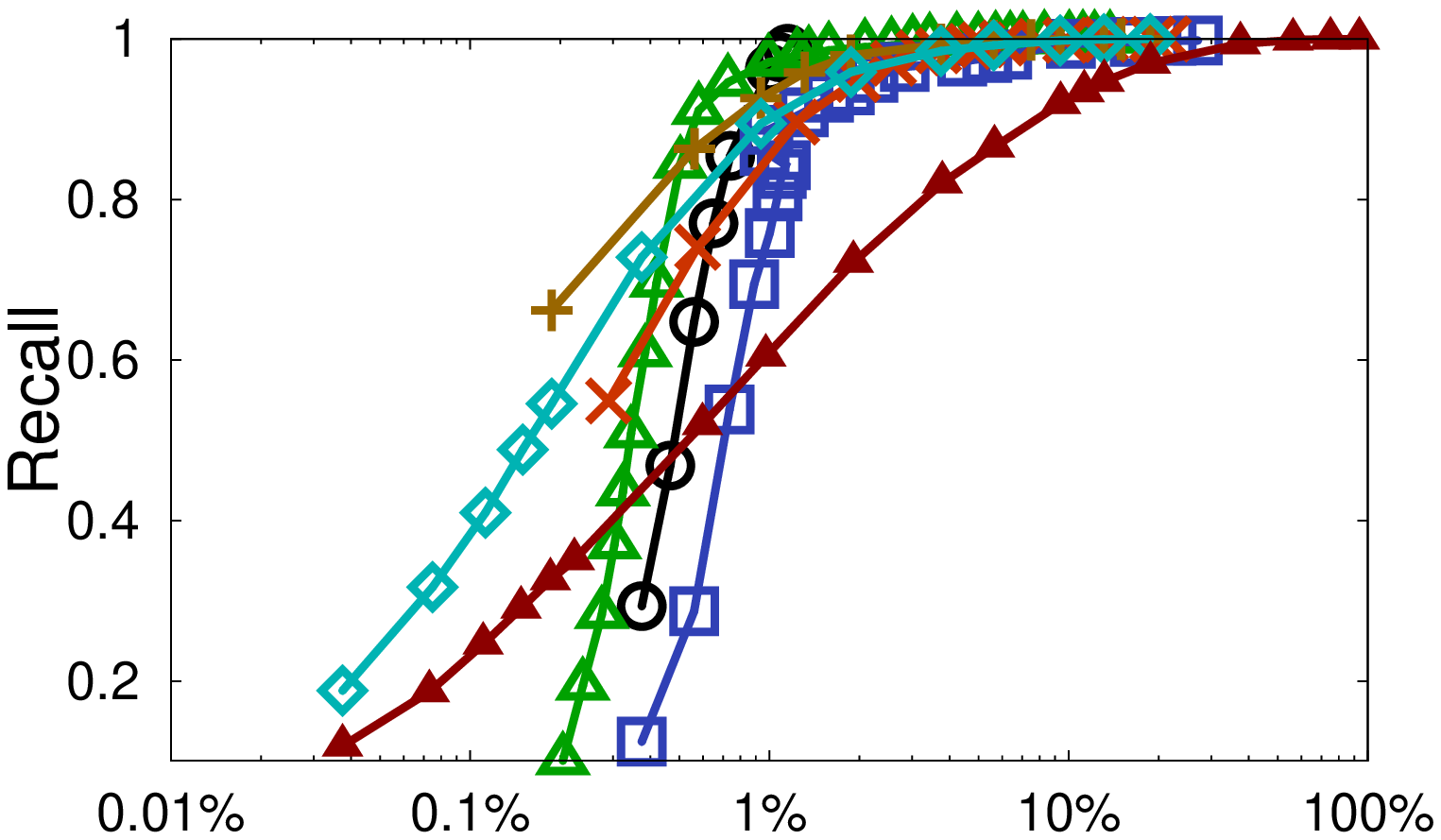}}
      \subfigure[\small cifar]{
      \label{fig:exp_recall_msong} 
      \includegraphics[width=0.23\linewidth]{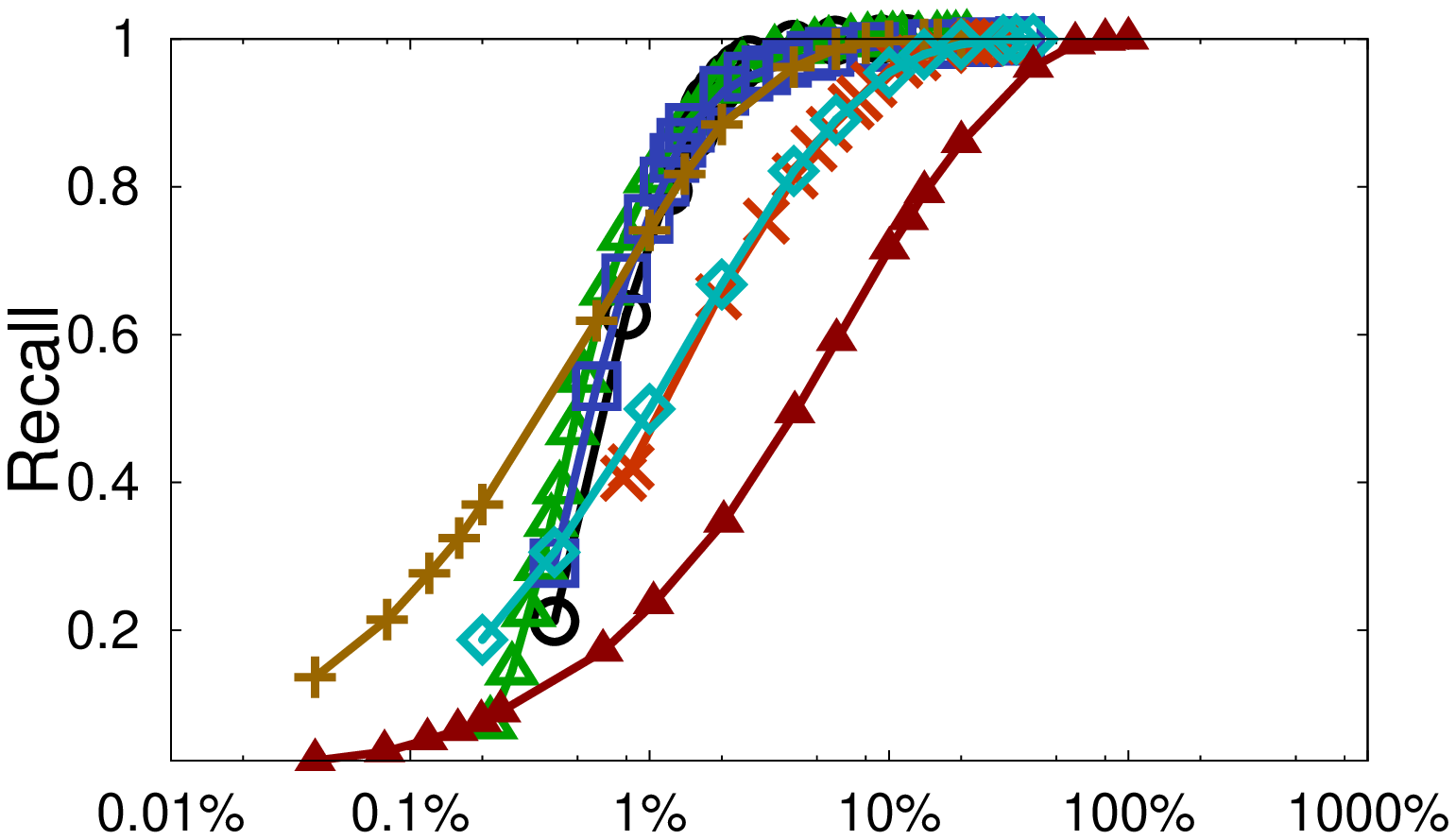}}
      \subfigure[\small enron]{
      \label{fig:exp_recall_sift} 
      \includegraphics[width=0.23\linewidth]{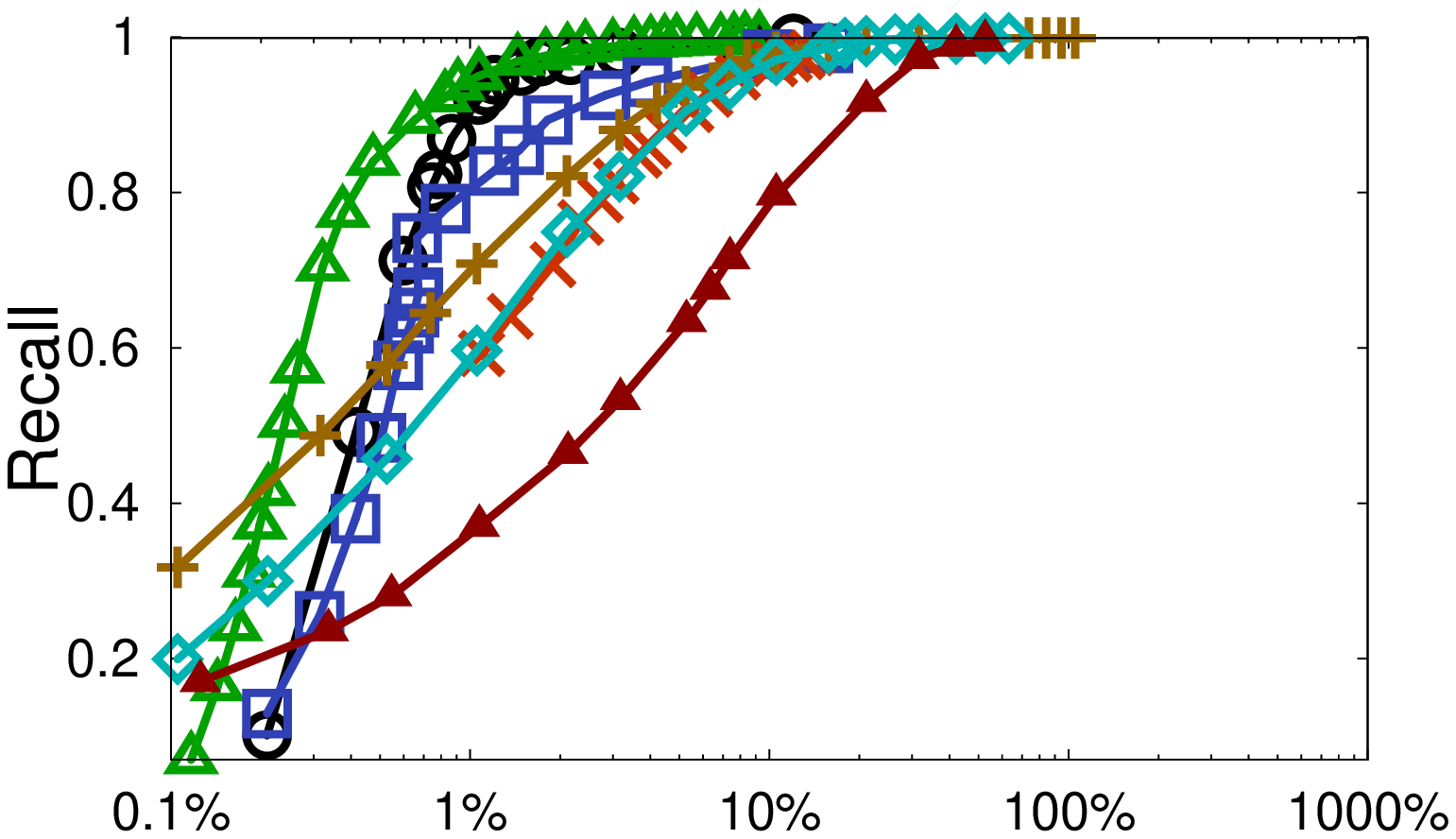}}
      \subfigure[\small imag ]{
      \label{fig:exp_recall_deep} 
      \includegraphics[width=0.23\linewidth]{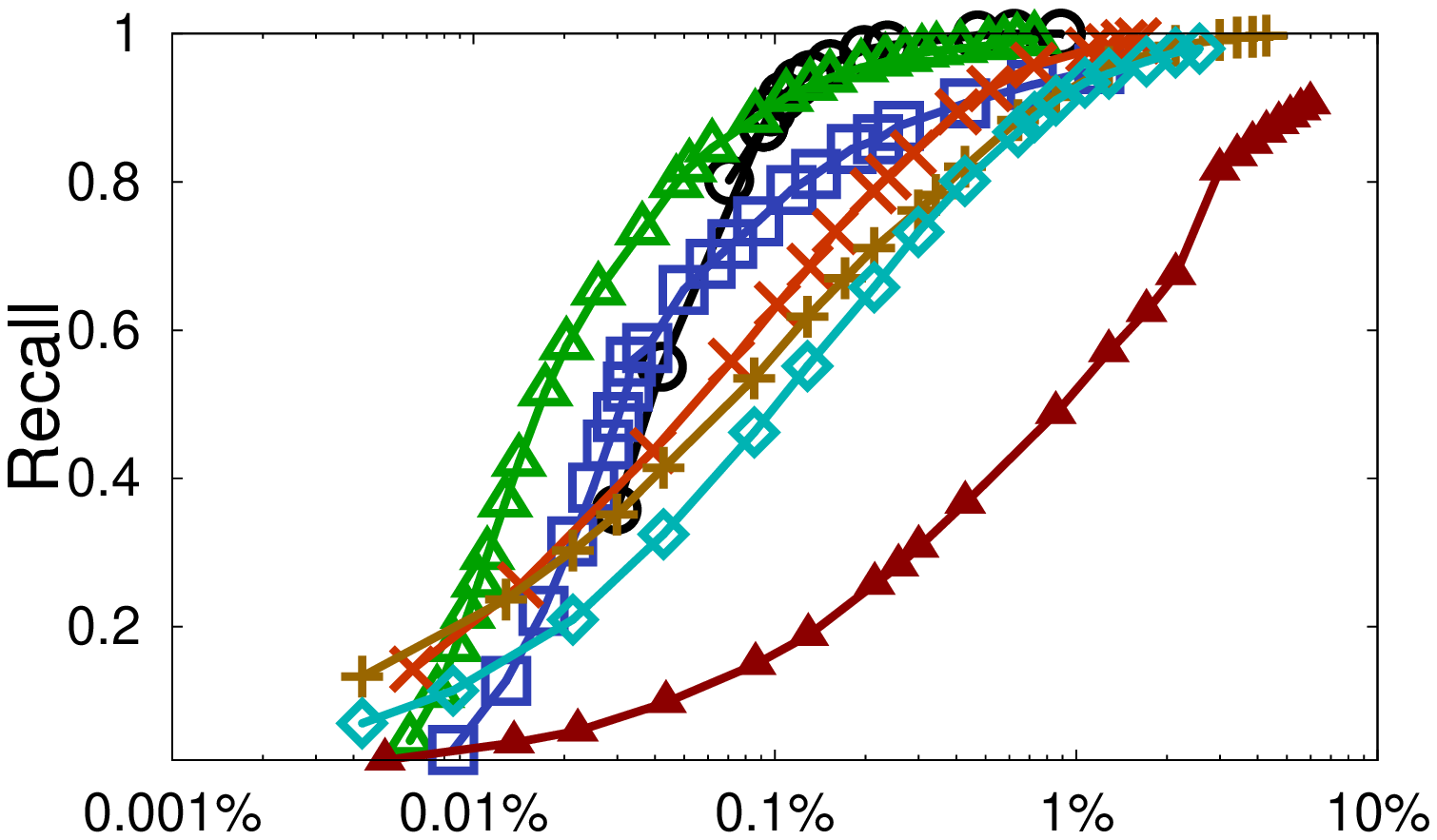}}
\end{minipage}%
\vspace{-1mm}
\begin{minipage}[t]{1.0\linewidth}
\centering
      \subfigure[\small Glove]{
      \label{fig:exp_recall_nus} 
      \includegraphics[width=0.23\linewidth]{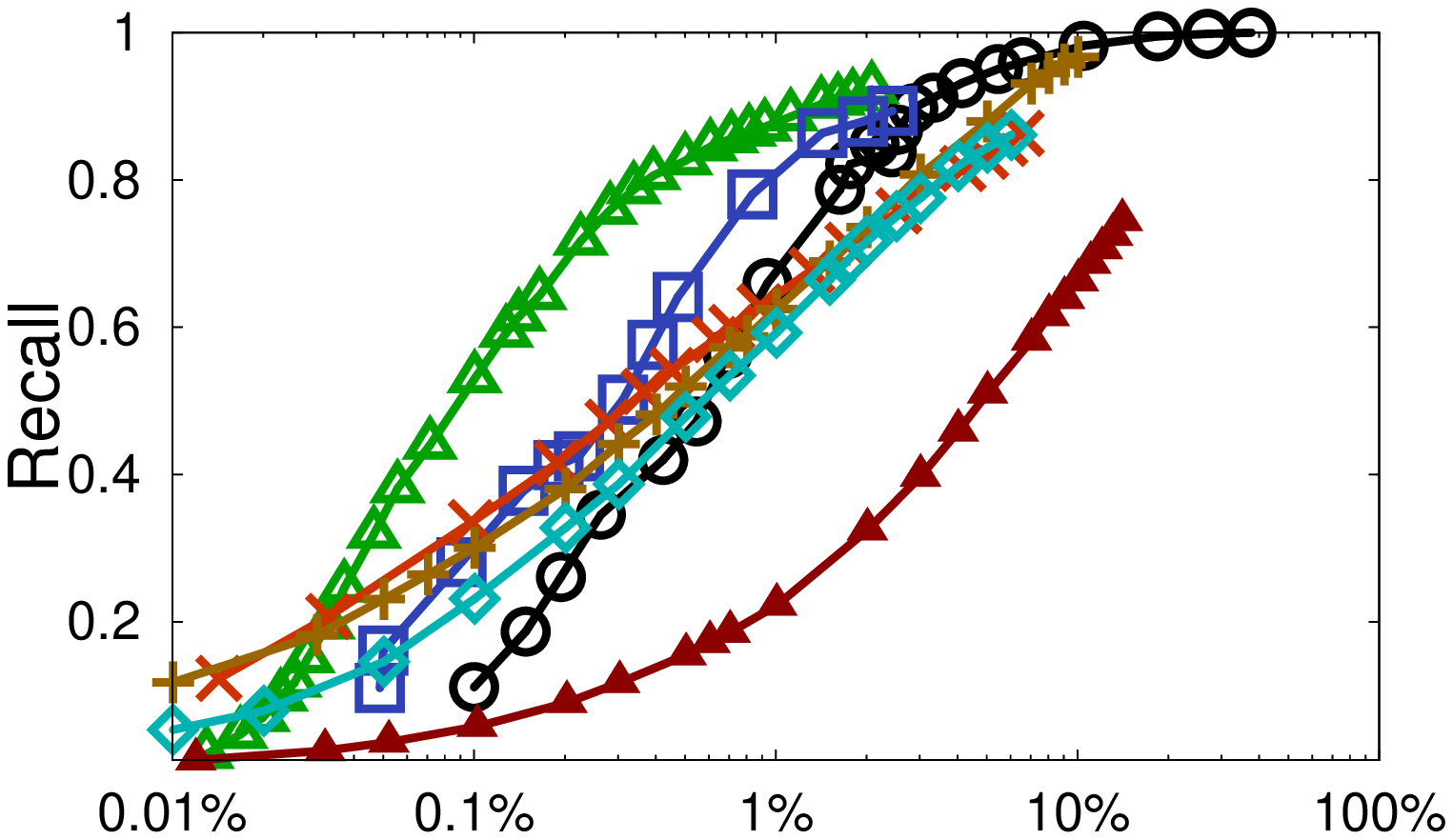}}
      \subfigure[\small Rand]{
      \label{fig:exp_recall_msong} 
      \includegraphics[width=0.23\linewidth]{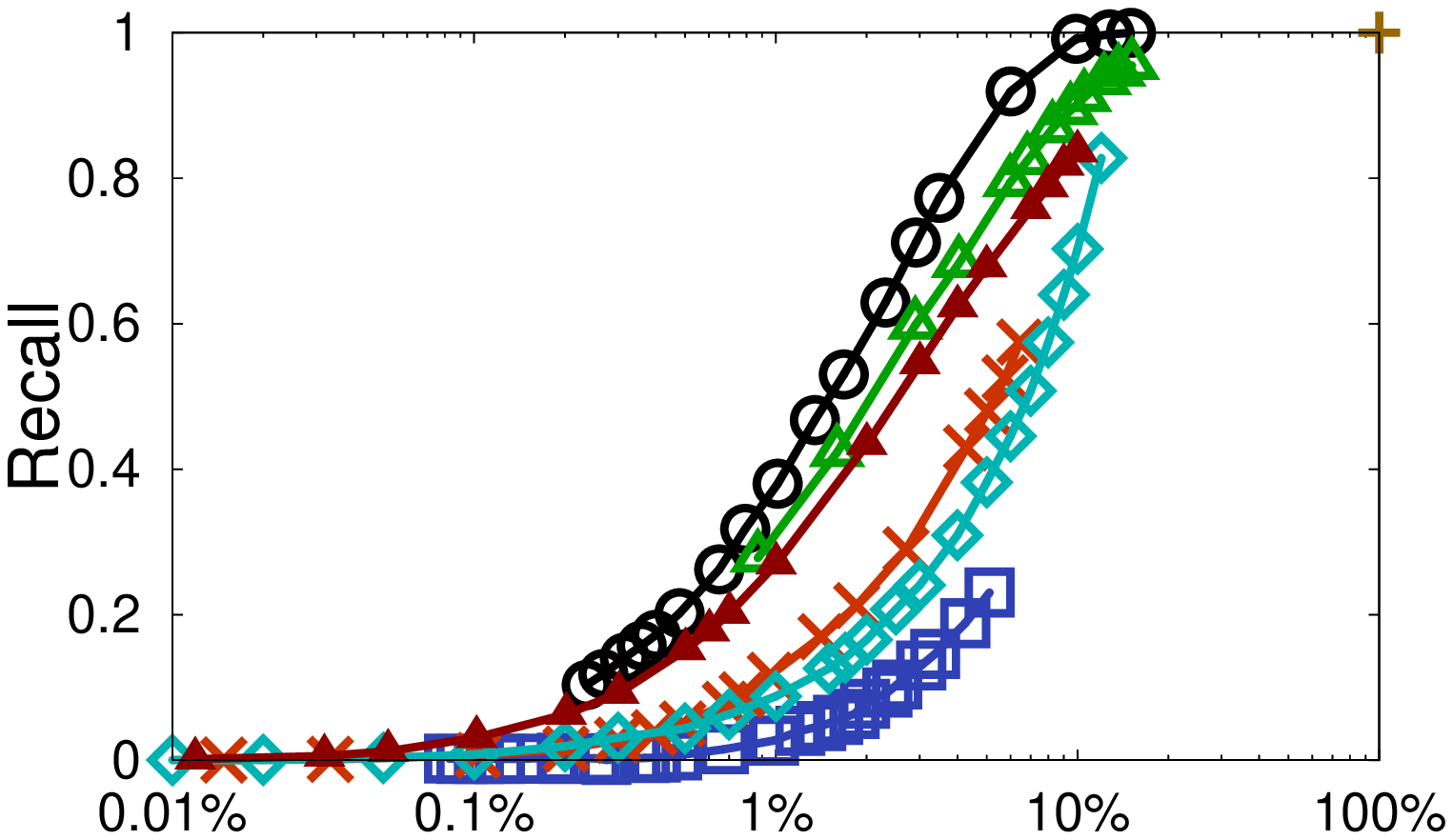}}
      \subfigure[\small Sift]{
      \label{fig:exp_recall_sift} 
      \includegraphics[width=0.23\linewidth]{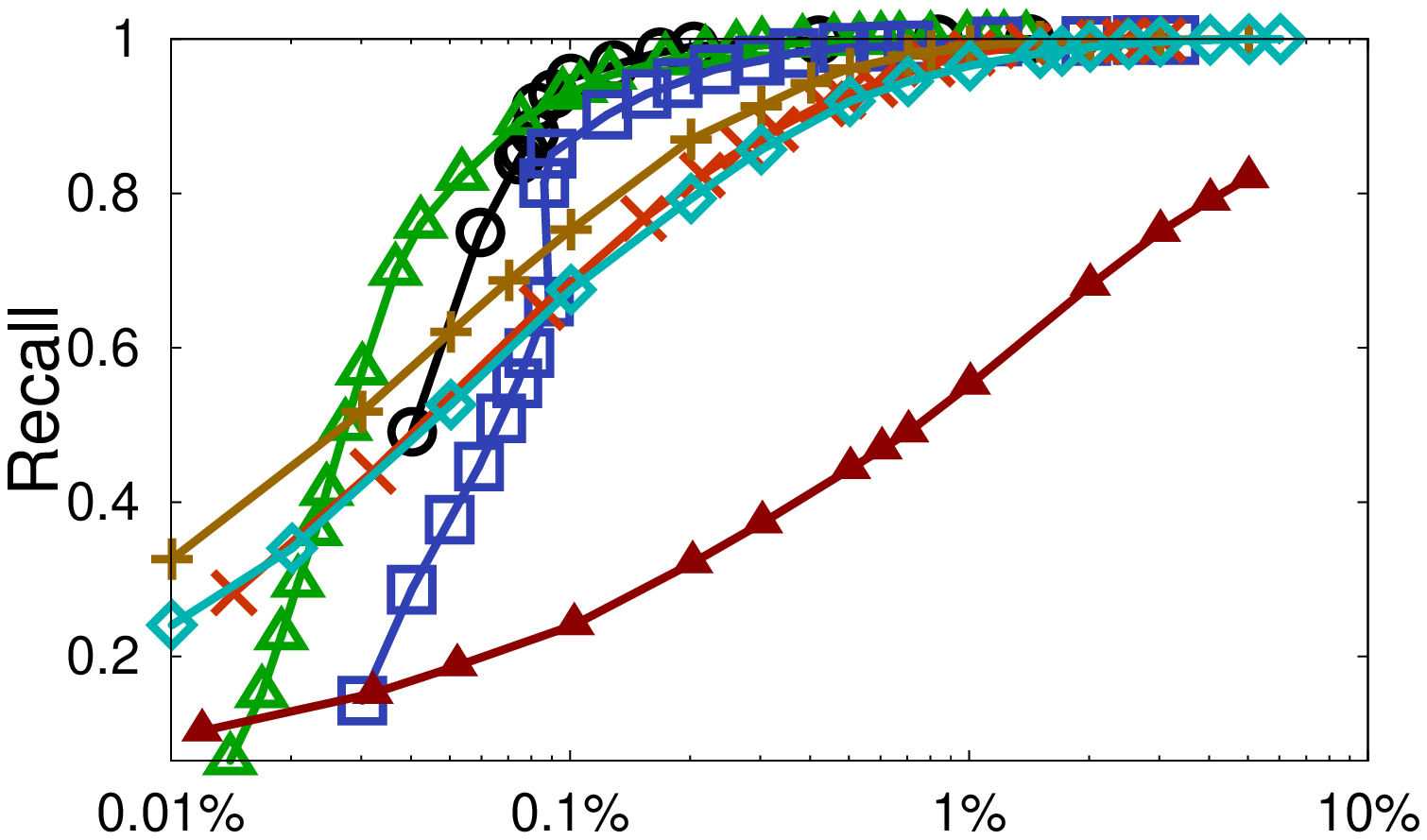}}
      \subfigure[\small Gauss ]{
      \label{fig:exp_recall_deep} 
      \includegraphics[width=0.23\linewidth]{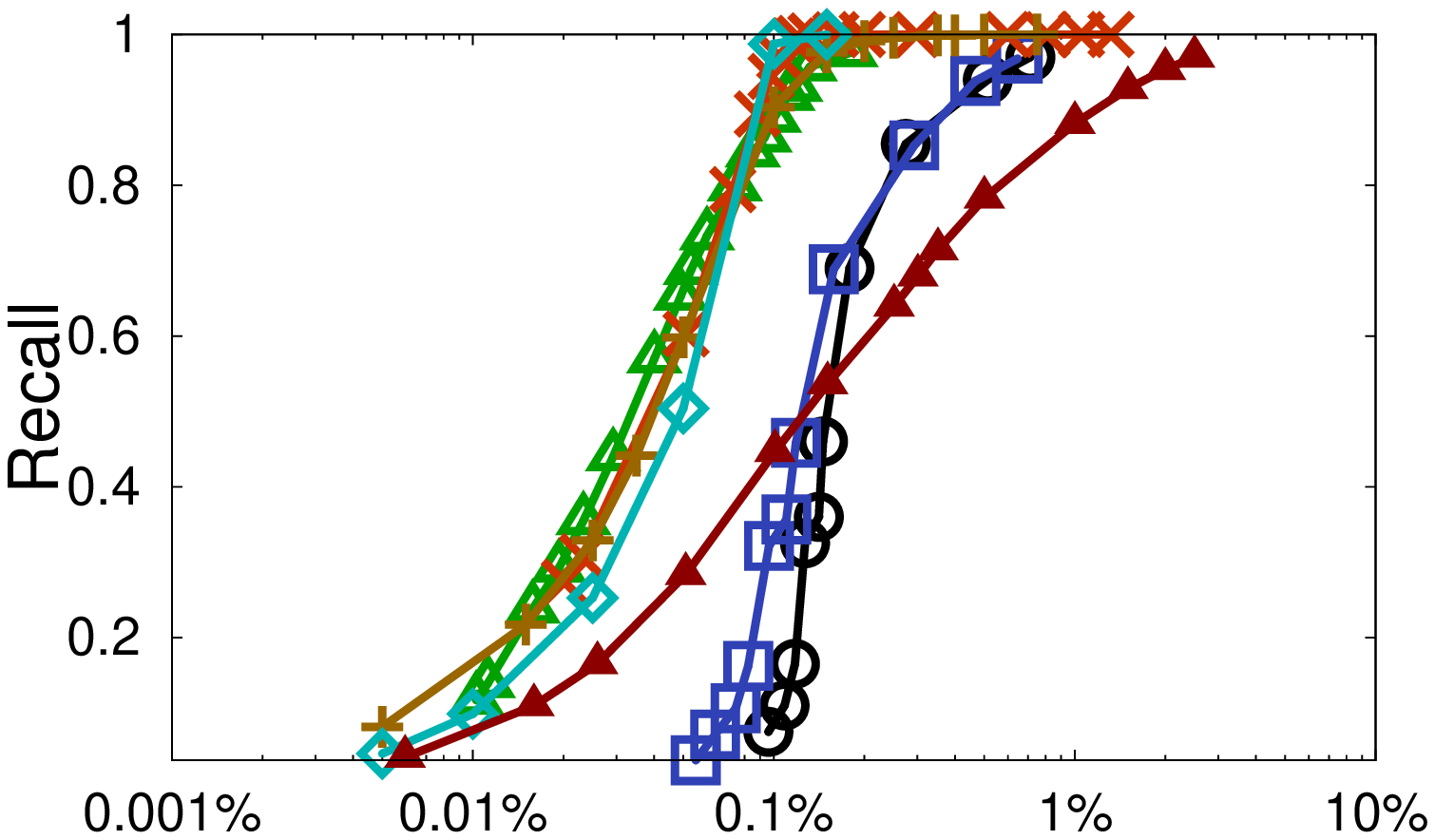}}
\end{minipage}%
\vspace{-1mm}
\begin{minipage}[t]{1.0\linewidth}
\centering
      \subfigure[\small Mnist]{
      \label{fig:exp_recall_nus} 
      \includegraphics[width=0.23\linewidth]{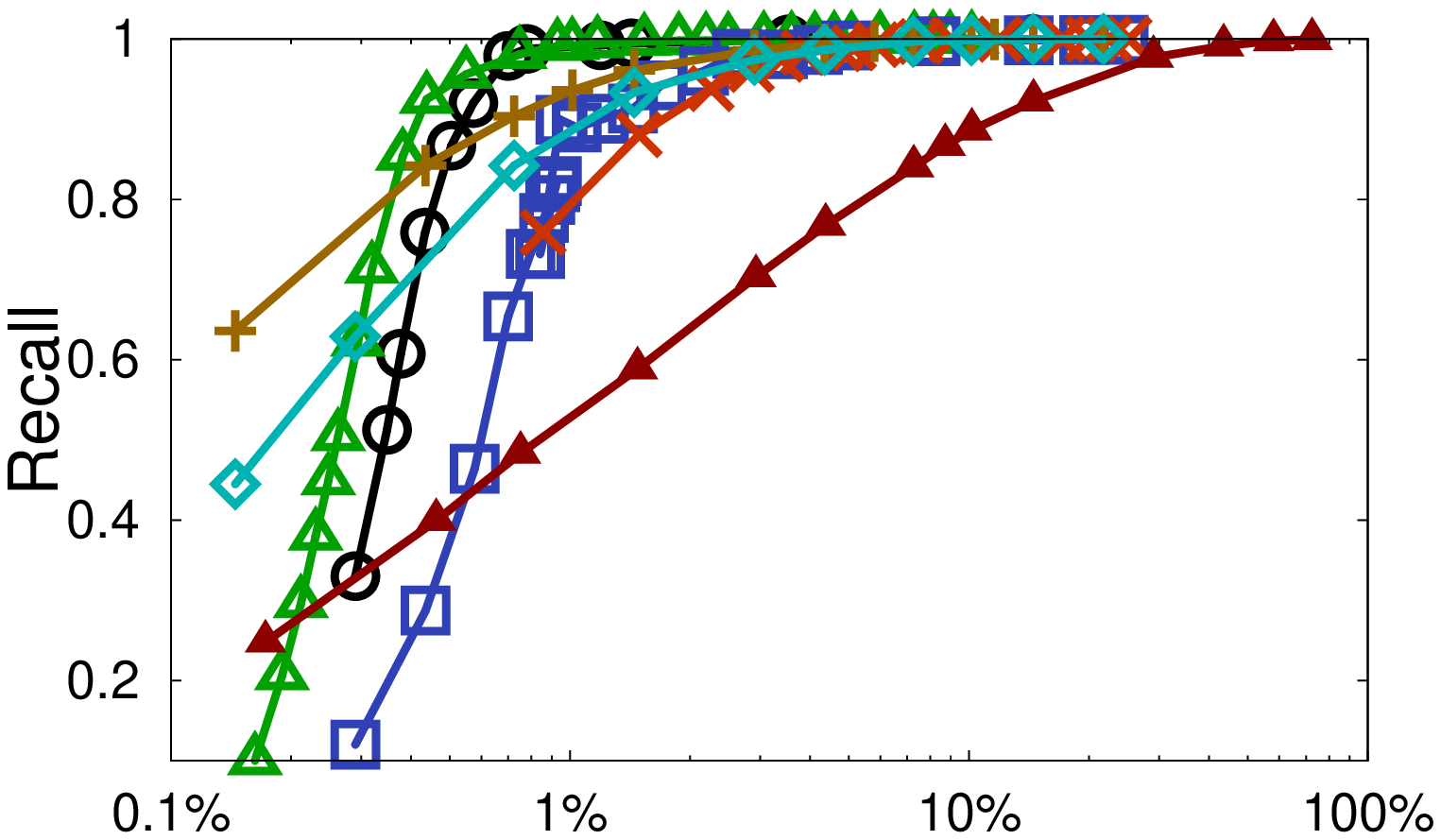}}
      \subfigure[\small Notre]{
      \label{fig:exp_recall_msong} 
      \includegraphics[width=0.23\linewidth]{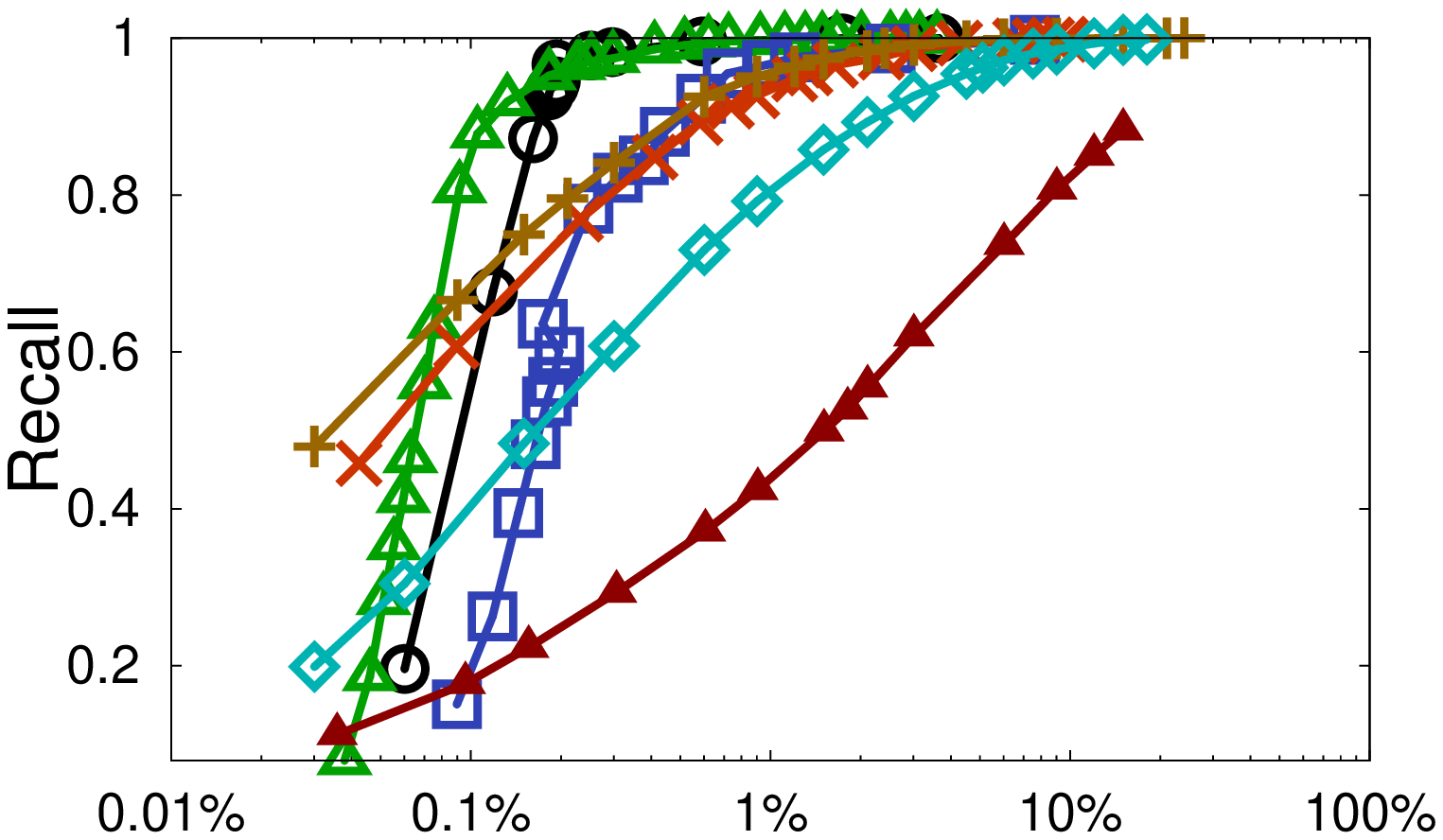}}
      \subfigure[\small Sun]{
      \label{fig:exp_recall_sift} 
      \includegraphics[width=0.23\linewidth]{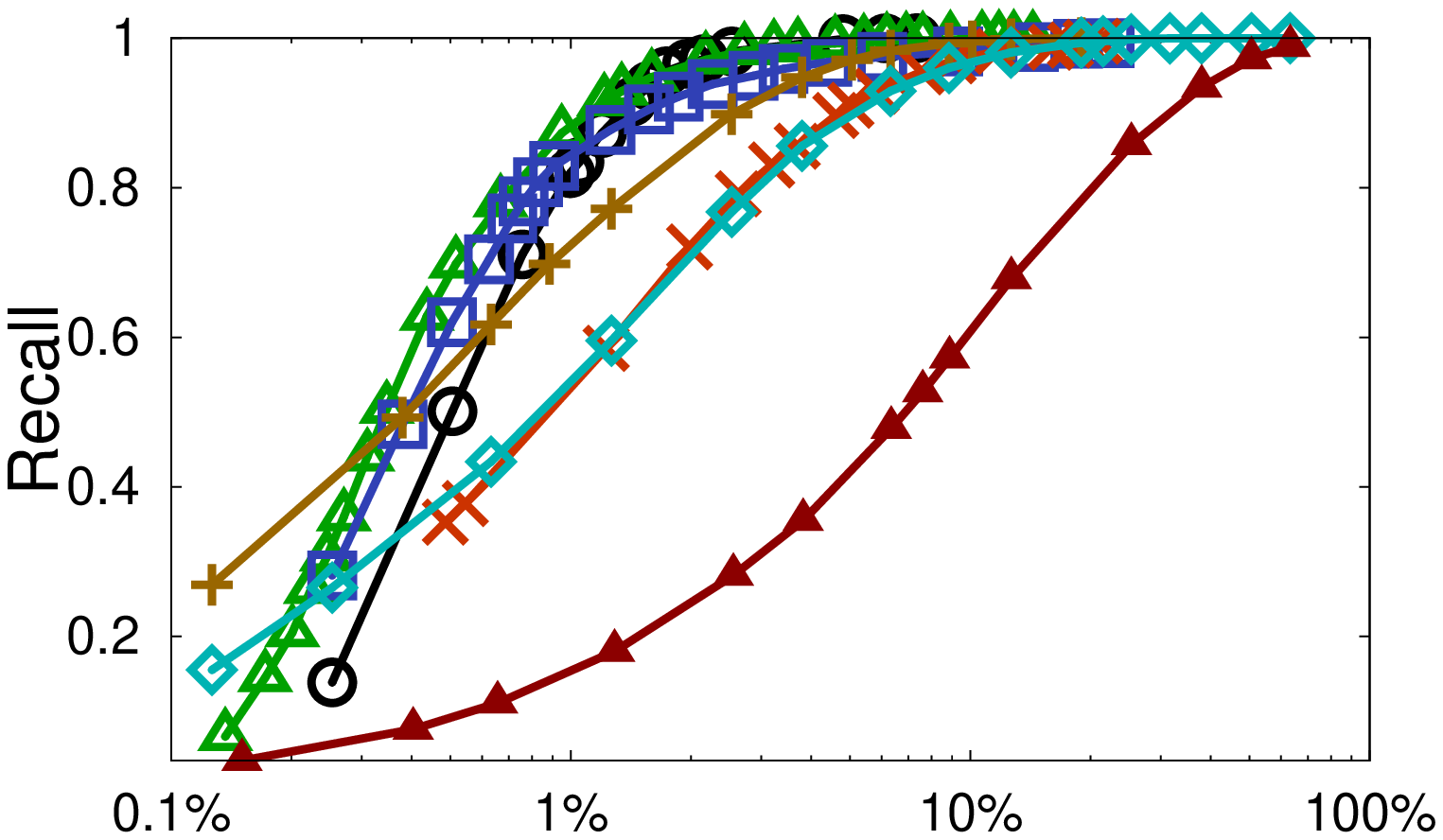}}
      \subfigure[\small Trevi ]{
      \label{fig:exp_recall_deep} 
      \includegraphics[width=0.23\linewidth]{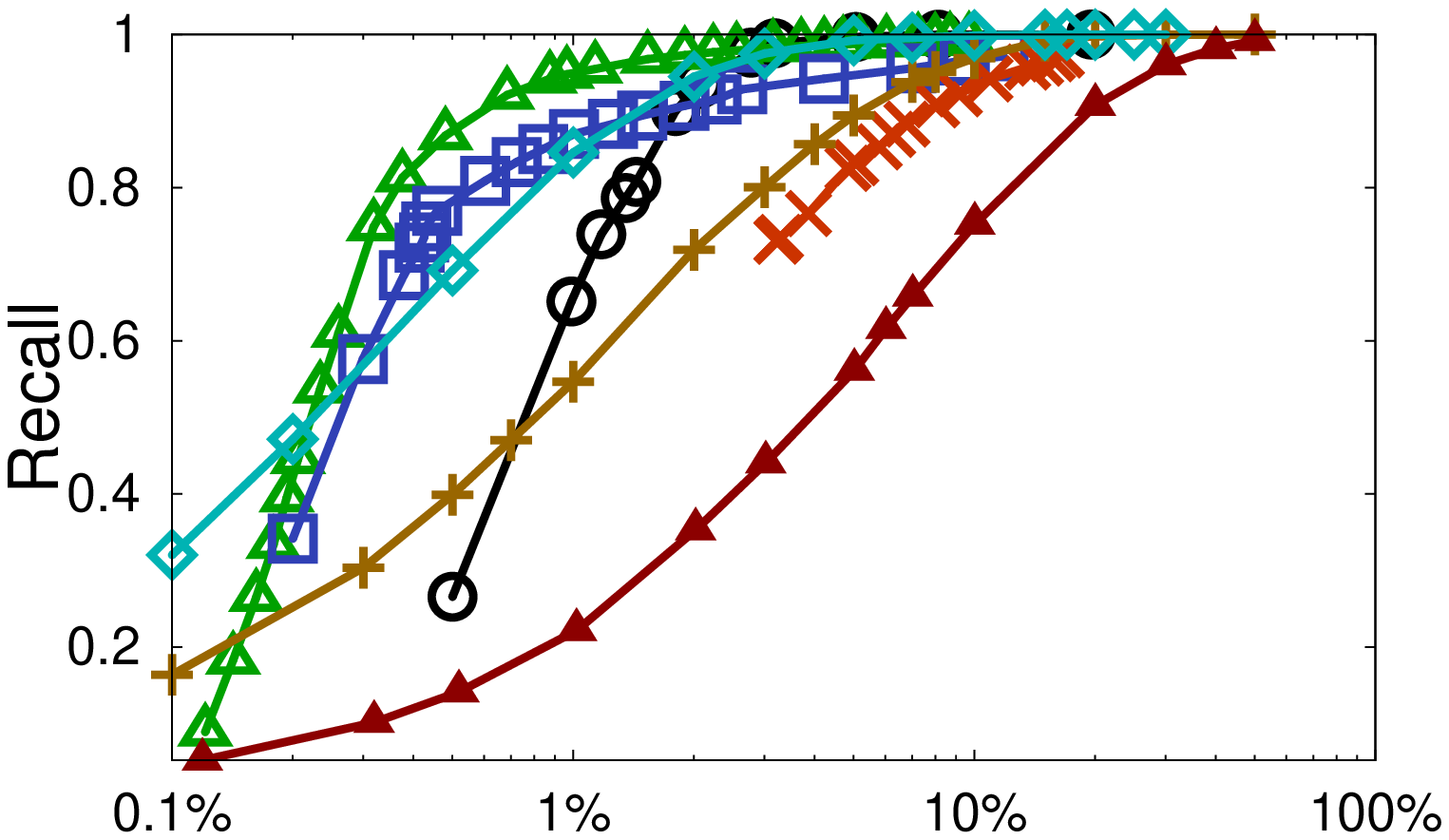}}
\end{minipage}%
\vspace{-1mm}
\begin{minipage}[t]{1.0\linewidth}
\centering
      \subfigure[\small Ben]{
      \label{fig:exp_recall_nus} 
      \includegraphics[width=0.23\linewidth]{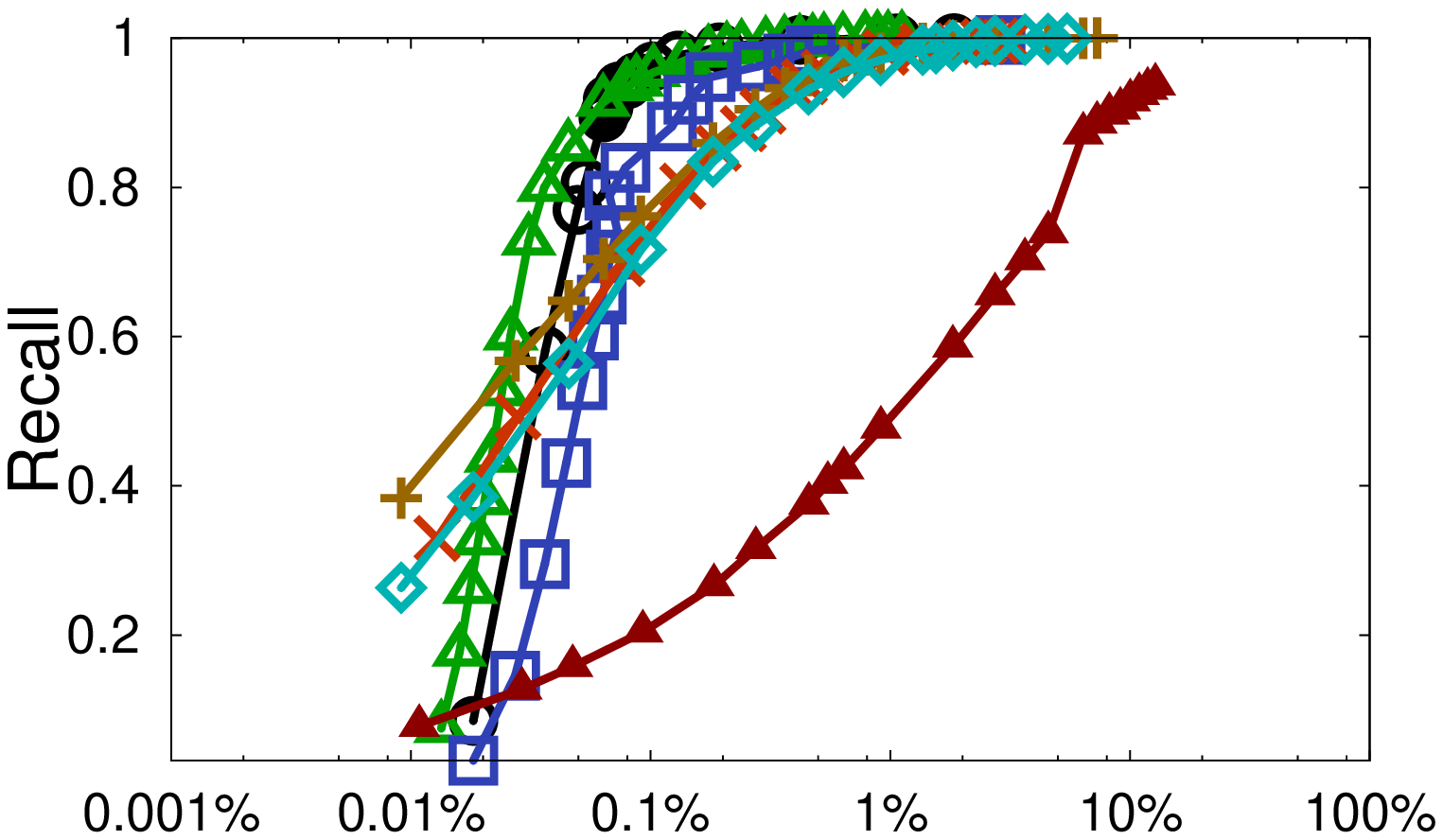}}
      \subfigure[\small UQ-V]{
      \label{fig:exp_recall_msong} 
      \includegraphics[width=0.23\linewidth]{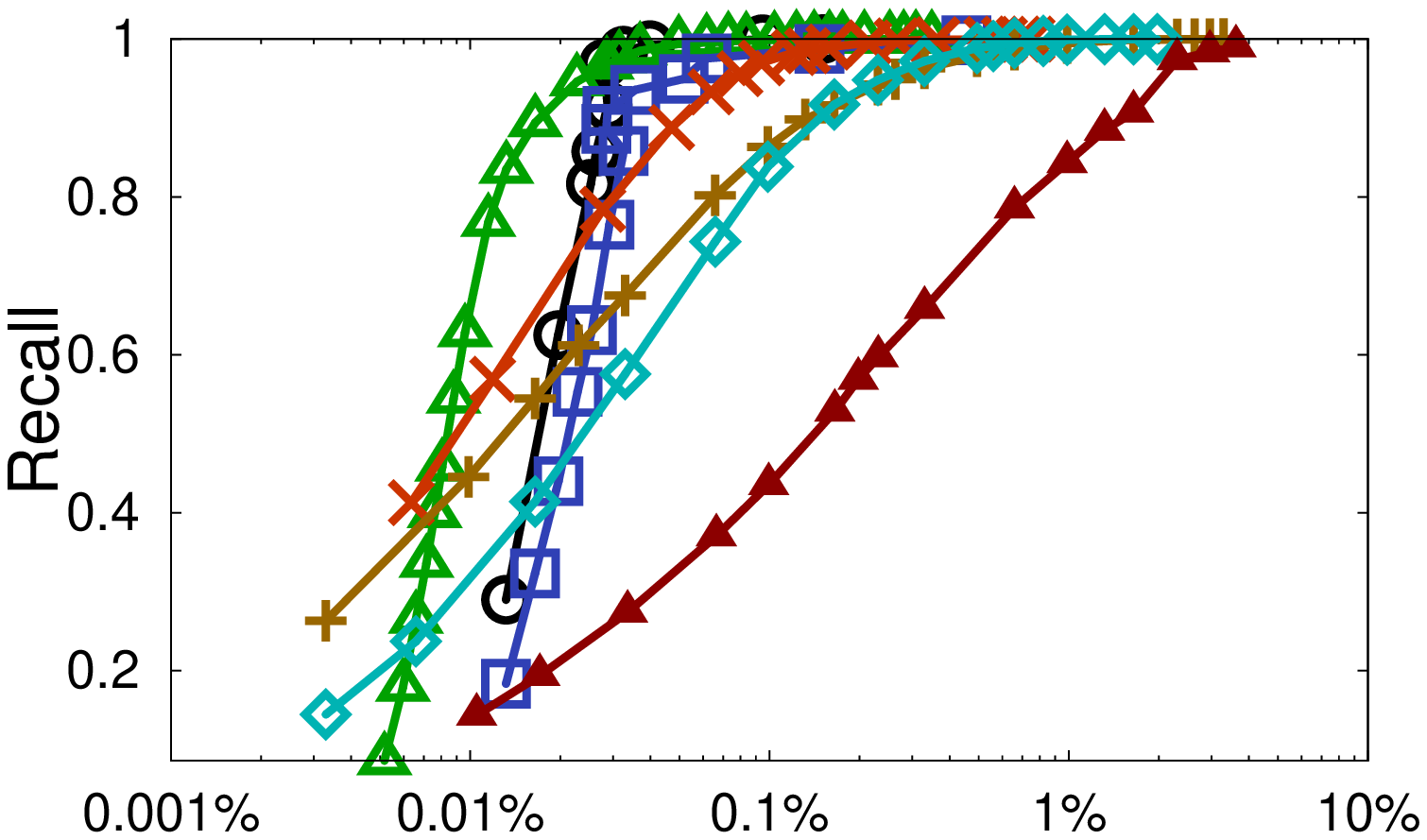}}
      \subfigure[\small Yout]{
      \label{fig:exp_recall_sift} 
      \includegraphics[width=0.23\linewidth]{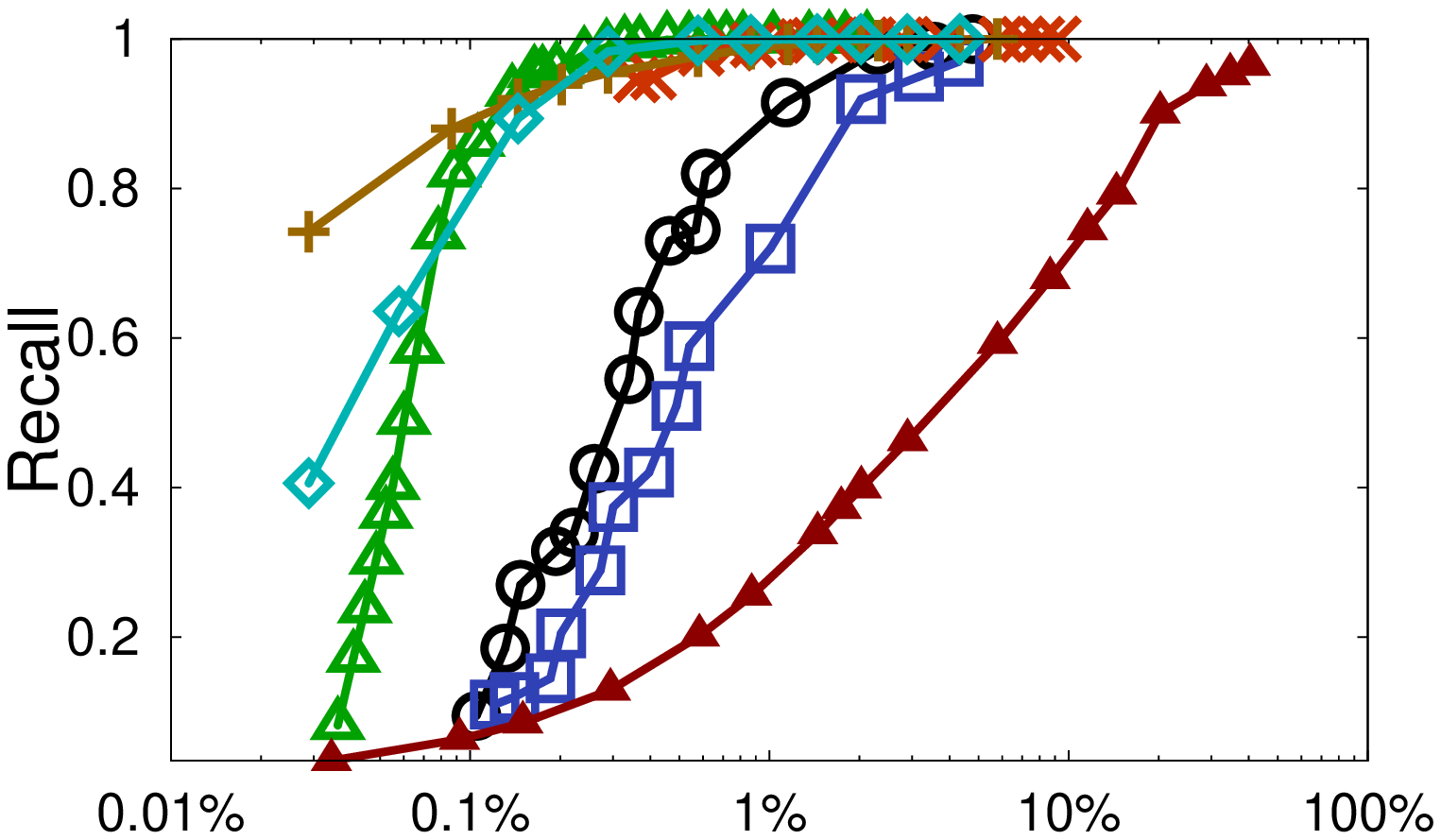}}
      \subfigure[\small BANN ]{
      \label{fig:exp_recall_deep} 
      \includegraphics[width=0.23\linewidth]{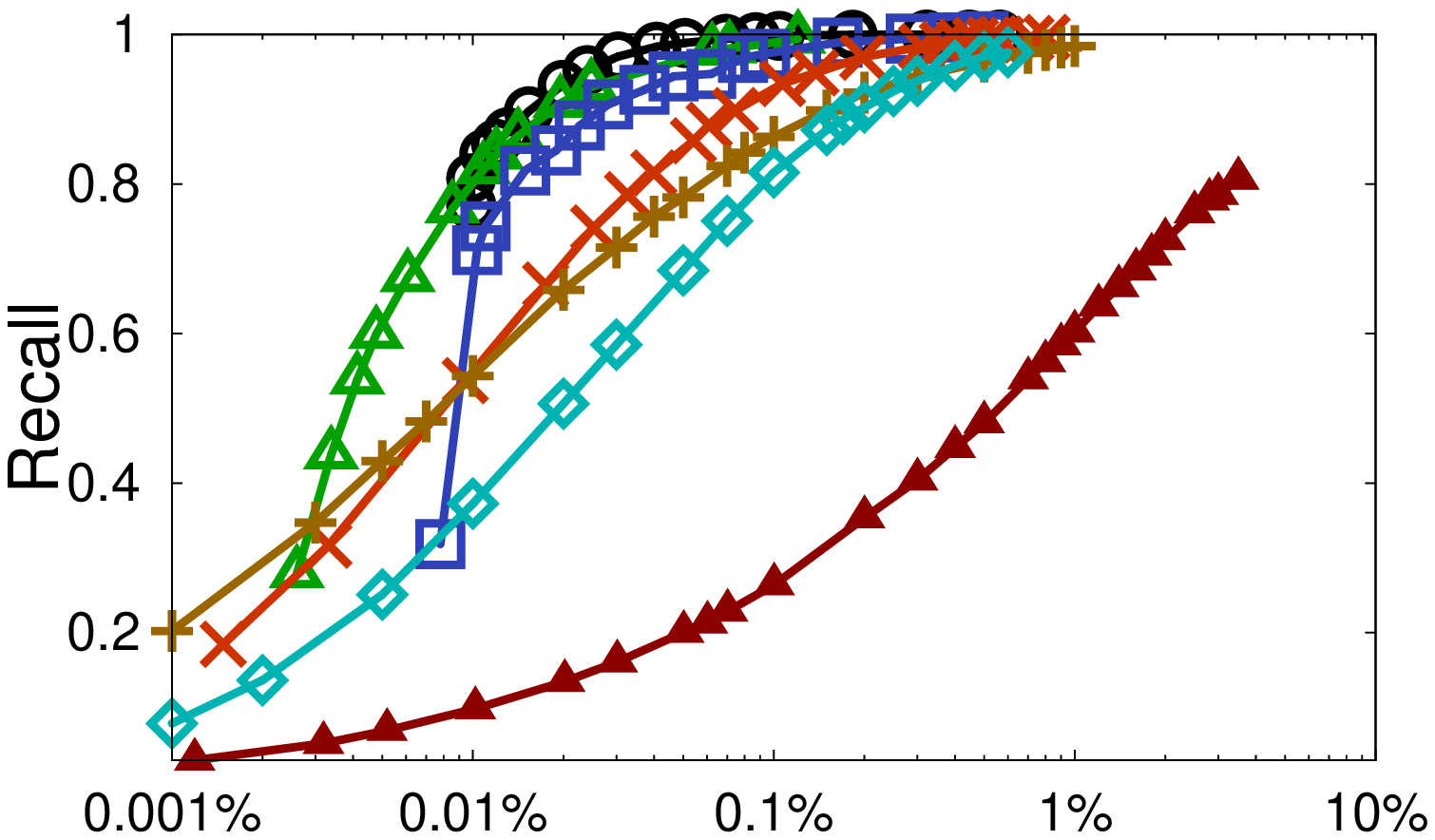}}
\end{minipage}%
\vspace{-1mm}
\caption{\small Recall vs Percentage of Data Points Accessed}
\label{fig:app_final_recall_N}
\end{figure*}

Figure \ref{fig:app_final_speedup_recall} and \ref{fig:app_final_recall_N} show the trade-off between search quality(Recall) and search time(Speedup and the percentage of data points to be accessed) for the remaining datasets(some have been shown in \ref{subsec:exp_final}).

\clearpage

\clearpage


\end{document}